Lecture Notes in Physics

# *Second Edition*
# Solar Energetic Particles II

## A Modern Primer on Understanding Sources, Acceleration and Propagation

**Donald V. Reames**

Institute for Physical Science and Technology
University of Maryland
College Park, MD
USA
dvreames@gmail.com

## Preface

In a field overflowing with beautiful images of the Sun, solar energetic particle (SEP) events are a hidden asset, perhaps a secret weapon, that can sample the solar corona and carry away unique imprints of its most bizarre and violent physics. Only recently have we found that the abundances of the elements in SEPs carry a wealth of data, not only on their own acceleration and history, but on plasma temperatures at their source, and on aspects of the genesis of the corona itself. SEPs are the tangible product of differing energetic outbursts at the Sun. They come in extremes. Little "impulsive" SEP events from magnetic reconnection in solar jets (also in flares), have most unusual 1000-fold resonant enhancements of $^3$He and of heavy elements like Au or Pb, while large "gradual" SEP events accelerated at shock waves driven by coronal mass ejections (CMEs), sample the composition of the corona itself, but also accelerate GeV protons that threaten Mars-bound astronauts with hazardous radiation. Direct SEP measurements plus solar images provide complimentary, "multi-messenger" data on high-energy physics at the Sun.

There have been new studies of abundances of chemical elements in SEPs and their ionization states, and of electrons that produce related radio emission; there is onset timing, and the ion streaming limit; we see evidence of resonant wave-particle interactions, delayed injection profiles, intensity dropouts, energy spectral shapes with spectral "knees", and quiescent particle "reservoirs", in addition to the associations with solar jets and CMEs. Spacecraft that measure SEPs have spread throughout the heliosphere and even dipped into the outer corona. All of this has sparked new understanding and new questions about the physics of SEPs and of the solar corona where they arise: the reasons for their composition, origin, acceleration, and distributions in time and space. This has become a rich new field.

Chapter 1 provides background on the Sun and an introduction to SEPs. Chapters 2 and 3 present the history and much of the physical evidence for the distinction of impulsive and gradual SEP events. Chapters 4 and 5 consider properties and physics of each of these classes individually. The later chapters focus on high energies and radiation hazards of SEPs (Chap. 6), on SEP measurement (Chap. 7), on the physics of element abundances in the solar corona and solar wind (Chap. 8), on the varied origins of protons and heavy ions (Chap. 9), and a Summary and Conclusions (Chap. 10).

This second edition has expanded material in all chapters and newly added chapters on the first-ionization-potential "FIP effect" of coronal element abundances and on the special role of H as a "shock indicator" in abundances. Design requirements for radiation storm shelters for astronauts in deep space are now discussed in Chapter 6. Connections of SEPs with radio bursts and gamma-ray lines have been expanded and new spatial distributions from STEREO have been included. In addition to new material, the discussions have been updated and expanded with new and improved figures. Updated references, all with titles and clickable "doi" references will help readers connect with the SEP literature.

College Park, MD                                                    Donald V Reames

## Acknowledgments

First, I would like to thank those scientists who have contributed their efforts to the progress of this field and those who have contributed the figures I have used to illustrate their discoveries.

Special thanks go to Louis Barbier, Daniel Berdichevsky, Ed Cliver, Steve Kahler, Mary Ann Linzmayer, Chee Ng, Ron Turner, and Gary Zank for reading and commenting on this manuscript and for helpful discussions leading to its preparation. I would especially like to thank Chee Ng for his assistance with the theory of particle transport, wave growth, and shock acceleration.

## About the Author

Born and raised in South Florida, Don Reames received his university education, leading in 1964 to a PhD in Nuclear Physics, at the University of California at Berkeley. He then joined a group at NASA's Goddard Space Flight Center in Maryland using sounding rockets and balloons to study galactic cosmic rays and energetic particles from the Sun. He subsequently used data from experiments on the *Gemini*, IMP, ISEE, *Helios*, *Voyager*, *Wind* and STEREO missions, as well as many related solar missions, to study those particles and their origins more extensively. He retired from NASA in 2003 to assume an Emeritus position, but also soon joined the University of Maryland in College Park to become a Senior Re-search Scientist in 2011. His honors include the 2012 George Ellery Hale Prize from the Solar Physics Division of the American Astronomical Society for his work on the composition and transport of solar energetic particles, and in 2001 he received Goddard's John C. Lindsay Memorial Award for Space Science for his work on solar $^3$He-rich events.

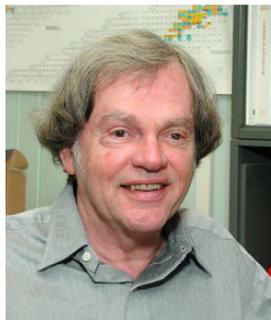

# Contents

# List of abbreviations

| | |
|---|---|
| BFO | Blood-forming organs (for radiation dose) |
| CNS | Central nervous system (for radiation dose) |
| CIR | Corotating interaction region |
| CME | Coronal mass ejection |
| DH | Decametric-hectometric (radio-emission frequencies) |
| DSA | Diffusive shock acceleration |
| EMIC | Electromagnetic ion cyclotron (plasma waves) |
| ESA | *European Space Agency* |
| ESP | Energetic storm particles (near shock) |
| FIP | First ionization potential |
| FSW | Fast solar wind ($> 500$ km s$^{-1}$) |
| GCR | Galactic cosmic ray |
| GLE | Ground-level event, Ground-level enhancement event |
| GSFC | Goddard Space Flight Center |
| ICME | Interplanetary CME |
| LSSA | Large-scale shock acceleration |
| NASA | *National Aeronautics and Space Administration* |
| NOAA | *National Oceanic and Atmospheric Administration* |
| PATH | Particle Acceleration and Transport in the Heliosphere (model) |
| QLT | Quasi-linear theory |
| SEP | Solar energetic particle |
| SPR | Solar particle release (time at the Sun) |
| SSW | Slow solar wind ($< 500$ km s$^{-1}$) |
| TAC | Time-to-amplitude converter |
| UV | Ultraviolet |

Instruments:

| | |
|---|---|
| AIA | *Atmospheric Imaging Assembly*, on SDO |
| AMS | *Alpha Magnetic Spectrometer,* on *International Space Station* |
| EIT | *Extreme-Ultraviolet Imaging Telescope*, on SOHO |
| EPS | *Energetic-Particle Sensor*, on GOES |
| LASCO | *Large-Angle and Spectrometric Coronagraph*, on SOHO |
| HEPAD | *High-Energy Proton and Alpha Detector*, on GOES |
| LEMT | *Low-Energy Matrix Telescope*, on Wind |

| SECCHI | *Sun Earth Connection Coronal and Heliospheric Investigation*, on STEREO |
| SIS | *Solar Isotope Spectrometer*, on ACE |
| SIT | *Suprathermal Ion Telescope,* on STEREO |
| STEP | *SupraThermal Energetic Particle*, on *Wind* |
| ULEIS | *Ultra Low-Energy Isotope Spectrometer*, on ACE |
| WAVES | *Radio and Plasma Wave Investigation*, on *Wind* |

Spacecraft:

| ACE | *Advanced Composition Explorer* |
| GOES | *Geostationary Operational Environmental Satellites* |
| IMP | *Interplanetary Monitoring Platform* |
| ISEE | *International Sun-Earth Explorer* |
| PAMELA | *Payload for Antimatter Exploration and Light-nuclei Astrophysics* |
| PSP | *Parker Solar Probe* |
| SDO | *Solar Dynamics Observatory* |
| SOHO | *Solar and Heliospheric Observatory* |
| SMM | *Solar Maximum Mission* |
| STEREO | *Solar Terrestrial Relations Observatory* |

# Chapter 1. Introducing the Sun and SEPs

**Abstract**   The structure of the Sun, with its energy generation and heating, creates convection and differential rotation of the outer solar plasma. This convection and rotation of the ionized plasma generates the solar magnetic field. This field and its variation spawn all of the solar activity: solar active regions, flares, jets, and coronal mass ejections (CMEs). Solar activity provides the origin and environment for both the impulsive and gradual solar energetic particle (SEP) events. This chapter introduces the background environment and basic properties of SEP events, time durations, abundances, and solar cycle variations.

We tend to think of the Sun as an image of its disk. Recent years have brought increasingly sophisticated images of that disk in the light of single spectral lines and images of active emissions from its surface and its corona with higher and higher spatial resolution. However, we have no such images of solar energetic particles (SEPs). In a photon-dominated discipline, SEPs are stealthy and obscure; they are invisible in the solar sky. While photons travel line-of-sight, SEPs are guided out to us along open magnetic field lines. We must measure, identify, and count SEPs directly one by one. Only in recent years have we overcome the limitations so our observations now begin to bear richer fruit. This is the story of that development.

Solar energetic particles (SEPs) come as bursts of high-energy particles from the direction of the Sun lasting for hours or sometimes days. The particle energies range from about 10 keV (kilo electron volts) to relativistic energies of several GeV, particle speeds 90% of the speed of light. In addition to the dominant protons and electrons, all of the other chemical elements from He through Au and Pb have now been measured. The relative abundances of these elements and their isotopes have been a powerful new resource in our quest for understanding the physical processes of acceleration and interplanetary transport of SEPs which alter those abundances in distinctive ways.

In this chapter we introduce properties of SEPs after reviewing some properties of the solar and interplanetary environment in which they are found.

## 1.1 The Structure of the Sun

With a mass of $1.989 \times 10^{33}$ g, the Sun dominates its neighborhood. It consists of gaseous, ionized plasma where the inner *core* (see Fig. 1.1) reaches temperatures of 15 million degrees Kelvin (MK) where some of the protons have enough energy to tunnel the Coulomb barrier of the nuclear charge. As they penetrate H, C, and N nuclei, they cause the nuclear reactions that catalyze the conversion of H into $^4$He. The energy released in this process is radiated and reabsorbed as it diffuses outward across the *radiative zone*, creating sufficient heat and pressure to balance the gravitational force trying to collapse the star.

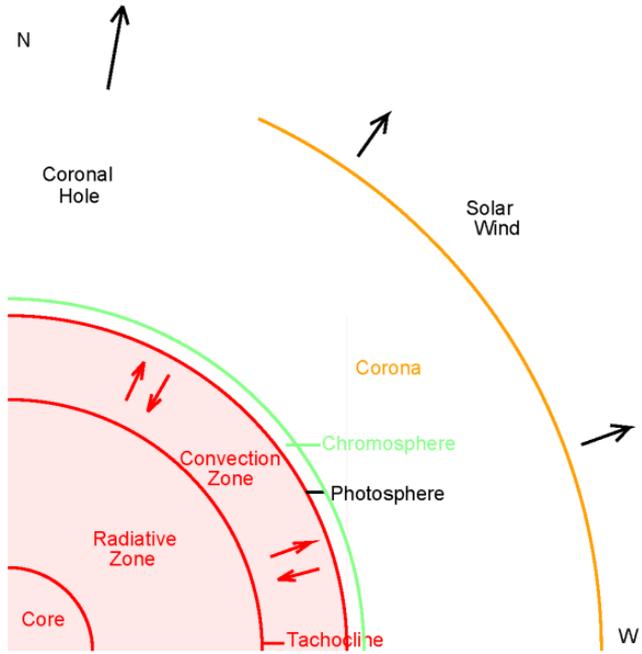

**Fig. 1.1** A cross section of the Sun shows its major radial structure from the core to the evaporating solar wind. (If we look at the Sun with North at the *top* and South at the *bottom*, West is to the *right* and East to the *left*. The solar *limb* is the edge of the visible disk.)

Circulation of the hot plasma across the *convection zone* brings energy to the *photosphere*, that surface where overlying material is too thin to absorb radiation or prevent its escape out into space. Here radiation of energy cools the photosphere to ~5800 K or to ~4500 K in *sunspots* which are sites of strong emerging magnetic field. At these temperatures, elements with a first ionization potential (FIP) below about 10 eV, just below that of H at 13.6 eV, remain ionized, while those with higher FIP can capture and retain electrons to become neutral atoms.

Above the photosphere lies the narrow *chromosphere* where the electron temperature $T_e$ remains about 6000 K over a height of about 2 Mm. At its upper boundary, the electron density $n_e$ suddenly falls from ~$10^{11}$ cm$^{-3}$ to $10^9$ cm$^{-3}$ and $T_e$ rapidly rises again to over 1 MK in the solar *corona* (e.g. Aschwanden 2005) which extends outward about another solar radius. The corona is heated either by numerous small sites of magnetic reconnection (nanoflares; Parker 1988) or by absorption of Alfvén waves, plasma waves created in the turbulent layers below, and is largely contained by rising closed magnetic loops. The outer layer of the corona evaporates to become the 400–800 km s$^{-1}$ *solar wind* which continues to blow past the Earth at 1 AU and far beyond the planets to nearly 100 AU. Properties of the solar wind were predicted by Parker (1963) before it was observed.

Inside the *tachocline*, which lies at the base of the convective zone, the Sun rotates (from East to West) like a rigid body, but throughout the convective zone the Sun rotates *differentially*, faster at the equator than at the poles. The sidereal period of solar rotation at the equator is 24.47 days but it is 25% longer at latitude 60°. Azimuthal surfaces of constant rotation-speed run radially through the convection zone forming conical shells about the rotation axis that extend inward only to the tachocline and not to their apex at the center of the Sun.

## 1.2 The Solar Magnetic Field

The Sun has a magnetic field that is generally dipolar in nature, although its origin is still not perfectly understood (see Parker 2009; Sheeley 2005). Magnetic fields, produced in the extreme rotational sheer at the tachocline, are buoyant and produce omega ($\Omega$) loops that rise through the convection zone and emerge through the photosphere to form *sunspots* and *active regions* (Fig. 1.2) as they are sheared and reconnected by the *differential rotation*. Clusters of magnetic field lines of one polarity tend to emerge from the photosphere at one sunspot and reenter at a nearby spot, leading or following it in the solar rotation. Magnetic fields in sunspots reach 2000–3000 G (0.2–0.3 T). Active regions tend to occur at mid-latitudes on the Sun where the effect of differential rotation on field generation is greatest. When oppositely directed fields reconnect in the largely collisionless regime of the corona, as much as half of the released magnetic energy can be converted to energy of SEPs, with especially copious electrons (Krucker et al. 2010). On *closed* magnetic loops, this can result in sudden heating and X-ray production in the denser loop footpoints, mainly by electron Bremsstrahlung (electron-ion scattering), which is seen as a solar *flare* (Fletcher et al. 2011). Heating trapped flare plasma to 10–40 MK causes the bright flash of softer radiation. Similar reconnection on *open* field lines, causing *jets* (Raouafi et al. 2016), can release electrons and ions into space, i.e. accelerate an impulsive SEP event, with minimal trapping or heating, as we shall see. As electrons stream out along open field lines they produce fast-drift type-III radio bursts at the local plasma frequency.

As we proceed to smaller and smaller flares, they become more and more numerous as a power law. Parker (1988) suggested that the magnetic reconnection in nanoflares actually provides the energy that heats the solar corona.

Fig. 1.2 shows an image of the Sun in ultraviolet (UV) light taken by the *Atmospheric Imaging Assembly* (AIA) on the NASA spacecraft *Solar Dynamics Observatory* (SDO; https://sdo.gsfc.nasa.gov/). Complex, bright areas in Fig. 1.2 are active regions while the large dark region on the solar image is a *coronal hole*. Coronal holes, often seen near the poles, are regions of *open* magnetic field lines extending into the outer heliosphere, stretched out by the plasma of the solar wind. The bright regions show locally *closed* field lines, i.e. loops, where any accelerated particles are contained and interact so that heating is greatly increased.

Of course, Maxwell's Equations tell us that *all* magnetic-field lines are *closed*. However, some field lines are drawn far out into the outer heliosphere by CMEs and the solar wind. For purposes of SEP flow, we describe those field lines as *open* if they can conduct charged particles out to an observer at or beyond Earth.

The direction of the solar dipolar magnetic field reverses in a cycle of one reversal in about 11 yr and solar activity increases as the field reverses. Solar minima occur when the field axis aligns with the solar rotation axis, in one polarity or the other, and the number and size of active regions decreases dramatically. Solar maxima occur during intermediate times and the Sun appears as in Fig. 1.2 late in 2013. During solar minimum the northern hemisphere contains nearly radial field lines of one polarity while the southern hemisphere contains the other; the hemi-spheres are separated by a plane (or wavy) current sheet, separating the opposite field polarities, extending out into interplanetary space from near the equator. High-speed solar wind (~700–800 km s$^{-1}$) emerges from coronal holes.

**Fig. 1.2.** An image of the Sun in 211 Å UV light, taken by the *Atmospheric Imaging Assembly* on the *Solar Dynamics Observatory*, shows brightening of magnetically-complex active regions and a large, dark coronal hole.

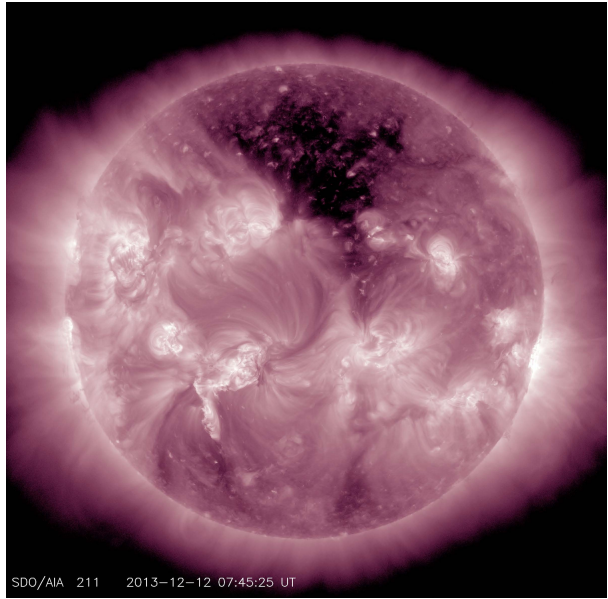

The plasma beta, $\beta_P = \rho kT/(B^2/8\pi)$, where $\rho$ is the density and $T$ the temperature, is the ratio of thermal to magnetic energy density. When $\beta_P < 1$, the field controls the plasma, $B$ is smooth and uniform, and particles are confined to magnetic flux tubes; when $\beta_P > 1$, the field becomes variable and distorted by plasma flow and turbulence. The internal structure of CMEs is dominated by magnetic field energy, with $\beta_P < 1$.

Most of the solar corona is controlled by magnetic fields with $\beta_P < 1$. Plasma can only flow along magnetic loops or flux tubes and cannot escape otherwise. Small neighboring flux tubes can have significantly different values of $T_e$ and $n_e$. However, $\beta_P$ increases with height in the corona and when $\beta_P > 1$, plasma is no longer trapped on magnetic loops; it can expand into space, drawing the magnetic fields outward into the solar wind. This tends to defines the "top" of the corona and typically occurs near 2 R$_S$ where $n_e \sim 10^6$ cm$^{-3}$.

## 1.3 Coronal Mass Ejections (CMEs)

Magnetic reconnection can lead to the ejection of large filaments containing $10^{14} - 10^{16}$ g mass and helical magnetic field with total kinetic energies of $10^{27} - 10^{32}$ ergs, carrying most of the energy in solar eruptions (Webb and Howard 2012). CME speeds can be as slow as the solar wind or can exceed 3000 km s$^{-1}$. Fig. 1.3 shows a large CME imaged by the *Large Angle and Spectrometric Coronagraph* (LASCO) on the *Solar and Heliospheric Observatory* (SOHO; https://sohowww.nascom.nasa.gov/) with a 304 Å image of the Sun from the *Extreme Ultraviolet*

*Imaging Telescope* (EIT) near the same time scaled onto the coronagraph occulting disk. CME theory and models have been reviewed by Forbes et al. (2006). CMEs only became visible when coronagraphs could block scattered light from the Sun which is $10^6$ times brighter.

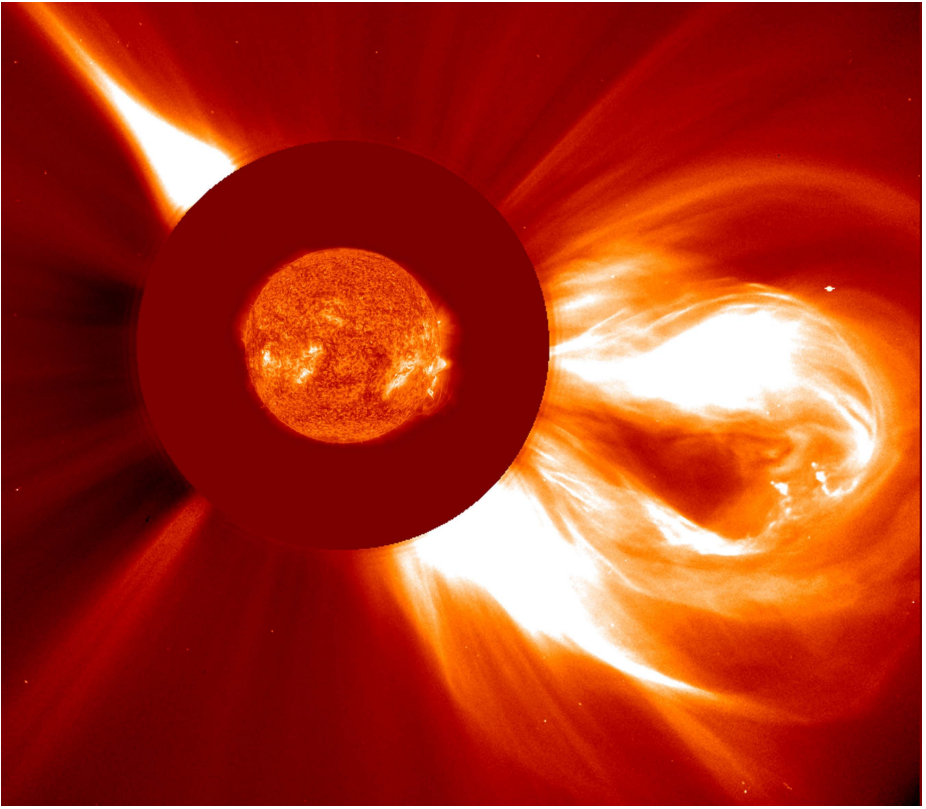

**Fig. 1.3.** A composite image from the EIT and LASCO telescopes on the NASA/ESA SOHO spacecraft shows a large CME being ejected toward the southwest.

*Filaments* are irregular linear structures of cool, dense, chromospheric plasma magnetically suspended in the corona lying parallel to the solar surface, supported at oppositely-directed magnetic fields beneath an arcade of coronal loops (Martin 1998). They appear dark in H$\alpha$ solar images and can hang above the photosphere for days. Filaments that project beyond the solar limb are called *prominences*. Filaments are often ejected as the core of CMEs. In some cases filaments that are present for many days, are suddenly ejected as a CME with no associated flare. These *disappearing-filament* events can drive shock waves and produce SEPs but they lack an associated flare (see Sect. 2.7). There is also a significant number of "stealth CMEs"; these also have no apparent effect on the lower corona but are also too slow to drive shock waves.

When the speed of a CME exceeds the speed of waves in the plasma of the corona or solar wind, it can drive a collisionless shock wave. We will see that fast, wide CMEs drive shock waves that are the primary source of acceleration of the largest SEP events (Kahler et al. 1984).

A bright *streamer* is seen in the upper left (northeast) corner of Fig. 1.3, opposite the CME. The streamer belt circles the Sun defining the neutral sheet between fields of opposite polarity. Streamers are the magnetic structures stretched behind CMEs after they move out into the heliosphere. As such, they represent newly opening field lines and may contribute to the slow (~400 km s$^{-1}$) solar wind, although the source of the slow solar wind is not fully resolved (e.g. Antiochos et al. 2011). Thus, out-flowing CMEs contribute to the average magnetic field in the heliosphere, which is larger following strong, active solar cycles than weak ones.

## 1.4 Interplanetary Space

The solar wind expands nearly radially outward from the Sun carrying plasma and magnetic field. The solar-wind speed remains approximately constant with distance from the Sun. As the Sun rotates, the field line connected to a given point on its surface is drawn into a spiral pattern, the Parker spiral. In the inner heliosphere, the plasma density and magnetic-field strength decrease approximately as $r^{-2}$ with distance $r$, from the Sun, and as $B \sim r^{-1.5}$ by 1 AU (Burlaga, 1995, 2001).

Near Earth the typical magnetic field $B$ is ~10 nT, the typical plasma density is ~10 particles cm$^{-3}$, and the electron plasma frequency, which varies with the electron density, $n_e$, as $n_e^{1/2}$, is ~30 kHz. The solar radius, $R_S = 6.96 \times 10^8$ m = 696 Mm, and the Earth-Sun distance, 1 AU, is $1.50 \times 10^{11}$ m = 216 $R_S$, often a useful number. In this spirit, plasma in the 400 km s$^{-1}$ solar wind takes 4.3 days to travel 1 AU, a shock wave with an average speed of 1700 km s$^{-1}$ takes one day, a 10 MeV proton or a 5 keV electron takes an hour, and a photon of light takes 8.3 min. Thus, it is not surprising that particles accelerated by a shock wave near the Sun arrive near Earth long before the arrival of the shock itself.

Alfvén waves propagate through plasma with correlated variations in $B$ and the plasma density $\rho$ with a speed $V_A = B/(4\pi\rho)^{1/2}$. In models of $V_A$ in the solar atmosphere above an active region (e.g. Mann et al. 2003), $V_A$ falls rapidly with height to a value of ~200–500 km s$^{-1}$ at $r \approx 1.5$ $R_S$, it then rises to a broad maximum of ~ 750 km s$^{-1}$ near 4 $R_S$ and finally decays approximately as $r^{-1}$ out toward Earth (Mann et al. 2003) where it is nominally 30 km s$^{-1}$. However, these values depend upon assumptions about the magnetic structure of an active region. The behavior of $V_A$ is important since the disturbance caused by a CME must exceed the speed of Alfvén waves to form a shock wave which can accelerate SEPs.

Large CMEs can be recognized in the solar wind when they pass Earth (often called ICMEs) and lists of them, with their associated coronagraphic origin, have been published (Richardson and Cane 2010). A class of particularly regular events called *magnetic clouds* is identified by a flux-rope magnetic field that spirals slowly through a large angle (Burlaga et al. 1981). Shock waves driven out by CMEs can also be observed near Earth and their properties can be determined (e.g. Berdichevsky et al. 2000). Lists of properties of interplanetary shock waves spanning many years (since 1995) are available for shocks observed at the *Wind* and ACE (*Advanced Composition Explorer*) spacecraft (https://www.cfa.har-vard.edu/shocks/). We will see examples of shock waves later in this book.

## 1.5 Solar Energetic Particles

Energetic charged particles must be accelerated in nearly collisionless plasma at high coronal altitudes so they do not lose all their new energy in Coulomb colli-sions. Particles accelerated at the tops of magnetic loops will soon scatter through the loss cone into the footpoints of those loops where they deposit their energy in much higher ambient densities, producing a solar flare. The rate of heating and production of photons generally depends upon the product of energetic-particle in-tensity times the ambient particle density. Thus SEP acceleration sites tend to be barely visible, high in the corona, while most of the photons are produced low in the corona. Looking only at photons, we do not see where these SEPs are "born"; we see where they "die". Especially troublesome is shock acceleration at ~2 solar radii and beyond, where SEPs in the acceleration site are completely invisible. This inconvenient fact has led to great confusion about the origin of SEPs.

The effort to understand the physical origin of SEP events finally led to the identification of two classes of SEP events, *impulsive* and *gradual* (or *long-dura-tion)* with the sources suggested by Fig. 1.4 (e.g. Reames 1999, 2013). The histo-ry of this journey will be discussed in Chap. 2 with further physical evidence in Chap. 3. Fortunately, the properties of the SEPs themselves carry the imprint of both their origin and their transport. Important differences lie in abundances of el-ements and isotopes, electron/ion ratios, energy spectra, onset timing, duration, angular distributions, and associations with visible phenomena, as we shall see.

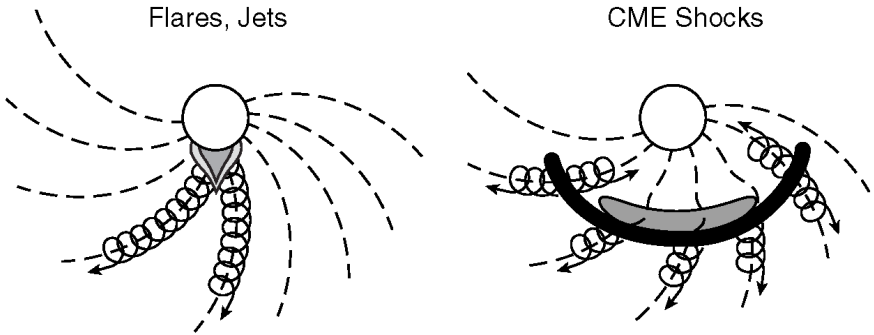

**Fig. 1.4.** Impulsive (*left*) and gradual (*right*) classes of SEP events are distinguished by the domi-nant sources of particle acceleration in each case (Reames 1999, © Springer). Impulsive SEP events are accelerated in magnetic-reconnection events on open field lines (i.e. jets) in the coro-na. Gradual SEP events are accelerated at shock waves (*solid black*) driven out from the Sun by CMEs (*gray*). Particle trajectories are shown as spirals along ***B*** (*dashed*).

The data base for many measurements from many spacecraft, including SEP in-tensities, from spacecraft where they were measured, is the NASA Coordinated

Data and Analysis Web site: https://cdaweb.gsfc.nasa.gov/sp_phys/. This web site has data from past and current space-physics missions.

### 1.5.1 Time Duration

While the terms impulsive and gradual did not originally refer to the SEP duration, this is often a reasonable characterization, as shown by the events in Fig. 1.5.

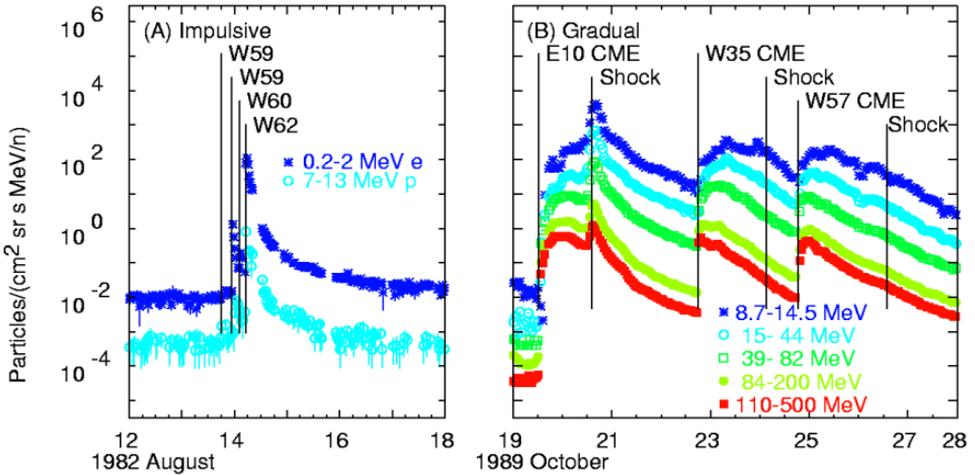

**Fig. 1.5**. Particle intensities are shown for a series of (**a**) impulsive and (**b**) gradual or long-duration SEP events at similar time and intensity scales. Flags labeled with the source longitude indicate the onset times of the events; also shown are the times of shock passage near Earth. Proton (or electron) energies are listed. It is difficult to obtain comparable proton energies because impulsive events are much less energetic (Reames 1999 © Springer).

### 1.5.2 Abundances

The abundances of elements and isotopes have been powerful indicators of the origin, acceleration, and transport of SEPs. It was found (Webber 1975; Meyer 1985) that the average element abundances, in events we now call large, gradual SEP events, were a measure of the corresponding solar *coronal* abundances. These differ from abundances in the photosphere by a factor which depends on the first ionization potential (FIP) of the element as shown in Fig. 1.6 and listed in Table 1.1 (Reames 1995, 2014). In the photosphere, low-FIP (<10 eV) elements are ionized while high-FIP elements are neutral atoms. Relative to the upward flow of neutral atoms, ions are also influenced by Alfvén waves during their transport across the solar chromosphere and into the corona (e.g. Laming 2009, 2015). Other measures of coronal abundances, such as in the solar wind (e.g. Geiss 1982; Bochsler 2009), show a FIP effect that is similar but not identical to SEPs (see Sect. 8.4; Schmelz et al. 2012; Reames 2018), reflecting a different origin. The SEP abundances in Table 1.1 can serve as reference abundances for discussion of "enhancements" throughout this book.

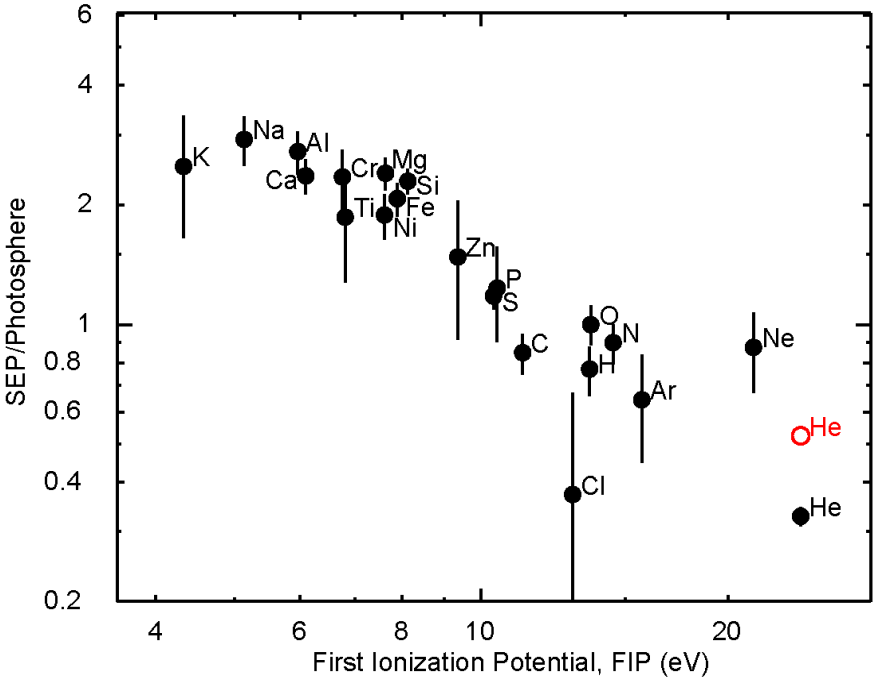

**Fig 1.6**. The average element abundance in gradual SEP events (Reames 1995, 2014, 2017), or reference abundance, relative to the corresponding abundance in the solar photosphere (Asplund et al. 2009) is plotted as a function of the FIP of the element (see text). Here He means ⁴He.

Table 1.1 lists the photospheric (Asplund et al. 2009) and the reference SEP (Reames 1995, 2014, 2020) abundances that we can use. A likely correction to the reference abundance of He (He/O = 91 rather than 57; Reames 2017), that will be discussed in Sect. 5.9, is shown as a red open circle in Fig. 1.6. Alternative photospheric abundances by Caffau et al. (2011), shown in Table 8.1, make some difference in the FIP plot as demonstrated by Reames (2015); the differences depend upon the choice of spectral lines used to obtain the photospheric abundance measurements (Chap. 8). Note that within this book, unspecified He means ⁴He.

The current theory of the "FIP effect" (Laming 2009, 2015, Laming et al. 2019) uses the ponderomotive force of Alfvén waves below the chromospheric-coronal boundary to preferentially boost low-FIP ions (e.g. Mg, Si, Fe) into the corona ahead of the slower evaporation of high-FIP neutral atoms (e.g. O, Ne, He). All elements become ionized in the 1-MK corona. The Alfvén waves can resonate with the loop length on closed magnetic loops and calculations show that the ion fractionation only occurs near the top of the chromosphere. Open fields lack this resonance so the fractionation is more extensive. Particles destined to become SEPs apparently arise mainly on closed loops in active regions and the elements P, S, and C are suppressed like high-FIP neutral atoms, while particles of the solar wind arise on open field lines where P, S, and C behave like low-FIP ions (see Sect. 5.9, Chap. 8, and Reames 2018, 2020). This means that SEPs are *not* merely accelerated solar wind, as originally noticed by Mewaldt et al. (2002) and Desai et al. (2003); SEPs and the solar wind are fundamentally different samples of coronal material. The FIP patterns of SEPs and the solar wind are compared with theory in Chap. 8 and the differing locations of their origin are discussed.

**Table 1.1.** Photospheric and SEP-reference abundances used in Fig. 1.6

| | Z | FIP [eV] | Photosphere | SEP Reference |
|---|---|---|---|---|
| H | 1 | 13.6 | $(2.04\pm0.05)\times10^6$ | $(\sim1.57\pm0.22)\times10^6$ |
| He | 2 | 24.6 | $(1.74\pm0.04)\times10^5$ | 57000±3000, 91000±13000 |
| C | 6 | 11.3 | 550±63 | 420±10 |
| N | 7 | 14.5 | 138±16 | 128±8 |
| O | 8 | 13.6 | 1000±115 | 1000±10 |
| Ne | 10 | 21.6 | 174±40 | 157±10 |
| Na | 11 | 5.1 | 3.55±0.33 | 10.4±1.1 |
| Mg | 12 | 7.6 | 81±8 | 178±4 |
| Al | 13 | 6.0 | 5.75±0.40 | 15.7±1.6 |
| Si | 14 | 8.2 | 66.1±4.6 | 151±4 |
| P | 15 | 10.5 | 0.525±0.036 | 0.65±0.17 |
| S | 16 | 10.4 | 26.9±1.9 | 25±2 |
| Cl | 17 | 13.0 | 0.65±0.45 | 0.24±0.1 |
| Ar | 18 | 15.8 | 5.1±1.5 | 4.3±0.4 |
| K | 19 | 4.3 | 0.22±0.14 | 0.55±0.15 |
| Ca | 20 | 6.1 | 4.47±0.41 | 11±1 |
| Ti | 22 | 6.8 | 0.182±0.021 | 0.34±0.1 |
| Cr | 24 | 6.8 | 0.89±0.08 | 2.1±0.3 |
| Fe | 26 | 7.9 | 64.6±6.0 | 131±6 |
| Ni | 28 | 7.6 | 3.39±0.31 | 6.4±0.6 |
| Zn | 30 | 9.4 | 0.074±0.009 | 0.11±0.04 |

Abundances also distinguish *impulsive* SEP events in a very different way (Mason 2007). The earliest of these was the greatly enhanced $^3$He/$^4$He ratio, which is $\sim5\times10^{-4}$ in the solar wind, but can be >1 in impulsive SEP events, as seen in the examples in Fig. 1.7. Enhancements of Fe/O by a factor of ~10 were subsequently observed and we now see these as part of an enhancement that is a power law in *A/Q* that becomes a 1000-fold enhancement for heavy elements up to Au and Pb, relative to He or O. These orderly systematic enhancements in impulsive SEP events will be discussed in Chap. 4. The observation of energy-dependent stripping of electrons from Fe ions after acceleration now suggests they were accelerated at or below 1.5 solar radii (DiFabio et al. 2008) in solar jets (see Sect 2.6).

**Fig. 1.7**. Intensities vs. time are shown in impulsive SEP event numbers 25 and 103 (shown in blue flags at event onsets) from the list in Reames, Cliver, and Kahler (2014 © Springer). [3]He exceeds [4]He in these events and Fe exceeds C and O. Flags in black preceding the SEP onsets are at the associated CME onset times and list the speed (km s[-1]), position angle (deg), and width (deg) of the CME. .

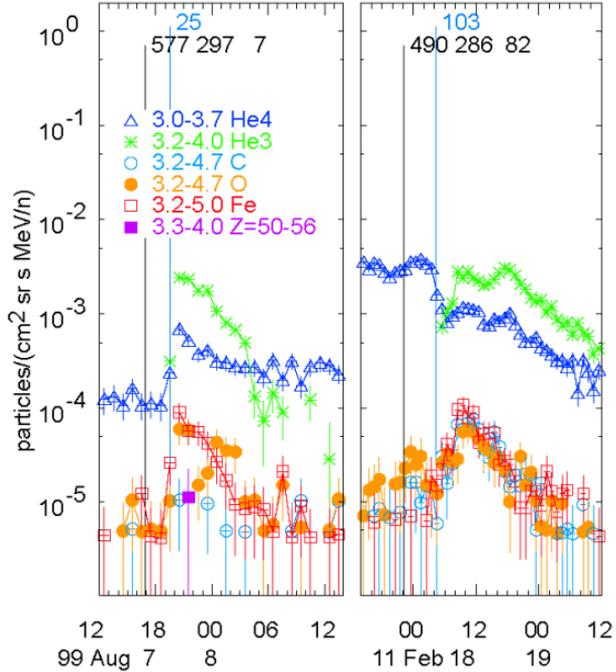

The two events in Fig. 1.7 have event-averaged Fe/O = 1.24±0.28 and 1.34±0.20, respectively, compared with the reference value of 0.131±0.006 in Table 1.1. Enhancements of even heavier elements (e.g. $Z > 50$) are much greater, on average, but are difficult to measure in single small events. These will be seen in Sect. 4.5.

Impulsive SEP events are associated with solar jets (Bučík 2020), i.e. magnetic reconnection on open field lines that allow the SEPs and narrow CMEs to escape (Kahler, Reames, and Sheeley 2001), especially those more-energetic jets from solar active regions (Bučík et al. 2018a, 2018b). Impulsive SEP events also have intense electron beams (Reames, von Rosenvinge, and Lin 1985) that emit type III radio bursts (Sect. 2.2; Reames and Stone 1986).

### 1.5.3 The Solar Cycle

SEP events do not precisely follow the solar activity level of sunspots, but they do have a definite solar cycle. The upper panel of Fig. 1.8 shows intensities of 120–230 MeV protons measured by the *Goddard Space Flight Center* telescope on the IMP-8 (*Interplanetary Monitoring Platform*) spacecraft. This telescope is sensitive to energetic protons of both solar and galactic origin and can thus observe the counter-cyclical behavior. When the Sun is active with SEP events, the greater ejection of CMEs increases the modulation that blocks and decreases the encroachment of galactic cosmic rays into the heliosphere. The monthly sunspot number is shown in the lower panel for comparison (see Hathaway 2010).

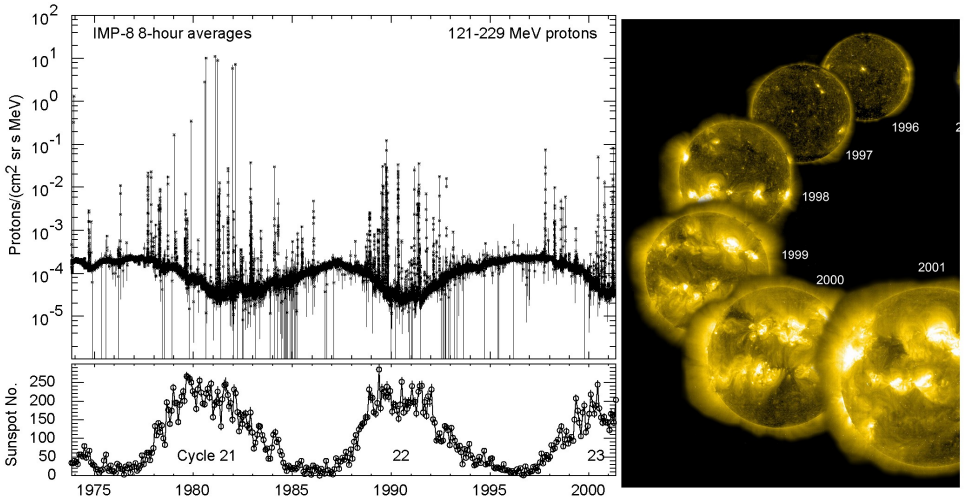

**Fig. 1.8**. Intensities of 120-230 MeV protons in 8-h averages from the Goddard IMP-8 telescope are shown over 27 years in the *upper panel*. Spikes from individual SEP events reach a factor of $10^5$ above a counter-cyclical baseline of galactic cosmic rays which the instrument also measures well. However, intensities in some SEP events during the later cycles are foreshortened because of increasingly frequent data gaps. The monthly international sunspot number is shown in the *lower panel* for comparison, and to the *right* are EIT images at 284 Å that show solar activity during each year from 1996 to 2001 during the rise of Solar Cycle 23.

### 1.5.4 Relativistic Kinematics

What we often call the particle "energy," *E,* commonly quoted as MeV amu$^{-1}$, is actually a measure of velocity $E = \mathcal{E}/A = M_u(\gamma - 1) \approx \frac{1}{2} M_u \beta^2$, where $\mathcal{E}$ is the total kinetic energy, *A* is the atomic mass, $M_u = m_u c^2 = 931.494$ MeV, $\gamma = (1 - \beta^2)^{-1/2}$, and $\beta = v/c$ is the particle velocity relative to the speed of light, *c*. Abundances of elements and isotopes are always compared at the same value of *E*. The total energy of a particle is $W = A M_u \gamma$ and the momentum is given by $pc = A M_u \beta\gamma$. The magnetic rigidity or momentum per unit charge is $P = pc/Qe = M_u \beta\gamma A/Q$ in units of MV. Note that the standard atomic mass unit (amu), 1/12 the mass of $^{12}$C, is close enough to nucleon masses that MeV nucleon$^{-1}$ is indistinguishable from MeV amu$^{-1}$ for SEP studies.

We can write the Lorentz force on a single particle in the form

$$m_u \frac{d}{dt}(\gamma \mathbf{v}) = \frac{Q}{A} e(\mathbf{E} + \mathbf{v} \times \mathbf{B}) \tag{1.1}$$

In a collisionless world where the electric and magnetic fields are independent of the nature of the particle, the only specific particle species dependence is $Q/A$. This will be the case for most of the wave-particle interactions we will encounter during particle acceleration and transport. The exception comes when the particle interacts with matter where the electric field **E** is that of the particle itself and depends upon $Q$ as it scatters electrons of the stopping material. This is the case in particle detectors (Chap. 7) where the species-dependence for energy loss becomes $Q^2/A$. Strong enhancements observed in elements with $76 \leq Z \leq 82$ in im-pulsive SEPs would have been suppressed by this dependence on $Q^2/A$ if the ions had traversed significant amounts of matter during acceleration or transport. Thus, acceleration and transport are primarily collisionless and depend upon $Q/A$, although, in some cases, ions may traverse enough material after acceleration to alter their ionization states (Sect. 2.6).

## 1.6 What Do We "See" at the Sun?

SEPs follow magnetic fields out from the Sun so the image of their source location is distorted and difficult to follow. Are there photons emitted that can help us locate SEP origins?

Many of the energetic photons we see from the Sun are produced by heating. Solar spectral lines that can image the Sun are a specific wavelength produced by ions of a specific element in a specific ionization state, i.e. temperature, of the ambient solar plasma. They show regions of heating and can indicate element abundances in the ambient plasma, but tell us little of the nature of any SEPs that may have caused the heating as they plunged into the denser plasma.

CMEs are imaged in white light reflected by ejected material. It is sunlight from the photosphere that has been Thomson-scattered by free electrons in the corona. The light intensity is a measure of electron density, and coronagraph images of fast CMEs can distinguish the location and evolution of shock waves and the driver gas (see Sect. 3.2).

Hard X-rays (> 20 keV) are produced by Bremsstrahlung (electron-ion scattering) of energetic electrons. Images show locations where energized (>20 keV) electrons encounter regions of denser plasma (Fletcher et al. 2011; Miller et al. 1997). Soft X-rays are produced in relatively dense regions by electron and ion heating as they stop and lose their energy to the plasma by Coulomb scattering. Thus flares are produced by SEP-heated plasma, but those SEPs don't get out. Magnetic trapping is the reason flares are so hot and bright.

Radio emission is produced by 10 – 100 keV electrons. Type III bursts are produced by electrons streaming out from the Sun along magnetic field lines while type II bursts are produced at interplanetary shock waves (see Sect. 2.2).

The only photons identified with accelerated ions are in the broad γ-ray lines produced by Doppler-shifted emission of interacting energetic ions of the "beam," while the narrow γ-ray lines are emitted from the ambient plasma in flares (see Sect. 4.9). The last measurements of solar γ-ray lines were made in the 1980s. These are the only measurable photons produced by the energetic ions.

It is also possible to trace field lines of direct measurements of SEP events back to the Sun to locate their source (Nitta et al. 2006; Wang, Pick, and Mason 2006; Ko et al. 2013).

Thus, the photons we measure are usually produced by electrons and rarely help us study accelerated ions, so, much of the SEP physics must be determined directly from measurements of the ions themselves. Associations are generally made from models or from similarities in the timing, although the photons usually relate to SEPs accelerated on closed field lines while the SEP ions and electrons we see in space come from related events on open field lines.

**Acknowledgements:** The author thanks the SOHO and SDO projects for figures used in this chapter.

# References

Antiochos, S.K., Mikić, S., Titov, V.S., Lionello, R., and Linker, J.A., A Model for the sources of the slow solar wind Astrophys. J. **731**, 112 (2011) doi: 10.1088/0004-637X/731/2/112

Aschwanden, M., *Physics of the solar corona*, Springer, Berlin, New York (2005) ISBN 3-540-30765-6

Asplund, M., Grevesse, N., Sauval, A.J., Scott, P., The chemical composition of the sun, Ann. Rev. Astron. Astrophys. **47**, 481 (2009) doi: 10.1146/annurev.astro.46.060407.145222

Berdichevsky, D.B., Szabo, A., Lepping, R.P., Vinas, A.F., Mariana, F., Interplanetary fast shocks and associated drivers observed through the 23rd solar minimum by Wind over its first 2.5 years, J. Geophys. Res., **105**, 27289 (2000) doi: 10.1029/1999JA000367; Errata in J. Geophys. Res., **106** 25133, (2001) doi: 10.1029/2001JA000074

Bochsler, P., Composition of matter in the heliosphere, *Proc. Int. Astron. Union Sympos.* **257**, 17 (2009) doi:10.1017/S1743921309029044.

Bučík, R, ³He-rich solar energetic particles: solar sources, Space Sci. Rev. **216** 24 (2020) doi: 10.1007/s11214-020-00650-5

Bučík, R., Innes, D.E., Mason, G.M., Wiedenbeck, M.E., Gómez-Herrero, R., Nitta, N.V., ³He-rich solar energetic particles in helical jets on the Sun, Astrophys. J. **852** 76 (2018a) doi: 10.3847/1538-4357/aa9d8f

Bučík, R., Wiedenbeck, M.E., Mason, G.M., Gómez-Herrero, R., Nitta, N.V., Wang, L., ³He-rich solar energetic particles from sunspot jets, Astrophys. J. Lett. **869** L21 (2018b) doi: 10.3847/2041-8213/aaf37f

Burlaga, L.F., *Interplanetary Magnetohydrodynamics*, Oxford University Press, Oxford, UK (1995) ISBN 13: 978-0-19-508472-6

Burlaga, L.F., Magnetic fields and plasmas in the inner heliosphere: Helios results, *Planet. Space Sci.* **49**, 1619 (2001) doi: 10.1016/S0032-0633(01)00098-8

Burlaga, L.F., Sittler, E., Mariani, F., Schwenn, R: Magnetic loop behind an interplanetary shock: Voyager, Helios, and Imp 8 observations, J. Geophys. Res., **86**, 6673 (1981) doi: 10.1029/JA086iA08p06673

Caffau, E., Ludwig, H.-G., Steffen, M., Freytag, B., Bonofacio, P., Solar chemical abundances determined with a CO5BOLD 3D model atmosphere, Sol. Phys. **268**, 255. (2011) doi: 10.1007/s11207-010-9541-4

Desai, M.I., Mason, G.M., Dwyer, J.R., Mazur, J.E., Gold, R.E., Krimigis, S.M., Smith, C.W., Skoug, R.M., Evidence for a suprathermal seed population of heavy ions accelerated by interplanetary shocks near 1 AU, Astrophys. J. **588**, 1149 (2003) doi: 10.1086/374310

DiFabio, R., Guo, Z., Möbius, E., Klecker, B., Kucharek, H., Mason, G. M., Popecki, M., Energy-dependent charge states and their connection with ion abundances in impulsive solar energetic particle events, Astrophys. J. **687**, 623.(2008) doi: 10.1086/591833

Fletcher, L., Dennis, B.R., Hudson, H.S., Krucker, S., Phillips, K., Veronig, A., Battaglia, M., Bone, L., Caspi, A., Chen, Q. et al., An observational view of solar flares, Space Sci. Rev. **159** 19 (2011) doi: 10.1007/s11214-010-9701-8

Forbes, T.G., Linker, J.A., Chen, J., Cid, C., Kóta, J., Lee, M.A., Mann, G., Mikic, Z., Potgieter, M.S., Schmidt, J.M., Syscoe, G.L., Vainio, R., Antiochos, S.K., Riley, P., CME theory and models, Space Sci. Rev. **123**, 251 (2006) doi: 10.1007/s11214-006-9019-8

Geiss, J., Processes affecting abundances in the solar wind, Space Sci. Rev. **33**, 201 (1982) doi: 10.1007/BF00213254

Kahler, S.W., Reames, D.V., Sheeley, N.R.,Jr., Coronal mass ejections associated with impulsive solar energetic particle events, Astrophys. J., **562**, 558 (2001) doi: 10.1086/323847

Kahler, S.W., Sheeley Jr., N.R., Howard, R.A., Koomen, M.J., Michels, D.J., McGuire R.E., von Rosenvinge, T.T., Reames, D.V., Associations between coronal mass ejections and solar energetic proton events, J. Geophys. Res. **89**, 9683 (1984) doi: 10.1029/JA089iA11p09683

Krucker S, Hudson H S, Glesener L, White S M, Masuda S, Wuelser J P, Lin R P. 2010. Measurements of the coronal acceleration region of a solar flare. Astrophys J, **714** 1108 doi: 10.1088/0004-637X/714/2/1108

Hathaway, D.H., The solar cycle, Living Rev. Sol. Phys. **7** 1 (2010) doi: 10.12942/lrsp-2010-1

Ko Y,-K., Tylka, A.J., Ng C.K., Wang Y.-M., Dietrich W.F., Source regions of the interplanetary magnetic field and variability in heavy-ion elemental composition in gradual solar energetic particle events, Astrophys. J. **776**, 92 (2013) doi: 10.1088/0004-637X/776/2/92

Laming, J.M., Non-WKB models of the first ionization potential effect: implications for solar coronal heating and the coronal helium and neon abundances, Astrophys. J. 695, 954 (2009) doi: 10.1088/0004-637X/695/2/954

Laming, J.M., The FIP and inverse FIP effects in solar and stellar coronae, Living Reviews in Solar Physics, **12**, 2 (2015) doi: 10.1007/lrsp-2015-2

Laming, J.M., Vourlidas, A., Korendyke, C., et al., Element abundances: a new diagnostic for the solar wind. Astrophys. J. **879** 124 (2019) doi: 10.3847/1538-4357/ab23f1 arXiv: 19005.09319

Mann, G., Klassen, A. Aurass, H., Classen, H.-T., Formation and development of shock waves in the solar corona and the near-Sun interplanetary space, Astron. Astrophys. **400**, 329 (2003) doi: 10.1051/0004-6361:20021593

Martin, S.F., Conditions for the formation and maintenance of filaments (invited review), Solar Phys, **182**, 126 (1998) doi: 10.1023/A:1005026814076

Mason, G.M., $^3$He-rich solar energetic particle events, Space Sci. Rev. **130**, 231 (2007) doi: 10.1007/s11214-007-9156-8

Mewaldt, R.A., Cohen, C.M.S., Leske, R.A., Christian, E.R., Cummings, A.C., Stone, E.C., von Rosenvinge, T.T., Wiedenbeck, M.E., Fractionation of solar energetic particles and solar wind according to first ionization potential, Advan. Space Res.**30** 79 (2002) doi: 10.1016/S0273-1177(02)00263-6

Meyer, J.P., The baseline composition of solar energetic particles, Astrophys. J. Suppl. **57**, 151 (1985) doi: 10.1086/191000

Miller, J.A., Cargill, P.J., Emslie, A.G., Holman, G.D., Dennis, B.R., LaRosa, T.N., Winglee, R.M., Benka, S.G., Tsuneta, S., Critical issues for understanding particle acceleration in impulsive solar flares, J. Geophys. Res. **102**, 14631 (1997) doi: 10.1029/97JA00976

Nitta, N.V., Reames, D.V., DeRosa, M.L., Yashiro, S., Gopalswamy, N., Solar sources of impulsive solar energetic particle events and their magnetic field connection to the earth, Astrophys. J. **650**, 438 (2006) doi: 10.1086/507442

Parker, E.N.: *Interplanetary Dynamical Processes,* Interscience, New York (1963).

Parker, E.N.: Nanoflares and the solar X-ray corona, Astrophys. J. **330** 474 (1988) doi: 10.1086/166485

Parker, E.N.: Solar magnetism: the state of our knowledge and ignorance, Space Sci. Rev. **144**, 15 (2009) doi: 10.1007/s11214-008-9445-x

Raouafi, N.E., Patsourakos, S., Pariat, E., Young, P.R., Sterling, A.C., Savcheva, A., Shimojo, M., Moreno-Insertis, F., DeVore, C.R., Archontis, V, et al., Solar coronal jets: observations, theory, and modeling, Space Sci. Rev. **201** 1 (2016) doi: 10.1007/s11214-016-0260-5 (arXiv:1607.02108)

Reames, D.V., Coronal abundances determined from energetic particles, Adv. Space Res. **15** (7), 41 (1995)

Reames, D.V.: Particle acceleration at the Sun and in the Heliosphere, Space Sci. Rev., **90**, 413 (1999) doi: 10.1023/A:1005105831781

Reames, D.V.: The two sources of solar energetic particles, Space Sci. Rev. **175**, 53 (2013) doi: 10.1007/s11214-013-9958-9

Reames, D.V., Element abundances in solar energetic particles and the solar corona, Sol. Phys., **289**, 977 (2014) doi: 10.1007/s11207-013-0350-4

Reames, D.V., What are the sources of solar energetic particles? Element abundances and source plasma temperatures, Space Sci. Rev. **194** 303 (2015) doi: 10.1007/s11214-015-0210-7

Reames, D.V., The abundance of helium in the source plasma of solar energetic particles, Sol. Phys. **292** 156 (2017) doi: 10.1007/s11207-017-1173-5 (arXiv: 1708.05034)

Reames, D.V., "The "FIP effect" and the origins of solar energetic particles and of the solar wind, Sol. Phys. **294** 47 (2018) doi: 10.1007/s11207-018-1267-8 (arXiv 1801.05840 )

Reames, D. V., Four distinct pathways to the element abundances in solar energetic particles, Space Science Rev. **216** 20 (2020) doi: 10.1007/s11214-020-0643-5 (arXiv 1912.06691)

Reames, D.V., Stone, R.G., The identification of solar He-3-rich events and the study of particle acceleration at the sun, Astrophys. J., **308**, 902 (1986) doi: 10.1086/164560

Reames, D.V., Cliver, E.W., Kahler, S.W., Abundance enhancements in impulsive solar energetic-particle events with associated coronal mass ejections, Sol. Phys. **289**, 3817, (2014) doi: 10.1007/s11207-014-0547-1

Reames, D.V., von Rosenvinge, T.T., Lin, R.P., Solar He-3-rich events and nonrelativistic electron events - A new association, Astrophys. J. **292**, 716 (1985) doi: 10.1086/163203

Richardson, I.G., Cane, H.V., Near-Earth interplanetary coronal mass ejections during solar cycle 23 (1996-2009): Catalog and summary of properties, Sol. Phys. **264**, 189 (2010) doi: 10.1007/s11207-010-9568-6

Schmelz , J.T., Reames, D.V., von Steiger, R., Basu, S., Composition of the solar corona, solar wind, and solar energetic particles, Astrophys. J. **755**:33 (2012) doi: 10.1088/0004-637X/755/1/33

Sheeley Jr., N.R., Surface evolution of the sun's magnetic field: a historical review of the flux-transport mechanism, Living Rev. Sol. Phys., **2**, 5 (2005) doi: 10.12942/lrsp-2005-5

Wang, Y.-M., Pick, M., Mason, G.M., Coronal holes, jets, and the origin of $^3$He-rich particle events, Astrophys. J. **639**, 495 (2006) doi: 10.1086/499355

Webb, D.F., and Howard, T.A., Coronal mass ejections: Observations, Living Rev. Sol. Phys. 9 3 (2012),doi 10.12942/lrsp-2012-3

Webber, W.R., Solar and galactic cosmic ray abundances - A comparison and some comments. *Proc. 14$^{th}$ Int. Cos. Ray Conf, Munich*, **5**, 1597 (1975)

# Chapter 2. A Turbulent History

**Abstract**   Large solar energetic-particle (SEP) events are clearly associated in time with eruptive phenomena on the Sun, but how? When large SEP events were first observed, flares were the only visible candidate, and diffusion theory was stretched to explain how the particles could spread through space, as widely as observed. The observation of coronal mass ejections (CMEs), and the wide, fast shock waves they can drive, provided better candidates later. Then small events were found with 1000-fold enhancements in $^3$He/$^4$He that required a different kind of source – should we reconsider flares, or their open-field cousins, solar jets? The $^3$He-rich events were soon associated with the electron beams that produce type III radio bursts. It seems the radio astronomers knew of both SEP sources all along. Sometimes the distinction between the sources is blurred when shocks reaccelerate residual $^3$He-rich impulsive suprathermal ions. Eventually, however, we would even begin to measure the source-plasma temperature that helps to better distinguish the SEP sources.

The first reported observation of a solar flare, that of 1118 GMT on 1 September, 1859, was published by a self-established astronomer Richard Carrington (1860) who saw the brightening of a white-light solar flare, which lasted over 5 min, while he was observing sunspots. The observation was confirmed by his friend Richard Hodgson. Carrington noted that the brightening did not seem to disrupt the underlying structure at all. However, apparently-associated geomagnetic effects were also noticed.

## 2.1 The First SEPs

Some 87 years later Scott Forbush (1946) reported the first SEPs as an increase in what we now call a ground-level event (GLE). Protons of GeV energies cause nuclear cascades through the atmosphere. Forbush was observing the intensities of similar secondary particles produced by galactic cosmic rays (GCRs) using ground-level ion chambers and especially the "Forbush decreases" now known to be caused by ejecta from the Sun whose shielding reduces the intensities of the GCRs. Three large solar events beginning in February and March 1942 produced sharp intensity *increases* from SEPs prior to the Forbush decreases. Since Forbush was unaware of CMEs and the shock waves they drive, it was natural for him to assume that the SEPs had come from the associated flares, which could even be *seen*.

The nuclear cascade from the large GLE of 23 February 1956 was measured by 6 neutron detectors widely spaced in geolatitude, and a balloon-borne detector which measured the atmospheric absorption mean free path of the solar protons (Meyer, Parker, and Simpson 1956). The SEP increase immediately preceded a Forbush decrease in GCRs that these authors regarded as a chance coincidence.

Since 1956, ground-level neutron monitors have held the promise of using the different geomagnetic cutoff rigidities at multiple sites to measure the high-energy proton spectra. Over 70 GLEs have been recorded in over 70 years (Cliver et al. 1982; Cliver 2006; Gopalswamy et al. 2012) but most of them barely rise above the GCRs. It is only recently that the neutron-monitor measurements, combined with satellite measurements have finally begun to yield rigidity spectra for 53 of the GLEs (Tylka and Dietrich 2009) as we will see in Sect. 6.1.

## 2.2 Solar Radio Bursts and Electrons

Much more sensitive ground-based evidence of SEPs was derived from the radio emission caused by streaming energetic electrons. As electrons of 10–100 keV stream out along magnetic fields from sources near the Sun, they excite Langmuir wave oscillations at the local plasma frequency. Since the plasma frequency depends upon the square root of the local plasma electron density, the emission, called a type III burst (e.g. Thejappa et al. 2012), drifts rapidly lower in frequency across the metric radio band as the electrons stream out from the Sun. At shock waves, electrons accelerated in the $V_S \times B$ electric field similarly excite local oscillations producing a type II burst (e.g. Ganse et al. 2012), but since the electrons are carried downstream of the shock soon after acceleration, the emission only drifts out with the shock speed, $V_S$, i.e. much more slowly.

In their review of the status of solar radio measurements Wild, Smerd, and Weiss (1963) identified two sites of acceleration near the Sun:

- Impulsive bursts of electrons were accelerated to produce type III radio bursts.

- Protons were accelerated at shock waves where accompanying electrons generated type II radio bursts.

After measurements in space became possible, Lin (1970, 1974) distinguished SEP events with 40 keV electrons that were associated with type III radio bursts, optical flares, and 20-keV X-ray bursts. These differed from the large proton events in which the accompanying electrons were mainly relativistic. Lin identified "pure" impulsive electron events, meaning events in which any accompanying ions were not yet detectible, at that time. The direct measurements of electrons by Lin supported the ideas of Wild, Smerd, and Weiss (1963).

An example of a modern space-based measurement of the dynamic radio spectra of type II and type III radio bursts from the *Wind*/WAVES instrument (Bougeret et al. 1995; https://ssed.gsfc.nasa.gov/waves/index.html) from a small event is shown in the lower panel of Fig. 2.1. The plasma frequency decreases with distance from the Sun as $\sqrt{n_e}$ where $n_e$ is the local plasma electron density. Ground-based radio instruments can measure only high frequencies produced near the Sun while space-based instruments cover sources moving from the Sun to the Earth. For type III bursts, frequencies drift rapidly, produced by 10–100 keV elec-trons streaming out from the Sun; frequencies in type II bursts drift outward from a source moving at the speed (~1000 km s⁻¹) of a shock wave.  In some events, type IV emission occupies the frequency region after and above the type II burst. Type IV emission may be produced by electrons accelerated on the sunward flanks of the shock or by reconnection regions in or behind the CME.

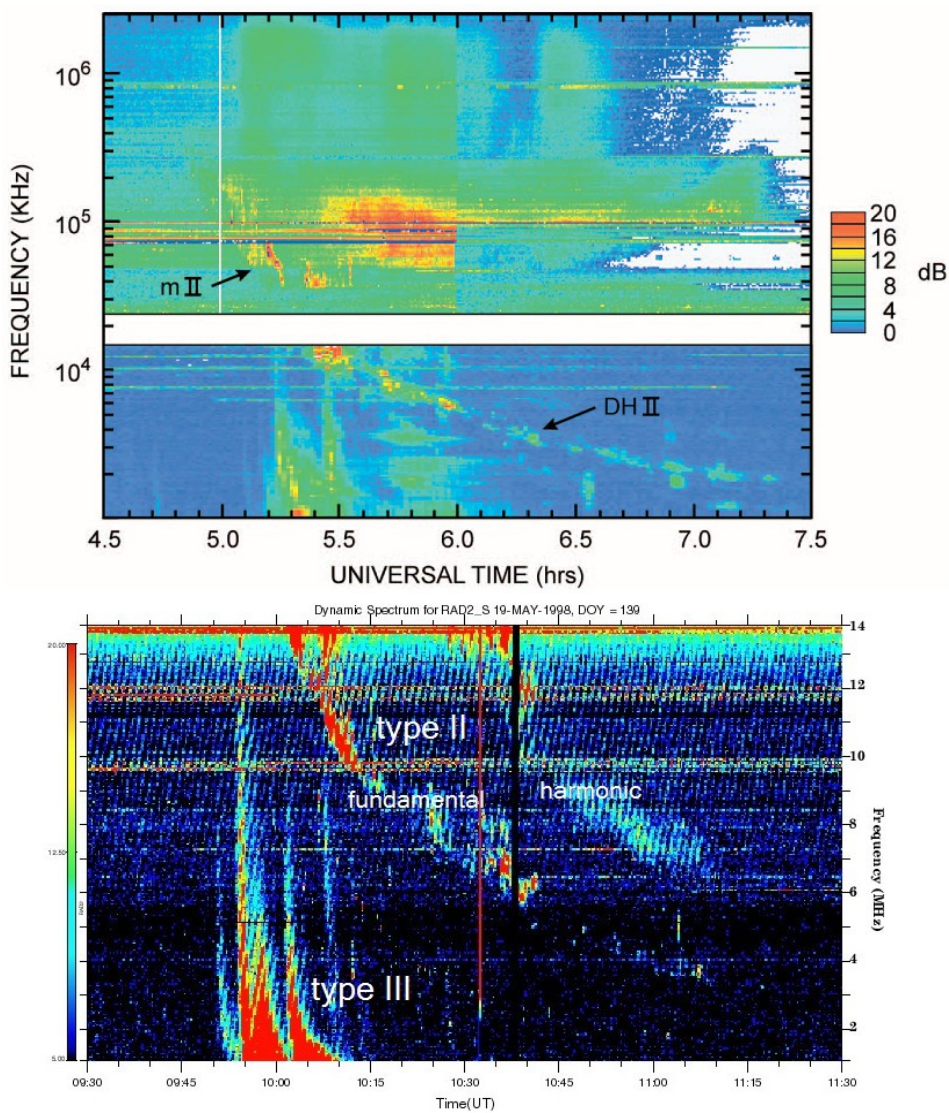

**Fig. 2.1** The *lower* panel shows the dynamic radio spectrum obtained by the *Wind*/WAVES instrument in May 1998 with type II and type III bursts.  The *upper panel* shows the type II burst associated with the large SEP event of 26 December 2001 as it propagates from the ground-based metric (≥ 25 MHz) Hiraiso station to the *Wind*/WAVES decametric-hectometric DH (≤ 14 MHz) regime (Cliver, Kahler, and Reames 2004, © AAS).

The upper panel in Fig. 2.1 shows a type II burst propagating from the ground-based metric (≥ 25 MHz) regime to the decametric-hectometric (DH, 1 – 14 MHz) regime at ~3 R$_S$ .  Cliver, Kahler, and Reames (2004) found a 90% association of DH type-II bursts with SEP events with 20 MeV protons but only 25% for metric type-II bursts without DH emission. This suggested shock acceleration that was strongest above ~3 $R_S$. Strong shocks that survive beyond ~3 $R_S$ are more likely to have expanded to a broad longitude extent.

Thus, prior to observations in space, there were observations of rare GLEs produced by GeV protons, and much-more-sensitive radio observations produced by ~10 keV electrons. Observers associated the GLEs with flares, but the radio observers provided the first evidence of two acceleration mechanisms and anticipated the importance of shock waves.

## 2.3 The Spatial Distribution

Parker (1963) not only described the continuous flow of the solar wind and the interplanetary magnetic field, he was aware of turbulence of that field and described the pitch-angle-scattering-induced diffusion of energetic particles flowing out along the field lines, an important description we still use often for ion transport along $B$ (see Sect. 5.1.1). However, in time we learned that all SEP features, such as their extreme longitudinal spread, do not arise solely from diffusion.

### 2.3.1 Lateral Diffusion and the Birdcage Model

"A man with only a hammer treats every problem like a nail." In early studies of large SEP events all the distributions seemed like they must be particle transport from a point-source flare, and diffusion theory was the transport tool of choice. The time dependence of the proton intensities had a smooth rise and a long, slow decay. Yet SEP events apparently associated with flares had such a wide span of solar longitudes, exceeding 180°. You could *see* the flares so they *must* be the source. Perhaps the particles from the flare diffused through the solar corona somehow and then out along the magnetic field lines toward Earth (Reid 1964).

In diffusion models, all of the physics of scattering is put into the diffusion coefficients, but when these coefficients are treated as adjustable parameters, their reality can become tenuous. Did the particles actually cross magnetic-field lines?

In fact, there was an early idea of a "fast propagation region" (Reinhard and Wibberenz 1974) of ≈60° in solar longitude after which particles diffused away more slowly. The authors did consider that the "fast propagation region" might actually be the surface of a shock wave, yet could not believe it to be the actual source of the acceleration. Shock waves were generally well known in 1974.

In the birdcage model (Newkirk and Wenzel 1978), arcades of coronal loops formed structures like wires of a birdcage, spreading particles across the corona. At the footpoints of the loops the fields were somehow connected to the next series of loops, and so on across the Sun. Transport through this grid was simply assumed to be diffusive and these diffusive transport models held sway for decades.

### 2.3.2 Large Scale Shock Acceleration and CMEs

A direct challenge to the birdcage model came first from Mason, Gloeckler, and Hovestadt (1984). They observed the abundances of low-energy H, He, C, O, and Fe ions over an extended time as connection longitudes drifted far (~120°) from the source.  Relative abundances of these ions representing different magnetic rigidities were not altered by their alleged complex journey through the coronal birdcage.  The authors suggested that the ions must actually result from large-scale shock acceleration (LSSA).  Shocks can easily cross magnetic field lines, accelerating particles locally across a broad surface, wherever they go.  LSSA also helped explain the long duration of the gradual events, especially at low energies, where the shocks continue acceleration as they come far out from the Sun.

In the same year Kahler et al. (1984) found a 96% correlation between the largest energetic SEP events and fast, wide CMEs.  This paper strengthened preliminary associations found during the earlier *Skylab* mission when CME observations began to become common.

It has long been known that particles are accelerated at shock waves.  SEP intensities often peak as a shock wave passes, indicating a local source from which particles diverge (Sect. 5.4).  These peaks were called "energetic storm particle" or "ESP events."  However, it has seemed difficult to communicate that the particles that arrived earlier than the ESP peak came from that same shock that was much stronger when it was nearer the Sun (e.g. Reames, Barbier, and Ng 1996).

### 2.3.3 The Longitude Distribution

When larger numbers of gradual SEP events had been accumulated, it became possible to organize them as a function of their apparent solar source longitude. Even today with multiple spacecraft available it is difficult to study many individual events by observing each of them with multiple spacecraft at several conveniently-spaced longitudes. Cane, Reames, and von Rosenvinge (1988) did the next best thing, studying 235 large events of >20-MeV protons observed on IMP and ISEE 3 by binning them as a function of their associated source longitudes. The authors concluded that the most important factor organizing the time profiles of large SEP events was the existence of an interplanetary-shock source and the curved Parker-spiral magnetic field which the particles were largely constrained to follow. Fig. 2.2 shows a version of their findings.

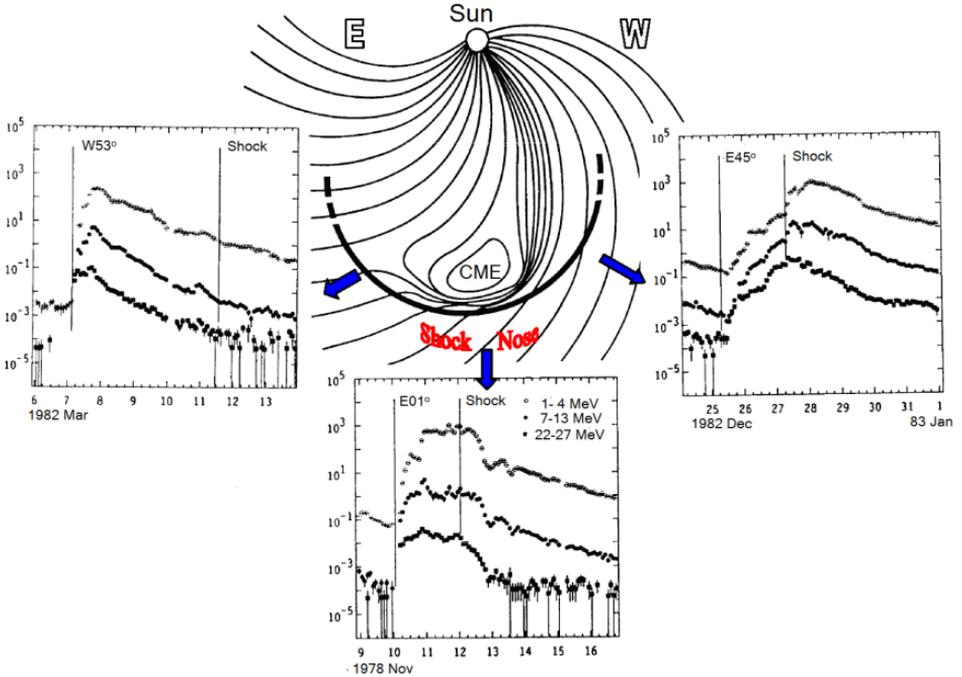

**Fig. 2.2** Variation of the appearance of typical SEP events is shown as viewed from three solar longitudes (see text; after Reames 1999 © Springer; see also Cane, Reames, and von Rosenvinge 1988, Reames, Barbier, and Ng 1996).

In Fig. 2.2, the three cases shown are described as follows:

1. A spacecraft on the East flank of the shock (a western solar-source longitude) sees a fast intensity increase early, when it is magnetically well-connected to the strongest source at the "nose" of the shock as it first appears near the Sun. At later times the intensity decreases as the magnetic connection point moves gradually around the shock toward its weaker eastern flank. When this flank of the shock would be expected to pass the spacecraft, the shock may be very weak or may have dissipated completely so far around from the nose.

2. A spacecraft observing a source near central meridian is magnetically connected far to the West of the shock nose early in the event but the intensity increases as the shock moves outward and the connection point approaches the nose. The connection to the shock nose occurs as the shock itself passes the spacecraft. Thereafter, the intensity may decline suddenly as the spacecraft passes inside the CME driving the shock.

3. A spacecraft on the West flank of the shock (an eastern source on the Sun) is poorly connected to the source but its connection and the observed intensities improve with time, reaching a maximum *behind* the shock when it encounters field lines that connect it to the nose of the shock *from behind*.

We will see that later observations of individual events from multiple space-craft generally supported the pattern seen in Fig. 2.2 (e.g. Fig. 5.16).

### 2.3.4 Scatter-Free Events

Does ambient turbulence in the interplanetary medium cause pitch-angle scatter-ing of the particles flowing out from the Sun? The classic Fig. 2.3 from Mason et al. (1989) provides an interesting answer especially late in an event.

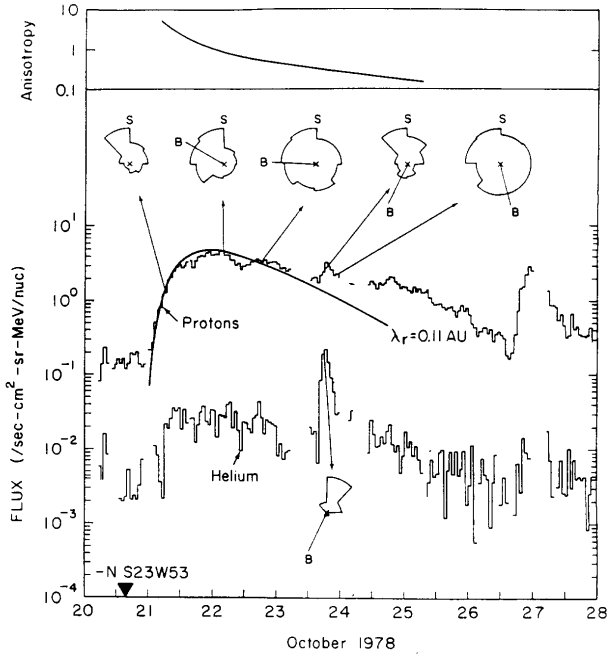

**Fig. 2.3**. Intensities and angular distributions of ~1 MeV amu$^{-1}$ H and He are shown for a large SEP event of 21 October, 1978 and for the newly anisotropic flow from a small $^3$He-rich event on 23 October. A diffusion fit, to the proton intensity is shown with a radial component of the scattering mean free path of 0.11 AU. How can scattering spread particles in time so much in the large event, but barely scatter those from the small event in its wake (Mason et al. 1989, © AAS)? The long duration of the large event comes from continu-ing acceleration, then from trap-ping behind the CME, not from scattering.

Mason et al. (1989) showed that most $^3$He-rich events (like that on 23 October 1978) actually propagate scatter free, i.e. with $\lambda \geq 1$ AU. We will see in Sect. 5.1.2 that in more intense events the streaming protons may be scattered early by self-amplified waves, but the slow decrease late in gradual events actually occurs when ions are adiabatically trapped in a magnetic *reservoir* (Sect. 5.7) behind the CME and shock. There is little scattering in the reservoir, but intensities decrease because the volume of the reservoir expands. Diffusion might be appropriate ear-lier in an event, but it *does not* produce the slow intensity decay of the large event, as the profile of the small scatter-free event on October 23 shows. Slow decays of SEPs are yet another misapplication of diffusion theory (see Sect. 5.7).

### 2.3.5 Field-line Random Walk

While particles do not easily cross field lines, and the field lines may not join for-tuitously, as suggested by the birdcage model, their footpoints do engage in a ran-dom walk which has the effect of spreading the longitude distribution of particles injected upon them (Jokipii and Parker 1969; Giacalone and Jokipii 2012). The footpoints of the open field lines are imbedded in turbulent velocity fields that cause adjacent lines of force to execute a random walk relative to each other in time, as each stage of the evolving field pattern is carried out by the solar wind. Field lines are also buffeted by turbulence from the passage of CMEs. Thus, even at quiet times, field lines from any small region on the Sun have a distribution that is spread about the Parker spiral so that particles from a compact impulsive SEP event have a Gaussian-like longitude (and latitude) distribution. In Fig. 2.4, this contributes to the longitude spread of the impulsive events shown in the right panel. In the left panel, the gradual events are also spread in longitude by the spatial extent of the shock-wave source. Recent more-sensitive instruments on STEREO see small, delayed echoes of impulsive events at longitudes >60° distant at reduced fluence (Wiedenbeck et al. 2013) vs. gradual events over nearly 360° (e.g. Reames, Barbier, and Ng 1996). The spread in impulsive SEPs includes variations in $V_{SW}$ and random walk of the field lines (Giacalone and Jokipii 2012).

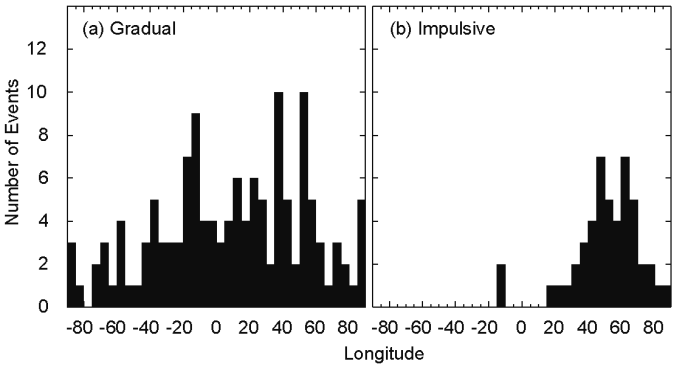

**Fig. 2.4.** Longitude distributions are shown for gradual SEPs (*left*) spread mainly because of the width of the shock source, and impulsive (*right*) SEP events spread by random walk of field lines and by variations in solar wind speed (Reames 1999 © Springer).

## 2.4 Shock Theory

Shock acceleration theory had an extensive history in GCR acceleration prior to its application to SEPs and that will not be repeated here. The plasma physics of shocks and shock acceleration has been reviewed by Jones and Ellison (1991; see also Lee 2005, Sandroos and Vainio 2007, Zank, Li, and Verkhoglyadova 2007; Verkhoglyadova, Zank, and Li 2015). Diffusive shock acceleration (DSA) occurs as ions are pitch-angle scattered back and forth across a shock wave, gaining an increment of velocity on each round trip. For an oblique shock wave, particles can gain additional energy in the $V_S \times B$ electric field of the shock (e.g. Decker 1983).

Acceleration of a particle at a quasi-parallel shock may be considered approximately as a series of frame transformations which are randomly required as a particle scatters from upstream to downstream to upstream of the shock. On each round trip the particle has gained a velocity related to the velocity difference between the upstream and downstream scattering centers; in this difference, the shock speed can be augmented by the Alfvén speed of the four possible wave modes – inward and outward along $B$ times two circular polarizations. The diffusion, the scattering against resonant waves, is determined by the particle's rigidity, but it gains in velocity. The rate of acceleration increases as the scattering mean free path decreases reducing the mean time between crossings. As accelerated particles stream away from the shock, they amplify resonant Alfvén waves of wave number $k \approx B/\mu P$, according to quasi-linear theory, where $P$ is the particle rigidity and $\mu$ its pitch-angle cosine. These amplified waves increase the resonant scattering of the ions that follow behind (see Sect. 5.1.2). Assuming $\mu \approx 1$ for simplicity, this traps particles of $P$ near the shock, increasing their energy to a higher value, say rigidity $P_2$, where they stream out, amplifying waves that resonate with $P_2$ which are trapped and accelerated to $P_3$, etc. Continuing this process indefinitely can lead to a power-law spectrum where the power depends upon the shock compression ratio. However, SEP spectra are produced by CME-driven shocks with finite lives and diminishing strength so they have spectral breaks or "knees" where spectra steepen, generally above 10 MeV (see Fig. 3.7), and produce complex behavior like that we will see in Figs. 2.9 and 2.10. The equations that control particle transport, scattering, and wave growth will be presented in Sects. 5.1.2 and 5.1.3.

Quasi-linear theory actually assumes that the energy density in wave turbulence is small with respect to the energy density in the field, $\delta B/B \ll 1$, a condition that is most likely violated at strong shocks which approach or even exceed the "Bohm limit" where the proton scattering mean free path equals its gyroradius. Lee (1983) applied equilibrium DSA theory to explain proton acceleration at interplanetary shocks assuming $\mu \approx 1$. At equilibrium, the growth of upstream waves that resonate with each rigidity is just sufficient to replace the waves being swept into the shock. In contrast, Zank, Rice, and Wu (2000) found that shock acceleration could produce GeV protons near the Sun, assuming that turbulence reaches the Bohm limit $\delta B/B \approx 1$ at the shock.

Ng, Reames, and Tylka (2003) considered the self-consistent, time-dependent particle transport with amplification of Alfvén waves, and Ng and Reames (2008) extended the calculation to the time-dependent shock acceleration of protons to energies of >300 MeV.

It is not uncommon to consider the action of two or more shocks on a population of particles (e.g. Gopalswamy et al. 2002). The equilibrium energy spectrum for two consecutive shocks is derived as Eq. 5.9 (Sect. 5.4) and is a power-law spectrum with the power dominated by the compression ratio of the strongest shock. It is appropriate to assume that the shocks contribute sequentially; pre-acceleration of seed ions may increase the probability of secondary acceleration. However, there is no stronger collaborative effect; shock acceleration occurs within a modest number of proton gyroradii of the shock and CME-driven shocks crossing each other spend a negligible time at such a small separation.

## 2.5 Element Abundances

The earliest observations of heavier elements in SEP events were made using nuclear-emulsion detectors on sounding rockets launched into large SEP events. Fichtel and Guss (1961) observed C, N, and O nuclei above 25 MeV amu$^{-1}$. The observations were extended to Fe by Bertsch, Fichtel, and Reames (1969). For the early measurements, the presence of SEPs was detected by a riometer, which measures radio absorption produced by ionization of the polar cap region produced by high intensities of SEPs. The riometer was used as an indication to fire sounding rockets above the atmosphere to measure SEP abundances from Ft. Churchill in northern Manitoba, Canada.

### 2.5.1 First Ionization Potential (FIP) and Powers of A/Q

Improving measurements led to comparison of element abundances in SEP events with those in the solar photosphere and corona (e.g. Webber 1975; Webber et al. 1975; Cook, Stone, and Vogt 1984). The measurements were summarized in the review of Meyer (1985). He found two factors that influenced element abundances in large SEP events ($^3$He-rich events were excluded). There was one component, present in all events that depended upon the first ionization potential (FIP) of the elements, and a second variable component that he called "mass bias" actually depending upon the mass-to-charge ratio $A/Q$ of the ions. The $A/Q$ dependence differed with time and from one event to another. The FIP dependence that was shown in Fig. 1.6 represents average abundances at the coronal origin of SEPs, relative to the corresponding photospheric abundances. Elements with FIP above about 10 eV are neutral atoms in the photosphere while lower-FIP elements are ionized. The ions are more rapidly swept up into the corona, as by Alfvén waves (e.g. Laming 2004, 2009) and thus have higher relative abundances there.

An increasing or decreasing power-law dependence on the $A/Q$ ratio of the ions was clearly found by Breneman and Stone (1985) and is shown in the left panel of Fig. 2.5. Brenenan and Stone (1985) used the newly available ionization-state measurements of Luhn et al. (1984) to determine $Q$.

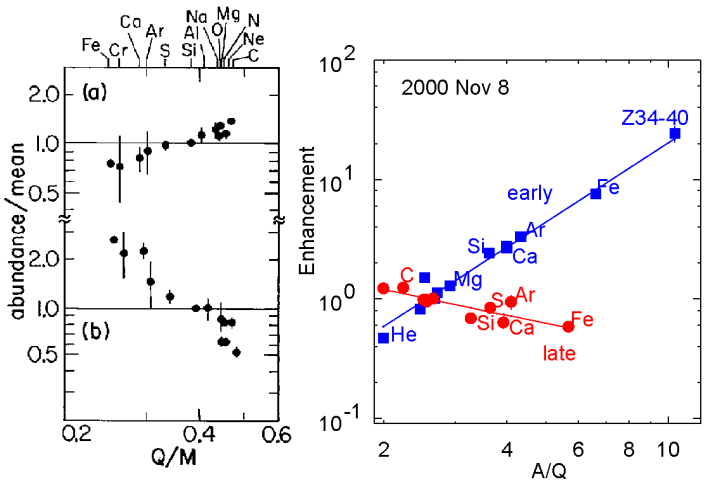

**Fig. 2.5.** The *left panel* shows the dependence of elemental abundances on the charge-to-mass ratio $Q/M$ (our $Q/A$) for two large SEP events (**a** and **b**) by Breneman and Stone (1985 © AAS). The *right panel* shows the $A/Q$ dependence early (*blue*) and late (*red*) in a large SEP event (Reames 2016; see Fig. 5.13 © Springer).

After languishing for over 30 years, these power-laws have gained renewed interest. The pattern of ionization states $Q$ depends upon the plasma temperature (see Fig. 5.11) and it has recently been shown (Reames 2016) that the pattern of the grouping of elements in enhancement vs. *A/Q* (Fig. 2.5, right panel) determine the source-plasma temperature (see Sect. 5.6).  In fact, grouping of elements C–Mg with similar enhancements and *A/Q* in the left panel of Fig. 2.5 suggests a temperature of about 1.5 MK, those in the right panel are both ~1.0 MK.  But we are getting ahead of our story (see Sect. 5.6).

### 2.5.2  $^3$He-rich Events

The first observation of $^3$He/$^4$He in SEP events (Hsieh and Simpson 1970) showed some evidence of enhancement which aroused interest because of the possibility that $^3$He could be produced in nuclear reactions in flares, but not when Serlemitsos and Balasubrahmanyan (1975) found $^3$He/$^4$He $= 1.52 \pm 0.10$ but $^3$He/$^2$H $> 300$. With no evidence of other reaction products, like $^2$H or $^3$H, it became clear that a new acceleration process was involved, since $^3$He/$^4$He $\approx 5 \times 10^{-4}$ in the solar wind. It also became apparent that there were other abundance enhancements, such as Fe/O that was ~10 times larger than in the solar wind (e.g. Gloeckler et al. 1975). However, there is still *no* evidence of nuclear-reaction secondaries, $^2$H, $^3$H, Li, Be, B, etc. in the SEPs; γ-ray and neutron measurements tell us they are produced in flare loops (Sect. 4.9), but are magnetically trapped there and cannot get out.

The next generation of measurements of $^3$He-rich events (Fig. 2.6) led to their association with non-relativistic electron events (Reames, von Rosenvinge, and Lin 1985) and with type III radio bursts (Reames and Stone 1986).  Thus Lin's (1970) "pure" electron events were actually $^3$He-rich or "impulsive" SEP events and were associated with the type III-burst electron events discussed by Wild, Smerd, and Weiss (1963) that we saw in Sect. 2.2.  While these events were also Fe-rich, Fe/O was not correlated with $^3$He/$^4$He (e.g. Mason et al. 1986), opening the possibility and the need for two different enhancement mechanisms.

The unique $^3$He enhancement suggested a resonant interaction with plasma waves. The earliest mechanism suggested was based upon the selective heating by absorption of ion-sound waves (Ibrigamov and Kocharov 1977; Kocharov and Kocharov 1978, 1984).  However, Weatherall (1984) found that this mechanism did not have the sensitivity to ion charge needed to account for the observed abundances. Fisk (1978) and Varvoglis and Papadopoulos (1983) suggested selective heating of $^3$He by absorption of electrostatic ion cyclotron waves at the $^3$He gyrofrequency; Winglee (1989) also invoked ion-ion streaming instability in an effort to enhance heavy ions. Riyopoulos (1991) considered electrostatic two-ion (H–$^4$He) hybrid waves.  Some of the mechanisms suggested during this period required high values of $^4$He/H in the source plasma, most required a second unspecified physical process for preferential acceleration of the pre-heated ions from the thermal distribution, such as a shock wave.

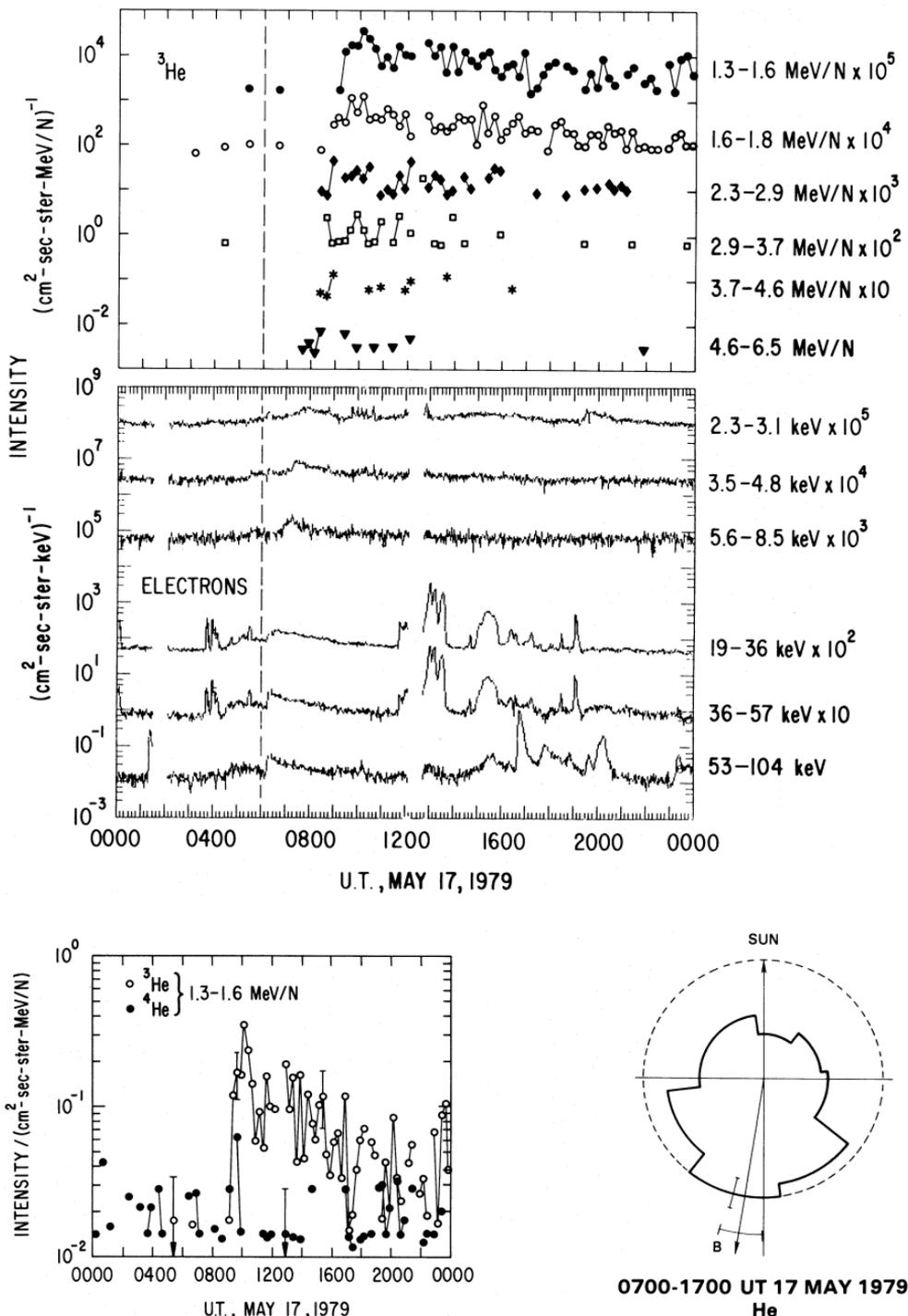

**Fig. 2.6** Intensities of ³He and electrons of various energies (*upper panels*) show velocity dispersion (i.e. fastest particles arrive first after traveling ~1 AU from the Sun) while intensities of ³He and ⁴He (*lower left*) show dominance of ³He, with ⁴He only at background, and the angular distribution (*lower right*) shows outward flow of ³He along the magnetic field **B**, all in a small ³He-rich SEP event of 17 May 1979 (Reames, von Rosenvinge, and Lin 1985, © AAS).

Temerin and Roth (1992; Roth and Temerin 1997) found that the streaming electrons that produce the type III bursts would generate electromagnetic ion cyclotron (EMIC) waves near the gyrofrequency of $^3$He. Ions mirroring in the converging magnetic field could be accelerated as they continue to absorb the waves, in analogy with the "ion conics" seen in the Earth's aurora. The strength of this model, with more robust waves, was that (i) it explained and used the strong association observed between type-III electrons and $^3$He-rich events, (ii) it produced $^3$He acceleration, not just preheating, and (iii) the streaming electrons provided a self-consistent source for the resonant waves that preferentially energized $^3$He. The authors suggested that heavier ions were accelerated through resonance with the second harmonic of their gyrofrequencies, but this required specific ionization states and did not produce the extreme and uniform increase in enhancement of the heavy elements with $Z > 50$ that was observed subsequently (e.g. Reames 2000; see also Reames, Cliver, and Kahler 2014a, b), as we shall see in Chap. 4. Miller, Viñas, and Reames (1993a, 1993b) considered electron beam generation of other wave modes such as sheer-Alfvén waves and their effect on heavy ions, and Steinacker et al. (1997) considered effects of broadened spectral lines produced by thermal damping in a hot (2.4–4.5 MK) plasma, producing the broadened "He valley" of damping which controls the wave regions left available for absorption and enhancement of various heavier ions.

Ho, Roelof, and Mason (2005) found that there was an upper limit to the fluence of $^3$He in events so that increasingly large impulsive events had decreasing $^3$He/$^4$He ratios. This agreed with an estimate by Reames (1999) that an impulsive event can accelerate and deplete most of the $^3$He in a typical flare (or jet) volume.

$^3$He-rich events were traced to their solar sources by Nitta et al. (2006) and by Wang, Pick, and Mason (2006) and there was a growing association with narrow CMEs that has become a clear association with solar jets (Kahler, Reames, and Sheeley 2001; Bučík et al. 2018; Bučík 2020; see also Reames, Cliver, and Kahler 2014a).

### 2.5.3. The Seed Population for Shocks

For a time, it seemed that impulsive and gradual events might be distinguished by their element abundances alone. Impulsive events were $^3$He-rich, weren't they? Then Mason, Mazur, and Dwyer (1999) found enhancements of $^3$He in large SEP events that clearly should otherwise be called gradual. In fact, there were even large $^3$He enhancements during relatively quiet times. Earlier evidence of this had been seen by Richardson et al. (1990). The mass distribution in Fig. 2.7 clearly shows $^3$He, and although the amount is small, it is 5 times the solar-wind abundance. The authors suggested that the $^3$He, and also Fe, are suprathermal remnants of previous impulsive SEP events. These *impulsive-suprathermal* ions contribute to the *seed population* for subsequent shock acceleration (see Tylka et al. 2001).

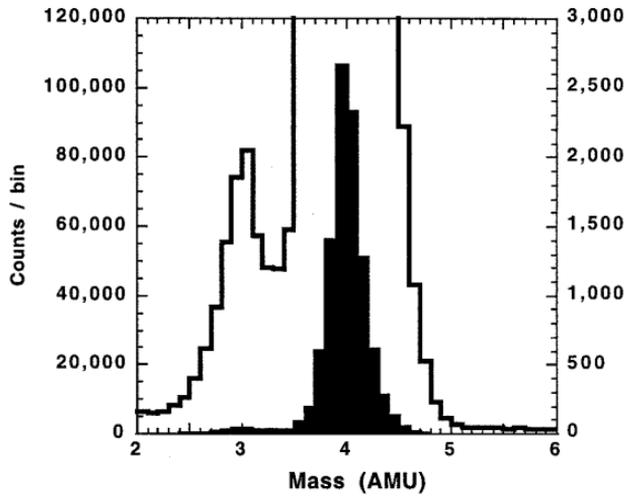

Exploring the seed population, Desai et al. (2001) found ³He intensity increases at shocks in 25 SEP events with enhancements of 3 to 600 relative to the solar wind ratio, and Desai et al. (2003) found Fe/O at the shock was correlated with Fe/O upstream. Fig. 2.8 shows intensities of ³He, ⁴He, O, and Fe before and during a strong shock event. The quiet period labeled A, "upstream" is both ³He-rich and has Fe/O >1 while the later period B on 24 June is just extremely Fe-rich. These strong Fe/O enhancements do not persist at the shock, but there clearly must be ³He in the seed population, suggesting that it contains suprathermal ions from earlier impulsive SEP events. The correlation of Fe/O at the shock with that upstream is consistent with that interpretation. Note, however, that *most* of the ions at this shock peak *do not* come from ³He-rich impulsive suprathermals. We will see that these ESP peaks represent ambient coronal material in most cases, although suprathermal ions may also contribute.

**Fig. 2.8.** (**a**) Intensities of 0.5 – 2.0 MeV amu$^{-1}$ $^3$He, $^4$He, O and Fe are shown during a large SEP event, with (**b**) a histogram of Fe arrivals, (**c**) the magnetic field B, and (**d**) the solar wind speed. $^3$He is clearly accelerated, peaking at the shock, S, but is not as strongly enhanced as in the $^3$He-rich period labeled **A**, "upstream." In quiet period **A** both $^3$He/$^4$He and Fe/O are enhanced; in **B** only Fe/O is enhanced. (Desai et al. 2003, © AAS). Quiet periods are frequently $^3$He- and Fe-rich (Bučík et al. 2104, 2015; Chen et al. 2015).

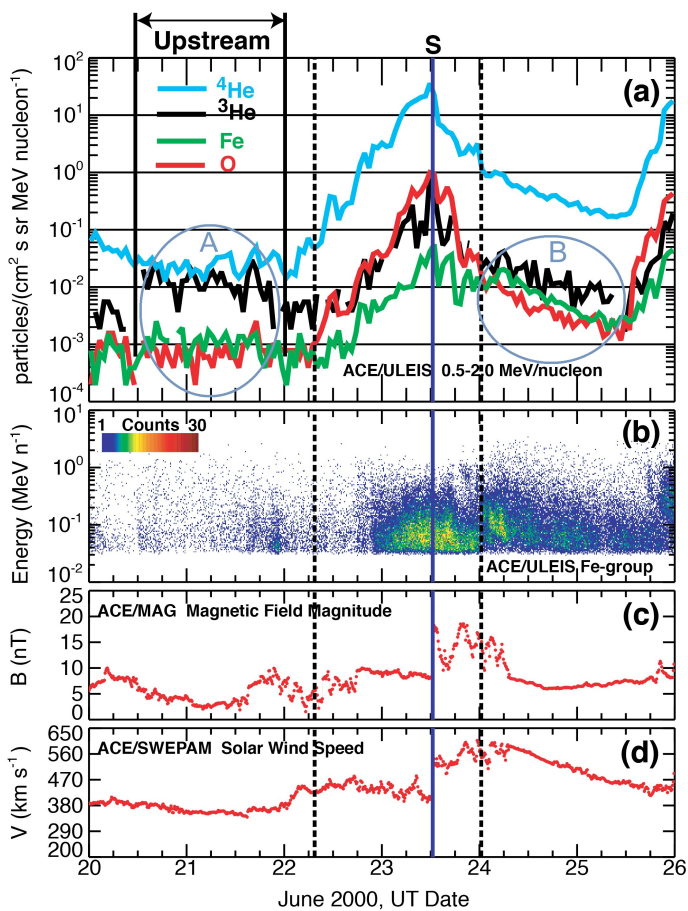

Tylka et al. (2005) found that in two otherwise-similar, large SEP events, the energy dependence of Fe/C above ~10 MeV amu$^{-1}$ suddenly increased in one event and decreased in the other, as shown in the left panel of Fig. 2.9. The authors considered the possible selection effect of impulsive suprathermal ions caused by differences in shock geometry. In quasi-perpendicular shock waves, with **B** perpendicular to the shock normal, injected ions may need a higher speed to reacquire the shock from downstream, so that pre-accelerated impulsive suprathermal ions would be preferentially selected as shown in the right panel of Fig. 2.9. Tylka and Lee (2006) calculated the effect different seed populations and shock geometries could have on the energy dependence of Fe/C. The higher-energy effects occur because the location of the high-energy "knee" (Sect. 3.4) where the power-law shock spectra roll downward, depends upon $Q/A$ of the ions and sec $\theta_{Bn}$, the angle between **B** and the shock normal. Coronal- and impulsive-suprathermal ions have different values of $Q$ and thus contribute differently above the spectral knee.

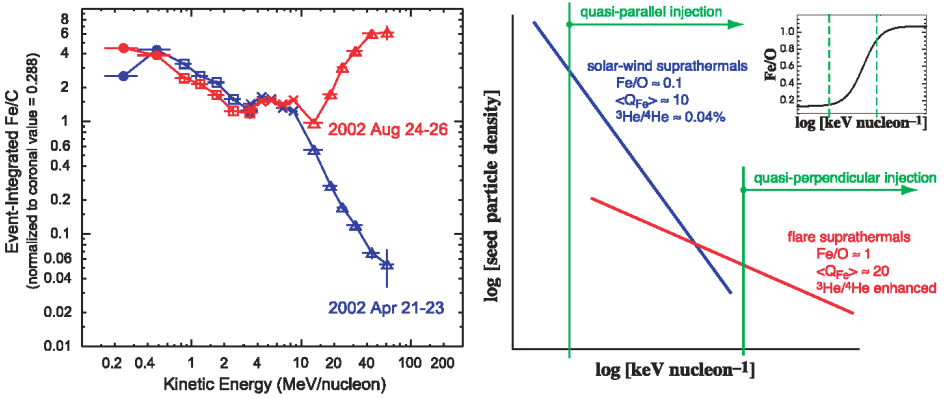

**Fig. 2.9** The *left panel* compares the energy dependence of Fe/C for two gradual events that are otherwise similar in their properties (Tylka et al. 2005 © AAS). The *right panel* shows hypothetical spectra of two sources of suprathermal ions where different injection thresholds will yield different abundance ratios (Tylka et al. 2005). Clearly, it would be unwise to use measurement of Fe/C above ~10 MeV amu$^{-1}$ in an attempt to distinguish impulsive and gradual SEP events.

Tylka and Lee (2006) assumed that the shock spectrum of species *i* varied as $j_i(E) = k_i E^{-\gamma} \exp(-E/E_{0i})$, a form originally suggested by Ellison and Ramaty (1985). Then letting $E_{0i} = E_0 \times (Q_i/A_i) \times (\sec \theta_{Bn})^{2/(2-\gamma)}$, where $E_0$ is the proton knee energy, a wide variety of energy dependence of Fe/O may be seen as in Fig. 2.10.

**Fig. 2.10**. The energy dependence of Fe/O is shown as a function of *R* which is the ratio, in the seed population, of O in impulsive suprathermal ions to coronal ions. The values of $\gamma$ and $E_0$ assumed for this case are shown in the lower left corner of the figure. (Tylka and Lee 2006 © AAS).

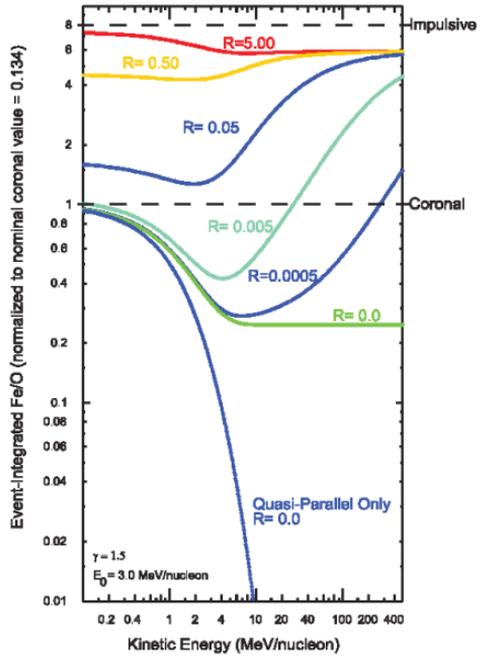

In fact, the seed population for shock acceleration can consist of ambient coronal material as well as residual suprathermal ions from previous impulsive and gradual SEP events. However, Giacalone (2005) noted that high turbulence near the shock with $\delta B/B \approx 1$ would allow oblique shocks better access to the low-energy seed population and diminish the selective dependence on $\theta_{Bn}$.

For further studies of the dependence of the spectral break, of its power-law dependence upon $Q/A$ and the variation with shock geometry, see Li et al. (2009) and Zhao, Zhang, and Rassoul (2016). The latter authors conclude that the energy of the spectral break depends upon $(Q/A)^\delta$ where $0.4 < \delta < 1.3$, with break energies varying between 10 and 120 MeV amu$^{-1}$. Note that abundances such as Fe/O will become affected when Fe is above its break but O is not.

## 2.6 Ionization States

Some of the earliest direct measures of SEP ionization states were the direct measurements at $0.34 – 1.8$ MeV amu$^{-1}$ for Fe (Luhn et al. 1984, 1987). They found an average of $Q_{Fe} = 14.2 \pm 0.2$ for gradual events, corresponding to a plasma temperature of ~2 MK, but a much higher value of $Q_{Fe} = 20.5 \pm 1.2$ for $^3$He-rich events. Either the $^3$He-rich events are much hotter, ~10 MK, or, as we now believe, the ions may be stripped in transit away from the impulsive sources which lie a little deeper in the corona. Subsequently Leske et al. (1995) used geomagnetic cutoffs to find the $Q_{Fe} = 15.2 \pm 0.7$ at 15–70 MeV amu$^{-1}$ in large events and Tylka et al. (1995) found $Q_{Fe} = 14.1 \pm 1.4$ at 200–600 MeV amu$^{-1}$.

More recently, DiFabio et al. (2008) found that the ionization states in impulsive SEP events increased with energy, suggesting that the ions had passed through enough material that electron stripping and capture were in equilibrium at each ion velocity. The authors suggested that the ions in *impulsive* events were accelerated below 1.5 $R_S$ where densities were higher, beginning at a temperature of 1–3 MK. It was once suggested that $^3$He-rich SEP events come from "high coronal flares", based upon their electron spectra, but the stripping of Fe associates them with the deepest known SEP sources. We will see in Sect. 3.1 that acceleration in *gradual* events does begin higher in the corona, at 2–3 $R_S$.

A different approach to determining ionization states in impulsive events was taken by Reames, Meyer, and von Rosenvinge (1994). They noted that in average impulsive SEP events, the elements $^4$He, C, N, and O showed *no* enhancement relative to reference coronal abundances, Ne, Mg, and Si were enhanced by a factor of ~2.5, and Fe by a factor of ~7. This suggested that, *at the time of acceleration*, C, N, and O were fully ionized like He, but that Ne, Mg, and Si were probably in a stable closed shell configuration with two orbital electrons. They suggested that this occurs in a temperature range of 3–5 MK. At higher temperatures, Ne would become stripped, have $Q/A = 0.5$ like lighter elements, and could not be enhanced relative to them. At lower temperatures, O could capture electrons and would no longer have $Q/A \approx 0.5$. More recent studies (Reames, Cliver, and Kahler 2014a, b) have lowered this range to 2–4 MK to account for i) more accurate measurements that showed Ne enhancements exceeding those of Mg, and Si, ii) O enhancements causing decreased He/O and C/O, and iii) a power-law fit in $A/Q$ extending to $(Z > 50)$/O (see Sect. 4.6). These values of 2–4 MK correspond to ambient electron temperatures in solar active regions where flares and jets occur. Thus we began to use abundances to measure temperatures.

The strong $A/Q$ dependence of the enhancements extending to a factor of ~1000 for (76 $\leq Z \leq$ 82)/O (e.g. Reames, Cliver, and Kahler 2014a, b) recently has been theoretically understood as occurring in collapsing islands of magnetic reconnection (e.g. Drake et al. 2009). These particle-in-cell simulations show that ions are Fermi-accelerated as they are reflected back and forth from the ends of the collapsing islands of magnetic reconnection (see Sect. 4.7).

While impulsive SEPs may have passed through the extremely small amount of matter required to attain equilibrium values of $Q$, they cannot have passed through enough material to lose significant energy, since the $Q^2/A$ dependence of the energy loss would destroy the strong ~1000-fold enhancement observed for heavy elements such as (76 $\leq Z \leq$ 82)/O.

Recent studies of the $A/Q$ dependence in *gradual* SEP events (Reames 2016) have found that most of these events (69%) have source-plasma temperatures $\leq$ 1.6 MK, consistent with shock acceleration of ambient coronal plasma (see Sect. 5.6). Only 24% of the events have active-region temperatures of 2.5 – 3.2 MK and thus include dominant enhancements from impulsive suprathermal seed ions.

Using the $A/Q$-dependence of abundance enhancements, with $Q$ vs. $T$ from atomic physics, these studies provide a new method of determining ionization states at the point of acceleration. This circumvents the effects of stripping that may be present in the ionization states measured later at 1 AU.

## 2.7 Disappearing-Filament Events

A "disappearing" filament occurs when a filament, which may have been visible in the corona for days, is suddenly destabilized and erupts within a CME, disappearing from its former position. An H$\alpha$ brightening may form a classic double-ribbon pattern along the filament channel with slight heating and soft X-ray emission, but no hard X-ray emission or flaring occurs. Such events can produce a fast CME, a shock wave, and a substantial gradual SEP event, without the need of a flare or even a solar active region.

An early association of SEPs with filament changes was made by Sanahuja et al. (1983) but a clear example was the SEP event of 5 December 1981, shown in Fig. 2.11, identified and discussed by Kahler et al. (1986). Cane et al. (1986) found six other disappearing-filament-associated SEP events with a CME and shock but no impulsive phase or flare, and Gopalswamy et al. (2015) have extended this study to recent large gradual SEP events. They conclude that fast CMEs that produce GLEs attain a high speed at 2 or 3 $R_s$ while those in filament eruptions begin slowly and accelerate, so that a shock wave is not produced until ~8 $R_s$. Thus the properties of the SEPs are controlled by properties of the CME and the shock. However, flares are not required for SEP acceleration.

**Fig. 2.11**. Intensities vs. time are shown for the disappearing-filament-associated SEP event of 5 December 1981. The peak in the low energy protons on 8 December occurs at the time of shock passage at 1 AU.

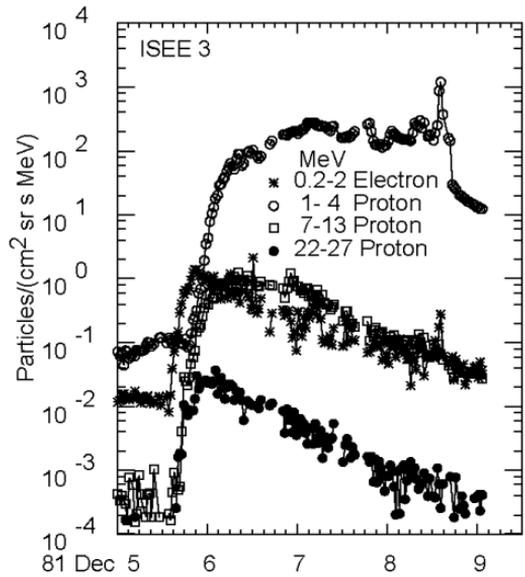

In contrast with these SEP events without flares, there are also "confined flares," X-class flares that have no CMEs or external SEPs (Gopalswamy, Akiyama, and Yashiro 2009).

## 2.8 "The Solar-Flare Myth"

By 1993, the idea of impulsive and gradual SEP events was fairly well documented, CMEs and CME-driven shocks had been studied for a decade in relation to SEPs, and ³He-rich events had been studied for two decades.  It became increasingly clear that the largest SEP events (and the only ones producing a significant radiation hazards) were gradual events related to CMEs and shocks, *not to flares*. The birdcage model (Sect. 2.3.1) was dead.  While reviews of this emerging paradigm were fairly common in invited talks at meetings, it was the publication of the review "The Solar Flare Myth" by Gosling (1993) that drew enormous criticism that surprised the SEP community.  This fairly straightforward review was thought to "wage an assault on the last 30 years of solar-flare research" (Zirin 1994) by a flare community that usually ignored SEPs entirely.  Apparently there was concern that if hazardous SEPs did not come from flares, flare research might be discontinued!  The sky was falling!  In hindsight, surely the last 25 years have proven such concerns to be unfounded.  Unfortunately, however, there is still some visceral reluctance to embrace the idea of shock acceleration of SEPs, especially the shocks that produce GLEs.

The controversy raised by the Gosling (1993) paper led to an invited discussion from three alternative viewpoints in *Eos* where Hudson (1995) argued that the term "flare" should include the CME, shock, and any related physics, Miller (1995) argued that flares, being more numerous, were a better subject for acceleration studies, and Reames (1995) argued for the separate study of the physics of both flare and shock acceleration of SEPs.  While the extension of the term "flare" has some philosophical merit, it is important for SEP studies to distinguish a point-source flare or now, a localized jet, from the acceleration source at a broadly-extensive, Sun-spanning, CME-driven shock wave, especially when they involve different physical mechanisms.

## 2.9 Wave Generation and the Streaming Limit

When intensities of particles streaming along $B$ are sufficiently great, they can amplify resonant Alfvén waves that exist or even generate them anew (Stix 1992; Melrose 1980). When in resonance, circularly polarized waves can maintain the orientation of their fields with respect to the velocity vector of the gyrating ions, maximizing the interaction. Systematic scattering of streaming ions reduces their energy only slightly but energy is conserved by amplifying the waves. These waves increase scattering and, in the vicinity of shock waves, increase acceleration. We have mentioned the early study of equilibrium wave growth and shock acceleration (Lee 1983). Here, waves are amplified upstream to compensate for those that are being swept into the shock. In fact, for simplicity, Lee assumed that $\mu = 1$ so that $k \approx B/P$, i.e. each wave vector couples to its own single particle rigidity. When we allow $k \approx B/P\mu$, the waves can couple particles of different rigidity, an extremely important factor for many phenomena we observe.

Reames (1990) observed that 3 – 6 MeV proton intensities early in large gradual events never seemed to exceed a plateau value of ~100 – 200 (cm² sr s MeV)$^{-1}$, subsequently called the "streaming limit," although intensities could rise much higher as the shock approached (see Fig. 5.3). Ng and Reames (1994) began by comparing transport with and without wave growth. They found that wave growth throttles the flow of particles, trapping them near the source, limiting their streaming. Ng, Reames, and Tylka (1999, 2003, 2012) extended these calculations showing how the scattering varied greatly in time and space, affecting H, He, O, and Fe differently. The wave generation modifies the "initial" abundances seen early in SEP events (e.g. Reames, Ng, and Tylka, 2000). Further observations extended the streaming limit to higher energies (Reames and Ng 1998) and showed how the low-energy spectra can be flattened, but only when sufficient intensities of streaming high-energy protons precede them (Reames and Ng 2010). Wave growth and the streaming limit will be considered in detail in Sects. 5.1.2 and 5.1.5.

## 2.10 SEP – CME Correlation

In his article on "the big-flare syndrome," Kahler (1982) pointed out that the fact that big SEP events are usually accompanied by big flares, does *not* mean that flares *cause* SEP events; rather, in larger events, *all* energetic phenomena may be more energetic or intense, including flares, CMEs, and SEPs. Flares were once incorrectly thought to cause CMEs. When there is a large rearrangement of the coronal magnetic field, much of the energy released is actually carried away by the CME (e.g. Emslie et al. 2004). Flares are not required to accompany CMEs or SEP events and are, in fact, a secondary phenomenon (Kahler 1992). When flares do accompany CMEs, the CME can precede the flare. Kahler (1992) asks "how did we form such a fundamentally incorrect view?" Probably, correlations of the other phenomena with familiar highly-visible flares were taken much too seriously.

While correlations do not necessarily imply a causal relationship, they are a starting point, and there is a steep dependence of peak particle intensity in large gradual SEP events on CME speed as shown in Fig. 2.12 (Kahler 2001). Two samples of events are shown in the figure i) SEPs measured on *Wind* and CMEs by SOHO/LASCO, both near Earth, and ii) SEPs measured on *Helios*, off the solar limbs, while the *Naval Research Laboratory's Solwind* coronagraph measured CMEs, from near Earth. The latter was an effort to correct for the projection effect in the direction of CME propagation. Of course the "peak intensity" is, in reality, a strong function of longitude, as expected from Fig. 2.2 (see also Fig. 5.16), as is the speed of the shock driven by the CME; these factors contribute to the spread of the measurement which, as we will see, may be reduced by using the measurements of multiple spacecraft in a single SEP event (see Fig. 3.4).

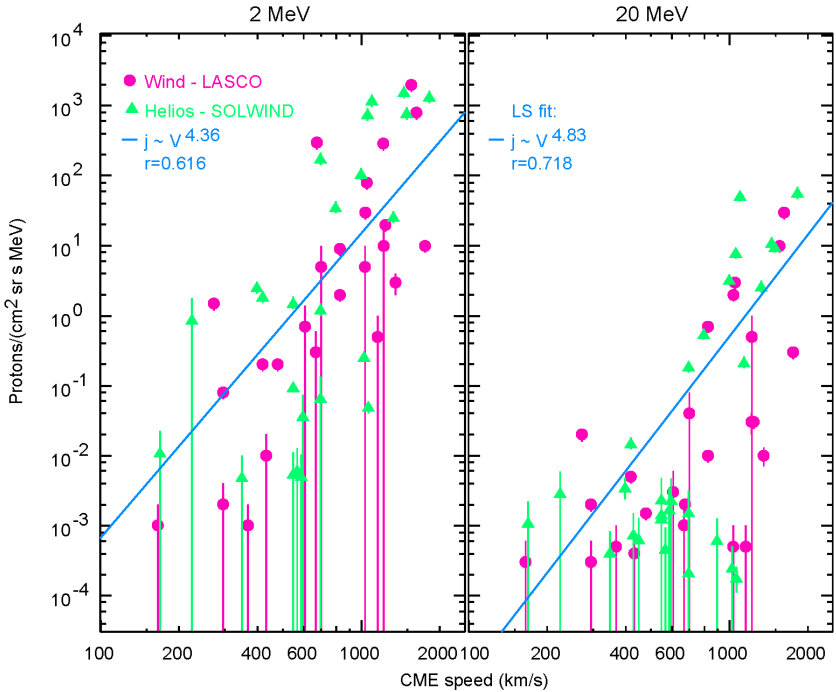

**Fig. 2.12.** Peak intensity is shown vs. CME speed for 2 MeV (*left*) and 20 MeV (*right*) protons for two event samples (see text). Power-law least-squares fits and correlation coefficients (*r*) are shown (see Kahler 2001).

Recently, Kouloumvakos et al. (2019) have greatly improved the CME-SEP relationship by modeling the full 3D geometry of shock waves using the three coronagraph images from SOHO/LASCO and STEREO A and B. The peak proton intensities in three energy intervals from 20–100 MeV, on the three spacecraft, were then correlated with shock properties at the base of their magnetic flux tubes. The best correlation was 40–60 MeV protons with the Alfvén Mach number, which exceeded 75%. Here a single 3D model of a shock predicted three SEP observations, each separated spatially by approximately 120° for a total of 84 observations.

The apparent dependence on CME speed in Fig. 2.12 is certainly quite steep, although there is no physical reason that the relationship should be a power law. Fast CMEs are surely required to produce significant SEP events as originally suggested by Kahler et al. (1984). However, this type of correlation is only a basis for further study, and must be tested and improved as we will see in Sect. 3.2. What variables, other than CME speed, contribute to SEP intensities? Much of the remaining spread must be due to differences in particle transport conditions that spread the particles in space and produce time variations and delays in reaching peak intensities.

## 2.11 SEPs Actually Cause Flares, Not the Reverse

A recent study of the global energy distribution in flares and CMEs (Aschwanden et al. 2019) found that, of the magnetic energy released in reconnection in loops, $51\pm17\%$ led to acceleration of electrons and $17\pm17\%$ led to acceleration of ions. Despite the unusual error, measurements of γ-rays and neutrons (Sect. 4.9) suggest that e/p ratios in flares are not unlike those in impulsive SEP events, and the essential point here is that ~half of the magnetic energy released is directly carried away by confined SEPs. These SEPs are scattered into the denser footpoints of the loops where they produced heating and evaporation of the plasma; this secondary thermal energy is dissipated as white light, UV, and soft X-ray emission, i.e. a hot, bright flare. Flares occur because the reconnection occurs on closed magnetic loops which the SEPs are unable to escape; when the reconnection involves open field lines, jets are produced and impulsive SEPs are seen in space.

We tend to think of SEPs as particles measurable in space, but the energetic particles accelerated in closed magnetic-reconnection sites certainly deserve the name SEPs, and are the intermediary between the magnetic islands of reconnection and the sudden burst of heat and light we call a flare.

## References

Aschwanden, M.J., Caspi, A., Cohen, C.M.S., Holman, G., Jing, J., Kretzschmar, M., Kontar, E.P., McTiernan, J.M., Mewaldt, R.A., O'Flannagain, A., Richardson, I.G. Ryan, D., Warren, H.P.. Xu, Y., Global energetics of solar flares and coronal mass ejections, J. Phys. Conf. Ser. **1332** 1 (2019) doi: 10.1088/1742-6596/1332/1/012002

Bertsch, D.L., Fichtel, C.E., Reames, D.V., Relative abundance of iron-group nuclei in solar cosmic rays, Astrophys. J. Lett. **157**, L53 (1969) doi: 10.1086/180383

Bougeret, J.-L., Kaiser, M.L., Kellogg, P.J., et al., WAVES: The radio and plasma wave investigation on the *Wind* spacecraft, Space Sci Rev, **71**, 5 (1995) doi: 0.1007/BF00751331

Breneman, H.H., Stone, E.C., Solar coronal and photospheric abundances from solar energetic particle measurements, Astrophys. J. Lett. **299**, L57 (1985) doi: 10.1086/184580

Bučík, R., ³He-rich solar energetic particles: solar sources, Space Sci. Rev. **216** 24 (2020) doi: 10.1007/s11214-020-00650-5

Bučík, R., Innes, D.E., Chen, N.H., Mason, G.M., Gómez-Herrero, R., Wiedenbeck, M.E., Long-lived energetic particle source regions on the Sun, J. Phys. Conf. Ser. **642**, 012002 (2015) doi: 10.1088/1742-6596/642/1/012002

Bučík,,R,. Innes, D.E., Mall, U., Korth, A., Mason, G.M., Gómez-Herrero, R., Multi-spacecraft observations of recurrent ³He-rich solar energetic particles, Astrophys. J. **786**, 71 (2014) doi: 10.1088/0004-637X/786/1/71

Bučík, R., Innes, D.E., Mason, G.M., Wiedenbeck, M.E., Gómez-Herrero, R., Nitta, N, ³He-rich solar energetic particles in helical jets on the sun, Astrophys. J. **852** 76 (2018) doi: 10.3847/1538-4357/aa9d8f

Chen N.H., Bučík R., Innes D.E., Mason G.M., Case studies of multi-day ³He-rich solar energetic particle periods,  Astron. Astrophys. **580**, 16 (2015) doi: 10.1051/0004-6361/201525618

Cane, H.V., Kahler, S.W., Sheeley Jr., N.R., Interplanetary shocks preceded by solar filament eruptions, J. Geophys. Res. **91**, 13321 (1986) doi: 10.1029/JA091iA12p13321

Cane, H.V., Reames, D.V., von Rosenvinge, T.T., The role of interplanetary shocks in the longitude distribution of solar energetic particles, J. Geophys. Res. **93**, 9555 (1988) doi: 10.1029/JA093iA09p09555

Carrington, R.C. Description of a singular appearance seen in the Sun on 1 September, 1859, Mon. Not. Roy. Astron. Soc. **20**, 13 (1860) doi: 10.1093/mnras/20.1.13

Cliver, E.W., The unusual relativistic solar proton events of 1979 august 21 and 1981 may 10, Astrophys. J. **639**, 1206 (2006) doi: 10.1086/499765

Cliver, E.W., Kahler, S.W., and Reames, D.V., Coronal shocks and solar energetic proton events, Astrophys. J. **605**, 902 (2004) doi: 10.1086/382651

Cliver, E.W., Kahler, S.W., Shea, M.A., & Smart, D.F., Injection onsets of 2 GeV protons, 1 MeV electrons, and 100 keV electrons in solar cosmic ray flares, Astrophys. J, **260**, 362 (1982) doi: 10.1086/160261

Cook, W.R., Stone, E.C., Vogt, R.E., Elemental composition of solar energetic particles, Astrophys. J. **279**, 827 (1984) doi: 10.1086/161953

Decker, R.B., Formation of shock-spike events at quasi-perpendicular shocks, J. Geophys. Res., **88**, 9959 (1983) doi: 10.1029/JA088iA12p09959

Desai, M.I., Mason, G.M., Dwyer, J.R., Mazur, J.E., Smith, C.W., Skoug, R.M., Acceleration of ³He Nuclei at Interplanetary Shocks", Astrophys. J. Lett., **553**, L89 (2001) doi: 10.1086/320503

Desai, M.I., Mason, G.M., Dwyer, J.R., Mazur, J.E., Gold, R.E., Krimigis, S.M., Smith, C.W., Skoug, R. M., Evidence for a Suprathermal Seed Population of Heavy Ions Accelerated by Interplanetary Shocks near 1 AU, Astrophys. J., **588**, 1149 (2003) doi: 10.1086/374310

DiFabio, R., Guo, Z., Möbius, E., Klecker, B., Kucharek, H., Mason, G.M., Popecki, M., Energy-dependent charge states and their connection with ion abundances in impulsive solar energetic particle events, Astrophys. J. **687**, 623.(2008) doi: 10.1086/591833

Drake, J.F., Cassak, P.A., Shay, M.A., Swisdak, M., Quataert, E.,A magnetic reconnection mechanism for ion acceleration and abundance enhancements in impulsive flares, Astrophys. J. Lett. **700**, L16 (2009) doi: 10.1088/0004-637X/700/1/L16

Ellison, D., Ramaty, R., Shock acceleration of electrons and ions in solar flares, Astrophys. J. **298**, 400 (1985) doi: 10.1086/163623

Emslie, A.G., et al., Energy partition in two solar flare/CME events, J. Geophys. Res. **109**, A10104 (2004) doi: 10.1029/2004JA010571

Fichtel, C.E., Guss, D. E., Heavy nuclei in solar cosmic rays, Phys. Rev. Lett. **6**, 495 (1961) doi: 10.1103/PhysRevLett.6.495

Fisk, L.A., ³He-rich flares - a possible explanation, Astrophys. J. **224**, 1048 (1978) doi: 10.1086/156456

Forbush, S.E., Three unusual cosmic ray increases possibly due to charged particles from the Sun, Phys. Rev. **70**, 771 (1946) doi: 10.1103/PhysRev.70.771

Ganse, U., Kilian, P., Vainio, R., Spanier, F., Emission of type II radio bursts - single-beam versus two-beam scenario, Solar Physics **280**, 551 (2012) doi: 10.1007/s11207-012-0077-7

Giacalone, J., Particle acceleration at shocks moving through an irregular magnetic field, Astrophys. J. **624**, 765 (2005) doi: 10.1086/429265

Giaclaone, J., Jokipii, J. R., The longitudinal transport of energetic ions from impulsive solar flares in interplanetary space, Astrophys, J. Lett. **751**, L33 (2012) doi: 10.1088/2041-8205/751/2/L33

Gloeckler, G., Hovestadt, D., Vollmer, O., Fan, C.Y., Unusual emission of iron nuclei from the sun, Astrophys, J. Lett. **200**, L45 (1975) doi: 10.1086/181893

Gopalswamy, N., Akiyama, S., Yashiro, S., Major solar flares without coronal mass ejections, in Proc. IAU Symp. **257**, 283, N. Gopalswamy, & D. F. Webb, eds (2009) doi: 10.1017/S174392130902941X

Gopalswamy, N., Mäkelä, P., Akiyama, S., Yashiro, S., Xie, H., Thakur, N., Kahler, S.W., Large solar energetic particle events associated with filament eruptions outside of active regions, Astrophys. J. **806**, 8 (2015) doi: 10.1088/0004-637X/806/1/8

Gopalswamy, N., Xie, H., Yashiro, S., Akiyama, S., Mäkelä, P., Usoskin, I.G., Properties of Ground level enhancement events and the associated solar eruptions during solar cycle 23, Space Sci. Rev. **171**, 23 (2012) doi: 10.1007/s11214-012-9890-4

Gopalswamy, N., Yashiro, S., Michalek, G., Kaiser, M.L., Howard, R.A., Reames, D.V., Leske, R., von Rosenvinge, T., Interacting coronal mass ejections and solar energetic particles, ,Astrophys. J. (Letters) **572**, L103 (2002) doi: 10.1086/341601

Gosling, J.T., The solar flare myth, J. Geophys. Res. **98**, 18937 (1993) doi: 10.1029/93JA01896

Ho, G.C., Roelof, E.C., Mason, G.M., The upper limit on ³He fluence in solar energetic particle events, Astrophys. J. Lett. **621**, L141 (2005) doi: 10.1086/429251

Hsieh, K.C., Simpson, J.A., The relative abundances and energy spectra of ³He and ⁴He from solar flares, Astrophys., J. Lett. **162,** L191 (1970) doi: 10.1086/180652

Hudson, H.S., Solar flares: No "myth", Eos Trans. AGU **76**(41), 405 (1995) doi: 10.1029/95EO00253

Ibragimov, I. A., Kocharov, G. E., Possible mechanism for enrichment of solar cosmic rays by helium-three and heavy nuclei, 15th Int. Conf. on Cosmic Rays (Plovdiv: Bulgarian Academy of Sciences), **11**, 340 (1977)

Jokipii, J.R., Parker, E.N., Stochastic aspects of magnetic lines of force with application to cosmic ray propagation, Astrophys J. **155**, 777 (1969) doi: 10.1086/149909

Jones, F.C., Ellison, D.E., The plasma physics of shock acceleration, Space Sci. Rev. **58**, 259 (1991) doi: 10.1007/BF01206003

Kahler, S.W., The role of the big flare syndrome in correlations of solar energetic proton fluxes and microwave burst parameters, J. Geophys. Res. **87**, 3439 (1982) doi: 10.1029/JA087iA05p03439

Kahler, S.W., Solar flares and coronal mass ejections, Annu. Rev. Astron, Astrophys. **30**, 113 (1992) doi: 10.1146/annurev.aa.30.090192.000553

Kahler, S.W., The correlation between solar energetic particle peak intensities and speeds of coronal mass ejections: Effects of ambient particle intensities and energy spectra, J. Geophys. Res. **106**, 20947 (2001) doi: 10.1029/2000JA002231

Kahler, S.W., Reames, D.V., Sheeley, Jr., N.R., Coronal mass ejections associated with impulsive solar energetic particle events, Astrophys. J., **562**, 558 (2001) doi: 10.1086/323847

Kahler, S.W., Sheeley Jr., N.R., Howard, R.A., Koomen, M.J., Michels, D.J., McGuire R.E., von Rosenvinge, T.T., Reames, D.V., Associations between coronal mass ejections and solar energetic proton events, J. Geophys. Res. **89**, 9683 (1984) doi: 10.1029/JA089iA11p09683

Kahler, S.W., Cliver, E.W., Cane, H.V., McGuire, R.E., Stone, R.G., Sheeley Jr., N.R., Solar filament eruptions and energetic particle events, Astrophys. J. **302** 594 (1986) doi: 10.1086/164009

Kocharov, G. E., Kocharov, L. G., Present state of experimental and theoretical investigations of solar events enriched by helium-3 , in Proc. 10th Leningrad Sympos. on Cosmic Rays (Leningrad: A. F. Yoffe Phys.-Tech. Inst.), p. 37 (1978)

Kocharov, G. E., Kocharov, L. G., ³He-rich solar flares, Space Science Rev. **38**, 89 (1984) doi: 10.1007/BF00180337

Kouloumvakos, A., Rouillard, A., Wu, Y., Vainio, R., Vourlidas,A., Plotnikov, I., Afanasiev, A., Önel, H., Connecting the properties of coronal shock waves with those of solar energetic particles, Astrophys. J., **876** 80 (2019) doi: [10.3847/1538-4357/ab15d7](10.3847/1538-4357/ab15d7)

Laming, J.M., A unified picture of the first ionization potential and inverse first ionization potential effects, Astrophys. J. **614**, 1063 (2004) doi: [10.1086/423780](10.1086/423780)

Laming, J.M., Non-WKB models of the first ionization potential effect: implications for solar coronal heating and the coronal helium and neon abundances, Astrophys. J. **695**, 954 (2009) doi: [10.1088/0004-637X/695/2/954](10.1088/0004-637X/695/2/954)

Lee, M.A., Coupled hydromagnetic wave excitation and ion acceleration at interplanetary traveling shocks, J. Geophys. Res., **88**, 6109. (1983) doi: [10.1029/JA088iA08p06109](10.1029/JA088iA08p06109)

Lee, M.A., Coupled hydromagnetic wave excitation and ion acceleration at an evolving coronal/interplanetary shock, Astrophys. J. Suppl., **158**, 38 (2005) doi: [10.1086/428753](10.1086/428753)

Leske, R.A., Cummings, J.R., Mewaldt, R.A., Stone, E.C., von Rosenvinge, T.T., Measurements of the ionic charge states of solar energetic particles using the geomagnetic field, Astrophys J. **452**, L149 (1995) doi: [10.1086/309718](10.1086/309718)

Li, G., Zank, G.P., Verkhoglyadova, O., Mewaldt, R.A., Cohen, C.M.S., Mason, G.M., Desai, M.I., Shock geometry and spectral breaks in large sep events, Astrophys. J. **702**, 998 (2009) doi: [10.1088/0004-637X/702/2/998](10.1088/0004-637X/702/2/998)

Lin, R.P., The emission and propagation of 40 keV solar flare electrons. I: The relationship of 40 keV electron to energetic proton and relativistic electron emission by the sun, Sol. Phys. 12, 266 (1970) doi: [10.1007/BF00227122](10.1007/BF00227122)

Lin, R.P., Non-relativistic solar electrons, Space Sci. Rev. **16**, 189 (1974) doi: [10.1007/BF00240886](10.1007/BF00240886)

Litvinenko, Y. E., On the formation of the helium-3 spectrum in impulsive solar flares, in R. Ramaty, N. Mandzhavidze, X.-M. Hua (eds), High Energy Solar Physics, AIP Conf. Proc. **374** (AIP Press, Woodbury, NY) p. 498. (1996) doi: [10.1063/1.50985](10.1063/1.50985)

Luhn, A., Klecker B., Hovestadt, D., Gloeckler, G., Ipavich, F.M., Scholer, M., Fan, C.Y. Fisk,L.A., Ionic charge states of N, Ne, Mg, Si and S in solar energetic particle events, Adv. Space Res. **4**, 161 (1984) doi: [10.1016/0273-1177(84)90307-7](10.1016/0273-1177(84)90307-7)

Luhn, A., Klecker, B., Hovestadt, D., Möbius, E., The mean ionic charge of silicon in He-3-rich solar flares, Astrophys. J. **317**, 951 (1987) doi [10.1086/165343](10.1086/165343)

Mason, G.M., Gloeckler, G., Hovestadt, D., Temporal variations of nucleonic abundances in solar flare energetic particle events. II - Evidence for large-scale shock acceleration, Astrophys. J. **280**, 902 (1984) doi: [10.1086/162066](10.1086/162066)

Mason, G.M., Mazur, J.E., Dwyer, J.R., $^3$He enhancements in large solar energetic particle events, Astrophys. J. Lett. 525, L133 (1999) doi: [10.1086/312349](10.1086/312349)

Mason, G.M., Ng, C.K., Klecker, B., Green, G., Impulsive acceleration and scatter-free transport of about 1 MeV per nucleon ions in $^3$He-rich solar particle events, Astrophys. J. **339**, 529 (1989) doi: [10.1086/167315](10.1086/167315)

Mason, G.M., Reames, D.V., Klecker, B., Hovestadt, D., von Rosenvinge, T.T., The heavy-ion compositional signature in He-3-rich solar particle events, Astrophys. J. **303**, 849 (1986) doi: [10.1086/164133](10.1086/164133)

Melrose, D.B., *Plasma Astrophysics*, Vol. 1, (New York: Gordon and Breach) (1980)

Meyer, J.P., The baseline composition of solar energetic particles, Astrophys. J. Suppl. **57**, 151 (1985) doi: [10.1086/191000](10.1086/191000)

Meyer, P., Parker, E.N., Simpson, J.A., Solar cosmic rays of February, 1956 and their propagation through interplanetary space, Phys. Rev. **104**, 768 (1956) doi: [10.1103/PhysRev.104.768](10.1103/PhysRev.104.768)

Miller, J.A., Much ado about nothing, Eos Trans. AGU **76** (41), 401 (1995) doi: [10.1029/95EO00246](10.1029/95EO00246)

Miller, J., Viñas, A., Reames, D.V., Selective $^3$He and Fe acceleration in impulsive solar flares, 23$^{rd}$ Intl. Cosmic-Ray Conf. (Calgary) **3** 13 (1993a)

Miller, J., Viñas, A., Reames, D.V., Heavy ion acceleration and abundance enhancements in impulsive solar flares, 23rd Intl. Cosmic-Ray Conf. (Calgary) **3** 17 (1993b)

Newkirk Jr., G, Wenzel, D.G., Rigidity-independent propagation of cosmic rays in the solar corona, J. Geophys. Res. **83,** 2009 (1978) doi: 10.1029/JA083iA05p02009

Ng, C.K., Reames, D.V., Focused interplanetary transport of approximately 1 MeV solar energetic protons through self-generated Alfven waves, Astrophys. J. **424**, 1032 (1994) doi: 10.1086/173954

Ng, C.K., Reames, D.V., Shock acceleration of solar energetic protons: the first 10 minutes, Astrophys. J. Lett. **686**, L123 (2008) doi: 10.1086/592996

Ng, C.K., Reames, D.V., Tylka, A.J., Modeling shock-accelerated solar energetic particles coupled to interplanetary Alfvén waves, Astrophys. J. **591**, 461 (2003) doi: 10.1086/375293

Ng, C.K., Reames, D.V., Tylka, A.J., Solar energetic particles: shock acceleration and transport through self-amplified waves, AIP Conf. Proc. **1436**, 212 (2012) doi: 10.1063/1.4723610

Nitta, N.V., Reames, D.V., DeRosa, M.L., Yashiro, S., Gopalswamy, N., Solar sources of impulsive solar energetic particle events and their magnetic field connection to the earth, Astrophys. J. **650**, 438 (2006) doi: 10.1086/507442

Reames, D.V., Acceleration of energetic particles by shock waves from large solar flares, Astrophys. J. Lett. **358**, L63 (1990) doi: 10.1086/185780

Reames, D.V., The dark side of the solar flare myth, Eos Trans. AGU **76** (41), 401 (1995) doi: 10.1029/95EO00254

Reames, D.V., Particle acceleration at the sun and in the heliosphere, Space Sci. Rev. **90**, 413 (1999) doi: 10.1023/A:1005105831781

Reames, D.V., Abundances of trans-iron elements in solar energetic particle events, Astrophys. J. Lett. **540**, L111 (2000) doi: 10.1086/312886

Reames, D.V., Temperature of the source plasma in gradual solar energetic particle events, Sol. Phys., **291** 911 (2016) doi: 10.1007/s11207-016-0854-9, arXiv: 1509.08948

Reames, D.V., Cliver, E.W., Kahler, S.W., Abundance enhancements in impulsive solar energetic-particle events with associated coronal mass ejections, Sol. Phys. **289**, 3817, (2014a) doi: 10.1007/s11207-014-0547-1

Reames, D.V., Cliver, E.W., Kahler, S.W., Variations in abundance enhancements in impulsive solar energetic-particle events and related CMEs and flares, Sol. Phys. **289**, 4675 (2014b) doi: 10.1007/s11207-014-0589-4

Reames, D.V. Ng, C.K., Streaming-limited intensities of solar energetic particles, Astrophys. J. **504**, 1002 (1998) doi: 10.1086/306124

Reames, D.V., Ng, C.K., Streaming-limited intensities of solar energetic particles on the intensity plateau, Astrophys. J. **722**, 1286 (2010) doi: 10.1088/0004-637X/723/2/1286

Reames, D.V., Stone, R.G., The identification of solar He-3-rich events and the study of particle acceleration at the sun, Astrophys. J., **308**, 902 (1986) doi: 10.1086/164560

Reames, D.V., Barbier, L.M., Ng, C.K., The spatial distribution of particles accelerated by coronal mass ejection-driven shocks, Astrophys. J. **466** 473 (1996) doi: 10.1086/177525

Reames, D.V., Meyer, J.P., von Rosenvinge, T.T., Energetic-particle abundances in impulsive solar flare events, Astrophys. J. Suppl. **90**, 649 (1994) doi: 10.1086/191887

Reames, D.V., Ng, C.K., Tylka, A.J., Initial time dependence of abundances in solar particle events, Astrophys. J. Lett. **531**, L83 (2000) doi: 10.1086/312517

Reames, D.V., von Rosenvinge, T.T., Lin, R.P., Solar He-3-rich events and nonrelativistic electron events - A new association, Astrophys. J. **292**, 716 (1985) doi: 10.1086/163203

Reid, G.C., A diffusive model for the initial phase of a solar proton event, J. Geophys. Res. **69**, 2659 (1964) doi: 10.1029/JZ069i013p02659

Reinhard, R., Wibberenz, G., Propagation of flare protons in the solar atmosphere, Sol. Phys. **36**, 473 (1974) doi: 10.1007/BF00151216

Richardson, I.G., Reames, D.V., Wenzel, K.-P., Rodriguez-Pacheco, J., Quiet-time properties of low-energy (less than 10 MeV per nucleon) interplanetary ions during solar maximum and solar minimum, Astrophys. J. Lett. **363**, L9 (1990) doi: 10.1086/185853

Riyopoulos, S., Subthreshold stochastic diffusion with application to selective acceleration of ³He in solar flares, Astrophys. J. **381** 578 (1991) doi: [10.1086/170682](10.1086/170682)

Roth, I., Temerin, M., Enrichment of ³He and Heavy Ions in Impulsive Solar Flares, Astrophys. J. **477**, 940 (1997) doi: [10.1086/303731](10.1086/303731)

Sanahuja, B., Domingo, V., Wenzel, K.-P., Joselyn, J.A., Keppler, E., A large proton event associated with solar filament activity, Sol. Phys. **84**, 321 (1983) doi: [10.1007/BF00157465](10.1007/BF00157465)

Sandroos, A., Vainio, R., Simulation results for heavy ion spectral variability in large gradual solar energetic particle events, Astrophys. J. **662**, L127 (2007) doi: [10.1086/519378](10.1086/519378)

Serlemitsos, A.T., Balasubrahmanyan, V.K., Solar particle events with anomalously large relative abundance of ³He, Astrophys. J. **198**, 195, (1975) doi: [10.1086/153592](10.1086/153592)

Steinacker, J, Meyer, J.-P., Steinacker, A., Reames, D.V., The helium valley: comparison of impulsive solar flare ion abundances and gyroresonant acceleration with oblique turbulence in a hot multi-ion plasma, Astrophys. J. **476** 403 (1997) doi: [10.1086/303589](10.1086/303589)

Stix, T.H., *Waves in Plasmas* (New York: AIP) (1992)

Temerin, M., Roth, I., The production of ³He and heavy ion enrichment in ³He-rich flares by electromagnetic hydrogen cyclotron waves, Astrophys. J. Lett. **391**, L105 (1992) doi: [10.1086/186408](10.1086/186408)

Thejappa, G., MacDowall, R.J., Bergamo, M., Papadopoulos, K., Evidence for the oscillating two stream instability and spatial collapse of Langmuir waves in a solar type III radio burst, Astrophys. J. Lett. **747**, L1 (2012) doi: [10.1088/2041-8205/747/1/L1](10.1088/2041-8205/747/1/L1)

Tylka, A.J., Boberg, P.R., Adams Jr., J.H., Beahm, L.P., Dietrich, W.F., Kleis, T., The mean ionic charge state of solar energetic Fe ions above 200 MeV per nucleon, Astrophys. J. **444**, L109 (1995) doi: [10.1086/187872](10.1086/187872)

Tylka, A.J., Cohen, C.M.S., Dietrich, W.F., Lee, M.A., Maclennan, C.G., Mewaldt, R.A., Ng, C.K., Reames, D.V., Shock geometry, seed populations, and the origin of variable elemental composition at high energies in large gradual solar particle events, Astrophys. J. **625**, 474 (2005) doi: [10.1086/429384](10.1086/429384)

Tylka, A.J., Cohen, C.M.S., Dietrich, W.F., Maclennan, C.G., McGuire, R.E., Ng, C.K., Reames, D.V., Evidence for remnant flare suprathermals in the source population of solar energetic particles in the 2000 bastille day event, Astrophys. J. Lett. **558**, L59 (2001) doi: [10.1086/323344](10.1086/323344)

Tylka, A.J., Dietrich, W.F., A new and comprehensive analysis of proton spectra in ground-level enhanced (GLE) solar particle events, in *Proc. 31ˢᵗ Int. Cos. Ray Conf*, *Lódz* (2009), [http://icrc2009.uni.lodz.pl/proc/pdf/icrc0273.pdf](http://icrc2009.uni.lodz.pl/proc/pdf/icrc0273.pdf)

Tylka, A.J., Lee, M.A., Spectral and compositional characteristics of gradual and impulsive solar energetic particle events, Astrophys. J. **646**, 1319 (2006) doi: [10.1086/505106](10.1086/505106)

Varvoglis, H., Papadopoulis, K., Selective nonresonant acceleration of He-3(2+) and heavy ions by H(+) cyclotron waves ,Astrophys. J. Lett. **270**, L95 (1983) doi: [10.1086/184077](10.1086/184077)

Verkhoglyadova, O., Zank, G.P., Li, G., A theoretical perspective on particle acceleration by interplanetary shocks and the Solar Energetic Particle problem, Phys. Reports **557**, 1 (2015) doi: [10.1016/j.physrep.2014.10.004](10.1016/j.physrep.2014.10.004)

Wang, Y.-M., Pick, M., Mason, G.M., Coronal holes, jets, and the origin of ³He-rich particle events, Astrophys. J. **639**, 495 (2006) doi: [10.1086/499355](10.1086/499355)

Weatherall, J., Turbulent heating in solar cosmic-ray theory, Astrophys. J. 281, 468 (1984) doi: [10.1086/162119](10.1086/162119)

Webber, W.R., Solar and galactic cosmic ray abundances - A comparison and some comments. *Proc. 14ᵗʰ Int. Cos. Ray Conf, Munich*, **5**, 1597 (1975)

Webber, W.R., Roelof, E.C., McDonald, F.B., Teegarden, B.J., Trainor, J., Pioneer 10 measurements of the charge and energy spectrum of solar cosmic rays during 1972 August, Astrophys. J. **199**, 482 (1975) doi: [10.1086/153714](10.1086/153714)

Wiedenbeck, M.E., Mason, G.M., Cohen, C.M.S., Nitta, N.V., Gómez-Herrero, R., Haggerty, D.K., Observations of solar energetic particles from ³He-rich events over a wide range of heliographic longitude,  Astrophys. J. **762** 54 (2013) doi: [10.1088/0004-637X/762/1/54](10.1088/0004-637X/762/1/54)

Wild, J.P., Smerd, S.F., Weiss, A.A., Solar Bursts, Annu. Rev. Astron. Astrophys., **1**, 291 (1963) doi: 10.1146/annurev.aa.01.090163.001451

Winglee, R.M., Heating and acceleration of heavy ions during solar flares, Astrophys. J. **343**, 511 (1989) doi: 10.1086/167726

Zank, G.P., Li, G., Verkhoglyadova, O., Particle Acceleration at Interplanetary Shocks, Space Sci. Rev. **130**, 255 (2007) doi: 10.1007/s11214-007-9214-2

Zank, G.P., Rice, W.K.M., Wu, C.C., Particle acceleration and coronal mass ejection driven shocks: A theoretical model, J. Geophys. Res., **105**, 25079 (2000) doi: 10.1029/1999-JA000455

Zhao, L., Zhang, M., Rassoul, H.K., Double power laws in the event-integrated solar energetic particle spectrum. Astrophys. J. **821**, 62 (2016) doi: 10.3847/0004-637X/821/1/62

Zirin, H., Solar storminess, Sky and Telescope, Nov., 9 (1994)

# Chapter 3. Distinguishing the Sources

**Abstract**   Our discussion of history has covered many of the observations that have led to the ideas of acceleration by shock waves or by magnetic reconnection in gradual and impulsive solar energetic particle (SEP) events, respectively. We now present other compelling observations, including onset timing, SEP-shock correlations, injection time profiles, high-energy spectral knees, e/p ratios, and intensity dropouts caused by a compact source, that have helped clarify these acceleration mechanisms and sources. However, some of the newest evidence now comes from source-plasma temperatures. In this and the next two chapters, we will find that impulsive events come from solar active regions at $\approx 3$ MK, controlling ionization states $Q$, hence $A/Q$, and, in most gradual events, shocks accelerate ambient coronal material from $\leq 1.6$ MK. When SEPs are trapped on closed loops they supply the energy for flares. In addition to helping to define their own origin, SEPs also probe the structure of the interplanetary magnetic field.

The history in Chap. 2 suggested how the flow of observations and ideas eventually led to credible evidence of two sites of SEP acceleration and the related physical mechanisms. While some observations have been described, some of the clearest evidence of origin has not yet been presented. In this chapter we continue the story of particle origin, showing where and when SEPs are accelerated, and measurements that allow us to compare impulsive and gradual events. There are many different lines of evidence that fit together to determine the most probable origins, and that evidence continues to grow.

## 3.1 SEP Onset Times

Even in relatively intense SEP events, it is likely that the earliest detectable particles at each energy will be those that were originally focused in the diverging magnetic field in the inner heliosphere and have scattered least, simply traversing along the magnetic field line from the source with an average pitch-angle cosine, $<\mu> \approx 1$, so that they arrive first. An example of the observed arrival times of particles of different energies is shown in Fig. 3.1. The rise of the intensities is clear and sharp and intensities rise by two or three orders of magnitude. If 1% or more of the ions in each energy interval have traveled with $<\mu> \approx 1$, we will be able to determine the scatter-free onset time with reasonable accuracy. The accuracy of this scatter-free approximation has been well studied and will be discussed below. As discussed in Sect. 2.3.4, impulsive events are all almost entirely scatter-free. The particle transit time $t = L/v$ where $L$ is the path length along the field line and $v$ is the particle velocity. By fitting the measurements we can determine both the path length and the time that the particles left the Sun, the so-called solar particle release (SPR) time.

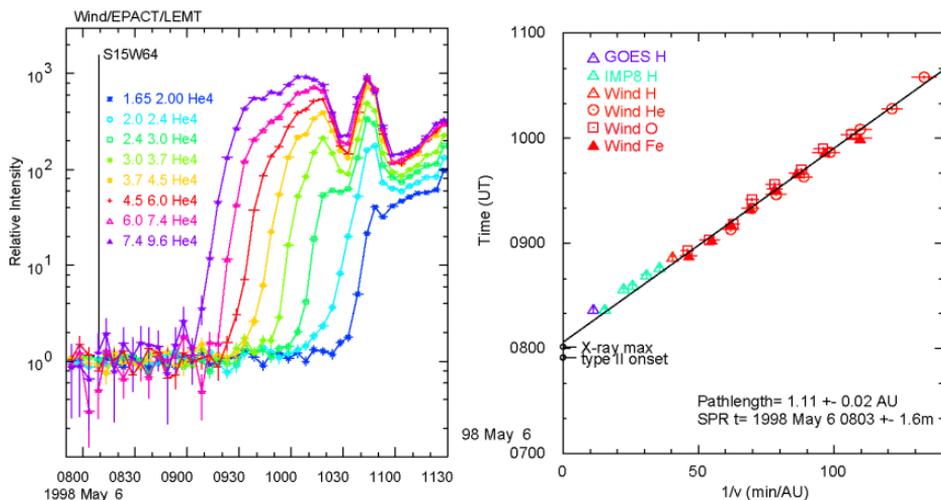

**Fig. 3.1** The *left panel* shows the arrival of $^4$He ions of the indicated MeV amu$^{-1}$ intervals at the *Wind* spacecraft near Earth. The *right panel* shows the onset time of these and other intervals vs. $v^{-1}$. For the fitted line, the slope is the pathlength and the intercept is the solar particle release (SPR) time at the Sun for this large gradual GLE event (Reames 2009a, © AAS).

Note that the SPR time is the release time *at the Sun*; to compare with photon observation times at Earth one should add 8.3 min to the SPR time. The path length of $1.11 \pm 0.02$ AU allows for some curvature of the Parker spiral, typically 1.1 – 1.2 AU. For large gradual events, including the ground level events (GLEs), generally the SPR times occur quite late in the event. Timing in impulsive (left) and gradual (right) events is compared in Fig. 3.2.

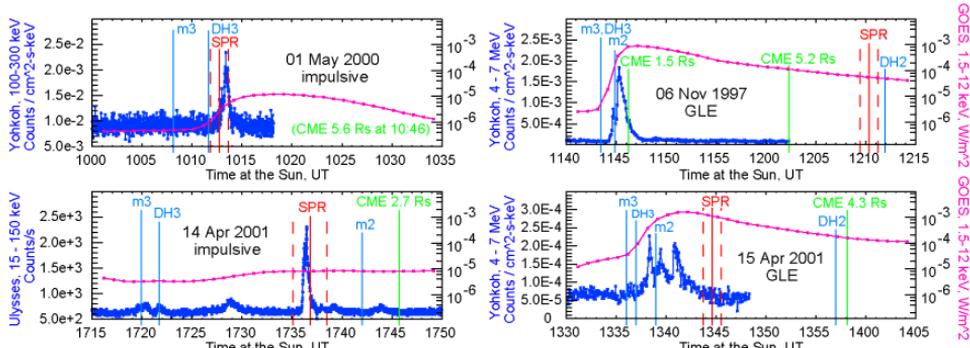

**Fig. 3.2** A comparison is shown of timing in two impulsive (*left*) and two gradual (*right*) SEP events. Solar particle release (SPR) times of the particles (*red* with *dashed* errors) are compared with hard X-ray (*dark blue, left*), γ-ray (*dark blue, right*) and GOES soft X-ray (*violet*) time profiles. Onset times of metric (m) and decametric-hectometric (DH; 1–14 MHz) type II and III radio bursts (*light blue*) and CME locations (*green*) are shown (adapted from Tylka et al. 2003).

For the impulsive SEP events in Fig. 3.2, the SPR times fall rather precisely on the hard-X-ray peak times (there are no measurable γ-rays in these events). For the GLEs, the SPR times often fall well after the γ-rays are over (by up to 30 min), but always after the metric type II onset indicates the formation of a shock wave. Rouillard et al. (2016) relate the SPR delay to the time required for the shock to become supercritical, i.e. Mach >3.

It is interesting to plot the height of the CME leading edge at SPR time as a function of longitude of the observer relative to that of the CME source as shown in the right panel of Fig. 3.3 for the GLEs. For a multi-spacecraft study of a single event see Reames and Lal (2010).

**Fig. 3.3** The *right panel* shows the height of the CME at SPR time vs. longitude for numbered GLEs. The cartoon on the *left* shows the CME and SPR location widening on the flanks (Reames 2009a, b © AAS). The height distribution is fit to a symmetric parabola for comparison; actually a height of 2 – 3 $R_S$ is fairly constant over ~70°. This could be the width of the source shock surface above closed loops that was once incorrectly called the "fast propagation region". SPR may differ on each local SEP field line, while type II onset occurs near the source longitude.

Clearly a correct estimate of the SPR time depends upon the intensities being sufficiently high that a small number of un-scattered ions are detectible. Gopalswamy et al. (2012) have simply assumed a path length of 1.2 AU in order to avoid the velocity-dispersion analysis. However, Rouillard et al. (2012, see Appendix) have calculated that the error in the SPR time from scattering should be less that 1 – 2 min., comparable with errors from the 5 min-averaged data used. If scattering delayed low energies more, the apparent SPR would be too *early*. Note also that the *impulsive* events on the left in Fig. 3.2 show *no* evidence of onset errors. Tan et al. (2013) have found that the SPR times and path lengths of the non-relativistic electrons agree with those of the ions, and Rouillard et al. (2012) have also shown lateral spreading of the shock wave as imaged by the coronagraphs.

High-energy (GeV) protons are often strongly beamed along the interplanetary magnetic-field **B**, so a particular neutron monitor on Earth sees a peak when its asymptotic look direction is aligned with **B**. As **B** varies, neutron monitors often see spiky increases or multiple peaks and valleys of intensity.

Surely there are a few GLEs where the SPR timing alone would permit some kind of (unspecified) acceleration at the time of the associated flare. However, these events may just have faster CMEs or a faster decrease in $V_A$ with radius that would permit earlier ion acceleration by the shock, or earlier arrival of the shock above closed magnetic loops. If the GLEs with late SPR times are clearly shock accelerated, why would we seek a new mechanism for those events with earlier SPR times which have equally strong shocks? Shock acceleration is able to account for SEP acceleration in all gradual events, including GLEs, especially in GLEs. No other mechanism is required, no other seems capable.

## 3.2 Realistic Shock-SEP Timing and Correlations

With recent measurements on the STEREO spacecraft, it has been possible to construct three-dimensional distributions of CMEs and shocks and compare them with SEPs, i.e. to compare the SEPs and the shock along the same single field line. (Rouillard et al. 2011, 2012, 2016). Fig. 3.4 shows aspects of this comparison.

The left-hand simulation in Fig. 3.4 reconstructs the way the CME and shock spread. The actual SPR time depends upon the time an active shock actually strikes the (dashed) field line to an observer. It would be a great improvement on the comparison in Fig. 3.3 if we could see the *local* shock as we can here. Some images of the shock are shown in the upper right panels of Fig. 3.4.

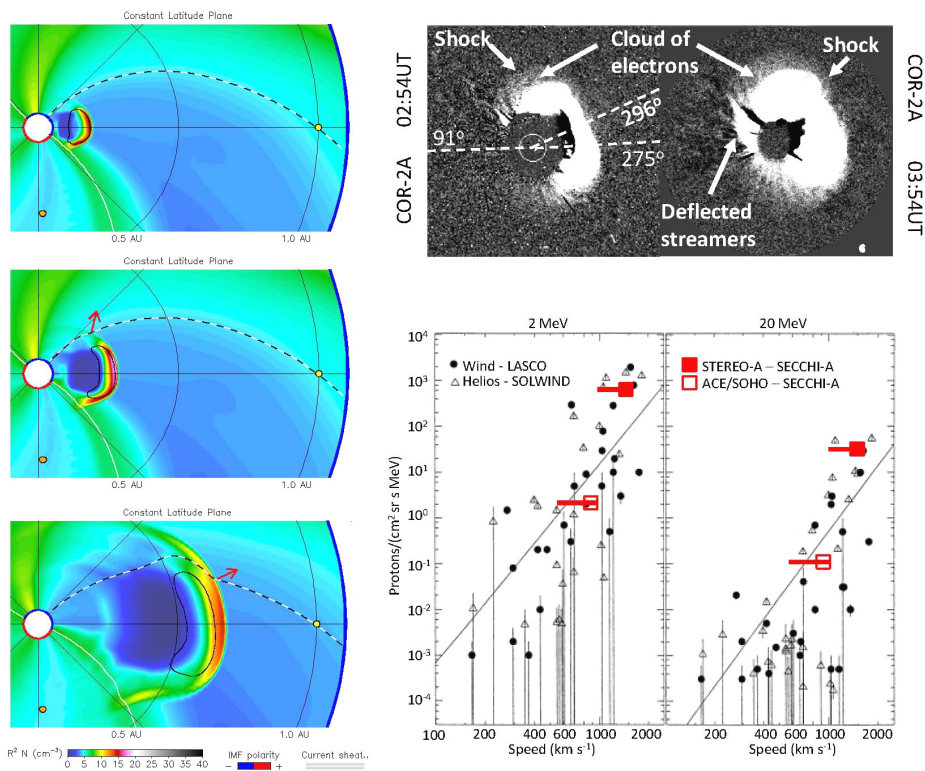

**Fig. 3.4** *Left-hand panels* show a STEREO simulation of the evolution and lateral spread of the CME and shock. *Upper-right panels* show actual images of the shock. *Lower right panels* show possible improvement in the intensity – CME-speed correlation (*red points*) when observed on single field lines at two different longitudes in an event (Rouillard et al. 2011, 2012 © AAS)

The lower-right panels in Fig. 3.4 show correlations of peak proton intensities at 2 and 20 MeV with CME speed from the earlier study by Kahler (2001) shown in Fig. 2.12. However, no single speed exists for any CME or shock, and there is no single peak proton intensity, since both vary strongly with longitude. The red points in the lower right panel of Fig. 3.4 compare intensity and CME speed on single magnetic flux tubes, apparently improving the correlation. Kouloumvakos et al. (2019) have extended the CME-SEP correlations by modeling the 3D geom-etry of shock waves using the three coronagraph images from SOHO/LASCO and STEREO A and B, as noted in Sect. 2.10.

More recently, Gopalswamy et al. (2013) studied the first GLE of Solar Cycle 24, GLE 71 on 17 May 2012, together with 6 other large, well-connected events with fast CMEs. The evolution of two of the CMEs is compared in Fig. 3.5.

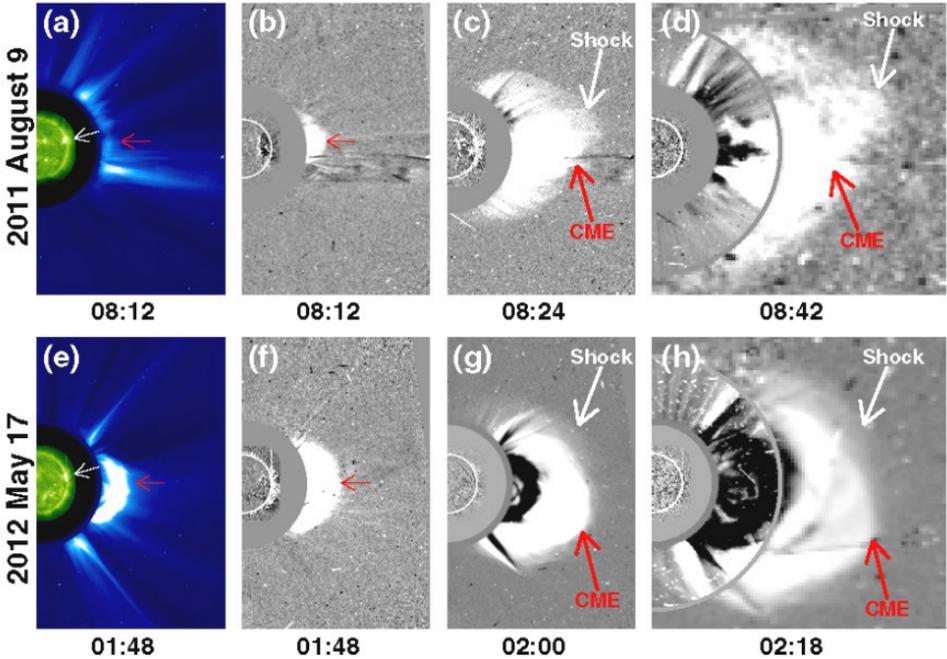

**Fig. 3.5** The time evolution of two CMEs, 9 August 2011 and the GLE on 17 May 2012, are shown from their first appearance on SOHO/LASCO. In panels (**a**) and (**e**), the SDO/AIA solar image at 195 Å shows the solar sources while the remaining difference images show evolution of the CMEs and shocks. *Red arrows* point to the CME nose. The shock remains closer to the CME in (**h**) than in (**d**), indicating a stronger shock (Gopalswamy et al. 2013 © AAS).

In the GLE, the shock formation height (type II radio burst onset) is at 1.38 R$_S$ and the observed CME height at the time of particle release was directly measured as 2.32 R$_S$. This is consistent with the findings from extrapolations of GLEs in Cycle 23. The authors concluded that the event of May 2012 was a GLE simply because it was better connected to Earth than the other large SEP events with similar or even faster CMEs. Note that the SPR time is the SEP onset on the local field-line longitude of the SEP observer, while the type II onset is at a longitude nearest the source; hence, the type II onsets do not show the quasi-parabolic behavior of the SPR height in Fig. 3.2.

Thakur et al. (2016; see also Tylka and Dietrich 2009) compared the >700 MeV proton channel on GOES as an alternate indicator of GLEs. They found two events that differed, one GLE with no increase at >700 MeV and one >700-MeV increase that was not a GLE. They ascribed the difference to the level of the background. They also found that GLEs were generally observed when the shocks form at 1.2 – 1.93 R$_S$ and when solar particle release (SPR) occurs between 2 – 6 $R_S$. Note that the electron acceleration that produces the type II burst could occur while the shock is still propagating within closed magnetic loops, but SPR time must occur where the shock is on open field lines, and at a local longitude.

Cliver, Kahler, and Reames (2004) found a strong (~90%) association of decametric-hectometric (DH; 1–14 MHz) type II radio emission produced at ~3 $R_S$ by SEP events with 20 MeV protons (see Fig. 2.1). The correlation was only 25% for lower-altitude metric type II's without DH, suggesting that shock acceleration is strongest at or above ~3 $R_S$.

## 3.3 Injection Profiles

Relating to SEP increases early in events, Kahler (1994) plotted the intensity of SEPs, not as a function of time, but as a function of the height of the CME, using the height-time plot for the CME, as shown in Fig. 3.6. Not only are the protons injected late, but their intensities continue to rise until $R > 6$ $R_S$, even at 21 GeV. A final peak in the Alfvén speed vs. height occurs at ~4 $R_S$ and $V_A$ has probably declined to about 600 km s$^{-1}$ at 6 $R_S$ (Mann et al. 2003; see also Sect. 1.4). The Alfvén-Mach number of the shock, $V_S/V_A$ remains at ~2 or greater above ~1.2 $R_S$ for these shock waves.

**Fig. 3.6** Injection profiles of high-energy protons are shown as a function of CME height for three GLEs in 1989: August 16, September 29, and October 24. The CME speeds for these events are 1377, 1828, and 1453 km s$^{-1}$, respectively (Kahler 1994, © AAS).

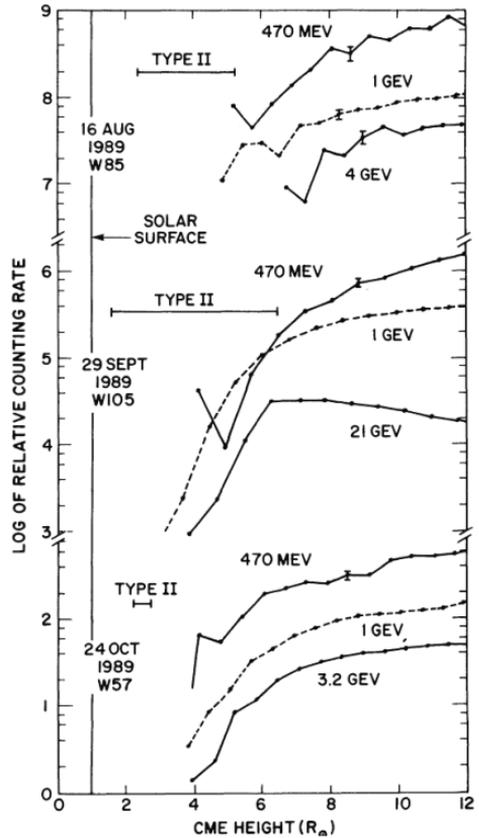

## 3.4 High-Energy Spectra and Spectral Knees

Are GLEs fundamentally different from other gradual SEP events? Is it likely that there is some new source of particles that can only be seen at energies above ~0.5 GeV? Much of the evidence connecting gradual SEPs to shock acceleration, especially element abundances and source-plasma temperatures, comes from energies below 100 MeV. Do the high-energy spectra come from the same source? Some spectra are shown in Fig. 3.7.

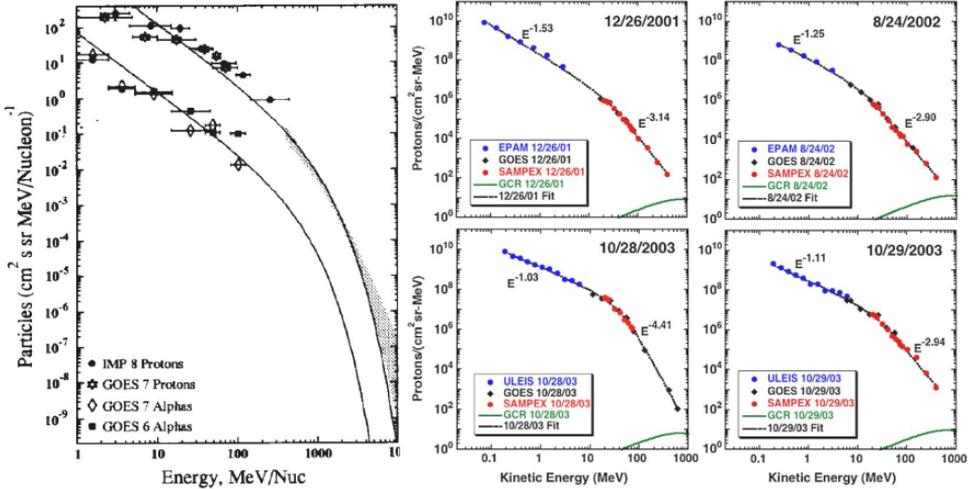

**Fig. 3.7** The *left panel* shows H and He in the large GLE of September 29, 1989 (Lovell, Duldig, and Humble 1998). The *shaded* region is the spectrum deduced from neutron monitors and the spectra are fit to the shock-spectral shape of Ellison and Ramaty (1985). The four *right-hand panels* show GLE fluence spectra that are typical of the 16 GLE spectra assembled by Mewaldt et al. (2012 © Springer) fit to double power-law spectra.

Mewaldt et al. (2012) studied spectra and element abundances of 16 GLEs. They found that the empirical double power-law spectral forms give a better fit than the power-law-times-exponential spectrum of Ellison and Ramaty (1985, see also Lee 2005) that models escape of high-energy particles from the shock. In any case, *none* of the 16 GLEs showed evidence of high-energy spectral *hardening* that might suggest the existence of a new source that could dominate higher energies.

Tylka and Dietrich (2009) have used the geomagnetic cutoff rigidities at neutron-monitor stations to develop integral rigidity spectra, using data from the world-wide neutron monitor network for 53 GLEs. The proton spectra are fit to double power laws in rigidity, decreasing with a power above 1 GV (430 MeV) in the range of 5 – 7 in 70% of the GLEs (see Sect. 6.1). None show hardening.

## 3.5 Intensity Dropouts and Compact Sources

When Mazur et al. (2000) plotted the energy of individual ions as a function of their arrival time, as seen in Fig. 3.8, they found that the pattern of velocity disper-sion that we described in Fig. 3.1 was sharply interrupted for time intervals when the spacecraft was simply not magnetically connected to the particle source. This was seen for impulsive SEP events and would be expected if magnetic flux tubes that were connected to a compact source were interspersed with others that were not, as confirmed by theory (Giacalone, Jokipii, and Mazur 2000). Gaps were not often seen in gradual events where a spatially extensive shock wave would be expected to populate all field lines with SEPs.

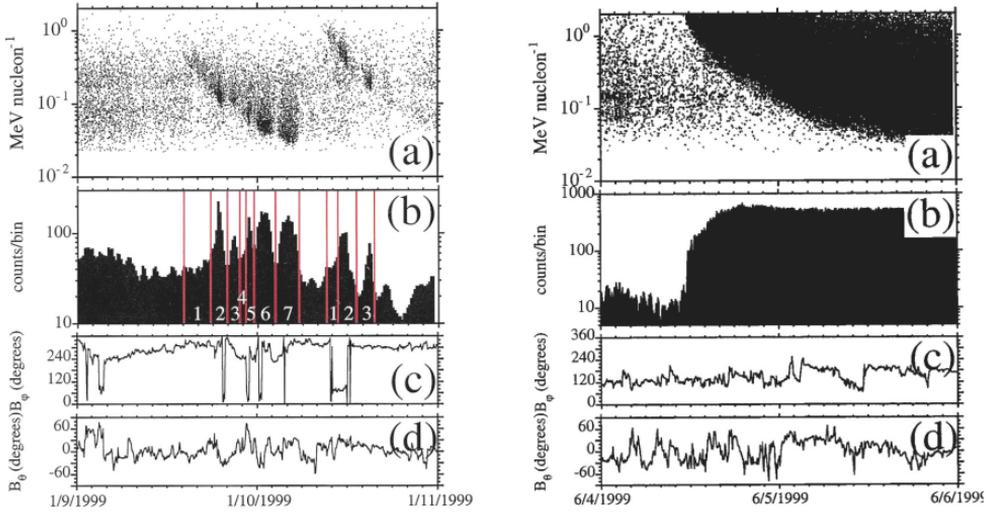

**Fig. 3.8** Panels (**a**) show the energy vs. the arrival time of individual ions from an impulsive (*left*) and a gradual (*right*) SEP event. Panels (**b**) show the corresponding ion count rates while (**c**) and (**d**) show the magnetic field direction. Particles or gaps occur because some flux tubes connect to the compact source of impulsive SEPs and others do not (Mazur et al. 2000, © AAS).

Subsequent observations (Chollet and Giacalone, 2011) found the boundaries between flux tubes with and without SEPs were extremely sharp. This indicated that there was little or no cross-field transport. The mixing of magnetic flux tubes that do and do not connect to any specific location on the Sun is expected from the random walk of their footpoints prior to the particle event (see Sect. 2.3.5).

More-recent observations (Tan and Reames 2016; Tan 2017) have shown that there are occasional dropouts during gradual SEP events as well. Differences in the scattering in some magnetic flux tubes can have an especially strong affect on the intensities and angular distributions of non-relativistic electrons. These profound sudden changes in particle intensities mostly occur when differently-connected flux tubes are sampled inside a passing CME, as measured by multiple spacecraft in the SEP event of 14 December 2006 (von Rosenvinge et al. 2009).

## 3.6 Abundances

Abundances of elements and isotopes were one of the earliest indications of the two different sources of SEPs:

1. The average abundances of the elements in gradual events, relative to those in the photosphere, measured a FIP pattern related to the abundances in the corona and solar wind (see Sects. 1.5.2 and 2.5.1). Since they were associated with fast, wide CMEs driving shock waves, this fit well with the idea of a shock wave sampling ambient coronal abundances.

2. The strong 1000-fold enhancements of $^3$He/$^4$He, and the associations with streaming electrons, and with the type III radio bursts they produce, were clearly related to an impulsive source at the Sun and soon connected with narrow CMEs and solar jets (see Kahler, Reames, and Sheeley 2001; Reames, Cliver, and Kahler 2014a; Bučík et al. 2018; see also Sects. 2.5.2 and 4.7).

In the next chapters (Sects. 4.6, 5.6) we will see that the pattern of the power-law dependence of abundance enhancements on $A/Q$ of the ions leads to a determination of $Q$ values and of the associated source-plasma temperature $T$. The results are:

1. Gradual events: ~69% of events $0.8 < T < 1.6$ MK, 24% of events $T = 2 – 4$ MK from re-accelerated impulsive SEP seed ions (Reames 2016)

2. Impulsive SEP events: $T = 2 – 4$ MK (Reames, Cliver, and Kahler 2014a, b)

Thus in 69% of gradual events, shocks sweep up material at ambient coronal temperatures. In 24%, shocks traverse active regions and re-accelerate some residual impulsive suprathermal ions diluted by some ambient active-region plasma, especially protons. We will develop the techniques for determining source temperatures in Sects. 4.6 and 5.6. The temperatures are strong evidence for shock acceleration of large gradual SEP events. Ambient coronal temperatures of SEPs would seem to be hard to explain for those who would like to accelerate gradual SEPs in hot flares, or even to *store* SEPs in hot flare loops. Furthermore, GLEs show a similar distribution of source-plasma temperatures and abundances as non-GLEs. We find nothing unique about the physics of GLEs; they just happen to have harder spectra and direct a few more high-energy particles toward Earth.

## 3.7 Electrons

In a review article, Ramaty et al. (1980) studied peak intensities of $0.5 – 1.1$ MeV electrons vs. those of 10 MeV protons. For sufficiently intense protons they found a correlation between the electrons and protons that they ascribed to common acceleration by a shock wave. Cliver and Ling (2007) revisited this study from the perspective of impulsive and gradual SEP events. Their interesting findings are shown in Fig. 3.9.

This clever study makes use of the fact that impulsive events are nearly all magnetically well-connected. The known impulsive events show no evidence of electron-proton intensity correlation. The shock-accelerated events span a much larger region of solar longitude and show electron-proton correlation where the impulsive events are absent, i.e. events that are poorly connected or those that have high proton intensities.

**Fig. 3.9** The panels each show peak 0.5-MeV electron intensity vs. peak 10-MeV proton intensity. Events in the *upper panel* are well connected (20W – 90W) while those in the *lower panel* are poorly connected. The events in blue are impulsive events that are ³He-rich with enhanced heavy elements. A proton-intense subset of well-connected events shows a strong correlation while essentially all of the poorly-connected events are correlated (Cliver and Ling 2007 © AAS).

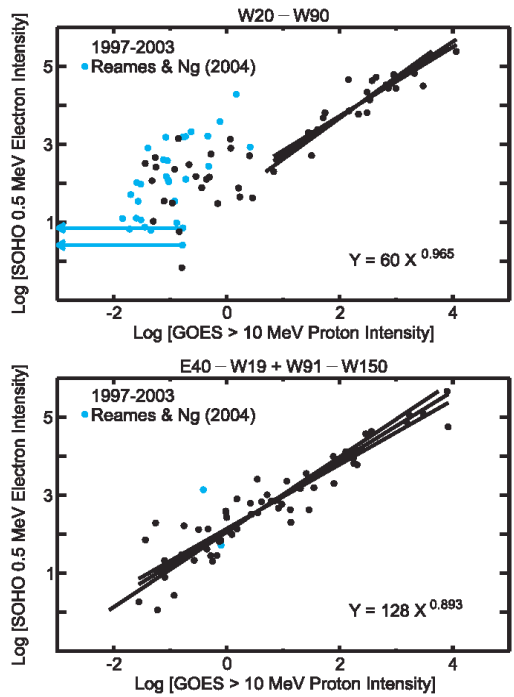

Are the 0.5-MeV electrons accelerated at the same shock as the 10-MeV protons? Apparently so. In general, it is difficult to know how to compare protons and electrons. Unfortunately, low-energy electrons do not resonate with Alfvén waves as low-energy protons do. Should they be compared at the same energy, rigidity, or velocity? Usually the available intervals are used that are neither; yet, despite the lack of an ideal variable for comparison, it is still usually possible to conclude that impulsive SEPs are electron rich, and that shocks in gradual events accelerate both protons and relativistic electrons.

Recently, an extensive list of over 1000 electron events spanning a solar cycle was prepared by Wang et al. (2012). The list includes radio, CME, X-ray, flare, ³He/⁴He, and 10-MeV proton data.

Electron acceleration at shock waves is a historical problem. There is even lore that shocks cannot accelerate electrons While ions can scatter back and forth against Alfvén waves at the quasi-parallel nose, where the shock is fastest, non-relativistic electrons cannot resonate with Alfvén waves and are not similarly accelerated. However, electrons can be accelerated in the $V_S \times B$ electric fields on the oblique flanks of the shock.

Are type-II radio bursts only emitted from the flanks of shocks? No. They come from the nose as well. Type-II emission from the shock nose and flanks can be distinguished because the nose is farther out from the Sun, at lower $n_e$, and therefore lower frequency; the frequencies can be correlated with the coronagraph image. Actually, type-II emission from the flanks of a shock goes by the special name: diffuse interplanetary radio emission (DIRE; e.g. Gopalswamy 2020).

Electron acceleration at the quasi-parallel nose of a shock probably occurs because real shocks are not planar but are very complex structures, varying in space and time. This complexity was even noted in Jones and Ellison (1991) and has been observed directly in interplanetary shocks with the *Cluster* spacecraft by Kajdič et al. (2019). The fluctuations can even involve $\theta_{Bn}$ so $V_S \times B$ electric fields may be available everywhere to accelerate electrons. Electrons are certainly accelerated at shocks, they produce type-II emission and they even manage to reach relativistic energies as seen in Fig. 3.9.

## 3.8 Why not Flares?

Impulsive SEP events in space are produced by solar jets, not flares. Solar flares exist *precisely* because the SEPs in them are magnetically trapped. Sites of electron acceleration in magnetic reconnection occur above the newly-forming closed magnetic loops at heights of >20 Mm and at densities of a few times $10^9$ cm$^{-3}$ (Krucker et al. 2010; Krucker and Battaglia 2014). Ion acceleration is likely to occur at the same location (e.g. Drake et al. 2009). These SEPs eventually scatter into the dense regions of the loop footpoints where they dump all their energy, causing heating and expansion of hot, bright plasma back up into the loops. Observations of γ-ray lines have shown that the energetic ions in flares are $^3$He-rich (Mandzhavidze, Ramaty, and Kozlovsky 1999; Murphy, Kozlovsky, and Share 2016) and Fe-rich (Murphy et al. 1991), just like the impulsive SEP events we see in space. However, these γ-rays arise from nuclear reactions that also produce secondary ions of $^2$H, $^3$H, and isotopes of Li, Be, and B. These reaction products are *not* seen in space. Serlemitsos and Balasubrahmanyan (1975) found $^3$He/$^4$He = $1.52 \pm 0.10$, compared with $(4.08 \pm 0.25) \times 10^{-4}$ in the solar wind (Gloeckler and Geiss 1998), but they also found $^3$He/$^2$H > 300. Limits on Be/O or B/O in large SEP events are $< 2 \times 10^{-4}$ (e.g. McGuire, von Rosenvinge, and McDonald 1979; Cook, Stone, and Vogt 1984). Neutral γ-rays and neutrons (Chupp et al. 1982; Evenson et al. 1983, 1990) from nuclear reactions escape, but SEPs that actually caused the flare never escape the closed magnetic loops. One cannot exclude a stray leak from a flare, but that must be swamped by SEPs from the shock source.

In Carrington's (1860) first observation of a flare, he was "surprised … at finding myself unable to recognize any change whatever as having taken place." The fields maintain their shape. Modern instruments allow measurement of the reconnection magnetic flux and a recent data base contains reconnection flux for 3137 solar-flare ribbon events (Kazachenko et al. 2017). Large reconnection events that lacked fast shock waves produce beautiful flares, but *no* SEPs (Kahler et al. 2017). Flares *require* containment by closed magnetic fields and disrupt the general field topology very little; the field energy transmitted to particles accelerated in flares is dissipated as heat in the footpoints and loops and we do not see these energetic particles in space.

Historically, it has been the neutron monitors that have been used most often to try to associate a flare source with GLEs. These are the *biggest* events, where *everything* happens, and are the most prone to the errors of causal association that result from "big flare syndrome" (Kahler 1982). Worse yet, neutron monitors do not measure particle abundances, their energy measurements are limited, and anisotropic intensities of streaming GeV protons are often modulated by variations in the direction of the interplanetary magnetic field relative to the asymptotic look direction of a site. At best, neutron monitors mainly measure timing, and many physical processes, including shock acceleration, can be fast.

We prefer to study SEPs measured in the *smallest* events, to identify sources. Jets (Sect. 4.7) *without* fast shocks give us impulsive SEP events, and disappearing-filament events (Sect. 2.7) drive fast shocks with no flares or jets, yet they give us gradual SEP events that depend upon the shock properties. These smaller, simpler single-source events better define SEP origins.

Flares are bright because they are *hot*, reaching temperatures if 10 – 40 MK. When we measure the source plasma temperatures of SEPs (see Sects. 2.6, 4.6 and 5.6), we find temperatures of 1 – 4 MK, i.e. ambient temperatures of coronal plasma. At 10 – 40 MK, even the elements Ne, Mg, and Si become fully ionized so they all have $A/Q = 2$, just like He, C, N, and O. This hot plasma would not support the observed enhancements seen in Ne/O or even Si/O (Chap. 4). SEPs from hot flare plasma are not seen. The observed properties of SEPs in space are not compatible with such hot flare plasma and closed-field regions do not just open to release them. The SEP sources we once attributed to specific flares most likely come from nearby jets where the SEPs escape rapidly with minimal heating.

To the extent that SEPs in flares are similar to those in jets, flare observations, especially γ-ray lines, could be a benefit in understanding the physics of impulsive acceleration mechanisms. It would be helpful to have more-extensive measurements of γ-ray lines (Sect. 4.9).

## 3.9 SEPs as Probes

As SEPs stream out along the interplanetary field they can map its structure. This was shown early by the radio mapping of the electron population in type III radio bursts. Knowing the direction to the center of the radio signal and its distance from the Sun determined by the frequency and models of the electron density vs. radius, the electrons could be followed, as seen in Fig. 3.10. Occasionally, trajectories of this kind can be made using triangulation from two or more spacecraft (Reiner et al. 1998; see also Li et al. 2016). While such electron trajectories generally follow the Parker spiral, it is important to realize that field lines are often distorted by variations in the solar-wind speed and by the passage of CMEs.

Low-energy (<100 keV) electrons are often seen passing Earth outbound then returning sunward from a magnetic reflection site beyond Earth (Kahler and Reames 1991; Tan et al. 2012, 2013) as shown in the example in Fig. 3.11. These electrons have highly scatter-free transport (Tan et al. 2011) and are thus excellent probes of the magnetic topology.

**Fig. 3.10** The trajectory of the electron population in the type III radio burst accompanying a ³He-rich SEP event is shown in three dimensions (**a**) and as a projection on the ecliptic (**b**) (Reames and Stone 1986 © AAS )

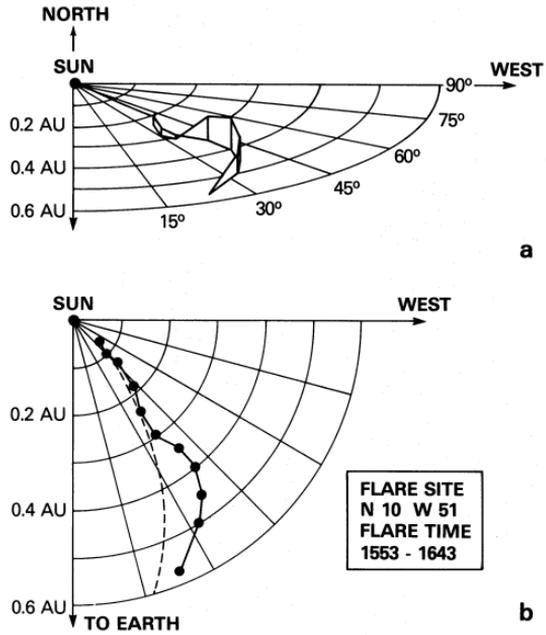

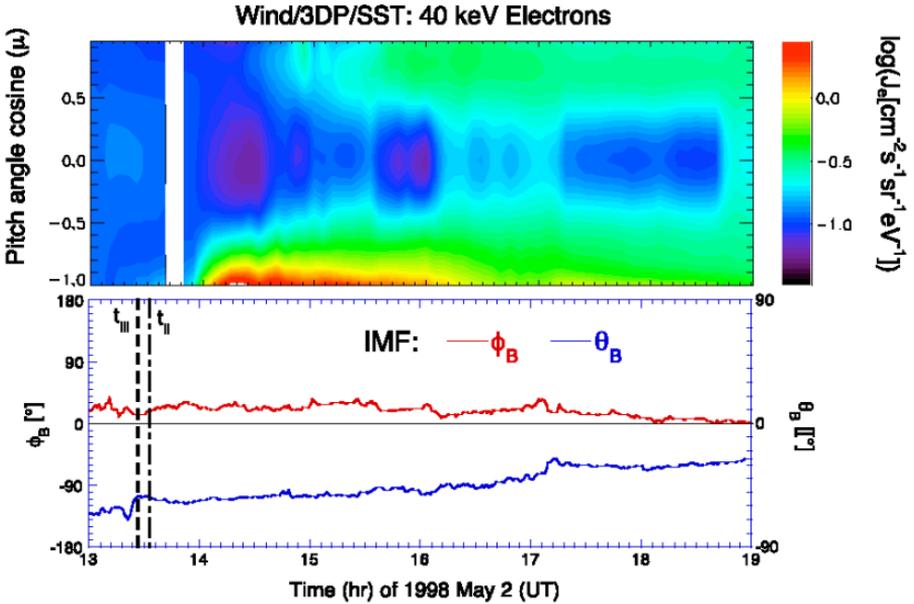

**Fig. 3.11.** The pitch-angle spectrogram of 40 keV electrons from the 2 May 1998 SEP event is shown in the *upper panel*. Electrons first appear from the Sun at $\mu \approx -1$ near 1400 UT and reflected electrons appear at $\mu \approx +1$ around 1500 UT. The *lower panels* show the field direction (Tan et al. 2013, © AAS).

Fig. 3.12 uses separate incident and reflected electron onset times to determine the release times and pathlengths for each, and also shows He ions, with no re-flected beam (Tan et al. 2012). Pathlengths for the incident ions and electrons are quite similar and near the length of the Parker spiral. Electrons depart about 13 min earlier. Reflected electrons travel ≈1 AU farther in this event.

**Fig. 3.12** The *upper panel* shows the onset times of both incident and reflected electrons vs. $v^{-1}$ with the fitted release time and pathlength for each. The *lower panel* shows om-ni-directional data for He (Tan et al. 2012, © AAS). The He SEPs are actually highly colli-mated for a day after onset; they show no reflection at the same time as the reflected elec-trons.

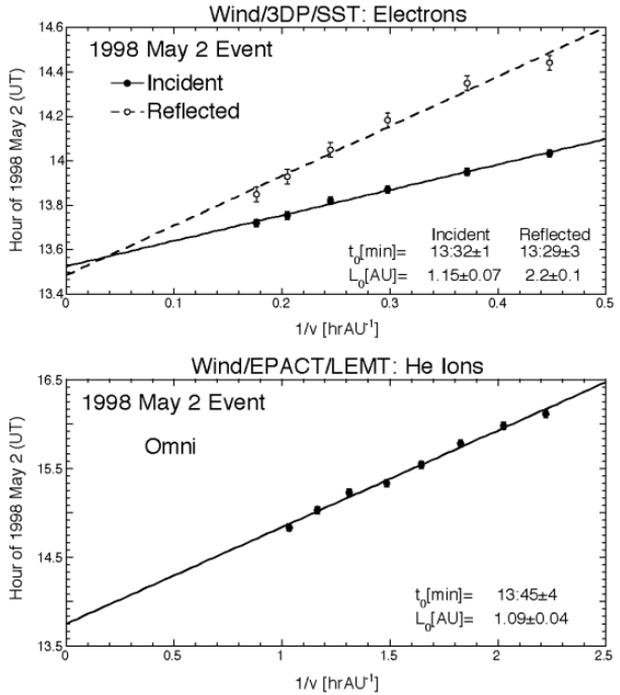

However, electrons are not the only particles affected by their journey, Reames and Ng (2002) found that in some SEP events Fe/O was higher for sunward bound ions than for those that were outward bound. Since Fe scatters less then O, the Fe more rapidly passed Earth to be reflected sunward than O, so the returning particles were more Fe-rich, i.e. they had simply traveled farther from the source.

There should be no surprise that the magnetic fields depart from the simple Parker spiral and become quite complex. Not only is there the random walk of field lines discussed in Sect. 2.3.5, but there is a constant progression of CMEs that disturb the field as suggested by Fig. 3.13.

The Sun can eject 2.5 CMEs day$^{-1}$ at solar maximum (Webb and Howard 1994). If each CME occupies one steradian and its typical speed is ~400 km s$^{-1}$, CMEs will be randomly spaced at radial distances of typically ~1 AU apart, one after the other, out into the heliosphere in any direction. While most of them would be too slow to generate shocks, they would have low $\beta_P$ and would carry magnetic flux ropes that contribute to magnetic distortions capable of reflecting particles.

**Fig. 3.13** The Sun can emit a progression of CMEs as suggested here. Particles accelerated at a shock preceding the smaller, newer CME can easily be reflected by converging field lines of the outer CME or by turbulence at the shock that precedes it.

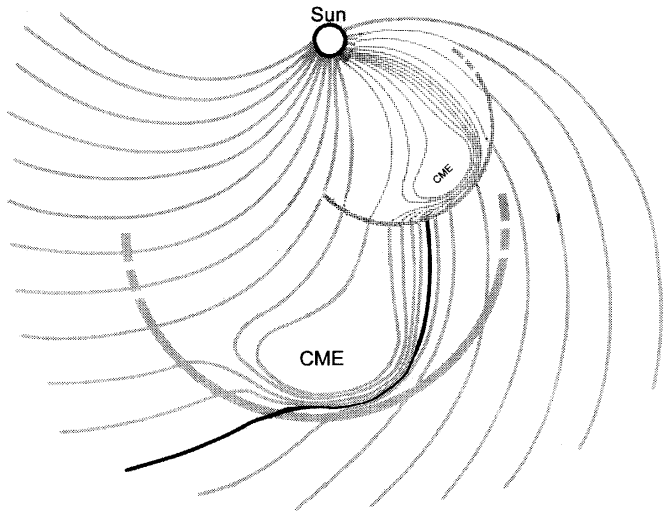

Bidirectional flows of protons or electrons are commonly seen within CMEs near 1 AU (e.g. Kahler and Reames 1991; Richardson and Reames 1993). These flows are often used as probes to study the topology of the magnetic fields, and even to identify CMEs. Their common occurrence also shows that pitch-angle scattering is minimal within CMEs.

## References

Bučík, R., Innes, D.E., Mason, G.M., Wiedenbeck, M.E., Gómez-Herrero, R., Nitta, N, ³He-rich solar energetic particles in helical jets on the sun, Astrophys, J. **852** 76 (2018) doi: 10.3847/1538-4357/aa9d8f

Carrington, R.C. Description of a singular appearance seen in the Sun on September, 1859, Mon. Not. Roy. Astron. Soc. **20**, 13 (1860) doi: 10.1093/mnras/20.1.13

Chollet, E.E., Giacalone, J., Evidence of confinement of solar-energetic particles to interplanetary magnetic field lines, Astrophys. J. **728**, 64 (2011) doi: 10.1088/0004-637X/728/1/64

Chupp, E.L., Forrest, D.J., Ryan, J.M., Heslin, J., Reppin, C., Pinkau, K., Kanbach, G., Rieger, E., Share, G.H., A direct observation of solar neutrons following the 0118 UT flare on 1980 June 21, Astrophys, J. Lett., **263**, L95 (1982) doi: 10.1086/183931

Cliver, E.W., Kahler, S.W., and Reames, D.V., Coronal shocks and solar energetic proton events, Astrophys. J. **605**, 902 (2004) doi: 10.1086/382651

Cliver, E.W., Ling, A.G., Electrons and protons in solar energetic particle events, Astrophys. J. 658, 1349 (2007) doi: 10.1086/511737

Cook, W.R., Stone, E.C., Vogt, R.E., Elemental composition of solar energetic particles, Astrophys. J. **279**, 827 (1984) doi: 10.1086/161953

Drake, J.F., Cassak, P.A., Shay, M.A., Swisdak, M., Quataert, E., A magnetic reconnection mechanism for ion acceleration and abundance enhancements in impulsive flares, Astrophys. J. Lett. **700**, L16 (2009) doi: 10.1088/0004-637X/700/1/L16

Ellison, D., Ramaty, R., Shock acceleration of electrons and ions in solar flares, Astrophys. J. **298**, 400 (1985) doi: 10.1086/163623

Evenson, P., Meyer, P., Pyle, K.R, Protons from the decay of solar flare neutrons, Astrophys. J. **274**, 875 (1983) doi: 10.1086/161500

Evenson, P., Kroeger, R., Meyer, P., Reames, D., Solar neutron decay proton observations in cycle 21, Astrophys. J. Suppl. **73** 273 (1990) doi: 0.1086/191462

Giacalone, J., Jokipii, J. R., Mazur, J. E., Small-scale gradients and large-scale diffusion of charged particles in the heliospheric magnetic field, Astrophys. J. Lett. **532** L78 (2000) doi: 10.1086/312564

Gopalswamy, N., Interplanetary radio emission: A summary of recent result, eprint (2020) arXiv:2008.09222

Gopalswamy, N., Xie, H., Akiyama, S., Yashiro, S., Usoskin, I.G., Davila, J.M, The first ground level enhancement event of Solar Cycle 24: Direct observation of shock formation and particle release heights, Astrophys. J. Lett. **765** L30 (2013) doi: 10.1088/2041-8205/765/2/L30

Gopalswamy, N., Xie, H., Yashiro, S., Akiyama, S., Mäkelä, P., Usoskin, I. G., Properties of Ground level enhancement events and the associated solar eruptions during solar cycle 23, Space Sci. Rev. **171**, 23 (2012) doi: 10.1007/s11214-012-9890-4

Jones, F.C., Ellison, D.C., The plasma physics of shock acceleration, Space Sci. Rev. **58**, 259 (1991) doi: 10.1007/BF01206003

Kahler, S.W., Injection profiles of solar energetic particles as functions of coronal mass ejection heights, Astrophys. J., **428**, 837 (1994) doi: 10.1086/174292

Kahler, S.W., The correlation between solar energetic particle peak intensities and speeds of coronal mass ejections: Effects of ambient particle intensities and energy spectra, J. Geophys. Res. **106**, 20947 (2001) doi: 10.1029/2000JA002231

Kahler, S.W., Kazachenko,M, Lynch, B.J., Welsch, B.T., Flare magnetic reconnection fluxes as possible signatures of flare contributions to gradual SEP events, J. Phys. Conf. Series **900** 012011 (2017) doi: 10.1088/1742-6596/900/1/012011

Kahler, S.W., Reames, D.V., Probing the magnetic topologies of magnetic clouds by means of solar energetic particles, J. Geophys. Res. **96**, 9419 (1991) doi: 10.1029/91JA00659

Kahler, S.W., Reames, D.V., Sheeley, Jr., N.R., Coronal mass ejections associated with impulsive solar energetic particle events, Astrophys. J., **562**, 558 (2001) doi: 10.1086/323847

Kajdič, P., Preisser, L., Blanco-Cano, X., Burgess, D., Trotta, D., First observations of irregular surface of interplanetary shocks at ion scales by *Cluster*, Astrophys. J. Lett. **874** L13 (2019) doi: 10.3847/2041-8213/ab0e84

Kazachenko, M.D., Lynch, B.J., Welsch, B.T., Sun, X., A database of flare ribbon properties from solar dynamics observatory I: Reconnection flux, Astrophys. J. **845** 49 (2017) doi: 10.3847/1538-4357/aa7ed6 arXiv:1704.05097.

Kouloumvakos, A., Rouillard, A., Wu, Y., Vainio, R., Vourlidas,A., Plotnikov, I., Afanasiev, A., Önel, H., Connecting the properties of coronal shock waves with those of solar energetic particles, Astrophys. J., **876** 80 (2019) doi: 10.3847/1538-4357/ab15d7

Krucker S., Battaglia M., Particle densities within the acceleration region of a solar flare. Astrophys J, **780**: 107 (2014) doi: 10.1088/0004-637X/780/1/107

Krucker S., Hudson H.S., Glesener L., White S.M., Masuda S., Wuelser J.P., Lin R.P., Measurements of the coronal acceleration region of a solar flare. Astrophys J, **714**: 1108 (2010) doi: 10.1088/0004-637X/714/2/1108

Lee, M.A., Coupled hydromagnetic wave excitation and ion acceleration at an evolving coronal/interplanetary shock, Astrophys. J. Suppl., **158**, 38 (2005) doi: 10.1086/428753

Li, B., Cairns, I.H., Gosling, J.T., Steward, G., Francis, M., Neudegg, D., Schulte in den Bäumen, H., Player, P.R., Milne, A.R., Mapping magnetic field lines between the Sun and Earth, J. Geophys. Res. **121**, 925 (2016) doi: 10.1002/2015JA021853

Lovell, J.L., Duldig, M.L., Humble, J.E., An extended analysis of the September 1989 cosmic ray ground level enhancement, J. Geophys. Res., **103**, 23,733 (1998) doi: 10.1029/98-JA02100

Mandzhavidze, N., Ramaty, R., Kozlovsky, B., Determination of the abundances of subcoronal $^4$He and of solar flare-accelerated $^3$He and $^4$He from gamma-ray spectroscopy, Astrophys. J. **518**, 918 (1999) doi: 10.1086/307321

Mann, G. Klassen, A. Aurass, H., Classen, H.-T., Formation and development of shock waves in the solar corona and the near-Sun interplanetary space, Astron. Astrophys. **400**, 329 (2003) doi: 10.1051/0004-6361:20021593

Mazur, J.E., Mason, G.M., Dwyer, J.R., Giacalone, J., Jokipii, J.R., Stone, E.C., Interplanetary magnetic field line mixing deduced from impulsive solar flare particles, Astrophys. J. Lett. **532**, L79 (2000) doi: 10.1086/312561

McGuire, R.E., von Rosenvinge, T.T., McDonald, F.B., A survey of solar cosmic ray composition, *Proc. 16th Int. Cosmic Ray Conf., Tokyo* **5**, 61 (1979)

Mewaldt, R.A., Looper, M.D., Cohen, C.M.S., Haggerty, D.K., Labrador, A.W., Leske, R.A., Mason, G.M., Mazur, J.E., von Rosenvinge, T.T., Energy spectra, composition, other properties of ground-level events during solar cycle 23, Space Sci. Rev. **171**, 97 (2012) doi: 10.1007/s11214-012-9884-2

Murphy, R.J., Kozlovsky, B., Share, G.H., Evidence for enhanced ³He in flare-accelerated particles based on new calculations of the gamma-ray line spectrum, Astrophys. J., **833** 196 (2016) doi: 10.3847/1538-4357/833/2/196

Murphy, R.J., Ramaty, R., Kozlovsky, B., Reames, D.V., Solar abundances from gamma-ray spectroscopy: Comparisons with energetic particle, photospheric, and coronal abundances, Astrophys. J. **371**, 793 (1991) doi: 10.1086/169944

Ramaty, R., Paizis, C., Colgate, S.A., Dulk, G.A., Hoyng, P., Knight, J.W., Lin, R.P., Melrose, D.B., Orrall, F., Shapiro, P.R., Energetic particles in solar flares, In: Solar flares: A monograph from Skylab Solar Workshop II. (A80-37026 15-92) Boulder, Colo., Colorado Associated University Press, p117 (1980)

Reames, D.V., Solar release times of energetic particles in ground-level events, Astrophys. J. **693**, 812 (2009a) doi: 10.1088/0004-637X/693/1/812

Reames, D.V., Solar energetic-particle release times in historic ground-level events, Astrophys. J. **706**, 844 (2009b) doi: 10.1088/0004-637X/706/1/844

Reames, D.V., Temperature of the source plasma in gradual solar energetic particle events, Sol. Phys., **291**, 911 (2016) doi: 10.1007/s11207-016-0854-9, arXiv: 1509.08948

Reames, D.V., Stone, R.G., The identification of solar He-3-rich events and the study of particle acceleration at the sun, Astrophys. J., **308**, 902 (1986) doi: 10.1086/164560

Reames, D.V., Cliver, E.W., Kahler, S.W., Abundance enhancements in impulsive solar energetic-particle events with associated coronal mass ejections, Sol. Phys. **289**, 3817, (2014a) doi: 10.1007/s11207-014-0547-1

Reames, D.V., Cliver, E.W., Kahler, S.W., Variations in abundance enhancements in impulsive solar energetic-particle events and related CMEs and flares, Sol. Phys. **289**, 4675 (2014b) doi: 10.1007/s11207-014-0589-4

Reames, D.V., Lal, N., A multi-spacecraft view of solar-energetic-particle onsets in the 1977 November 22 event, Astrophys. J., **723**, 550 (2010) doi: 10.1088/0004-637X/723/1/550

Reames, D.V., Ng, C.K., Angular distributions of Fe/O from wind: new insight into solar energetic particle transport, Astrophys, J. Lett. 575, L37 (2002) doi: 10.1086/344146

Reames, D.V., Ng, C.K., Heavy-element abundances in solar energetic particle events, Astrophys. J. **610**, 510 (2004) doi: 10.1086/421518

Reiner, M.J., Fainberg, J., Kaiser, M.L., Stone, R.G., Type III radio source located by Ulysses/Wind triangulation, J. Geophys. Res. **103**, 1923. (1998) doi: 10.1029/97JA02646

Richardson, I.G., Reames, D.V., Bidirectional ~1 MeV amu⁻¹ ion intervals in 1973--1991 observed by the Goddard Space Flight Center instruments on IMP 8 and ISEE 3/ICE, Astrophys. J. Suppl. **85** 411 (1993) doi: 10.1086/191769

Rouillard, A.C., Odstrčil, D., Sheeley, N.R. Jr., Tylka, A.J., Vourlidas, A., Mason, G., Wu, C.-C., Savani, N.P., Wood, B.E., Ng, C.K., et al., Interpreting the properties of solar energetic particle events by using combined imaging and modeling of interplanetary shocks, Astrophys. J. **735**, 7 (2011) doi: 10.1088/0004-637X/735/1/7

Rouillard, A.P., Plotnikov, I., Pinto, R.F., Tirole, M., Lavarra, M., Zucca, P., Vainio, R., Tylka, A.J., Vourlidas, A., De Rosa, M.L. Linker, J., Warmuth, A., Mann, G., Cohen, C.M.S., Mewaldt, R.A., Deriving the properties of coronal pressure fronts in 3D: application to the 2012 May 17 ground level enhancement, Astrophys J. **833** 45 (2016) doi: 10.3847/1538-4357/833/1/45

Rouillard, A., Sheeley, N.R., Jr., Tylka, A., Vourlidas, A., Ng, C.K., Rakowski, C., Cohen, C.M.S., Mewaldt, R.A., Mason, G.M., Reames, D., et al., The longitudinal properties of a solar energetic particle event investigated using modern solar imaging, Astrophys. J. **752**, 44 (2012) doi: 10.1088/0004-637X/752/1/44

Serlemitsos, A.T., Balasubrahmanyan, V.K., Solar particle events with anomalously large relative abundance of ³He, Astrophys. J. **198**, 195, (1975) doi: 10.1086/153592

Tan, L.C., Electron-ion intensity dropouts in gradual solar energetic particle events during solar cycle 23, Astrophys. J. **846** 18 doi: 10.3847/1538-4357/aa81d1

Tan, L.C., Reames, D.V., Dropout of directional electron intensities in large solar energetic particle events, Astrophys. J. **816** 93 doi: 10.3847/0004-637X/816/2/93

Tan, L.C., Reames, D.V., Ng, C.K., Shao, X., Wang, L., What causes scatter-free transport of non-relativistic solar electrons?, Astrophys J. **728**, 133 (2011) doi: 10.1088/0004-637X/728/2/133

Tan, L.C., Malandraki, O.E., Reames, D.V., Ng, C.K., Wang, L., Dorrian, G. Use of Incident and Reflected Solar Particle Beams to Trace the Topology of Magnetic Clouds, Astrophys. J. **750**, 146 (2012) doi: 10.1088/0004-637X/750/2/146

Tan, L.C., Malandraki, O.E., Reames, D.V., Ng, C.K., Wang, L., Patsou, I., Papaioannou, A., Comparison between path lengths traveled by solar electrons and ions in ground-level enhancement events, Astrophys J. **768**, 68 (2013) doi: 10.1088/0004-637X/768/1/68

Thakur, N., Gopalswamy, N., Mäkelä, P., Akiyama, S., Yashiro, S., Xie, H., Two exceptions in the large SEP events of Solar Cycles 23 and 24, Solar. Phys. **231** 519 (2016) doi: 10.1007/s11207-015-0830-9

Tylka, A.J., Cohen, C.M.S., Dietrich, W.F., Krucker, S., McGuire, R. E., Mewaldt, R.A., Ng, C.K., Reames, D.V, Share, G. H., Onsets and release times in solar particle events, Proc. 28th Int. Cosmic Ray Conf., 3305 (2003)

Tylka, A.J., Dietrich, W.F., A new and comprehensive analysis of proton spectra in ground-level enhanced (GLE) solar particle events, in Proc. 31st Int. Cos. Ray Conf , Lódz (2009), http://icrc2009.uni.lodz.pl/proc/pdf/icrc0273.pdf

von Rosenvinge, T.T., Richardson, I.G., Reames, D.V., Cohen, C.M.S., Cummings, A.C., Leske, R.A., Mewaldt, R.A., Stone, E.C., Wiedenbeck, M.E., The solar energetic particle event of 14 December 2006, Sol. Phys. **256** 443 (2009) doi: 10.1007/s11207-009-9353-6

Wang, L., Lin, R.P., Krucker, S., Mason, G.M., A statistical study of solar electron events over one solar cycle, Astrophys. J. **759**, 69 (2012) doi: 10.1088/0004-637X/759/1/69

Webb, D.F., Howard, R.A., The solar cycle variation of coronal mass ejections and the solar wind mass flux, J. Geophys. Res. **99**, 4201 (1994) doi: 10.1029/93JA02742

# Chapter 4. Impulsive SEP Events (and Flares)

**Abstract**   $^3$He-rich, Fe-rich, and enriched in elements with $Z > 50$, the abundances of solar energetic particles (SEPs) from the small impulsive SEP events stand out as luminaries in our study. The $^3$He is enhanced by resonant wave-particle interactions. Element abundances increase 1000-fold as the ~3.6 power of the mass-to-charge ratio $A/Q$ from He to heavy elements like Au or Pb, enhanced during acceleration in islands of magnetic reconnection in solar jets, and probably also in flares. This power-law of enhancement vs. $A/Q$ implies $Q$ determined by a source temperature of $2.5 – 3.2$ MK, typical of jets from solar active regions where these impulsive SEPs occur. However, a few small events are unusual; several have suppressed $^4$He, and rarely, a few very small events with steep spectra have elements N or S greatly enhanced, perhaps by the same resonant-wave mechanism that enhances $^3$He. Which mechanism will dominate? The impulsive SEP events we see are associated with *narrow* CMEs, from solar jets where magnetic reconnection on open field lines gives energetic particles and CMEs direct access to space. Gamma-ray lines tell us that the same acceleration physics may occur in flares.

Impulsive SEP events were first identified by their unusual enhancements of $^3$He/$^4$He, with ~1000-fold increases over the abundance $^3$He/$^4$He $\approx 5 \times 10^{-4}$ in the solar wind, frequently with $^3$He/$^4$He $> 1$, and occasionally with $^3$He/H $> 1$. Next we found enhancements of Fe/C or Fe/O of ~10, which were more-stable indicators of impulsive events, since $^3$He/$^4$He varies widely. Then ~1000-fold increases in elements with $(76 \leq Z \leq 82)$/O were added to the unusual picture.

Despite the huge enhancements of $^3$He, the isotopes $^2$H and $^3$H are *not* observed in SEPs (<1% of $^3$He according to Serlemitsos and Balasubrahmanyan 1975). Observations of γ-ray lines and neutrons show the presence of nuclear reactions in the low corona during flares (e.g. Ramaty and Murphy 1987), but isotopes of Li, Be, and B have never been observed in SEPs. Limits on Be/O or B/O in large SEP events are $< 4 \times 10^{-4}$ (e.g. Cook, Stone, and Vogt 1984) Reaction secondaries are trapped on flare loops and cannot escape, and the $^3$He we see is *not* a nuclear-reaction product, it is enhanced by resonant wave-particle reactions (e.g. Temerin and Roth 1992; see Sect. 2.5.2). In fact, as we have seen, only particles accelerated on *open* field lines, e.g. in solar jets (or at shock waves), can ever escape.

## 4.1 Selecting Impulsive Events

Many years ago, Reames (1988) examined the distribution of all daily abundance averages with measurable Fe/O ratios during 8.5 years, and found a bimodal distribution with peaks near Fe/O $\approx 0.1$ and Fe/O $\approx 1.0$. The technique was free from bias related to individual event selection, although long-duration events were certainly more heavily sampled. Periods with Fe/O near 0.1 had unremarkable abundances of other elements, but those near 1.0 also had enhancements in $^3$He/$^4$He, $^4$He/H, and e/p ratios. While the two distributions of Fe/O did have an overlap region, largely because the poor statistics available at that time spread the distributions, the results showed that Fe/O at about $2 - 5$ MeV amu$^{-1}$ was more reliable for selecting candidate periods (e.g. Fig. 4.1) for impulsive SEP events than $^3$He/$^4$He, which depends strongly upon energy (e.g. Fig. 4.4).

A more-recent version of the bimodal abundance study is the two-dimensional histogram shown in Fig. 4.1, based upon much more accurate data. Here 8-h measurements of Ne/O vs. Fe/O are binned for a 19-year period, and this time we have the luxury of requiring 20% accuracy to prevent excessive spreading of the distributions. Of course, it is still true that gradual events occupy many more 8-h periods and impulsive events, with lower intensities, are less likely to achieve 20% accuracy, but the presence of two peaks is clear.

**Fig. 4.1** Measured relative enhancements in Ne/O vs. Fe/O for 8-h periods during 19 years are binned for all periods with errors of 20% or less. (Reames, Cliver, and Kahler 2014a © Springer)

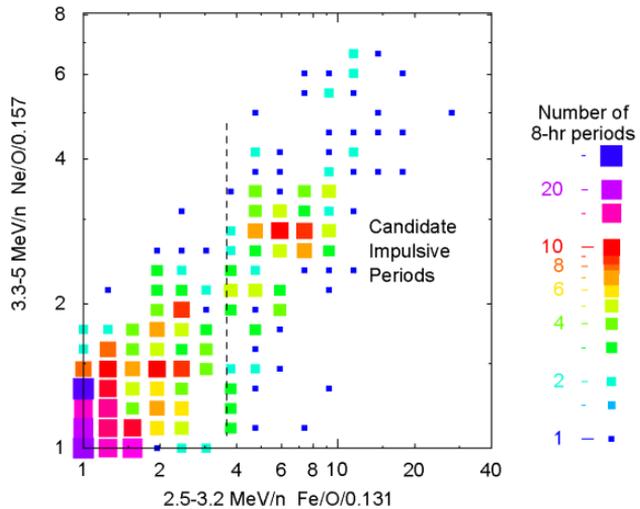

Time periods near coordinates (1, 1) in Fig. 4.1, occur during large gradual SEP events for which the normalization was chosen. The peak near (6, 3) in the figure represents impulsive events, but the Ne/O value was not actually used for selection of candidate periods for defining impulsive SEP events.

## 4.2 Sample Impulsive Events

Fig. 4.2 shows intensities of several particle species in a sample of impulsive SEP events with various properties. In events 1, 2, and 5, we have $^3$He/$^4$He $\gg$ 1, and in event 1, $^3$He > H. In events 1 and 2 the O, which may seem high relative to Fe, is actually background from anomalous cosmic-ray O and is present at almost the same rate before and after the events. In events 5 and 6, O is closer to $^4$He than in other events, these "He-poor" events have low $^4$He/O. Events $6 - 13$ on the bottom row are an order of magnitude larger in $^4$He, O, or Fe than those in the first row, and heavy elements begin to appear; these larger events are also not as strongly $^3$He-rich. Instrument limitations: only groups of elements are resolved above $Z = 34$ and, when $^3$He/$^4$He $< 0.1$, $^3$He is poorly resolved and is not plotted.

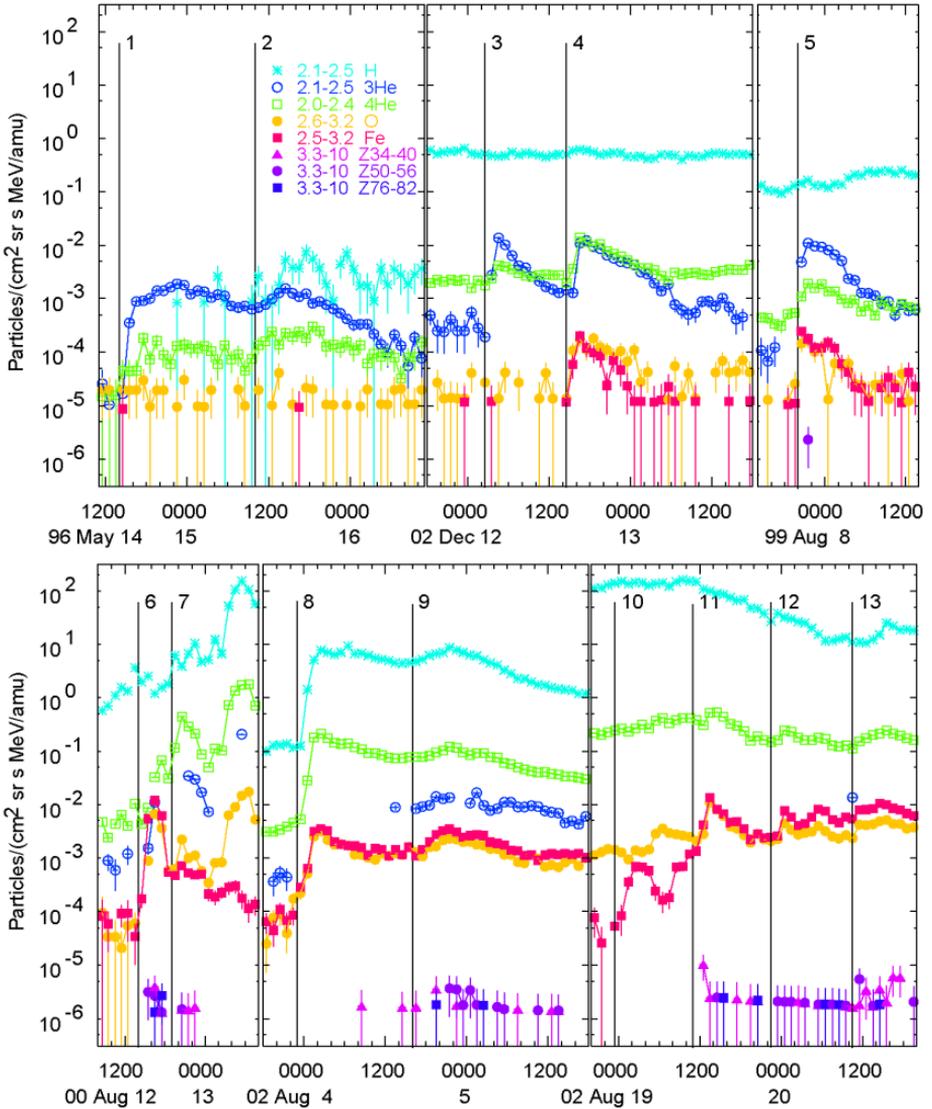

**Fig. 4.2** Intensities of H, $^3$He, $^4$He, O, Fe, and heavy elements are shown as a function of time during 13 impulsive SEP events (see text, Reames and Ng 2004 © AAS).

## 4.3 Energy Dependence

Some sample energy spectra in $^3$He-rich events are shown in Fig. 4.3.

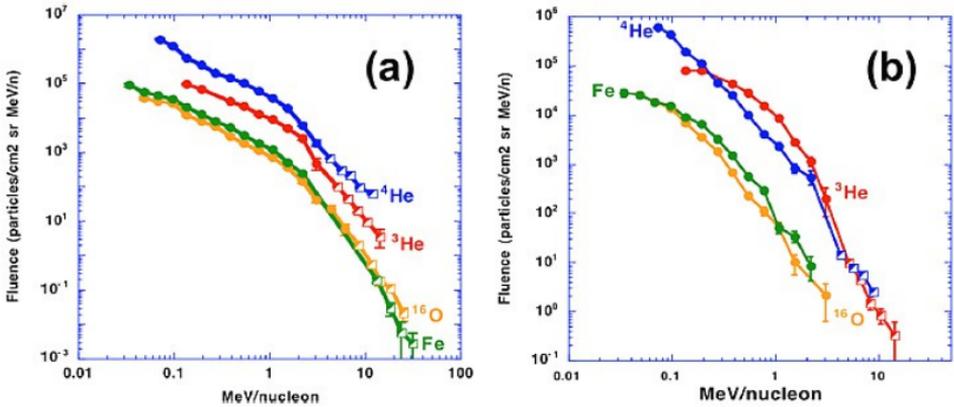

**Fig. 4.3.** Spectra of $^3$He, $^4$He, O and Fe are shown in the (**a**) 9 September 1998 and the (**b**) 21 March 1999 events (Mason et al. 2002b, © AAS).

The spectra on the left in Fig. 4.3 appear as broken power-law spectra while those on the right are more curved and show large energy variations in $^3$He/$^4$He as seen in Fig. 4.4. Abundance ratios of Fe/O show much less spectral variation.

**Fig. 4.4** Energy dependence is shown for $^3$He/$^4$He ratios. Red and blue are for events shown in Fig. 4.3. (**a**) and (**b**), respectively. The *green* event is 27 September 2000 (Mason et al. 2002b © AAS).

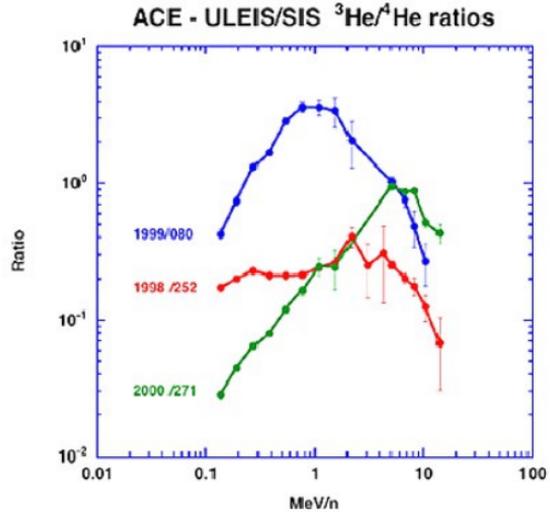

The $^3$He/$^4$He variations shown in Fig. 4.4 make it difficult to characterize an event by this ratio, which seems to peak in the region of $1 - 10$ MeV amu$^{-1}$. Fe/O at a few MeV amu$^{-1}$ is a better-behaved alternative for defining impulsive events, as suggested in Fig. 4.1.

Liu, Petrosian, and Mason (2006) have been able to fit the complex spectra of $^3$He and $^4$He with a model of stochastic acceleration by a power-law spectrum of plasma-wave turbulence, presumably associated with magnetic reconnection. This work follows the tradition of stochastic acceleration involving the general transfer of energy from waves to particles (see reviews: Miller et al. 1997; Miller 1998).

These models have difficulty explaining the strong *A/Q*-dependent enhancements extending to heavy elements that we will discuss in Sects. 4.5 and 4.6.

## 4.4 Abundances for Z ≤ 26

Given the spectra and variations of $^3$He we have seen, it is not surprising that the $^3$He/$^4$He ratio is uncorrelated with other abundance ratios as seen in Fig. 4.5. This was known to Mason et al. (1986) and Reames, Meyer, and von Rosenvinge (1994) and is often taken as evidence that the mechanism of $^3$He enhancement is different from that causing enhancement of Fe/O and heavy elements.

**Fig. 4.5** Cross plots of Fe/C vs. $^3$He/$^4$He at 1.3-1.6 MeV amu$^{-1}$ in impulsive SEP events shows little evidence of correlation (Reames 1999; adapted from Reames, Meyer, and von Rosenvinge 1994 © AAS).

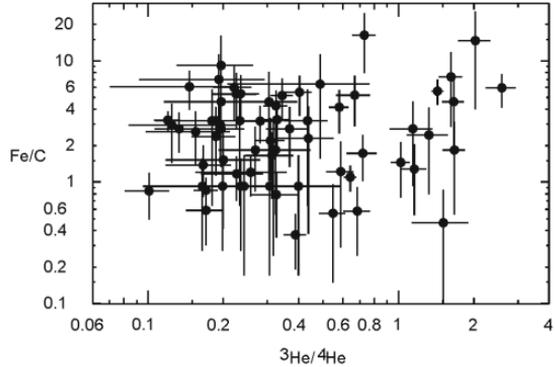

The average enhancements of the elements from $^4$He through Fe were summarized by Reames (1995, 1999) as seen in Fig. 4.6.

**Fig. 4.6** Average abundance enhancements of elements in impulsive SEP events vs. *Q/A* at 3.2 MK as of 1995 as shown by Reames (1999 © Springer).

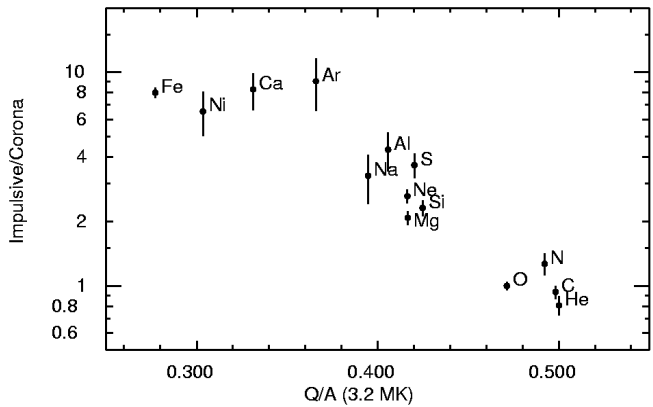

It was suggested by Reames, Meyer, and von Rosenvinge (1994) that the grouping of the enhancements of $^4$He, C, N, and O occurs because C, N, and O, are fully ionized, like $^4$He, and thus have $Q/A = 0.5$. Ions in the group from Ne – S have closed shells of 2 orbital electrons and $Q/A \approx 0.4$. This occurs at a temperature of about $3 – 5$ MK, as we shall see. The observation that $^4$He and C are the same supports the idea that $A/Q = 2$ for both, and matter traversal is excluded since the dependence cannot be $A/Q^2$.

## 4.5 Abundances for 34 ≤ Z ≤ 82

Beginning with the launch of the *Wind* spacecraft late in 1994, abundances of elements in the remainder of the periodic table well above Fe started to become available on a regular basis (Reames 2000). Although resolution of individual elements was not possible, the pattern of enhancement of element groups gave a new perspective to the term "enhancement" as the high-*Z* elements approached 1000-fold enhancements, comparable with those of $^3$He. Subsequently, two completely different instrument techniques yielded: i) the abundances vs. *A* at 0.1 – 1.0 MeV amu$^{-1}$ up to *A* = 200 (Mason et al. 2004) and ii) the abundances vs. *Z* at 3.3 – 10 MeV amu$^{-1}$ up to *Z* ≈ 82 (Reames and Ng 2004). Both are seen in Fig. 4.7.

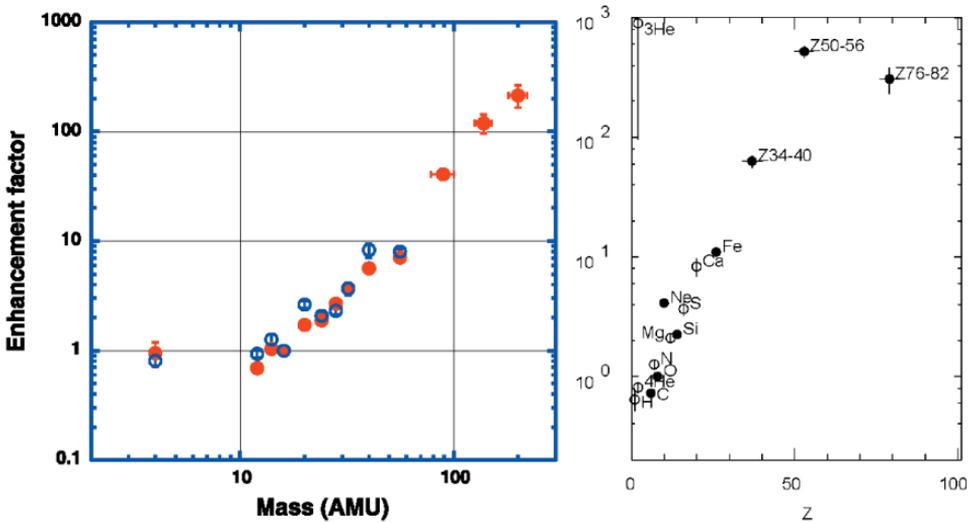

**Fig. 4.7** Enhancements relative to solar system and coronal abundances are extended to high masses at 0.1 – 1.0 MeV amu$^{-1}$ in the *left panel* (*red*, Mason et al. 2004 © AAS) and to high *Z* at 3.3 – 10 MeV amu$^{-1}$ in the *right* (Reames and Ng 2004 © AAS). Open symbols are from Reames (1995).

Reference abundances used for Fig. 4.7 are solar system abundances for the red symbols in the left panel and solar system abundances corrected for FIP (see Chap. 8) to simulate coronal abundances in the right panel.

## 4.6 Power-Law Enhancements in A/Q: Source-Plasma Temperatures

The atomic physics that describes ionization states as a function of plasma temperature *T* has been studied for many years. Fig. 4.8 shows *A/Q* vs. *T*, based on *Q* vs. *T* from atomic physics, which was used by Reames, Cliver, and Kahler (2014a) to determine the appropriate value of *T* for the power-law fit for impulsive SEP events shown in Fig. 4.9. Values of *Q* vs. *T* below Fe are from Arnaud and Rothenflug (1985), Fe is from Arnaud and Raymond (1992) and elements in the high-*Z* region from Post et al. (1977).

**Fig. 4.8.** *A/Q* is shown as a function of equilibrium temperature for various elements (*left panel*) and is enlarged for low *Z* (*right panel*). Data on elements below Fe are from Arnaud and Rothenflug (1985), Fe from Arnaud and Raymond (1992) and sample element data in the high-*Z* region from Post et al. (1977). The region used for Fe-rich impulsive SEP events is *shaded*.

**Fig. 4.9.** The mean enhancement in the abundances of elements in impulsive SEP events relative to reference gradual SEP events is shown as a function of *A/Q* of the element at ~3 MK. For the least-squares fit line shown in the figure the enhancement varies as the 3.64 ± 0.15 power of *A/Q* (Reames, Cliver, and Kahler 2014a, © Springer).

Reames, Cliver, and Kahler (2014a) found a temperature that was somewhat lower than in earlier work. They noted that *A/Q* for Ne is higher than that for Mg or Si in this region $T \leq 3.0$ MK and would help explain the observation that, in the impulsive event averages, Ne/O > Mg/O > Si/O (see right panel of Fig. 4.8). Also, *A/Q* for O was beginning to approach $\approx 2.2$ in the region; this would help explain the "He-poor" events observed with low $^4$He/O. Finally, *A/Q* values in the 2.5 – 3.2-MK region fit the enhancements in the elements with $Z \geq 34$ quite well.

It is also possible to determine a best-fit temperature and a power-law fit for individual impulsive SEP events. Each impulsive event has measured enhancements for the elements and each temperature in a region of interest has its own pattern of *A/Q*. We fit the enhancements vs. *A/Q* for *each* temperature, note the values of $\chi^2$ for each fit, and then choose the fit, and temperature, with the minimum $\chi^2$. Values of $\chi^2$ vs. *T* are shown for 111 impulsive events in Fig. 4.10. The number of events with minima at each temperature is listed along the *T* axis. The right panel shows the spread and magnitude of the enhancements.

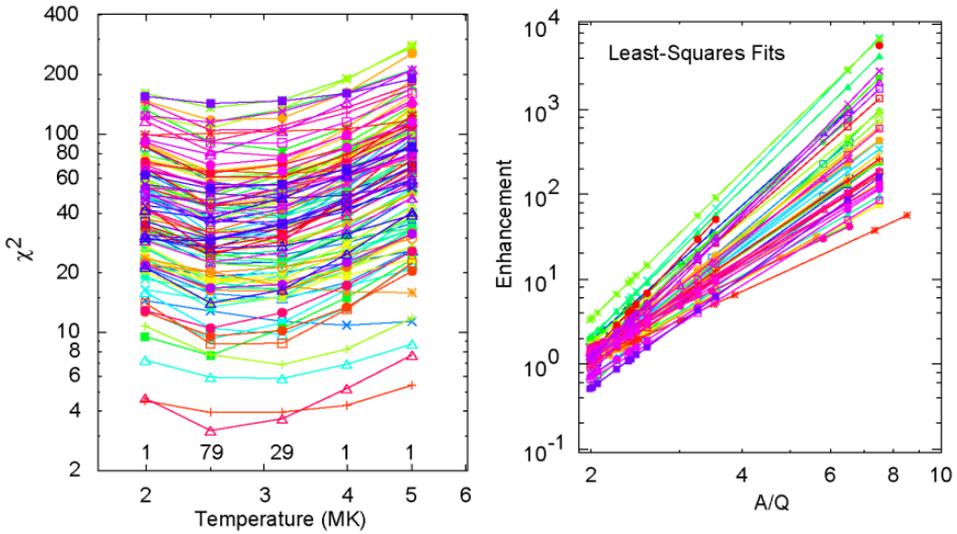

**Fig. 4.10.** The *left panel* shows $\chi^2$ vs. *T* for all 111 impulsive SEP events using different *colors* and *symbols* for each event. The number of events with $\chi^2$ minima at each temperature is shown along the *bottom* of the panel. The *right panel* shows the distribution of the fits (Reames, Cliver, and Kahler 2014b, © Springer)

It is noteworthy that temperatures of 2.5 – 3.2 MK we find for impulsive SEP events are quite appropriate for solar jets from active regions (see Fig. 14 of Raouafi et al. 2016). The larger jets are associated with active regions.

Fig. 4.11 shows fits of enhancement vs. *A/Q* obtained for six individual impulsive SEP events. The event numbers listed with the onset time in each panel correspond to the numbers in the published list of Reames, Cliver, and Kahler (2014a) to provide continuity. In performing the least-squares fits for these impulsive events, errors used to determine weighting factors have been increased by 20% of the values, in quadrature with the statistical errors (Reames, Cliver, and Kahler 2014b). The physical origin of this 20% variation observed in the abundances, presumably local, is not entirely understood, but its necessity is clear.

Deviations of individual abundances of Ne, Mg, S, and Ca above the fit lines can be seen sometimes in Fig. 4.11. In these cases, C, O, and Fe may fall below to balance the fit. In some other events larger excursions are seen, especially for Ne.

These variations are not understood, but resonant enhancements at specific values of *A/Q* are a possibility or, more likely, just local abundance variations of the solar corona. The observed variations do not appear to be variations in the FIP bias of the underlying coronal material (see Chap. 8) but may be local variations in individual element abundances.

Note that, if these ions at this energy of $3 - 5$ MeV amu$^{-1}$ had traversed a significant amount of material during or prior to acceleration, all the elements from He up to Ne, Mg, and Si would be fully ionized with $A/Q \approx 2.0$, which is not consistent with observed power-law enhancements.

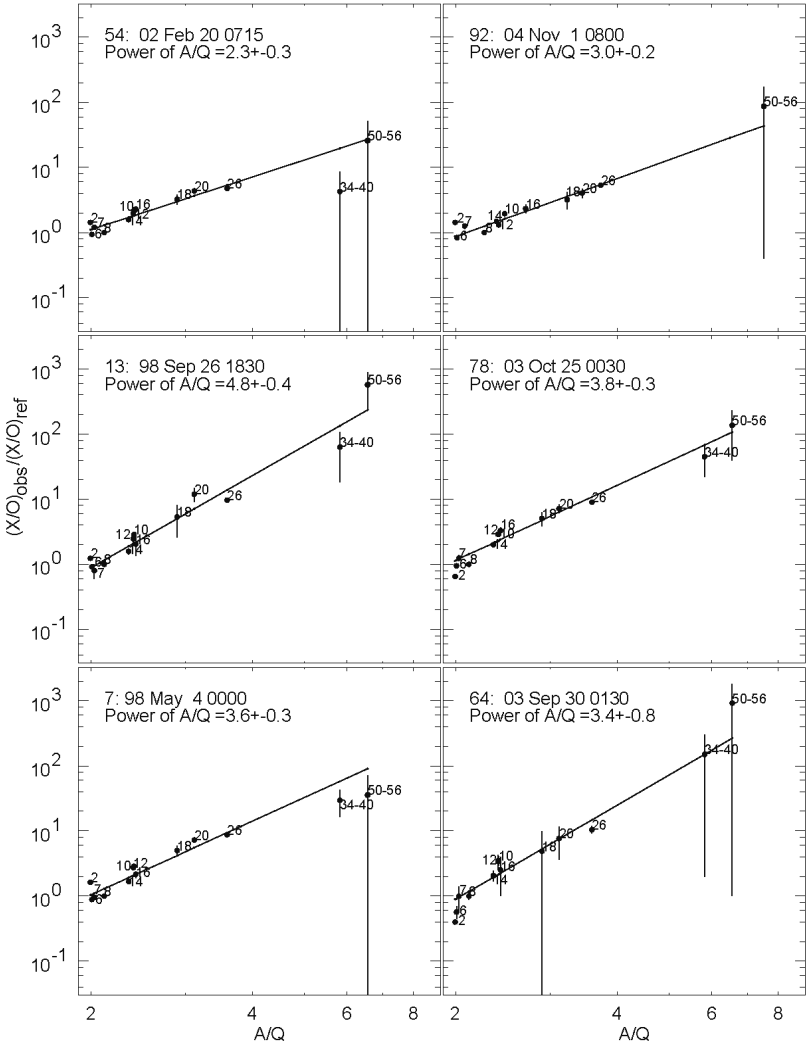

**Fig. 4.11.** Observed element enhancements relative to corresponding reference values vs. best-fit values of *A/Q* (at 2.5 or 3.2 MK) are shown for six individual impulsive SEP events with event numbers (see text), onset times, and power of *A/Q* shown. Individual points in the panels are labeled with the atomic number *Z* of the element (Reames 2019, © Springer).

For the most part, abundance variations in successive SEP events in a sequence do not seem to be correlated, as seen in the example in Fig. 4.12. However, the He abundance may be an exception.

**Fig. 4.12**. The *lower panel* shows the time history of several species during two events, 34 and 35, in March 2000. The *upper panels* show fits to the relative abundances vs. *A/Q*. Excursions, such as Ne in the first event and Ar in the second are not shared. However, a large suppression of He is seen in both events denoted by the arrows (Reames 2019 © Springer).

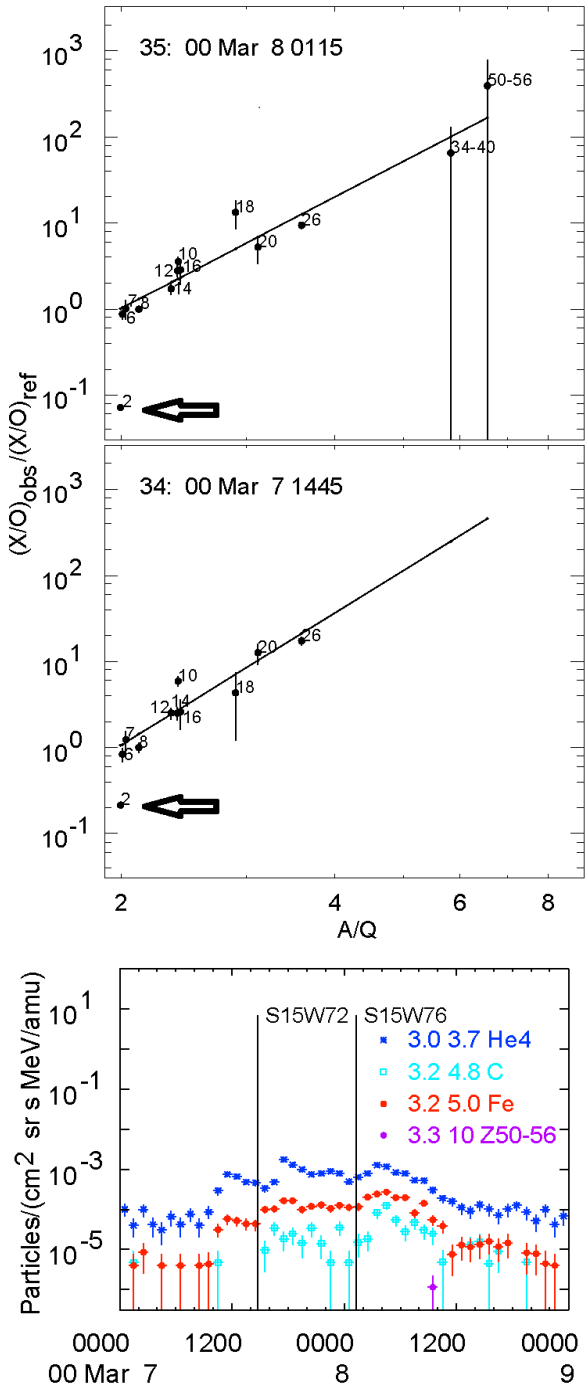

The element He, with the highest value of FIP = 24.6 eV, may be the last element to become ionized on its transit across the solar chromosphere and into the corona. Thus its ion abundance may lag behind those of other high-FIP elements. Presumably this condition could apply throughout a region where multiple impulsive SEP events occur. However, for the events shown in Fig. 4.12, He/C (or He/O) is suppressed by a factor of 5 in the earlier event and a factor of 10 in the later one. Except for coincidence, there is no known reason that the suppression of He should increase with time. However, in these "He-poor" events He/C often decreases with energy and He/O can be as low as $\approx$2 at $\approx$8 MeV amu$^{-1}$ (Reames 2019). $^3$He/$^4$He is often high in these He-poor events, perhaps partly because the denominator is reduced, but, of course, $^3$He and $^4$He have the same value of FIP. He-poor events are not well understood but they may be the only FIP-dependent variation from event to event (see Chap. 8). If the He-poor SEP events are caused by incomplete FIP processing, which occurs in the chromosphere, then plasma from any CME from the same jet might also be He-poor. The pattern of enhancements for $Z \geq 6$ suggests (minimum $\chi^2$) a source temperature of $\approx$3 MK where He and C both have $A/Q = 2$, so how else can we decrease He/C so greatly?

Are there any impulsive SEP events outside the region 2 – 4 MK? Reames, Cliver, and Kahler (2015) could find only a few new events outside 2.5 – 3.2 MK by relaxing the requirement for high Fe/O, and none elsewhere. However, Mason, Mazur, and Dwyer (2002a) did find a small $^3$He-rich event with enhanced N that may have had a temperature of < 1.5 MK, but this event was not even visible above 1 MeV amu$^{-1}$ so it must have been a very small event with a very steep energy spectrum. Thus, impulsive SEP events outside solar active regions are rare and very small (but see also Sect. 4.8).

## 4.7 Associations: CMEs, Flares, and Jets

While gradual SEP events are associated with fast, wide CMEs, impulsive SEP events are associated with smaller, slower, and especially *narrow* CMEs. Are these just extremes on a continuum? Probably not, since narrow CMEs from solar jets involve plasma motion along $B$ and may be less likely to produce shocks, while wide ones are from extensive eruptive events that can drive plasma perpendicular to $B$ and produce strong shocks (e.g. Vršnak and Cliver 2008). Nearly 70% of impulsive Fe-rich SEP events in a recent study (Reames, Cliver, and Kahler 2014a) have associated CMEs with the properties shown in Fig. 4.13. We often associate impulsive SEP events with flares with the proper timing; flares identify the location of an active region. Flares are carefully tabulated while nearby jets, where the SEPs actually originate, are not. Narrow, slower CMEs are, in fact, associated with jets (Kahler, Reames, and Sheeley 2001; Bučík et al. 2018a, 2018b). Sources of $^3$He-rich events have been reviewed recently by Bučík (2020).

**Fig. 4.13.** Properties of the impulsive-SEP-associated CMEs and flares are as follows: flare longitude (*top*), CME width and speed, and the CME-SEP delay (Reames, Cliver, and Kahler 2014a © Springer). The median speed is 597 km s$^{-1}$ vs. 408 km s$^{-1}$ for all CMEs and 1336 km s$^{-1}$ for gradual SEP events (Yashiro et al., 2004). The average transport delay from CME launch to SEP onset is 2.7 h. From type III onset to SEP onset is 2.3 h, corresponding to a path length of ~1.4 AU, which suggests average pitch cosine, $<\mu> \approx 0.8$ or a complex path like that in Fig. 3.10.

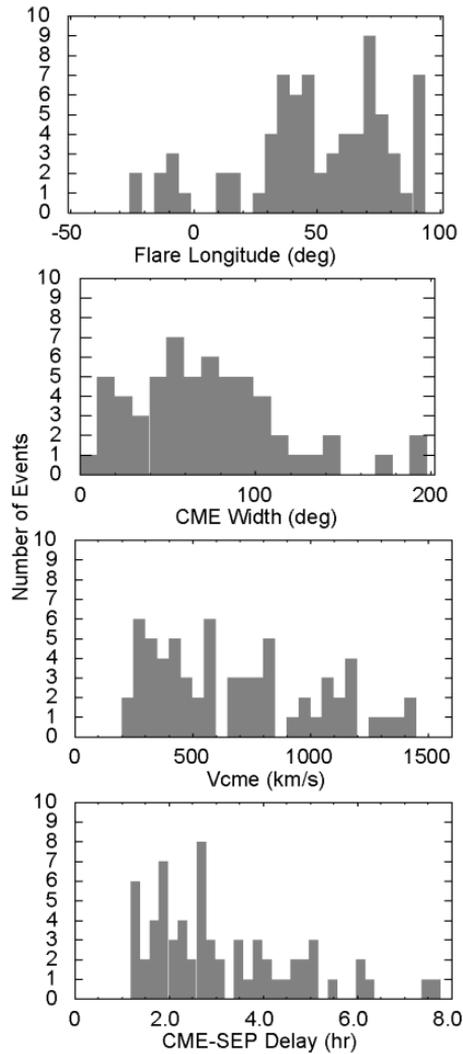

Any correlation between impulsive-SEP abundances and CMEs is difficult to quantify, but Fig. 4.14 shows enhancements, relative to He, vs. $Z$ for individual events with measured $Z \geq 50$ ions, where the symbols denote the CME width. The events with the greatest enhancements have small, narrow CMEs or no visible CME. Yashiro et al. (2004) previously examined small $^3$He-rich SEP events with no CME and found associated brightness changes and "coronal anomalies" that were probably small CMEs that were too faint to qualify to be cataloged – i.e. they were below threshold. The evidence in Fig. 4.14 suggests that the smaller the CME the greater the enhancements. The smallest events tend to have both suppressed He/O and enhanced ($Z \geq 50$)/O.

**Fig. 4.14.** Enhancements relative to He are shown vs. Z for impulsive SEP events. Symbol sizes indicate the associated CME width (Reames, Cliver, and Kahler 2014a © Springer).

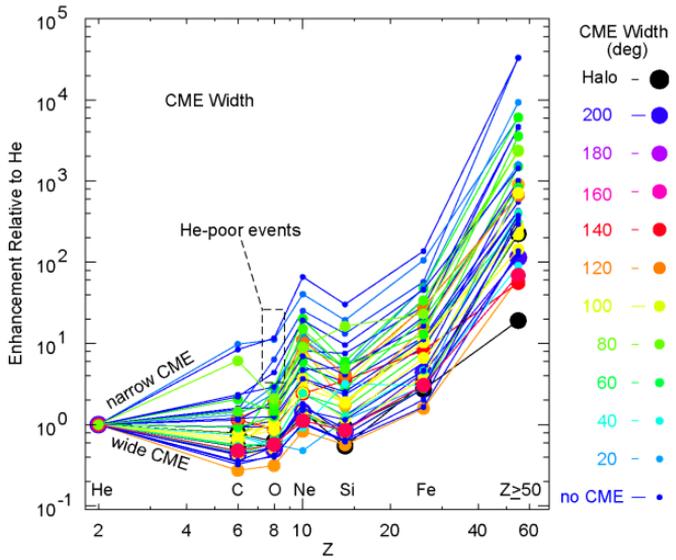

When abundance enhancements are displayed as a function of the GOES soft X-ray peak intensity as in Fig. 4.15, the smaller B- and C-class events have steeper abundance variations than the brighter M- and X-class X-ray events. The smaller events are also more likely to have large $^3$He/$^4$He ratios (not shown, see Reames, Cliver, and Kahler, 2014b).

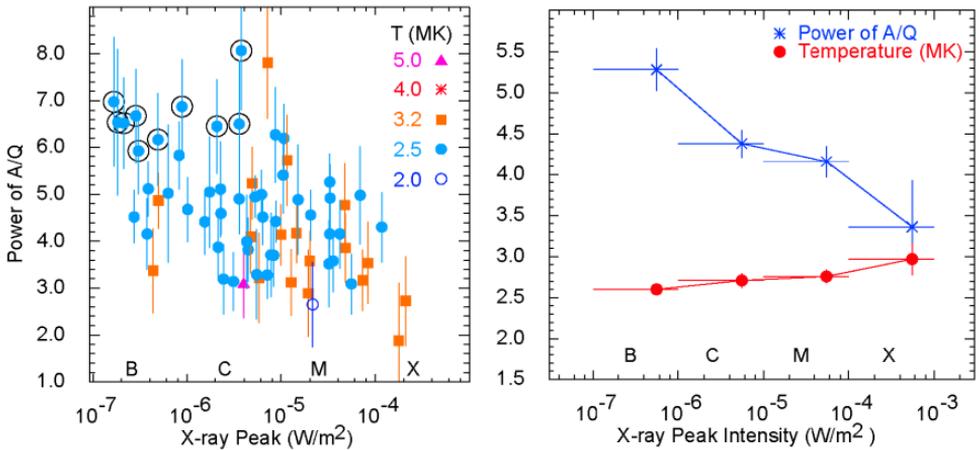

**Fig. 4.15.** The *left panel* shows the power of $A/Q$ vs. the GOES soft X-ray peak intensity (and "CMX" class) for individual SEP events with temperature as a *symbol* and *color*. *Circled* events are "He-poor" with low $^4$He/O caused only partly by increased $A/Q$ for O. The *right panel* shows variation of the mean temperature and power of $A/Q$ within each soft X-ray class (after Reames, Cliver, and Kahler 2014b © Springer).

As we have said often, flares involve hot *closed* loops that do not produce SEPs in space. Jets are reconnection events involving *open* field lines so that plasma and any energetic particles accelerated will be ejected in the diverging field. Thus, it is not at all clear why there should be *any* correlation of SEP properties with flare heating in the neighboring closed loops of an associated flare, let alone the inverse correlation seen in Fig. 4.15. Apparently it suggests that more-modest reconnection in smaller jets, which may also have minimal flaring, are more likely to produce stronger and steeper abundance enhancements. There is flaring from closing field lines in jets and evaporation of hot plasma from the base of field lines; X-rays may come from hotter regions that differ from the reconnection and SEP source.

The cartoon in Fig. 4.16 illustrates the basic mechanism behind a jet produced when new magnetic flux (blue) emerges, pressing into oppositely directed open magnetic field (black). The time evolution of an isolated jet is shown in Fig. 4.17.

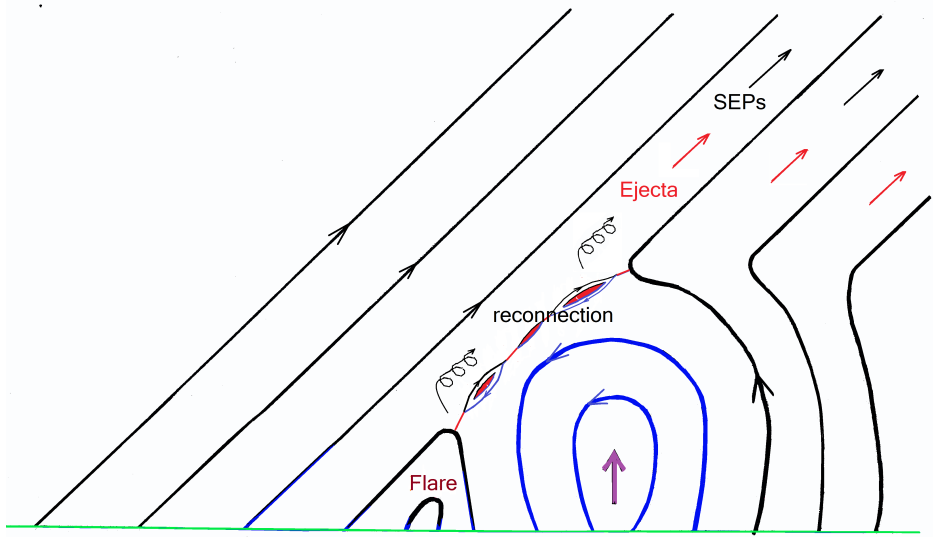

**Fig. 4.16.** A jet is produced when newly emerging magnetic flux (*blue*) reconnects with oppositely directed field (*black*) in the *red* region. The reconnection region is not a uniform surface but forms islands of reconnection. Energetic particles and plasma can escape toward the *upper right*; an enclosed flaring region forms at the *lower left*. (see Reames 2002). Real jets can be much more complex, involving twisted fields, etc. (see review by Raoufi et al. 2016)

The drawing in Fig. 4.16 shows a reconnection region formed when emerging magnetic flux is oppositely directed from the overlying flux. The magnetic reconnection region is non-uniform and forms islands of reconnection. Oppositely-directed fields do not perfectly cancel, and a small out-of-plane "guide field" may stabilize the reconnection. Particle-in-cell simulations show Fermi acceleration of ions reflected back and forth from the ends of the collapsing islands of reconnection (Drake et al. 2009). If the distribution of the widths $w$ of islands of reconnection is $P(w) \sim w^{-\alpha}$, then the rate of production of energetic ions with mass $A$ and charge $Q$ is

$$dN/dt \sim w_{th}^{3-\alpha} \sim (A/Q)^{\alpha-3} \tag{4.1}$$

where $w_{th}$ is the threshold for ion heating which typically occurs when $A/Q > 5$ (Drake et al. 2009). However, we see enhancements all the way from $A/Q = 1$.

The theory of ion (other than $^3$He, covered in Sect. 2.5.2) and electron acceleration in flares is extensive (see Ramaty 1979; Steinacker, Jackel, and Schlickeiser 1993; Miller et al. 1997; Miller 1998). Stochastic acceleration of ions by an arbitrary wave spectrum is common and the balance between acceleration and Coulomb losses (e.g. Mason and Klecker 2018) has been considered for nearly 40 years (Ramaty 1979). However, the observation that $^4$He and C are both unenhanced and have $A/Q = 2$ at $T \approx 3$ MK seems to argue against a dependence upon $A/Q^2$ that would be appropriate for traversal of material. We tend to favor the particle-in-cell results (Drake et al. 2009) because they seem more directly related to reconnection and jets.

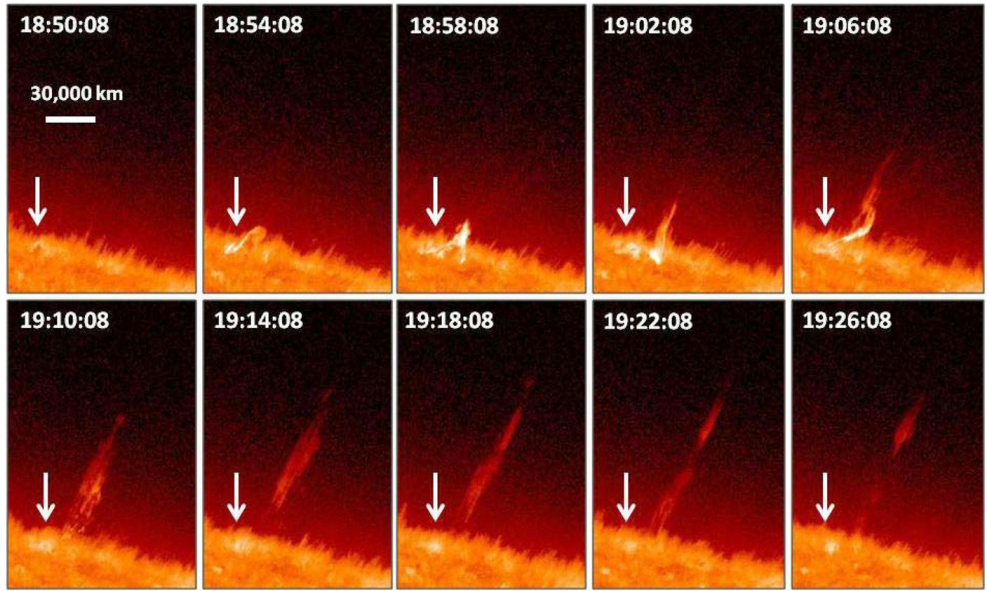

**Fig. 4.17** Eruption of an isolated jet on 11 August 2010 observed as a function of time on SDO/AIA of He II at 304 Å by Moore, Sterling, and Falconer (2015 © AAS)

X-ray properties of jets were described by Shimojo and Shibata (2000) and they were associated with impulsive SEP events by Kahler, Reames, and Sheeley (2001). X-ray jets were also previously associated with type III radio bursts which provide the streaming electrons that may generate the EMIC waves needed for $^3$He enhancements (Temerin and Roth 1992; Roth and Temerin 1997). The narrow CMEs associated with two impulsive SEP events are shown in Fig. 4.18.

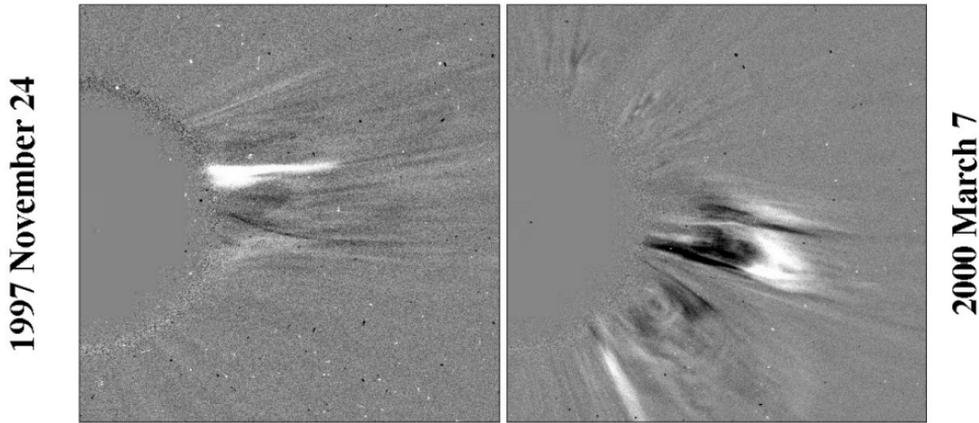

**1997 November 24**

**2000 March 7**

**Fig. 4.18**. Difference images show change in intensity in the LASCO C2 coronagraph for narrow CMEs associated with impulsive SEP events (Kahler, Reames, and Sheeley 2001 © AAS)

It has also been possible to trace the magnetic field lines from Earth to the Sun to locate the sources of events (Wang, Pick, and Mason 2006; Nitta et al. 2006). Much more sophisticated models of jets have also evolved that show reconnection associated with untwisting of axial field lines and generation of Alfvén waves (Moore et al. 2013; Lee, Archontis and Hood 2015). However, such models do not yet include any related particle acceleration (see review of solar jets by Raouafi et al. 2016) although observations now clearly associate $^3$He-rich SEP events directly with helical jets (Bučík et al. 2018a, 2018b; Bučík 2020). These SEP events are often associated with active-region coronal holes (Innes et al. 2016), i.e. open field lines extending out of an active region. Active-region jets tend to have temperatures of ~3 MK (see Fig. 14 of Raoufi et al. 2016) like those we deduce from impulsive SEP events (Fig. 4.10).

Recently, Paraschiv and Donea (2019; Paraschiv 2018) have associated recurrent solar jets with coronal "geysers", magnetic structures that are observed to generate as many as a dozen individual jets over a period of a day or so. These jets are found to produce the electron beams that generate type-III radio bursts (Sect. 2.2) which we associate with impulsive SEP events (e.g. Sect. 2.5.2). Geysers may be involved in the long periods of $^3$He-rich, Fe-rich suprathermal ions that are observed (Fig. 2.8). It is well known that multiple impulsive SEP events commonly occur together on this time scale (see e.g. Fig. 4.2). Magnetic flux emergence is still regarded as an important trigger mechanism for solar jets (Paraschiv, Donea, and Leka 2020).

The studies of ion acceleration in islands of magnetic reconnection come from particle-in-cell simulations discussed above (Drake et al. 2009; Knizhnik, Swizdak, and Drake 2011; Drake and Swizdak 2012). This acceleration produces a strong power-law dependence on $A/Q$ and provides the most promising explanation of the element abundance enhancements in impulsive SEP events appropriate to solar jets. However, $^3$He enhancements seem to require a separate explanation involving resonant waves (e.g. Temerin and Roth 1992).

## 4.8. Can We Have it Both Ways?

It seems suspicious to derive ³He enhancements from one mechanism and heavy-element enhancements from another – in the same SEP event. Do both really contribute?

Recently, Mason et al. (2016) found 16 ³He-rich events in 16 years with extremely high S/O abundances in the 0.4 – 1.0 MeV amu⁻¹ interval. Most of these events are too small and their spectra too steep to be measurable above 1.0 MeV amu⁻¹, the few we can measure show no significant anomalies. Properties of the most extreme event of 16 May 2014 are shown in Fig. 4.19.

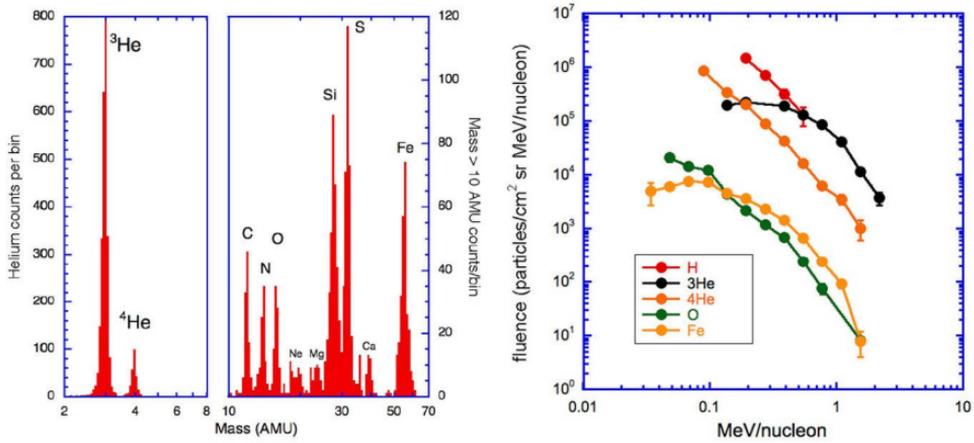

**Fig. 4.19.** Mass histograms of the 16 May 2014 ³He-rich event are shown in the *left two panels* and some corresponding energy spectra are shown in the *right panel* (Mason et al. 2016 © AAS).

This event has ³He/⁴He = 14.88±1.36 and S/O = 1.14±0.12, although the spectra show that these abundances are not constant but vary with energy. The shapes of the spectra of Si, S, and Fe are similar to that of ³He, strongly suggesting that these elements are all accelerated or modified by the same mechanism.

Mason et al. (2016) consider a wide range of plasma temperatures above 0.4 MK. However, most models of resonant wave-particle acceleration (e.g. Fisk 1978; Roth and Temerin 1997) suggest that heavy elements might resonate with the same waves as ³He, but through the second harmonic of their gyrofrequency. Since $A/Q$ =1.5 for ³He, naively we might expect Si and S to be accelerated when they have $A/Q \approx 3.0$, if the resonance is broad enough for us to use an average value of $A/Q$. For S, this average value occurs at 2.0 MK and for Si near 1.5 MK, both reasonable temperatures that might exist near the fringes of active regions. Solar jets near active regions have temperatures around 3 MK while those from coronal holes are nearer 1.5 MK (see Fig. 14 of Raouafi et al. 2016). Jets from coronal holes might also be based upon different FIP-dependent coronal abundances like those in the solar wind where both C and S behave like a low-FIP elements (see Sects. 8.4 and 8.5; Reames 2020). This increased factor of ~4 in S/O would only partly explain the observations, but the elevated C/O may be a clue.

However, the greatly enhanced S does not seem to be supported at 2.0 MK by Roth and Temerin (1997) either. Present theory of $^3$He involves resonance of ions with electron-beam-generated EMIC waves; direct wave generation in the reconnection region is also a possibility that has not yet been explored.

The enhancements of other elements with less-rounded spectra might result from the power-law in *A/Q* more-commonly produced at higher energies by magnetic reconnection (e.g. Drake et al. 2009). It seems possible that the two mechanisms may compete to dominate different species at different energies in different impulsive SEP events, but a clear picture of the physics is still elusive and the degree of enhancement of all species may not be easily accommodated.

The understanding of the relative roles of the reconnection and the resonant wave mechanisms is the largest outstanding problem in the physics of flares, jets, and impulsive SEP events. Nature has tempted us with two huge 1000-fold enhancements, in $^3$He and in heavy elements. Neither one is subtle. Yet they seem to be unrelated and we are unable to incorporate their explanations into a single physical model that can tell us which will dominate and when.

## 4.9 Nuclear Reactions: Gamma-ray Lines and Neutrons

It may seem incongruous to discuss γ-ray-line events in a chapter on impulsive SEP events since γ-ray lines have not been observed in small events or jets. Line emission is observed in large flares from the de-excitation of nuclei produced in nuclear reactions that occur when ions, accelerated on closed coronal loops, are scattered into the loss cone and plunge into the higher-density corona (e.g. Ramaty and Murphy 1987; Kozlovsky, Murphy, and Ramaty 2002). Hard (>20 keV) X-rays, common in flares, are produced by non-relativistic electrons but X-rays tell us nothing about accelerated ions; only γ-ray lines can help here. Protons, undergoing nuclear reactions with C, O, and Fe, for example, produce narrow γ-ray spectral lines in the region of ~0.5 – 7 MeV, whose relative intensities can be used to measure abundances of elements in the corona. Energetic heavy ions in the "beam" interacting with protons in the corona produce Doppler-broadened spectral lines that measure abundances in the accelerated beam, i.e. the flare equivalent of our SEPs. Fig. 4.20, from Murphy et al. (1991) compares observed and calculated γ-ray spectra in the large event of 27 April 1981.

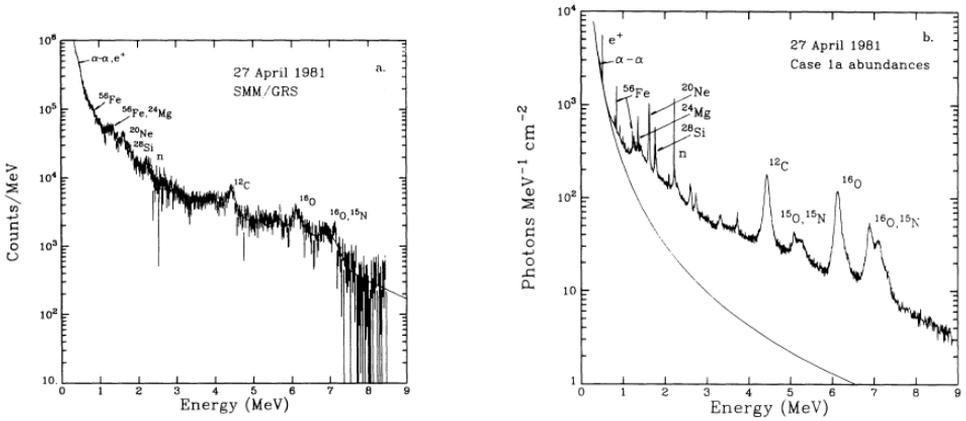

**Fig. 4.20.** An observed γ-ray spectrum (**a**) is compared with a calculated spectrum (**b**). Relative strengths of the lines are determined by element abundances in the corona and in the accelerated "beam" (Murphy et al. 1991 © AAS).

These lines suggest that the accelerated ions in large flares are both $^3$He-rich (Mandzhavidze, Ramaty, and Kozlovsky 1999) and Fe-rich (Murphy et al. 1991). The γ-ray lines excited by nuclear reactions of a $^3$He beam differ from those produced by a $^4$He beam, allowing measurement of both isotopes. The presence of enhanced $^3$He/$^4$He in the accelerated "beam" in these events has been further confirmed recently by Murphy, Kozlovsky, and Share (2016). Thus it seems that flares and jets might accelerate ions in the same way and γ-ray-line spectroscopy could contribute to our understanding of that process. Unfortunately, the masks used to improve position resolution in modern spectroscopy missions, which operate rather like a pin-hole camera, block too many γ-rays to permit γ-ray-line spectroscopy on modern spacecraft.

Nuclear reactions in the corona also produce $^2$H, $^3$H, positrons, π-mesons, and isotopes of Li, Be, and B as inferred from the γ-ray-line spectroscopy. However, as we re-emphasize, none of these secondary products (except γ-rays and neutrons) has been observed into space. Apparently the secondary ions are magnetically trapped on the loops where they are created, suggesting that the primary ions that produced them were similarly trapped.

Evidence of a spatially extended γ-ray source and long-duration γ-ray events are discussed in Sect. 5.7. We will see that the >100 MeV γ-rays are produced by >300 MeV protons accelerated by CME-driven shock waves (Plotnikov, Rouillard, and Share 2017; Share et al. 2018).

Neutrons are also produced in nuclear reactions in solar flares and 50 – 300 MeV neutrons have been observed directly in space (Chupp et al. 1982; Chupp 1984). Neutrons decay into a proton, electron, and neutrino with a 10-min half life and neutron-decay protons of 5 – 200 MeV have also been measured, allowing a neutron spectrum to be calculated (e.g. Evenson et al. 1983, 1990). Neutron-decay protons are best measured for eastern solar events where they can be measured on field lines that are much less accessible to protons directly from the shock source, which slowly increase later.

## 4.10. Open Questions

This section suggests open questions that might be addressed by future research.

1. If most $A/Q$ enhancements come from magnetic reconnection and $^3$He enhancements come from wave-particle interactions, where and when do these two acceleration mechanisms fit into models of solar jets? What parameters control their relative contributions?

2. Why do the impulsive SEP events with the lowest intensities, smallest associated flares and narrowest CMEs have the greatest enhancements of $^3$He/$^4$He and the heavy-element abundance enhancements with the steepest power of $A/Q$? To what extent can all the $^3$He event observations be explained by source depletion of the enhanced species? Surely not Fe-enhancements?

3. Measurements by a spacecraft near the Sun, like PSP, could improve SEP onset timing by removing the blurring effect of scattering during transport. What is the duration and time profile of impulsive events at 10 s to 1 min resolution and how does it compare with X-ray, γ-ray, and type-III-burst timing? Note that intensities may vary as $\sim r^{-3}$, providing greatly improved statistics nearer the Sun for smaller events, but anisotropies may be high. Do the electron and ion sources differ? What is the relative timing of $^3$He and $^4$He?

4. For a spacecraft near the Sun, the $\sim r^{-3}$ intensity increase would allow observation of many more small impulsive SEP events from smaller jets. What is the size distribution? Are there microjets and nanojets supplying many impulsive suprathermal ions during times that are otherwise quiet?

5. Discrete ionization states affect the assignment of source-plasma temperatures. $^{12}$C$^{+5}$ is enhanced but $^{12}$C$^{+6}$ is not; treating $Q$ as 5.5 is approximate. $A/\langle Q \rangle$ is not the same as $\langle A/Q \rangle$. Then there is $^{13}$C which is always enhanced. Can we improve the estimates of $T$?

6. Can $^{22}$Ne enhancement explain the occasional extra enhancements of Ne?

7. What causes the large enhancements of S in some small impulsive events? Are the events connected to active regions?

8. What causes the occasional large suppression in the He-poor events?

## References

Arnaud, M., Raymond, J., Iron ionization and recombination rates and ionization equilibrium, Astrophys. J. **398**, 394 (1992) doi: 10.1086/171864

Arnaud, M., Rothenflug, R., An updated evaluation of recombination and ionization rates, Astron. Astrophys. Suppl. **60**, 425 (1985)

Bučík, R, $^3$He-rich solar energetic particles: solar sources, Space Sci. Rev. **216** 24 (2020) doi: 10.1007/s11214-020-00650-5

Bučík, R., Innes, D.E., Mason, G.M., Wiedenbeck, M.E., Gómez-Herrero, R., Nitta, N, [3]He-rich solar energetic particles in helical jets on the sun, Astrophys. J. **852** 76 (2018a) doi: 10.3847/1538-4357/aa9d8f

Bučík, R., Wiedenbeck, M.E., Mason, G.M., Gómez-Herrero, R., Nitta, N.V., Wang, L., [3]He-rich solar energetic particles from sunspot jets, Astrophys. J. Lett. **869** L21 (2018b) doi: 10.3847/2041-8213/aaf37f

Chupp, E.L., High-energy neutral radiations from the sun, Ann. Revs Astron. Astrophys. **22**, 359 (1984) doi: 10.1146/annurev.aa.22.090184.002043

Chupp, E.L., Forrest, D.J., Ryan, J.M., Heslin, J., Reppin, C., Pinkau, K., Kanbach, G., Rieger, E., Share, G.H., A direct observation of solar neutrons following the 0118 UT flare on 1980 June 21, Astrophys, J. Lett., **263**, L95 (1982) doi: 10.1086/183931

Cook, W.R., Stone, E.C., Vogt, R.E., Elemental composition of solar energetic particles, Astrophys. J. **279**, 827 (1984) doi: 10.1086/161953

Drake, J.F., Cassak, P.A., Shay, M.A., Swisdak, M., Quataert, E., A magnetic reconnection mechanism for ion acceleration and abundance enhancements in impulsive flares, Astrophys. J. Lett. **700**, L16 (2009) doi: 10.1088/0004-637X/700/1/L16

Drake, J.F., Swisdak, M., Ion heating and acceleration during magnetic reconnection relevant to the corona, Space Sci. Rev. **172**, 227 (2012) doi: 10.1007/s11214-012-9903-3

Evenson, P. Kroeger, R., Meyer, P., Reames, D., Solar neutron decay proton observations in cycle 21, Astrophys. J. Suppl. **73**, 273 (1990) doi: 0.1086/191462

Evenson, P., Meyer, P., Pyle, K.R, Protons from the decay of solar flare neutrons, Astrophys. J. **274**, 875 (1983) doi: 10.1086/161500

Fisk, L.A., [3]He-rich flares - a possible explanation, Astrophys. J. **224**, 1048 (1978) doi: 10.1086/156456

Innes, D.E., Bučík, R., Gao, L.-J., Nitta, N., Observations of solar X-ray and EUV jets and their related phenomena, Astronomische Nachrichten **337**, 1024 (2016) doi: 10.1002/asna.201612428

Kahler, S.W., Reames, D.V., & Sheeley, Jr., N.R., Coronal mass ejections associated with impulsive solar energetic particle events, Astrophys. J., **562**, 558 (2001) doi: 10.1086/323847

Knizhnik, K., Swisdak, M., Drake, J.F., The acceleration of ions in solar flares during magnetic reconnection, Astrophys. J. Lett. **743**, L35 (2011) doi: 10.1088/2041-8205/743/2/L35

Kozlovsky, B., Murphy, R.J., Ramaty, R., Nuclear deexcitation gamma-ray lines from accelerated particle interactions, Astrophys. J. Suppl. **141** 523 (2002) doi: 10.1086/340545

Lee, E. J., Archontis, V., Hood, A. W., Helical blowout jets in the sun: untwisting and propagation of waves, Astrophys. J. Lett. **798**, L10 (2015) doi: 10.1088/2041-8205/798/1/L10

Liu, S., Petrosian, V., Mason, G.M., stochastic acceleration of [3]He and [4]He in solar flares by parallel-propagating plasma waves: general results, Astrophys. J. **636**, 462 (2006) doi: 10.1086/497883

Mandzhavidze, N., Ramaty, R., Kozlovsky, B., Determination of the abundances of subcoronal [4]He and of solar flare-accelerated [3]He and [4]He from gamma-ray spectroscopy, Astrophys. J. **518**, 918 (1999) doi: 10.1086/307321

Mason, G.M., [3]He-rich solar energetic particle events, Space Sci. Rev. **130**, 231 (2007) doi: 10.1007/s11214-007-9156-8

Mason, G.M., Klecker, B., A possible mechanism for enriching heavy ions in [3]He-rich solar energetic particle events, Astrophys. J. **862** 7 (2018) doi: 10.3847/1538-4357/aac94c

Mason, G.M., Mazur, J.E., Dwyer, J.R., A new heavy ion abundance enrichment pattern in [3]He-rich solar particle events, Astrophys. J. Lett. **565**, L51 (2002a) doi: 10.1086/339135

Mason, G.M., Mazur, J.E., Dwyer, J.R., Jokipii, J.R., Gold, R.E., Krimigis, S.M., Abundances of heavy and ultraheavy ions in [3]He-rich solar flares, Astrophys. J. **606**, 555 (2004) doi: 10.1086/382864

Mason, G.M., Nitta, N.V., Wiedenbeck, M.E., Innes, D.E., Evidence for a common acceleration mechanism for enrichments of [3]He and heavy ions in impulsive SEP events, Astrophys. J. **823**, 138 (2016) doi: 10.3847/0004-637X/823/2/138

Mason, G.M., Reames, D.V., Klecker, B., Hovestadt, D., von Rosenvinge, T.T., The heavy-ion compositional signature in He-3-rich solar particle events, Astrophys. J. **303**, 849 (1986) doi: 10.1086/164133

Mason, G.M., Wiedenbeck, M.E., Miller, J.A., Mazur, J.E., Christian, E.R., Cohen, C.M.S., Cummings, A.C., Dwyer, J.R., Gold, R.E., Krimigis, S.M., Leske, R.A., Mewaldt, R.A., Slocum, P.L., Stone, E.C., von Rosenvinge, T.T., Spectral properties of He and heavy ions in ³He-rich solar flares, Astrophys, J. **574** 1039 (2002b) doi: 10.1086/341112

Miller, J.A., Particle acceleration in impulsive solar flares, Space Sci. Rev. **86**, 79 (1998) doi: 10.1023/A:1005066209536

Miller, J.A., Cargill, P.J., Emslie, A.G., Holman, G.D., Dennis, B.R., LaRosa, T.N., Winglee, R.M., Benka, S.G., Tsuneta, S., Critical issues for understanding particle acceleration in impulsive solar flares, J. Geophys. Res. **102**, 14631 (1997) doi: 10.1029/97JA00976

Moore, R.L., Sterling, A.C., Falconer, D.A., Robe, D., The cool component and the dichotomy, lateral expansion, and axial rotation of solar X-ray jets, Astrophys. J., **769**, 134 (2013) doi: 10.1088/0004-637X/769/2/134

Moore, R.L., Sterling, A.C., Falconer, D.A., Magnetic untwisting in solar jets that go into the outer corona in polar coronal holes, Astrophys. J. **806** 11 (2015) doi: 10.1088/0004-637X/769/2/134

Murphy, R.J., Kozlovsky, B., Share, G.H., Evidence for enhanced ³He in flare-accelerated particles based on new calculations of the gamma-ray line spectrum, Astrophys. J., **833** 196 (2016) doi: 10.3847/1538-4357/833/2/196

Murphy, R.J., Ramaty, R., Kozlovsky, B., Reames, D.V., Solar abundances from gamma-ray spectroscopy: Comparisons with energetic particle, photospheric, and coronal abundances, Astrophys. J. **371**, 793 (1991) doi: 10.1086/169944

Nitta, N.V., Reames, D.V., DeRosa, M.L., Yashiro, S., Gopalswamy, N., Solar sources of impulsive solar energetic particle events and their magnetic field connection to the Earth, Astrophys. J. **650**, 438 (2006) doi: 10.1086/507442

Paraschiv, A.R., On long timescale recurrent active region coronal jets: the coronal geyser structure, Ph. D. thesis, Monash Univ. (2018) doi: 10.26180/5bc9d76627396

Paraschiv, A.R., Donea, A., On solar recurrent coronal jets: coronal geysers as sources of electron beams and interplanetary type-III radio bursts, Astrophys. J. **873**, 110 (2019) doi: 10.3847/1538-4357/ab04a6

Paraschiv, A.R., Donea, A., Leka, K.D., The trigger mechanism of recurrent solar active region jets revealed by the magnetic properties of a coronal geyser site, Astrophys, J., **891**, 149 (2020) doi: 10.3847/1538-4357/ab7246

Post, D.E., Jensen, R.V., Tarter, C.B., Grasberger, W.H., Lokke, W.A., Steady-state radiative cooling rates for low-density, high temperature plasmas, At. Data Nucl. Data Tables **20**, 397 (1977) doi: 10.1016/0092-640X(77)90026-2

Plotnikov, I., Rouillard, A., Share, G., The magnetic connectivity of coronal shocks to the visible disk during long-duration gamma-ray flares, Astron. Astrophys. **608** 43 (2017) doi: 10.1051/0004-6361/201730804 (arXiv:1703.07563)

Ramaty, R., Energetic particles in solar flares, AIP Conf. Proc. **56**, 135 (1979) doi: 10.1063/1.32074

Ramaty, R., Murphy, R.J., Nuclear processes and accelerated particles in solar flares, Space Sci. Rev. **45**, 213 (1987) doi: 10.1007/BF00171995

Raouafi, N.E., Patsourakos, S., Pariat, E., Young, P.R., Sterling, A.C., Savcheva, A., Shimojo, M., Moreno-Insertis, F., DeVore, C.R., Archontis, V, et al., Solar coronal jets: observations, theory, and modeling, Space Sci. Rev. **201** 1 (2016) doi: 10.1007/s11214-016-0260-5 (arXiv:1607.02108)

Reames, D.V., Bimodal abundances in the energetic particles of solar and interplanetary origin, Astrophys. J. Lett. **330**, L71 (1988) doi: 10.1086/185207

Reames, D.V., Coronal Abundances determined from energetic particles, Adv. Space Res. **15** (7), 41 (1995)

Reames, D.V.: Particle acceleration at the Sun and in the Heliosphere, Space Sci. Rev., **90**, 413 (1999) doi: 10.1023/A:1005105831781

Reames, D.V., Abundances of trans-iron elements in solar energetic particle events, Astrophys. J. Lett. **540**, L111 (2000) doi: 10.1086/312886

Reames, D.V., Magnetic topology of impulsive and gradual solar energetic particle events, Astrophys. J. Lett. **571**, L63 (2002) doi: 10.1086/341149

Reames, D.V., Helium suppression in impulsive solar energetic-particle events, Sol. Phys. 294 32 (2019) doi: 10.1007/s11207-019-1422-x  (arXiv: 1812.01635)

Reames, D. V., Four distinct pathways to the element abundances in solar energetic particles, Space Science Rev. **216** 20 (2020) doi: 10.1007/s11214-020-0643-5

Reames, D.V., Ng, C.K., Heavy-element abundances in solar energetic particle events, Astrophys. J. **610**, 510 (2004) doi: 10.1086/421518

Reames, D.V., Cliver, E.W., Kahler, S.W., Abundance enhancements in impulsive solar energetic-particle events with associated coronal mass ejections, Sol. Phys. **289**, 3817, (2014a) doi: 10.1007/s11207-014-0547-1 (arXiv: 1404.3322 )

Reames, D.V., Cliver, E.W., Kahler, S.W., Variations in abundance enhancements in impulsive solar energetic-particle events and related CMEs and flares, Sol. Phys. **289**, 4675 (2014b) doi: 10.1007/s11207-014-0589-4

Reames, D.V., Cliver, E.W., Kahler, S.W., Temperature of the source plasma for impulsive solar energetic particles, Sol. Phys. 290, 1761 (2015) doi: 10.1007/s11207-015-0711-2

Reames, D.V., Meyer, J.P., von Rosenvinge, T.T., Energetic-particle abundances in impulsive solar flare events, Astrophys. J. Suppl. **90**, 649 (1994) doi: 10.1086/191887

Roth, I., Temerin, M., Enrichment of $^3$He and heavy ions in impulsive solar flares, Astrophys. J. **477**, 940 (1997) doi: 10.1086/303731

Serlemitsos, A.T., Balasubrahmanyan, V.K., Solar particle events with anomalously large relative abundance of $^3$He, Astrophys. J. 198, **195**, (1975) doi: 10.1086/153592

Share, G.H., Murphy, R.J., White, S.M., Tolbert, A.K., Dennis, B.R., Schwarz, R.A., Smart, D.F., Shea, M.A., Characteristics of late-phase >100 MeV γ-ray emission in solar eruptive events, Astrophys. J. **869** 182 (2018) doi: 10.3847/1538-4357/aaebf7

Shimojo, M., Shibata, K., Physical parameters of solar X-ray jets, Astrophys. J. **542**, 1100 (2000) doi: 10.1086/317024

Steinacker, J., Jaeckel, U., Schlickeiser, R., Ion acceleration in impulsive solar flares, Astrophys J. **415**, 342 (1993) doi: 10.1086/173168

Temerin, M., Roth, I., The production of $^3$He and heavy ion enrichment in $^3$He-rich flares by electromagnetic hydrogen cyclotron waves, Astrophys. J. Lett. **391**, L105 (1992) doi: 10.1086/186408

Vršnak, B., Cliver, E.W., Origin of Coronal Shock Waves, Sol. Phys. **253** 215 (2008) doi: 10.1007/s11207-008-9241-5

Wang, Y.-M., Pick, M., Mason, G.M., Coronal holes, jets, and the origin of $^3$He-rich particle events, Astrophys. J., **639**, 495 (2006) doi: 10.1086/499355

Yashiro, S., Gopalswamy, N., Cliver, E.W., Reames, D.V., Kaiser, M., Howard, R., Association of coronal mass ejections and type II radio bursts with impulsive solar energetic particle event, In: Sakurai, T., Sekii, T. (eds.) *The Solar-B Mission and the Forefront of Solar Physics*, ASP Conf. Ser. 325, 401 (2004)

# Chapter 5. Gradual SEP Events

**Abstract**   Gradual solar energetic-particle (SEP) events are "big proton events" and are usually much more "gradual" in their decay than in their onset.   As their intensities increase, particles streaming away from the shock amplify Alfvén waves that scatter subsequent particles, increasing their acceleration, eventually limiting ion flow at the "streaming limit."   Waves generated by higher-speed protons running ahead can also throttle the flow of lower-energy ions, flattening spectra and altering abundances in the biggest SEP events.   Thus, we find that the $A/Q$-dependence of scattering causes element-abundance patterns varying in space and time, which define source-plasma temperatures $T$, since the pattern of $Q$ values of the ions depends upon temperature.   Differences in $T$ explain much of the variation of element abundances in gradual SEP events.   In nearly 70% of gradual events, SEPs are shock-accelerated from ambient coronal plasma of ~0.8–1.6 MK, while 24% of the events involve material with $T \approx 2$–4 MK re-accelerated from residual impulsive-suprathermal ions with pre-enhanced abundances.   This source-plasma temperature can occasionally vary with solar longitude across the face of a shock.   Non-thermal variations in ion abundances in gradual SEP events reaccelerated from the 2–4 MK impulsive source plasma are reduced, relative to those in the original impulsive SEPs, probably because the accelerating shock waves sample a pool of ions from multiple jet sources.   Late in gradual events, SEPs become magnetically trapped in a *reservoir* behind the CME where spectra are uniform in space and decrease adiabatically in time as the magnetic bottle containing them slowly expands.   Finally, we find variations of the He/O abundance ratio in the source plasma of different events.

We begin by showing proton intensities in the classic large gradual SEP event of 4 November 2001 in Fig. 5.1.   This event, from a source longitude of W17 on the Sun, has the typical time profile of a centrally located event (see Sect. 2.3.3).   The figure lists phases of the event along the abscissa, which we will study in approximate time order, although onsets were discussed previously in Sect. 3.1.

In impulsive SEP events, most particles traveled to us scatter free so we had little need to discuss transport.   With increased intensities, protons from gradual SEP events generate or amplify their own spectrum of resonant Alfvén waves for pitch-angle scattering, which complicates their transport more and more as intensities increase.   In fact, it is the resonant waves, generated by the out-flowing particles, which scatter subsequent particles back and forth across the shock, incrementally increasing ion velocity, driving particles to higher and higher energy.

For recent reviews of gradual SEP events see Desai and Giacalone (2016), and Lee, Mewaldt, and Giacalone (2012).   For theoretical background see Parker (1963) and Jones and Ellison (1991).

**Fig. 5.1** Proton intensities vs. time from the NOAA/GOES satellite are shown for the large gradual SEP event of 4 November 2001 at solar longitude W19 (compare Fig. 2.2). Distinctive event phases are listed along the *abscissa* (Reames 2013 © Springer).

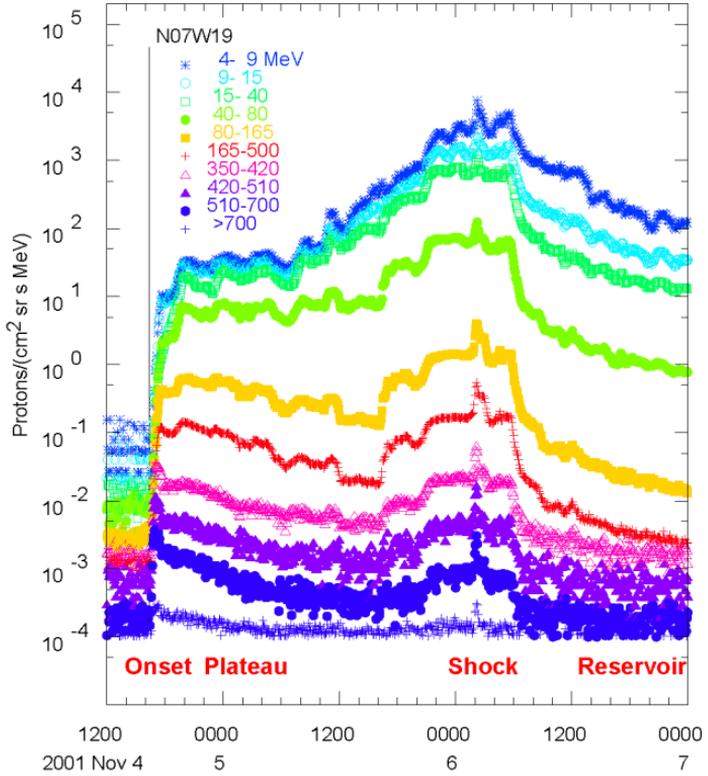

## 5.1 Parallel Transport

### 5.1.1 Diffusive Transport

The diffusion of particles of type X and velocity $v$ by pitch-angle scattering with scattering mean free path $\lambda_X$ with a power-law dependence on radial position $r$ as $\lambda_0 r^\beta$ varies as (Parker 1963; see Equation C1 in Ng, Reames, and Tylka 2003)

$$n_X(r,t) = \frac{1}{4\pi\Gamma(\varepsilon)}\left(\frac{\varepsilon}{3}\right)^{2\varepsilon-1}\left(\frac{3}{\lambda_0 v t}\right)^\varepsilon \exp\left[\frac{-3r^{(2-\beta)}}{(2-\beta)^2 \lambda_0 v t}\right] \qquad (5.1)$$

where $\varepsilon = 3/(2-\beta)$ and $\beta$ must be less than 2.

If we examine the ratio of species X and Y, where $\lambda$ is a power of rigidity $P$ and where $L = \lambda_X / \lambda_Y = R^\alpha = ((A_X/Q_X)/(A_Y/Q_Y))^\alpha$, as a result of the rigidity dependence of $\lambda$ and $\tau = 3r^{2-\beta}/[\lambda_Y(2-\beta)^2 v]$, remembering that log x $\approx 1-1/x$, it can easily be shown (e.g. Reames 2016a, b) that

$$X/Y = L^{-\varepsilon}\exp[(1-1/L)\tau/t] \approx L^{\tau/t-\varepsilon} \qquad (5.2)$$

The ratio in Eq. 5.2 is the enhancement or suppression relative to that ratio at the SEP source and does not include any pre-enhanced impulsive suprathermal ions, although those are also power-law in form. Thus, relative abundances vary approximately as *a power of A/Q*. This will prove to be important in determining source-plasma temperatures (Sect. 5.6). If the ratio $R > 1$, as for Fe/O, the abundance ratio, X/Y begins at infinity and falls asymptotically to $R^{-\alpha\varepsilon}$. Ratios begin at infinity because diffusion does not account for the particle transit time at the onset. Breneman and Stone (1985) observed that element abundance enhancements were power laws in *A/Q*, rising with *A/Q* in some SEP events and falling in others as we saw in Fig. 2.5 in Sect. 2.5.1. In standard diffusion theory, scattering does *not* change with time; thus, the waves affect the particles, but the particles have no affect on the waves (defying energy conservation).

### 5.1.2 Wave Growth

The amplification of Alfvén waves by streaming protons has been discussed in textbooks on plasma physics for many years (e.g. Stix 1962, 1992; Melrose 1980; see also Ng, Reames, and Tylka 2003; Rice, Zank, and Li 2003; Li, Zank, and Rice 2005). In quasi-linear theory, ions, streaming along the magnetic field *B*, resonate with Alfvén waves of wave number *k*:

$$k \approx \frac{B}{\mu P} \tag{5.3}$$

*in the rest frame of the waves*. Here $P=pc/Qe$ is the rigidity of a particle of charge *Qe*, and momentum *p*, and μ is the cosine of its pitch angle relative to *B*.

Equation 5.3 results from quasi-linear theory (QLT) where particles are assumed to orbit the unperturbed field and the electric field vector of the resonant circularly-polarized Alfvén wave rotates so as to maintain its phase relative to the direction of rotation of the gyrating particle. This resonance maximizes the transfer of energy between the wave and the particle, seen as pitch-angle scattering in the rest frame of the wave, or wave frame, approximately the plasma rest frame.

The growth rate of the σ polarization mode of Alfvén waves (see Ng, Reames, and Tylka 2003; Stix 1992; Melrose 1980) produced by protons is clearest and simplest in the wave frame, where it is given by

$$\gamma_\sigma(k) = 2\pi^2 g_\sigma e^3 c V_A \iint d\mu dP \frac{P^3}{W^2} R_{\mu\mu}^\sigma \frac{\partial f_H^\pm}{\partial \mu} \tag{5.4}$$

where $g_\sigma = \pm 1$ for outward (inward) wave direction and $f_H^\pm$ is the proton phase-space density in each corresponding wave frame. Here *W* is the total proton energy, and $R_{\mu\mu}^\sigma$ is the resonance function (see Ng and Reames 1995; Ng, Reames and Tylka 2003) that imposes the resonance condition (Eq. 5.3) while allowing for resonance broadening near $\mu \approx 0$. Resonance broadening overcomes the limitation of QLT that prevents scattering across $\mu = 0$ (Ng and Reames 1995). If we can ig-nore the effects of slow propagation of the waves, then the wave intensity of the $\sigma$ mode, $I_\sigma(k,r,t)$ obeys the simple equation

$$\frac{\partial I_\sigma(k,r,t)}{\partial t} = \gamma_\sigma(k,r,t) I_\sigma(k,r,t) \tag{5.5}$$

also in the wave frame, where we have explicitly shown the dependence upon space $r$ and time $t$, which may be quite significant. We will see that the pitch-angle diffusion coefficient for protons depends linearly upon the intensity of resonant waves (Sect. 5.1.3). Equation 5.5 was used by Ng and Reames (1994) to study time-dependent wave growth during proton transport that was quantitatively consistent with the streaming limit as we will see in Sect. 5.1.5.

Thus, streaming protons grow the waves, and those waves scatter the subsequent protons to reduce the streaming and the wave growth. This causes the scattering mean free paths to vary in both time and space (Ng, Reames, and Tylka 1999, 2003). While the wave growth caused by heavier ions is negligible, they respond to the waves in ways that are not always obvious, a priori. Waves, grown by protons at a particular value of $\mu P$, resonate with other energies and other species with the same value of $\mu P$, as shown in Eq. 5.3.

Wave growth is commonly combined with quasi-parallel shock acceleration, where scattering is especially important. However, wave growth is entirely a transport phenomenon, its dependence upon the particles is only through $\partial f_H^\pm/\partial\mu$; it is otherwise completely independent of the nature of the proton source. Wave growth will also be important near quasi-perpendicular shocks when streaming intensities of protons become large. This point is sometimes overlooked by students.

Working in the wave frame is illustrative but inconvenient when both inward and outward waves are present and when the Alfvén speed $V_A$ decreases as $r^{-1}$ with distance. Transforming to the plasma frame introduces terms of order $(V_{sw}+g_\sigma V_A)/v$ (see e.g. Ng, Reames, and Tylka 2003)

### 5.1.3 Particle Transport

The equation of particle transport may be simplified in the fixed inertial frame where $f$ is the phase-space density of a given particle species averaged over gyrophase (Roelof 1969; Ng and Reames 1994; Ng, Reames, and Tylka 1999)

$$\frac{\partial f}{\partial t} + \mu v \frac{\partial f}{\partial r} + \frac{1-\mu^2}{r} v \frac{\partial f}{\partial \mu} - \frac{\partial}{\partial \mu}\left( D_{\mu\mu} \frac{\partial f}{\partial \mu} \right) = G \tag{5.6}$$

The third term in Eq. 5.6 represents focusing of the particles in the diverging magnetic field while the fourth term represents pitch-angle scattering with the diffusion coefficient $D_{\mu\mu}$. Here $v$ is the particle speed, $\mu$ is its pitch angle cosine, and the term $G$ on the right-hand side of the equation represents particle sources, for example, it might be a power-law energy spectrum times a delta-function at the radial location of a shock wave.

The diffusion coefficient $D_{\mu\mu}$ is given by

$$D_{\mu\mu} = \frac{v^2}{4P^2} \sum_{\sigma} \int dk I_{\sigma} R_{\mu\mu}^{\sigma} \qquad (5.7)$$

where $P$ is the particle rigidity and $\sigma$ runs over wave modes. The wave intensity $I_{\sigma}$ and the resonance function $R_{\mu\mu}^{\sigma}$ were discussed in the previous section.

The set of Eqs. 5.4–5.7 completely describe the evolution of both particles and waves and their coupling. Equation 5.4 shows that the growth of waves is controlled by the streaming particles and Eq. 5.7 relates the particle scattering to the intensity of waves. *Scattering causes wave growth as a direct consequence of energy conservation* (Ng, Reames, and Tylka 2003, Appendix B)

### 5.1.4 Initial Abundance Ratios

We noted above that in diffusion theory, when $\lambda$ has a power-law dependence on rigidity, hence upon $A/Q$, ratios like Fe/O or He/H begin with large enhancements that decrease with time. While this occurs for small gradual SEP events, Fig. 5.2 shows that He/H can reverse in large SEP events where wave growth becomes important. This is an example of a case where waves that control the arrival of protons depend upon the proton's velocity, but those that affect He, for example, depends upon protons of a much higher velocity, at their common value of $\mu P$; those protons already arrived and generated their waves much earlier.

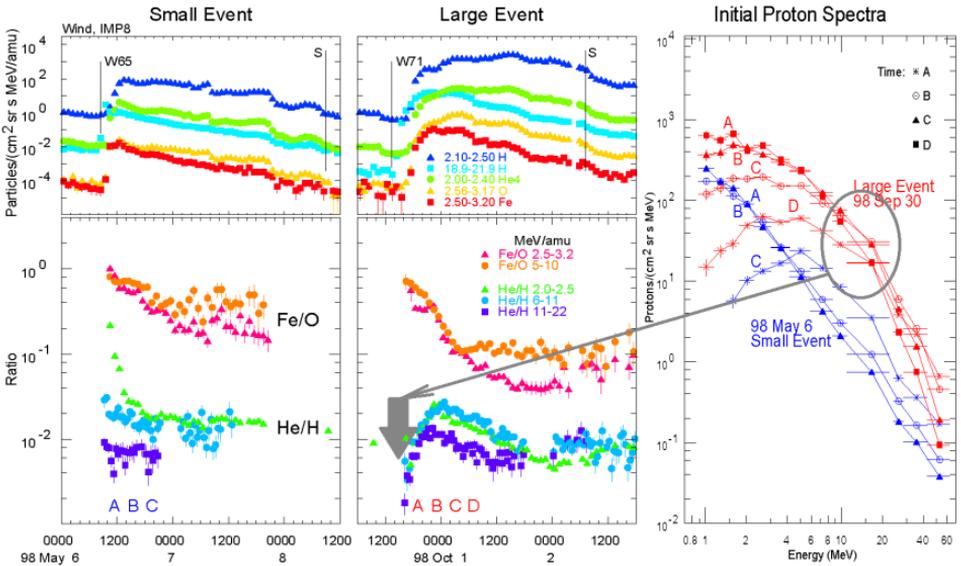

**Fig. 5.2** Particle intensities and abundance ratios are shown for small (*left*) and large (*center*) gradual SEP events (Reames, Ng, and Tylka 2000 © AAS). Initially He/H ratios fall in the small event but rise in the large one. Proton spectra at times A, B, C, and D, (shown in the *right panel*) are much more intense (*gray circle*) in the large October event (*red*), their greater wave generation causing initial suppression of He/H (*gray arrow*) since He in the ratio resonates with waves produced by copious higher-energy protons that arrived much earlier than the H in the ratio.

Why does the initial behavior reverse for He/H in the large event? The early ions stream out into space from the event with $\mu \approx 1$ with few resonant Alfvén waves and little scattering. The H at 2 MeV, for example, has suffered little scattering and is only beginning to make its own resonant waves. He at 2 MeV amu$^{-1}$, however, is scattered by waves that were amplified by 8-MeV protons (same rigidity) that came out much earlier. If the intensity of 8-MeV protons is high (i.e. a big event), they arrive earlier and generate waves so the newly arriving 2-MeV amu$^{-1}$ He will be trapped near the shock since it is scattered much more than the 2-MeV H. Similar logic applies to He/H at higher energies. This effect does not occur for Fe/O since both species are scattered by earlier proton-generated waves. Waves that scatter Fe are coupled to protons of quite high energy, which are less intense, so they actually increase Fe/O initially. The progression of enhancements is modeled by Ng, Reames, and Tylka (2003).

### 5.1.5 The Streaming Limit

In a study of large SEP events observed at the *Helios* spacecraft in solar orbit, Reames (1990) noticed that there was an early plateau period (see Fig. 5.1) during large SEP events near 1 AU, where the proton intensities seemed to have an upper limit of intensity as shown in Fig. 5.3.

**Fig. 5.3** Initial intensities of 3–6 MeV protons are shown overlapped for six large SEP events, all near 1 AU. Intensities do not seem to exceed ~200 (cm$^2$ sr s MeV)$^{-1}$ early in the events, but can become much higher later when shock peaks arrive (Reames 1990 © AAS).

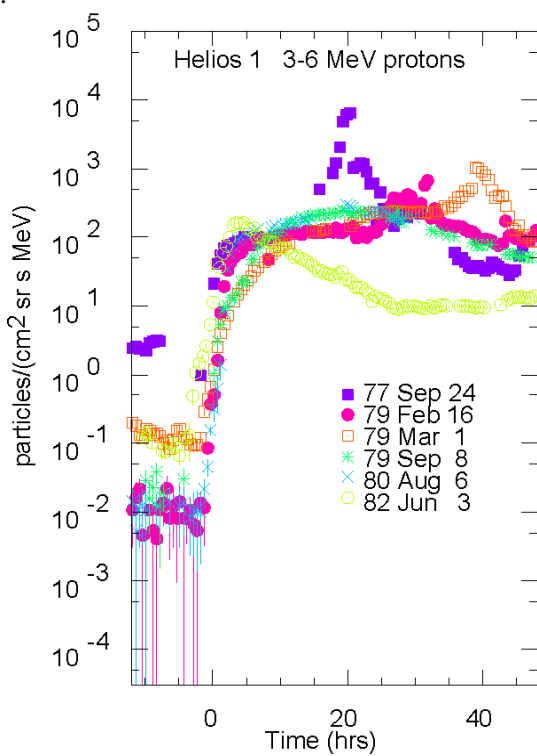

The intensities can rise much higher at the shock peaks, which are at the particle source, because particles at the shock have no net streaming. The streaming limit is a transport phenomenon.

   Imagine an experiment that slowly increases the SEP injection intensity at a source near the Sun. At first, the intensity at 1 AU would increase proportionally. Then, at higher source intensities, wave growth would begin to scatter and trap the particles, with most wave growth near the source where intensities are highest. Eventually, further increasing flow from the source would increase the wave growth and scattering so much that the intensity at 1 AU would no longer increase. This is the "streaming limit" that also emerges from theoretical transport models that include wave growth (e.g. Lee 1983, 2005; Ng and Reames 1994; Ng, Reames, and Tylka 1999, 2003, 2012). The intensity behavior at 1 AU vs. that at the source near the Sun is shown in the left panel of Fig. 5.4 while the right panel shows the spatial dependence caused by increasing injection levels at the source.

   Note that the wave growth depends upon the *absolute* value of the streaming intensity and the parameters shown in Eq. 5.4; there are *no arbitrarily adjustable parameters*. The peak intensity in the left panel of Fig. 5.4 is just over 200 (cm² sr s MeV)⁻¹, similar to the value observed in Fig. 5.3.

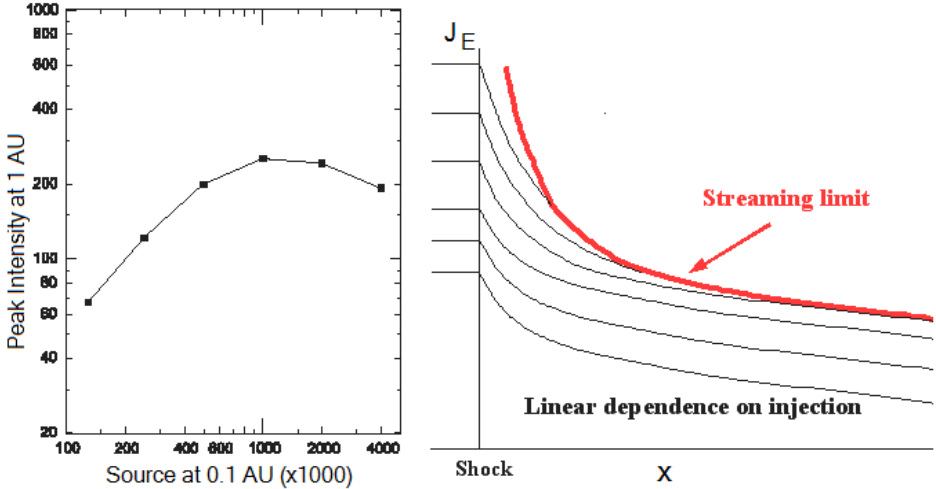

**Fig. 5.4**. The *left panel* shows intensity at 1 AU vs. that at 0.1 AU. The *right panel* shows the spatial variation as the source intensity level is increased with linear behavior at low intensities (see Ng and Reames 1994; Ng, Reames, and Tylka 2003, 2012).

   However, the plateau intensities in the largest gradual SEP events can involve more than just waves that are self-generated by particles of a single energy. They can involve waves generated by higher-energy protons that contribute to the scattering of lower-energy ions by coupling through the μ dependence of Eq. 5.3. These waves preferentially retard the lower-energy particles and flatten the power-law source spectra on the plateau as seen in the left panel of Fig. 5.5. Intense protons of 10–100 MeV stream out early, generating waves as they scatter toward smaller μ. Waves generated at high *P* and low μ resonate with ions of low *P* and μ ≈ 1 which are coming behind more slowly. Thus, waves amplified by protons of 10 MeV at $\mu \approx 0.5$ will scatter protons at 2.5 MeV and $\mu \approx 1$, retarding their flow and thus flattening their spectrum at 1 AU.

Some proof of this mechanism in given by its absence in the 2 May 1998 SEP event; its plateau proton spectrum is shown in the right panel of Fig. 5.5. The spectrum in this event remains a power law since the intensity of 10 MeV protons is two orders of magnitude smaller than that in 28 October 2003. The low intensities of 10–100 MeV protons do not generate enough waves to suppress the low-energy spectrum in the May event. The theoretical fits to these spectra, shown in Fig. 5.5, support this explanation. Wave growth can control spectral shape.

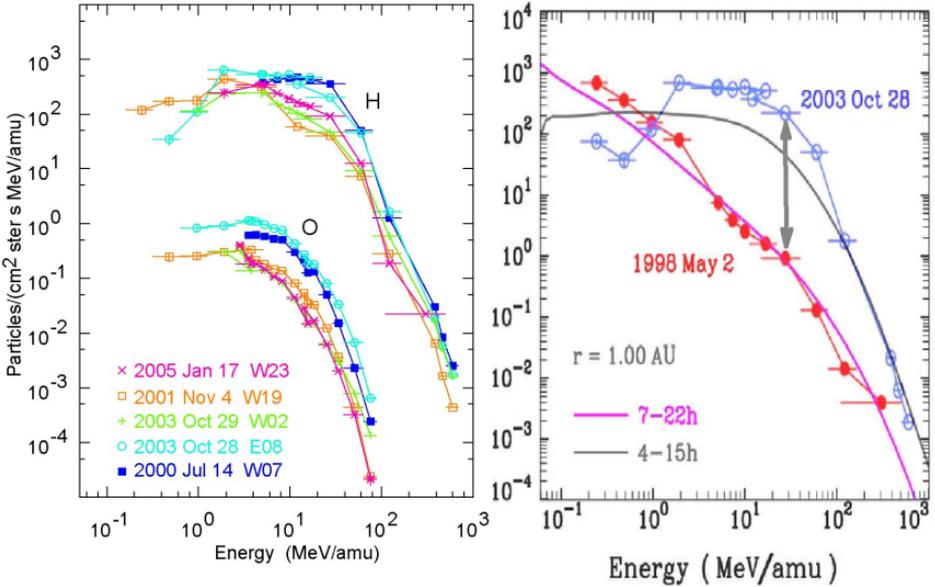

**Fig. 5.5**. The *left panel* shows energy spectra of H and O in five large gradual SEP events (all GLEs) that are flattened at low energies (Reames and Ng 2010 © AAS). The *right panel* shows that the small event of 2 May 1998, with greatly reduced H intensities at 10–100 MeV (*gray arrows*), cannot generate enough waves to suppress lower energies. Model fits to the spectra are shown in *gray* and *purple* curves (see Ng, Reames, and Tylka 2012).

### 5.1.6 Electron Transport

Non-relativistic electrons cannot resonate with Alfvén waves, so they do not participate in much of the physics we have just described. Low-energy electrons usually propagate scatter free with highly-anisotropic angular distributions mainly because of absorption by the solar wind of 0.1–1 Hz frequencies that would resonate with these electrons. Electron spectra often show a break in the ~100-keV energy region. Above the break the spectrum steepens and the width of the angular distribution broadens as scattering becomes much more important (see Tan et al. 2011). It is sometimes erroneously concluded that 1 MeV electrons are accelerated much later than those at 20–50 keV in SEP events; this apparent delay could result from transport rather than acceleration (Strauss and le Roux 2019).

## 5.2 Angular Distributions

Not surprisingly, angular distributions also show the effects of increased scattering when high proton intensities amplify waves. This is seen in the angular distributions of H and He ions in large and small SEP events as shown in Fig. 5.6. The particle intensities remain clustered along the field direction around 180° for more than a day in the angular distributions of the small event on the left in Fig. 5.6 but, in the more intense event on the right, the angular distributions begin to spread in only a few hours.

Of course, the scattering and the wave growth depend upon the initial wave intensity. However, small impulsive and gradual events usually remain scatter-free and angular distributions rapidly isotropize in more-intense gradual events and especially in GLEs (see Reames, Ng, and Berdichevsky 2001). Most SEP events begin nearly scatter free at energies above a few MeV amu$^{-1}$, but not at low energies where $\mu$-coupling shown in Fig. 5.5 applies and traps ions with energies below a few MeV amu$^{-1}$ near their source.

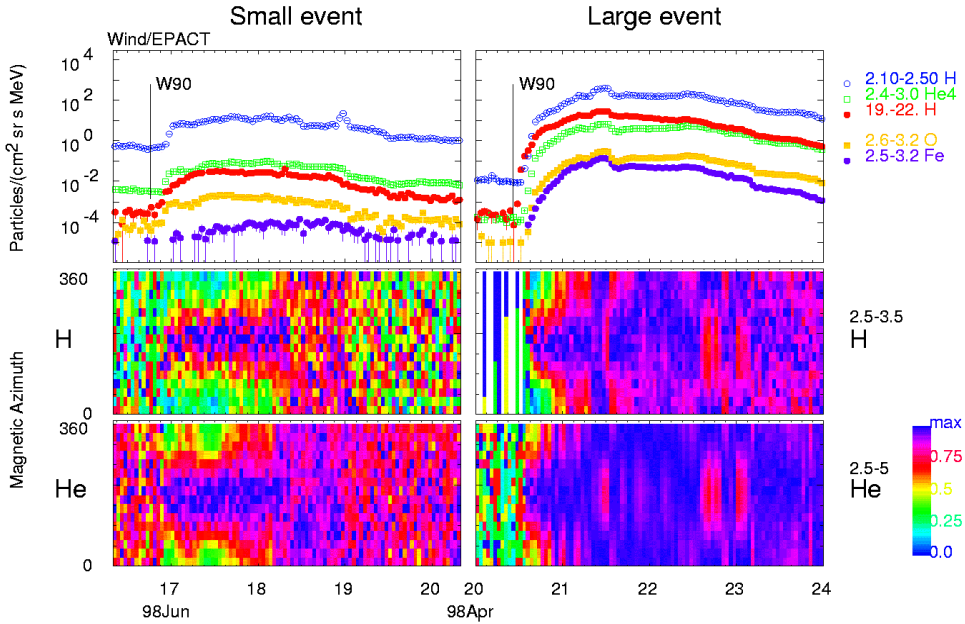

**Fig 5.6**. Intensities (*top*) and angular distributions, relative to **B**, for H (*middle*) and $^4$He (*bottom*) are shown for small (*left*) and large (*right*) gradual SEP events. Note the much higher intensity of the (*red*) 19–22 MeV protons in the *upper right panel*.

## 5.3 Models and Shock Acceleration

General information about shock formation and acceleration may be found in comprehensive review articles (Jones and Ellison 1991; Lee, Mewaldt, and Giacalone 2012; Desai and Giacalone 2016). However, there is such compelling experimental evidence of wave growth in the larger gradual SEP events that we focus on models that include it.

The earliest time-equilibrium model of shock acceleration with self-consistent treatment of particles and waves was the work of Bell (1978a, b) on GCRs, which was subsequently adapted to interplanetary shocks by Lee (1983). Shock models were applied to acceleration of GeV protons in the corona by Zank, Rice and Wu (2000, see also Lee 2005, Sandroos and Vainio 2007, Zank, Li, and Verkhoglyadova 2007; Afanasiev, Batterbee, and Vainio, 2016, 2018).

The time-dependent self-consistent model of particle transport with wave amplification (Ng, Reames, and Tylka 2003) was applied to shock acceleration by Ng and Reames (2008) resulting in modeling of the time-evolution of the proton spectra at the shock shown in Fig. 5.7 along with the evolution of the radial dependence of the intensity upstream of the shock for a given energy proton. A streaming limit soon forms within 0.1 $R_S$ of the shock as seen in the right panel.

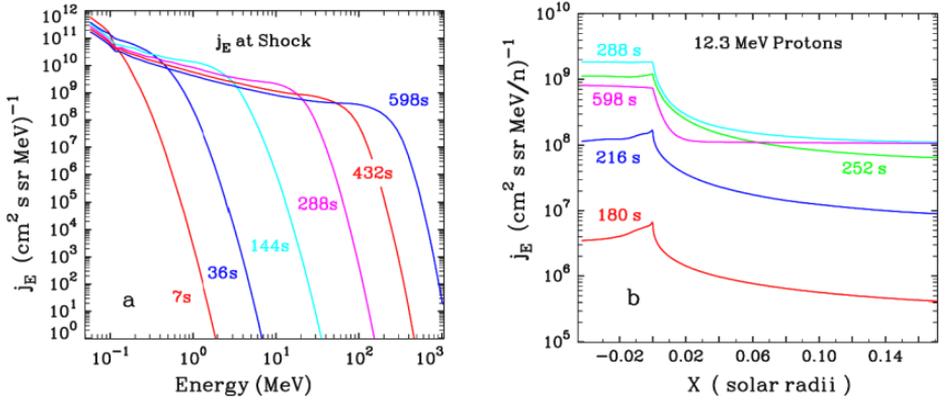

**Fig. 5.7**. The *left panel* shows the time evolution of the proton energy spectrum at the shock for the first ~10 min. The *right panel* shows the time evolution of the spatial distribution of 12.3 MeV protons upstream of the shock. Once accelerated at ~3 min, 12.3 MeV proton intensities form a streaming limit within 0.1 $R_S$ of the shock at ~4.2 min (Ng and Reames 2008 © AAS).

An interesting feature of the time-dependent numerical acceleration calculations is the growth of waves as the proton spectrum grows to higher energy. With the growth of waves that resonate with particles of the highest energy $E_1$ and rigidity $P_1$, some protons will begin to be accelerated to still-higher energy $E_2$ and rigidity $P_2$. Initially, the only waves that can trap ions at $E_2$ are those that resonate with protons with $\mu_2 < P_1/P_2$, i.e. only at small $\mu_2$ can ions at the new energy find resonant waves generated by lower-energy protons. Thus at each new energy the particles begin with a pancake distribution at small $\mu$ (Ng and Reames 2008).

The Ng and Reames (2008) model prevents the scattering from approaching the Bohm limit by requiring that the scattering mean free path be more than three times the particle gyroradius, so that the quasi-linear approximation remains valid. This makes the maximum energy lower and the acceleration rate slower than that in the calculation of Zank, Rice, and Wu (2000), who assumed the more-likely Bohm limit where scattering mean free path equals the particle gyroradius, i.e. $\delta B/B \approx 1$, as has been observed in strong shocks (Lario and Decker 2002; Terasawa et al. 2006). It is also true that an oblique shock, where ions gain energy in the $V_S \times \boldsymbol{B}$ electric field, can affect the acceleration time and maximum energy by increasing the particle energy gained on each traversal of the shock. We would speculate that a fast shock traversing a sufficiently dense seed population should have no trouble accelerating GeV protons in a minute or so.

Ng and Reames (2008) solved the self-consistent time-dependent wave-particle interaction while Zank, Rice, and Wu (2000) used a series of instantaneous steady-state solutions in their PATH (Particle Acceleration and Transport in the Heliosphere) model. The latter simplicity gave a better determination of maximum particle energies of GeV and allowed the inclusion of more-realistic two-dimensional versions of the CME and shock in the PATH models, including oblique shocks (Li et al. 2012) and perpendicular diffusion of the particles caused by random walk of the magnetic fields (Hu et al. 2017, 2018).

It can not escape our attention that it is much easier for theoreticians to work in a universe where particle scattering is constant in time, and waves never grow. Quasi-perpendicular shocks need no change in scattering to increase acceleration, only a small change in $\theta_{Bn}$. Such approximations are useful in making tractable solutions to explore specific functional dependences. However, observations show that wave growth dominates the largest SEP events. Further realistic studies that include it could help advance our understanding of these important events.

## 5.4 Shock Acceleration In Situ

Traveling interplanetary shock waves near Earth are the local continuation of the CME-driven shock waves that produce gradual SEP events. These shocks provide an opportunity to directly measure, in situ, the properties of accelerated particles together with the characteristics of the shock and its driver under an extremely wide variety of shock conditions (see e.g. Berdichevsky et al. 2000). Desai et al. (2003) showed that low-energy ion abundances near the shock peak were much more closely related to ambient abundances of those ions upstream of the shock than to the abundances of the corresponding elements in the solar wind, as might be expected from our discussion of the seed population in Sect. 2.5.3. Desai et al. (2004) found that energy spectra at the shocks were better correlated with the spectra upstream than with those expected from the shock compression ratio. Especially for low-energy ions, shock acceleration persists far out from the Sun and tends to reaccelerate ions from the same population that was accelerated earlier.

The choice of a location to measure the ambient, background, or reference abundances and spectra upstream of the shock is difficult. If it is chosen prior to the time that shock leaves the Sun, perhaps ~2–3 days before the shock arrival, then solar rotation insures that background is sampled at a longitude of 26-40 degrees to the west of the longitude sampled at the shock peak. If it is chosen hours prior to the shock arrival, background will be dominated by particles accelerated earlier by the same shock. Neither choice is ideal.

In effect, the re-acceleration of ions from the seed population found in the reservoir of an earlier event evokes the classical two-shock problem considered, for example, in the review by Axford (1981) and subsequently by Melrose and Pope (1993). Here, the integral equilibrium distribution function $f(p)$ of momentum $p$ of accelerated particles from a shock with compression ratio $s$ is

$$f_a(p) = ap^{-a} \int_0^p dq \ q^{a-1} \phi(q) \tag{5.8}$$

where $a=3s/(s-1)$ and $\phi(p)$ is the injected distribution. If we take $\phi(p)$ as a delta function at $p_0$ we find a power-law spectrum $f_a(p) \sim (p/p_0)^{-a}$ after the first shock. If we reapply Eq. 5.8, injecting $f_a(p)$ into a shock with compression ratio $s'$ and let $b=3s'/(s'-1)$, we find that integrating the power law gives

$$f_{a,b}(p) = \frac{kab}{p_0(b-a)} \left[ \left(\frac{p}{p_0}\right)^{-a} - \left(\frac{p}{p_0}\right)^{-b} \right] \quad \text{for } a \neq b. \tag{5.9}$$

The corresponding intensity is $j(E)=p^2 f(p)$.

Note that Eq. 5.9 is symmetric in the powers $a$ and $b$, and will be dominated by the shape of the hardest, flattest spectrum, either the background (i.e. $a$) or the new shock, $b$. Thus, it is no surprise that one finds local-shock spectra that are dominated by the shape of the upstream background spectrum (Desai et al. 2004; Reames 2012) produced earlier when the shock was stronger. It is quite often that ions with properties from an earlier epoch seem to be only "trapped" at a shock when their intensities are actually being increased by that shock. A further complication occurs when we include a spectral knee with a factor like exp ($-E/E_0$) (e.g. Ellison and Ramaty 1985; see also Mewaldt et al. 2012) to allow for the finite acceleration time. At energies above the knee, observers will find spectra that are much steeper than either the background or the expected equilibrium spectra.

**Fig. 5.8**. Particle intensities are shown vs. time in the *upper left panel* with plasma parameters *below* for shock number 83. Spectra of the shock and background are shown to the *right* with spectral slopes indicated (Reames 2012 © AAS).

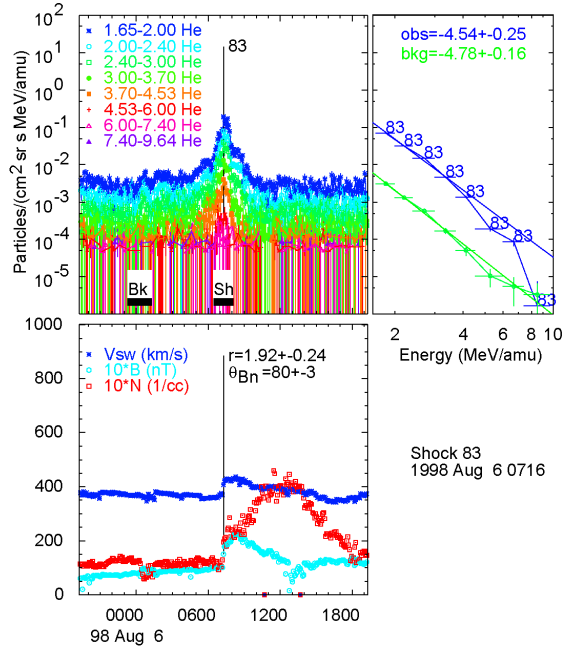

These possibilities for spectral shapes were considered in the observations of Reames (2012), who studied $^4$He spectra of ~1–10 MeV amu$^{-1}$ in 258 in situ interplanetary shocks observed near Earth by the *Wind* spacecraft (https://www.c-fa.harvard.edu/shocks/).  The purpose of this study was to determine which shock parameters were important to produce measurable particle acceleration and which were not.  Fig. 5.8 shows well-defined shock event No. 83.

Particle intensities in Fig. 5.8 are shown in the upper left panel, the plasma parameters: solar-wind speed $V_{SW}$, magnetic field intensity $B$, and density $N$, in the lower left panel, and the shock and background spectra in the upper right panel.  The times over which the two spectra are taken are shown in the upper left panel (Bk and Sh).  This is a quasi-perpendicular shock with the angle between $B$ and the shock normal, $\theta_{Bn} = 80^\circ \pm 3^\circ$

Fig. 5.9 compares properties of the shocks in this study.  The left panel shows a histogram of the shock speed distribution for all of the shocks and for the subset that showed measurable particle acceleration.  High shock speed was the strongest determinant for measurable acceleration, followed by high shock compression ratio, and large $\theta_{Bn}$.  High background intensity was also important: more input produced more output.  Measurable acceleration was more than twice as likely for shocks with $\theta_{Bn} > 60^\circ$ as for those with $\theta_{Bn} < 60^\circ$.  Quasi-parallel shocks, i.e. small $\theta_{Bn}$, may have been more likely to have knee energies below the energy of observation.  Recently, Zank et al. (2006) have suggested that "higher proton energies are achieved at quasi-parallel rather than highly perpendicular interplanetary shocks within 1 AU."  The in situ observations (Reames 2012) show the opposite; quasi-perpendicular shocks are favored; this difference likely occurs because ample pre-accelerated seed populations were available for these real shocks.

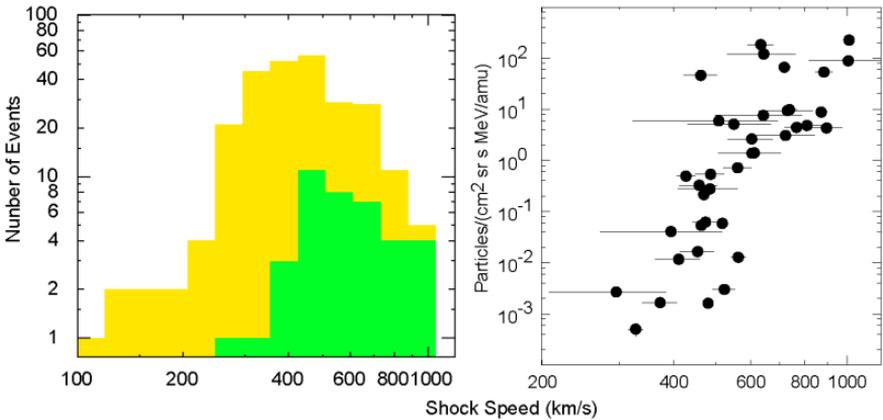

**Fig. 5.9**.  The *left panel* shows the distribution of shock waves at 1 AU with measurable acceleration of >1 MeV amu$^{-1}$ $^4$He vs. shock speed (*green*) within the distribution of all 258 shock waves vs. shock speed (*yellow* and *green*) observed by the *Wind* spacecraft.  The *right panel* shows the background-corrected peak intensity of 1.6–2.0 MeV amu$^{-1}$ $^4$He versus shock speed for the shocks in situ.  Shock speed is the strongest determinant of accelerated intensity for local shocks; this mirrors the correlated behavior of peak intensity versus CME speed in Fig. 2.12 (adapted from Reames 2012 © AAS, 2013).

The right panel in Fig. 5.9 shows the background-corrected peak shock intensity of 1.6–2.0 MeV amu$^{-1}$ $^4$He as a function of shock speed. The shock speed has a correlation coefficient of 0.80 with intensity. This correlation for in situ shocks mirrors the correlation of peak proton intensity with CME speed in Fig. 2.12 as modified by Rouillard et al. (2012) and shown in the lower right-hand panel of Fig. 3.4.

Particle intensities peak at the time of shock passage in nearly all of the events in the Reames (2012) study. However, sometimes intensities peak before or after shock passage when a spacecraft encounters magnetic flux tubes that connect it to a stronger part of the shock nearby, perhaps even one with a different value of $\theta_{Bn}$.

Absolute intensities of accelerated particles are not directly predicted by acceleration theories that omit wave growth. The rate of injection of seed particles is treated as an adjustable parameter – more input results in more output, and this is the case for in situ events. However, streaming protons and increasing wave intensities can trap particles near the source. At a few powerful shock waves, such as 20 October 1989, it has been observed that the energy in energetic particles exceeds that in the plasma and magnetic field (Lario and Decker 2002). Those authors suggested that the peak intensities of particles up to 500 MeV are simply trapped in a region of low density and low magnetic field near a shock. Maybe, but, how did they get there? Surely they were accelerated there near the peak intensity even though their spectrum may reflect shock properties from an earlier time (see Eq. 5.9). Perhaps the wave-trapped particles are in the process of destroying (i.e. pushing apart $B$ at) the shock that accelerated them. Another shock where the particle energy exceeds the magnetic energy is that of 6 November 2001, in Fig. 5.1 (C. K. Ng, private communication), where the sharp proton peak up to 700 MeV shows a shock that is still clearly intact. This is the issue of "cosmic-ray-mediated" shocks discussed by Terasawa et al. (2006) for two additional interplanetary shocks. This is a fascinating process that can be observed, in situ, at some interplanetary shocks.

## 5.5 Averaging SEP Abundances

We began by discussing the reference abundances in Chap. 1 and comparing them with the solar photospheric abundances as a function of first ionization potential (FIP) in Fig. 1.6 (see also Chap. 8). The reference abundances are obtained by averaging over many gradual SEP events. Since the transport of particles varies as a power of $A/Q$ (see Eq. 5.2), different species such as Fe and O will be distributed differently in space and time, but these particles are likely to be conserved. If we can successfully average over time or space we will recover the source abundances. If this assumption is correct and our averaging is representative, the reference abundances will approach the coronal abundances. Evidence for the space-time distribution is shown in Fig. 5.10.

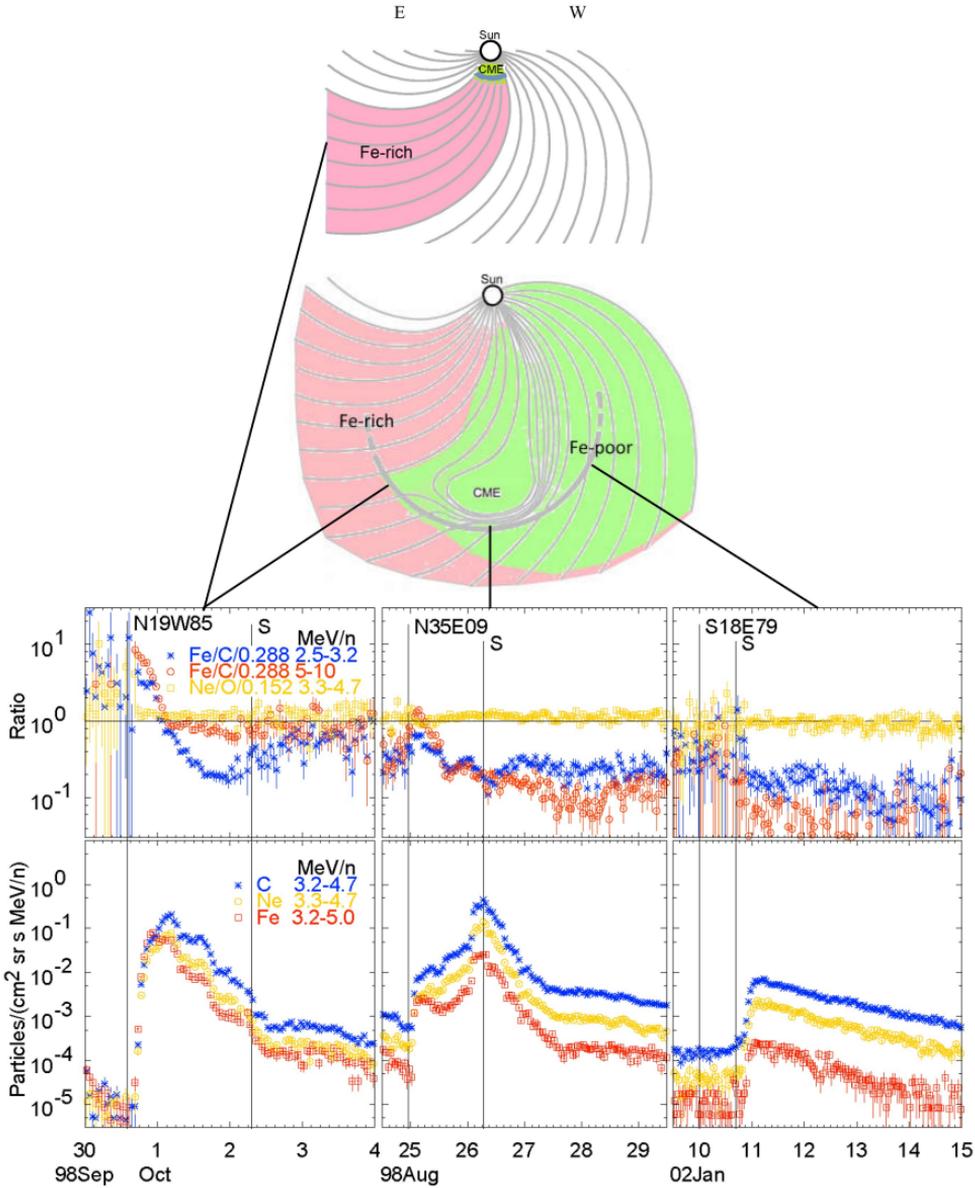

**Fig. 5.10**. Intensities of C, Ne, and Fe are shown for three gradual SEP events at different solar longitudes in the *lower panels*, relative abundances in the *middle panels*, and the location and evolution of a CME *above* (after Reames 2014 © Springer).

The SEP event on the East flank of the CME (W85 source, on the left in Fig. 5.10), shows enhancement of Fe/O early then suppression later, since Fe, with higher *A/Q*, scatters less than O. Ne/O, involving similar values of *A/Q*, varies little. Solar rotation and the Parker spiral translates this time variation into a spatial one and the events toward the West flank of the CME show mainly depleted Fe/O.

## 5.6 Source-Plasma Temperatures

Since particle transport of gradual SEPs varies as a power of $A/Q$, and $Q$ varies with $T$, we can use this power law to find the source-plasma temperature $T$ that gives the best-fit pattern vs. $A/Q$, just as we did for impulsive events. Fig. 5.11 (similar to Fig. 4.8) shows $A/Q$ vs. $T$ with $Q$ derived from the atomic physics.

**Fig. 5.11.** *A/Q* is plotted as a function of the theoretical equilibrium temperature for the elements named along each curve. Points are spaced 0.1 units of $\log_{10} T$ from 5.7 to 6.8. Bands produced by closed electron shells with 0, 2, and 8 orbital electrons are indicated, He having no electrons at this *T*. Elements tend to move from one closed-shell group to another as the temperature changes. (Data for $Z \leq$ 28 from Mazzotta et al. 1998, for $Z > 28$ from Post et al. 1977)

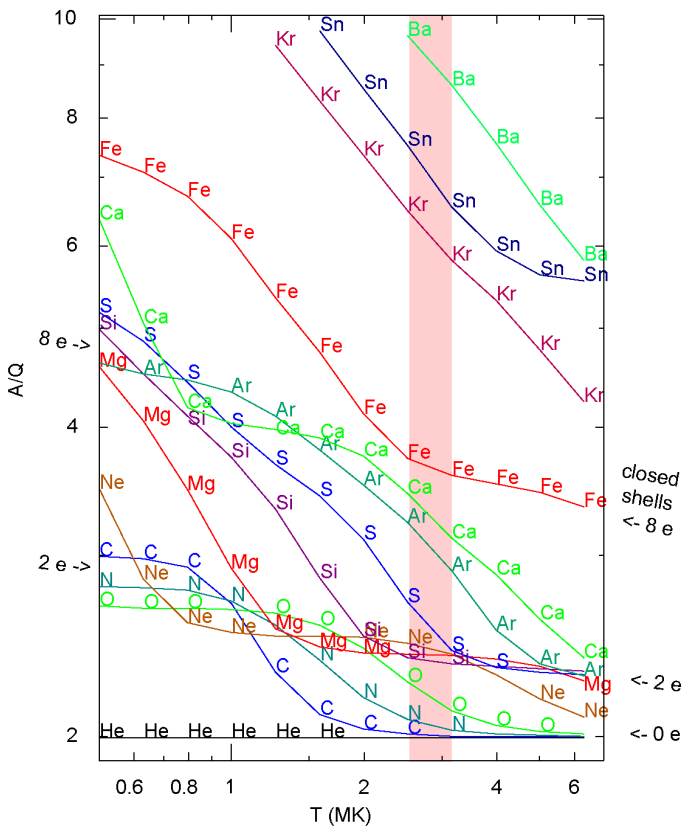

The red shaded region in Fig. 5.11 is 2.5–3.2 MK, corresponding to active region temperatures that we found for the impulsive SEP events that reflect the temperature of active-region jets where the SEPs originate (see Sect. 4.6). As we decrease $T$ below this region, O, then N, then C move from the 0-electron to the 2-electron closed shells. Meanwhile, Ca, then Ar, then S, then Si, then Mg move from the 2-electron to the 8-electron shells. Thus, we can tell the temperature by the pattern of abundance enhancements. We need only notice which elements are in which group; which elements have no enhancement like He; which elements are in the group with Ne; which are in the group with Ar.

Fig. 5.12 compares the observed pattern of enhancements early in a large gradual SEP event (on the left) scaled to the pattern of $A/Q$ (on the right). The patterns match best near $T \approx 0.6$ MK, an unusually low temperature for an SEP event. Note that C, N, and O have moved well above He to the 2-electron shell with Ne, while Mg, Si, and S have moved up to the 8-electron shell close to Ar and Ca. Patterns of enhancement in other SEP events are shown in Reames (2016a).

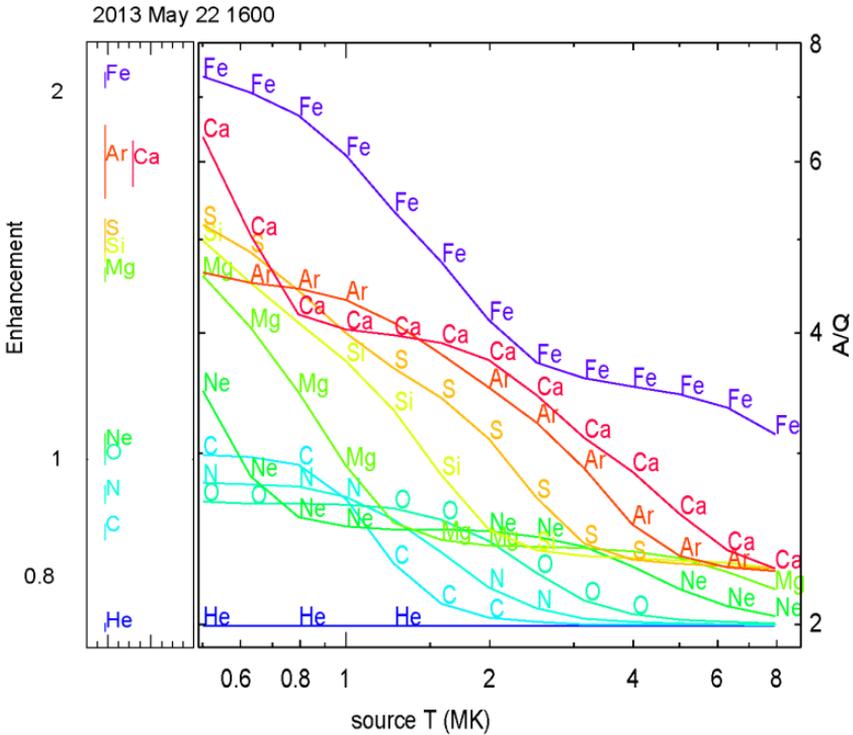

**Fig. 5.12**. The *left panel* shows the abundance enhancements at ~3–5 MeV amu$^{-1}$ observed early in the 22 May 2013 SEP event. The *right panel* compares a section of the *A/Q* vs. *T* plot from Fig. 5.11. The patterns match best at about 0.6 MK (Reames 2016a © Springer)

For the LEMT telescope on the *Wind* spacecraft, 8-hr intervals during a large SEP event will provide adequate statistics for the rarer elements to determine enhancement patterns. For each 8-hr period we can calculate least-squares fits of enhancement vs. *A/Q(T)* for all values of *T* in the range of interest and plot $\chi^2$ of the fit vs. *T* (upper-right panel in Fig. 5.13). The minimum value of $\chi^2$ gives the best-fit temperature and power of *A/Q* for that time interval. This process gives the source-plasma temperature as a function of time during an event, as shown in the upper-left panel of Fig. 5.13 for the event of 8 November 2000. For this event we find temperatures near 1 MK for all time periods with either abundance enhancements or suppressions. For two of the time periods, the best fits to enhancement vs. *A/Q* are shown in the lower-right panel of Fig. 5.13. However, for time periods when enhancements in the abundances are flat, neither enhanced nor suppressed relative to the coronal reference abundances, we cannot measure *T*, since any *A/Q* values will fit and $\chi^2$ has no minimum. Larger enhancement or suppression of the abundances produces clearer minima in $\chi^2$ and smaller errors in *T*.

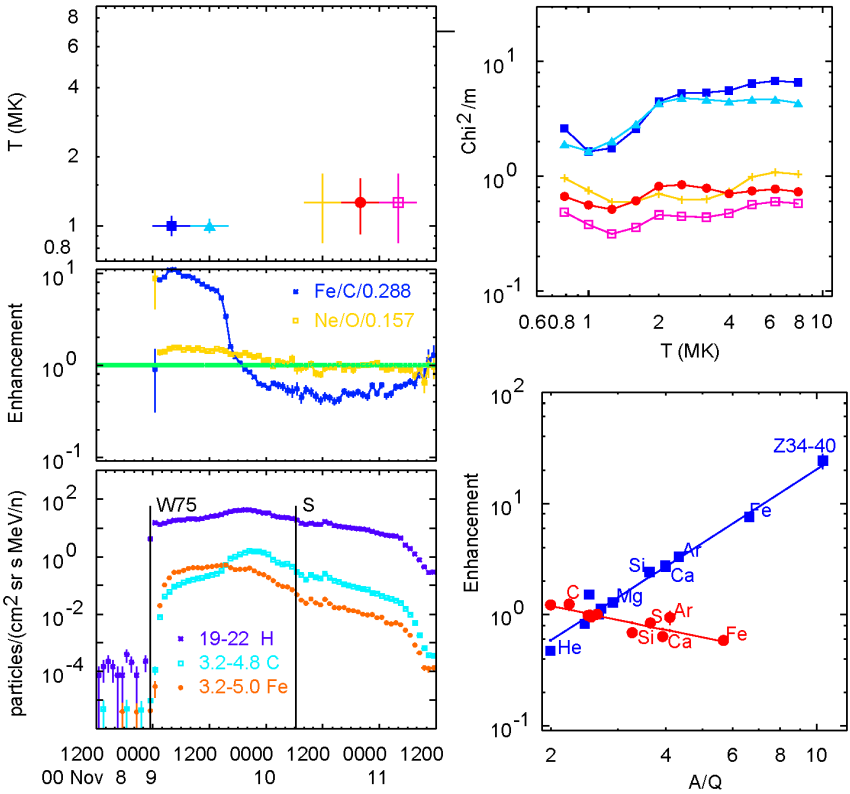

**Fig. 5.13.** Clockwise from the *lower-left panel* are the intensities of H, C, and Fe during the 8 November 2000 SEP event, the enhancements in Fe and Ne during the event, the best-fit temperatures in *color-coded* 8-hr intervals, values of $\chi^2/m$ vs. $T$ for each time interval (where $m$ is the number of degrees of freedom), and fits of enhancements from samples (*blue filled square* and *red filled circle*), early and late in the event, relative to O, vs. $A/Q$ (Reames 2016a © Springer).

For 45 gradual SEP events that had reasonably well-defined temperatures, Reames (2016a) found:

- 69% (31 events) showed ambient coronal temperatures $T \leq 1.6$ MK
- 24% (11 events) had $2.5 \leq T \leq 3.2$ MK active region temperatures, like impulsive SEP events where $T \approx 3$ MK is the typical temperature in solar jets associated with active regions (see Fig. 14 in Raouafi et al. 2016).

Some (11) of the events with ambient coronal temperatures showed a second minimum at the upper limit of $T$ in $\chi^2$ vs. $T$. These were originally attributed to the possible presence of stripped ions, but have been subsequently found to be artificially induced by variations in the source abundances of He (Reames 2017b; see Sect. 5.9). Subsequent studies omit H and He from fits (see Chap. 9)

While the gradual-event temperatures and fit parameters are not strongly correlated with any particular properties of the accelerating CME or shock, Fig. 5.14 shows $T$ vs. CME speed. The unweighted correlation coefficient is -0.49 for these events. Events that happen to be GLEs are identified in the figure; their tempera-ture distribution and other properties are similar to those of the other gradual SEP events.

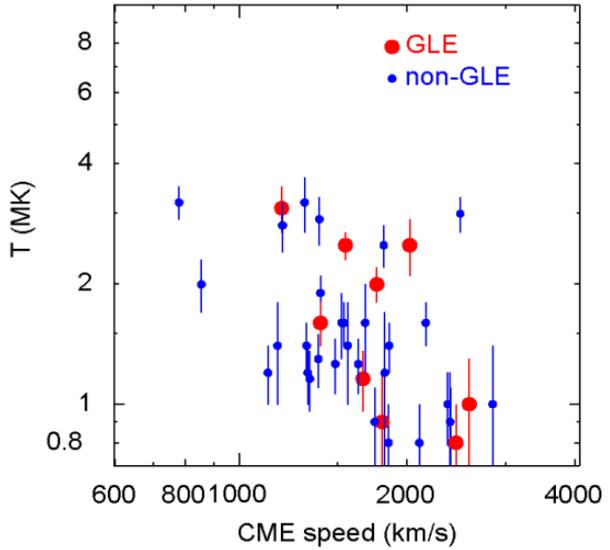

**Fig 5.14**. Source-plasma temperature is shown as a function of associated CME speed for gradual SEP events with GLE events identified (data from Reames 2016a).

We now realize that attempts to study abundance cross-correlations in gradual SEP events were often ineffective because most variations were caused by temperature differences that previously were not known. For example, the average value of Fe/O is a factor of ~10 higher in gradual events with $T = 3.2$ MK than in those with $T = 1.5$ MK. This is shown in Fig. 5.15 which plots normalized Fe/O vs. C/He, for intervals during the gradual SEP events, in both panels, with $T$ as symbols in the lower panel and power of $A/Q$ as symbols in the upper.

The region of abundances showing active-region temperatures $T \geq 2$ MK is immediately distinguishable, clustering in the upper left of the lower panel of Fig. 5.15. These events, which get their enhanced Fe/O from their impulsive seed population which also has enhanced He (see Sects. 5.8 and 5.9), are distinguished as open circles in the upper panel as well. Points during large events accelerated from specific temperatures of ambient coronal plasma stretch from upper right, with steep positive $A/Q$ enhancements early in the events, toward the lower center, where the $A/Q$ slopes are reduced or negative, late in the same events.

The events beginning in the large gray circle in the upper right corners of the panels in Fig. 5.15 are all high-fluence, $T < 2$ MK events with >30 MeV proton fluences above $10^7$ protons $cm^{-2}$ $sr^{-1}$. The intense streaming protons early in these events are likely to generate waves that increase particle scattering and steepen the positive $A/Q$ dependence early in the events (e.g. Fig. 9.18), enhancing both Fe/O and C/He. Smaller events with $T < 2$ MK lack large enhancements of Fe/O. The events with $T > 2$ MK are also smaller but they get their Fe/O enhancement from their impulsive seed population. Thus, gradual SEP events can become Fe-rich either by reaccelerating Fe-rich impulsive suprathermal ions or by having protons intense enough that self-generated waves preferentially transmit Fe and retard O.

**Fig. 5.15.** Normalized abundance ratios Fe/O vs. He/C is plotted in both panels with *symbol size* and *color* representing $T$ (*lower panel*) and power of $A/Q$ (*upper panel*). (Reames 2016b © Springer). The *large circle* surrounds "big events" with high fluences of >30 MeV protons.

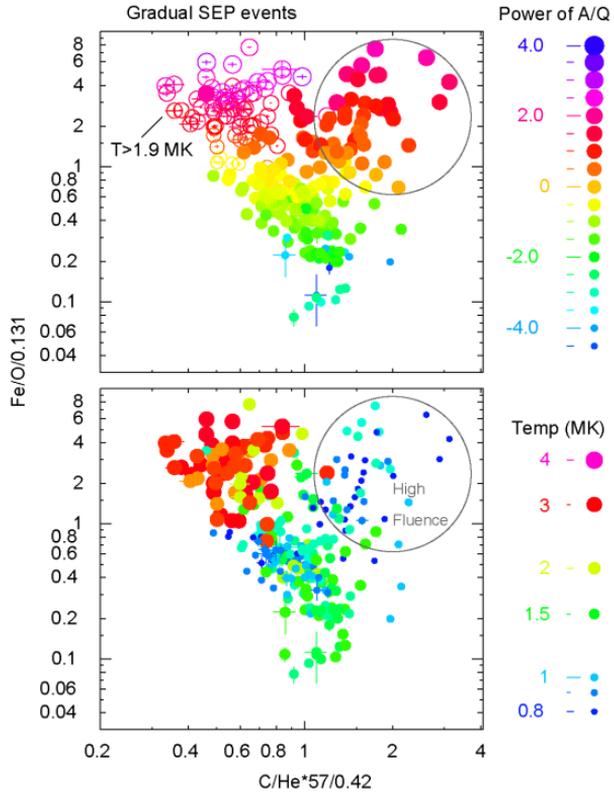

One might well ask: why do we use theoretical values of $Q$ vs. $T$ when there are actually direct measurements of $Q$ in some events (see Sect. 2.6)? Direct measurements can determine distributions of $Q$ which may not be isothermal and may differ somewhat from element to element. Primarily, direct measurements are rare, especially above 1 MeV, where deflection in the Earth's field is usually required (e.g. Leske et al. 1995). Also, in impulsive events, for example, $Q_{Fe}$ measured at 1 AU, is observed to increase with energy at low energies (e.g. DiFabio et al. 2008), suggesting that the ions have traversed enough material after acceleration to strip them to equilibrium charges that depend only upon their velocity, i.e. they are stripped to high $Q$. The theoretical charges, obtained from abundance measurements, are more likely to be appropriate earlier, i.e. at the time of acceleration. In addition, the theoretical charges from atomic physics are available for essentially all elements we measure and abundance measurements are often available at places and times where direct measurements are not. Thus we can measure the average source plasma temperatures wherever abundance measurements are available, but, unfortunately, these measurements do not determine charge distributions, distinguish multiple sources, or assess small temperature variations among elements, as direct measurements can.

An alternative method suggested for measuring ionization states in large SEP events involves fitting the $Q/A$-dependent shapes of the intensity-time profiles us-ing diffusion theory (Sollett et al. 2008). The comparison of this diffusive method with the power-law techniques is presently somewhat limited, and, unfortunately, we will see in the next section that the time decay profile of SEP events is actually controlled much less by diffusion than by the expansion of a magnetic trapping volume called the *reservoir*.

## 5.7 Spatial Distributions and the Reservoir

As spacecraft began to probe more-distant areas of the heliosphere, it became possible to view spatial distributions of SEPs, and their time variations, within a single SEP event. While spatial gradients were expected, it was rather surprising when equal intensities of ~20 MeV protons were found over long distances of solar longitude of ~180° on the *Pioneer* spacecraft by McKibben (1972). Twenty years later equal intensities were found in radius late in large events between *Ulysses* at 2.5 AU and IMP 8 near Earth by Roelof et al. (1992) who named the regions "reservoirs." Reservoirs extend to *Ulysses* at heliolatitudes up to >70°, N and S (Lario 2010), and they are also seen in other electron observations (Daibog et al. 2003).

The two spacecraft of the *Helios* mission provided another opportunity to measure the evolution of SEP events at different longitudes confirming that the longitude distribution of Fig. 2.2 was appropriate for each individual event. Fig. 5.16 shows that, at widely separated spacecraft, the intensities merge with that at *Helios 1* as each of the other spacecraft, *Helios 2* and IMP 8, joins it in the reservoir. Spectra are identical throughout the reservoir but intensities decrease adiabatically with time as the volume of this "magnetic bottle" expands. (The drawing in the lower panel of Fig. 5.16 shows the spacecraft penetrating the CME; in reality, of course, the spacecraft are nearly stationary as the CME expands past them, but that version would be much more difficult to draw.)

If there were significant leakage from the reservoir, one would expect the highest-energy protons to leak first, since they are faster, scatter less, and encounter the boundary most often, but this would steepen the spectrum with time and is *not* observed. Thus the leakage must be minimal.

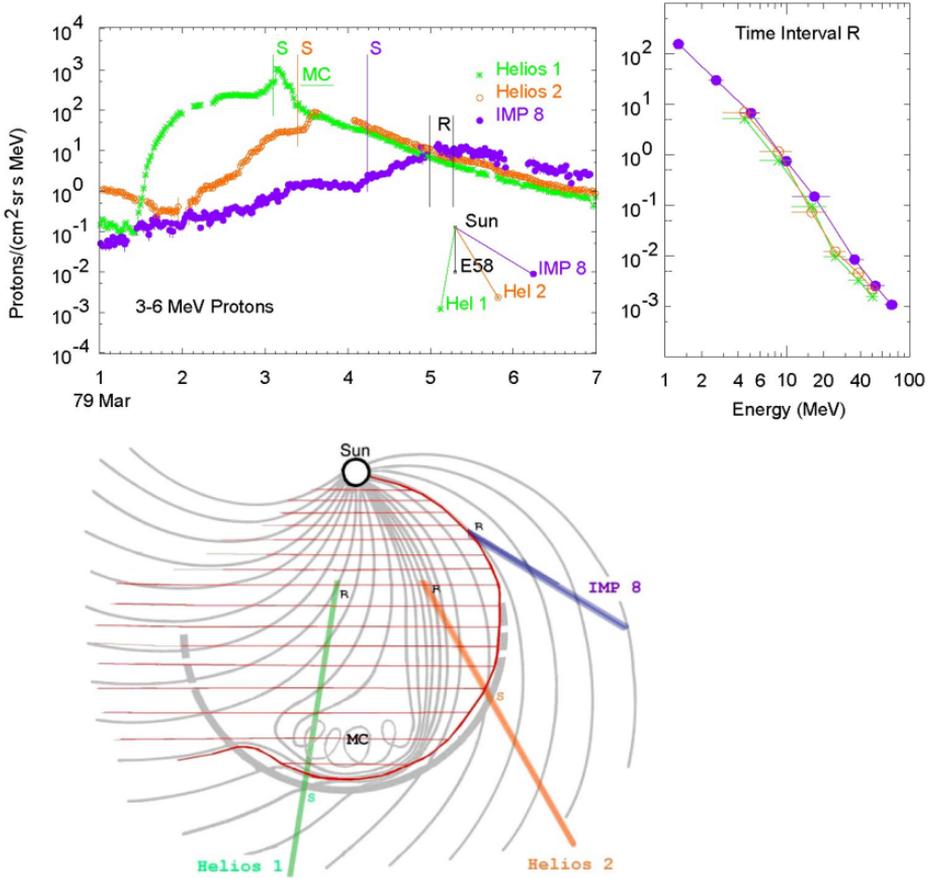

**Fig. 5.16**. The *upper left* panel shows the intensities of 3–6 MeV protons at three spacecraft vs. time. The paths of the spacecraft into the expanding CME are shown *below* as they penetrate into the reservoir region (*red hashing*) behind the shock and CME where all intensities and spectra (*upper right*) are equal spatially, though they decrease with time as the trapping volume expands (Reames 2013; after Reames, Kahler, and Ng 1997b © AAS).

One common, but rather poor, way of comparing spatial variations is to plot peak intensity at each of three spacecraft vs. longitude and fit the three points with a parabola. Does this measure particle spread in longitude? Suppose we made such a plot with the data shown in Fig. 5.16. The intensity at *Helios 1* peaks at the time of local shock passage. The intensity at *Helios 2* peaks when it enters the reservoir, where it has the same intensity as *Helios 1*. The intensity at IMP 8 peaks when it enters the reservoir later, where *all three intensities are equal*. What does the parabola defined by these three peak intensities measure? Is it the spread of the particles or the spread in the trapping volume behind the CME with time? The peaks all occur at different times and that *essential* timing information is lost when plotting only peak intensities vs. longitude. Isn't it more important to note that all intensities are equal when the intensity at IMP 8 peaks? It seems more productive to try to *distinguish* spatial and temporal effects rather than combining them.

For a single spacecraft, one way to show that spectra do not change their shape in time is to normalize the intensity-vs.-time plots at one point in time. If they stay normalized subsequently, then the spectral shapes are invariant. This is shown for two gradual SEP events in Fig. 5.17. This technique demonstrates invariance even when the spectra do not have power-law form. Multiple spacecraft at different locations can be included or abundance variations can be compared similarly.

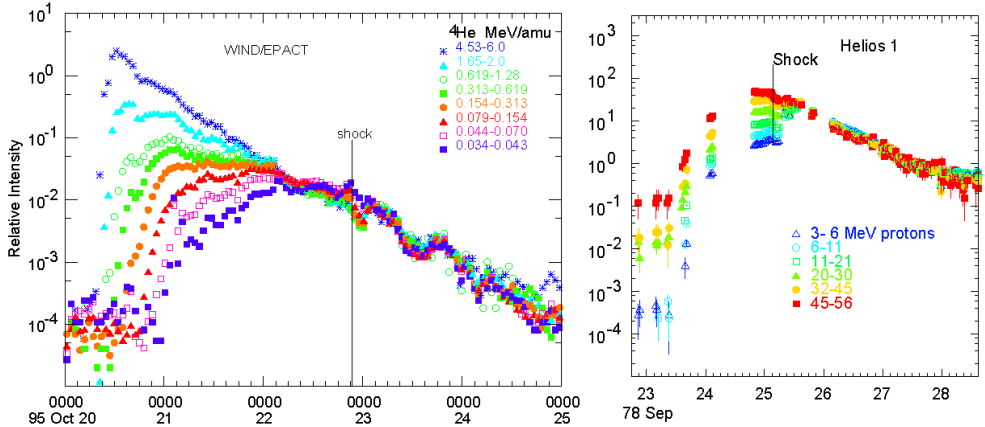

**Fig. 5.17**. In invariant spectral regions, particle intensities at different energies maintain the same relative normalization as a function of time, as shown for different species in two different events (Reames 2013; after Reames et al. 1997a and Reames, Kahler, and Ng 1997b © AAS).

Note that a reservoir can sometimes extend upstream of the CME and shock on the East flank, as seen in the left panel of Fig. 5.17; here the particles may be partly contained by self-amplified waves from earlier streaming or by other preexisting magnetic boundaries.

The realization that the slow decline in a gradual SEP event results from expansion of a reservoir is most important because it displaces the previous idea that slow particle spatial diffusion detained the ions to explain the decay phase of events. Actually, reservoirs are scatter free, as shown by the striking example of the little scatter-free $^3$He-rich event from Mason et al. (1989) shown as Fig. 2.3 in Sect. 2.3.4. A whole literature of fitting SEP events to diffusion theory had emerged, leading to the "Palmer (1982) consensus" that "$\lambda_\parallel$ = 0.08-0.3 AU over a wide range of rigidity." This is yet another example of the misapplication of diffusion theory; the intensity decline comes from the expansion of a magnetic bottle in time, not inefficient transport through space. There are no significant spatial gradients within reservoirs.

In some events we may use abundances to map variations of source plasma temperatures across the face of the shock and back into the reservoir. Fig 5.18 shows a temperature analysis that compares conditions at the STEREO B and the *Wind* spacecraft during an SEP event on 31 August 2012.

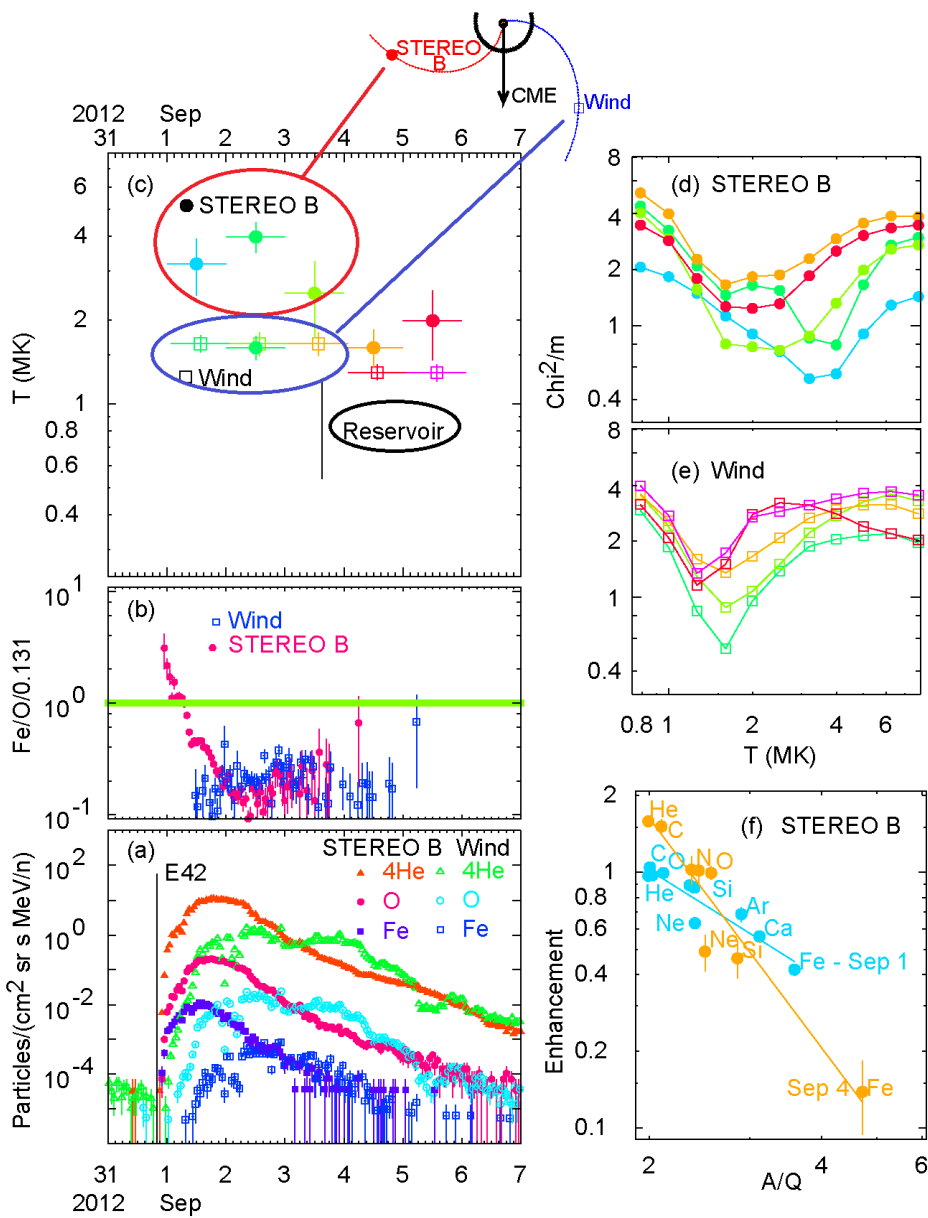

**Fig. 5.18** Time variations of (**a**) intensities, (**b**) Fe/O, and (**c**) derived source plasma temperatures are compared at STEREO B (*closed symbols*) and *Wind* (*open symbols*) for the SEP event of 31 August 2012 where spacecraft locations are shown above. Probable entry of the spacecraft into a reservoir occurs sometime on 3 September as noted in (**c**). Curves of $\chi^2/m$ vs. $T$ used to select the best-fit temperatures are shown in (**d**) and (**e**) and fits to enhancements vs. $A/Q$ are shown at STEREO (**f**) before (September 1) and after (September 3) reservoir entry (Reames 2017a © Springer).

The event in Fig. 5.18 shows the presence of hot (~3–4 MK) source plasma at the well-connected spacecraft (STEREO B), but not an enhancement from impul-sive seed particles, and ambient coronal plasma (~1.6 MK) on the flanks of the CME (at *Wind*). This condition persists until both spacecraft seem to have entered a reservoir, where similar intensities (Fig. 5.18a) and lower-temperature plasma (Fig. 5.18c) prevails.

Similar temperature variations across the face of the shock are possible, but less dramatic, in other events that have been measured (Reames 2017a). In one event, 23 January 2012, a spectral break disrupts the abundances, but we successfully revisit this event in Sect. 9.7. Spectral breaks are likely to make it difficult to derive temperatures from abundances; breaks are more likely above ~10 MeV amu$^{-1}$. At low energy (< 1 MeV amu$^{-1}$), shock waves accelerate ions far out into the heliosphere, sampling a widely varying seed population that may blur temperatures. In addition, strong suppression of low-energy spectra by the streaming limit (see Fig. 5.5) may disrupt temperature measurements at energies below 1 MeV amu$^{-1}$. Thus, ion temperatures for gradual SEP events are best measured in the range of 1–10 MeV amu$^{-1}$.

### 5.7.1 Reservoirs, Loops, and Long-Duration γ Rays

It is important to recognize that reservoirs trap energetic ions in an expanding volume above the solar surface for a long period of time. While this population of particles tends to be mirrored by the converging magnetic fields above the corona, some undoubtedly scatter into the loss cone and plunge into the corona to produce γ-rays (just as the particles in flaring loops must do on a much faster time scale). Vestrand and Forrest (1993) observed γ-ray production spanning over ≈30° of the Sun's surface in the large GLE of 29 September 1989. More recently, Plotnikov, Rouillard, and Share (2017) found three events for which CMEs from *behind the solar limb* drove quasi-perpendicular shock waves around to the footpoints of field lines on the visible disk, producing long-duration emission of >100 MeV γ-rays.

Ryan (2000) discussed long-duration γ-ray events lasting an hour or more while the flare-associated X-rays died away rapidly. See, also, the long-duration γ-ray observations by Ackermann et al. (2014) and Ajello et al. (2014). Share *et al.* (2018) studied 29 long-duration events with >100 MeV γ-rays. They concluded that the γ-rays resulted from the decay of neutral πmesons produced by shock-accelerated >300 MeV SEP protons trapped in the reservoirs behind these events. Reservoirs provide an invariant spectrum of shock-accelerated ions that can bombard a large area of the solar corona with energetic SEP proton intensities that slowly decrease over many hours or days. The SEPs trapped in a reservoir may have difficulty directly penetrating the magnetic fields above active regions, and the higher-energy SEPs may begin to reach the lower solar corona only as the trapping volume widens laterally beyond the active region. In some cases studied by Share et al. (2018), the >100 MeV γ-ray intensity rises slowly for several hours before it eventually begins to decrease, suggesting a slow increase in the area of the reservoir footprint on the Sun.

Gopalswamy et al. (2019) found a strong correlation of the duration of events with sustained γ-ray emission with the frequency and duration of associated type II radio emission. They found that the tail of the γ-ray event "seems to last until the end of the associated type II burst." Evidently both types of emission were associated with shock acceleration, as would not be too surprising if CMEs drive shocks and trap reservoirs. However, Gopalswamy et al. (2020) found an unusual geometry in the 1 September 2014 SEP event, where a shock wave was driven along an east-west flux rope, accelerating protons that plunged into the corona at the end of the flux rope, producing π⁰ mesons and their energetic γ-ray decays. Thus, some γ-ray events are produced directly by shock-driven proton events.

Those few events where the γ-rays seem to exceed estimated production by the observed proton spectra can be explained by poor magnetic connection. For example, a case where the nose of the shock is seen to be driven toward high latitude above the ecliptic so the Earth only samples an inadequate soft proton spectrum from the flank of the shock. All of the γ-ray events can be reasonably explained by shock-accelerated protons (Gopalswamy et al. 2019, 2020).

## 5.8 Non-thermal Variations: Impulsive vs. Gradual SEPs

Knowing the source-plasma temperatures allows us to compare impulsive and gradual SEP events from the same temperature source – i.e. impulsive SEP ions and impulsive seed particles. Fig. 5.19 compares the normalized abundances of O/C vs. C/He for impulsive and gradual SEP events plotted at the same scale. The impulsive events have been limited to those with modest <20% statistical errors in the ratios and the gradual events at $T \geq 2$ MK come from impulsive seed ions.

**Fig. 5.19.** Enhancements of O/C vs. C/He are compared, for gradual events with $T \geq 2.0$ MK (*upper panel*) and impulsive events with < 20% errors so that the spread is not from statistical errors (*lower panel*). Both panels are plotted at the same scale and $T$ is indicated by the *size* and *color* of the symbols. (1) The distribution is much smaller for the gradual events. (2) The median of the distribution of C/He for the gradual events, shown as a dashed line, implies a reference value for He/O of 91 rather than 57 (Reames 2016b © Springer).

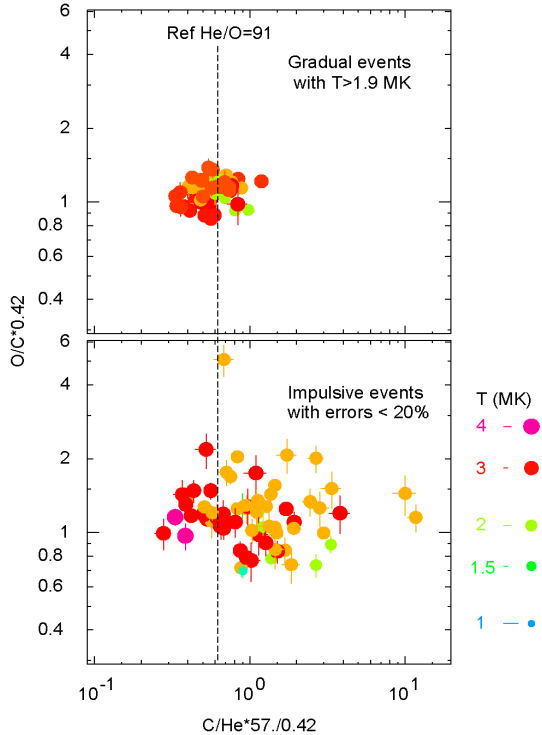

Especially at a temperature of 3.2 MK (red symbols in Fig. 5.19) the elements He and C are likely to be fully ionized with $A/Q = 2$ and O is nearly so (as seen in Fig. 5.11). Thus the ratios should be unaltered source abundances for both populations. However the dashed line also shows that the normalization is wrong for C/He since the central mean should be at 1.0 if properly normalized. This suggests that the reference abundance He/O should be 91 rather than 57. This would bring He in somewhat better alignment with other high-FIP elements on a FIP plot (Fig. 1.6) and is shown as a red open circle on that figure.

More significantly, the spread in the distribution of gradual events is much smaller than that of impulsive events in Fig. 5.19. The spread in the impulsive events must come from non-thermal abundance variations in the local plasma of jets where the magnetic reconnection is occurring. However, neither wave-particle interactions nor magnetic reconnection can alter C/He when both elements have $A/Q = 2.0$. If the shock wave of a gradual SEP event were accelerating only suprathermal ions from a single impulsive source, we would expect the same non-thermal distribution for gradual events that we see for impulsive events. This is *not* the case.

As the shock in a gradual event passes over an active region, it must average contributions (i) from impulsive suprathermal ions, which have enhancements in Fe/O and ${}^3$He/${}^4$He, for example, and perhaps (ii) from ions in the ambient ~3 MK plasma, which have no such enhancements. Ko et al. (2013) found that Fe-rich gradual SEP events were commonly connected to active regions. The result of the two contributions is to reduce the enhancements, as observed, and somewhat reduced distributions in the spread of abundance ratios, more like those in the upper panel of Fig. 5.19.

However, if we really expect to reduce the spread of the distributions as seen in the gradual events, we need to average over *several* small jets producing impulsive SEP events rather than only one; $n$ events will reduce the spread by a factor of $\sqrt{n}$. Thus a pool of 10 contributing events with 30% abundance variations would reduce the average variation of the seed population below 10%. It is likely that the number of small impulsive SEP events in an active region increases as the event size decreases, contributing a fairly steady flow of impulsive suprathermal ions; each temporarily contributes to the potential seed population before it diminishes. Based on the increasing number of flares with decreasing size, Parker (1988) proposed that a large number of small nanoflares could actually heat the corona. We need only a small increase in the number of jets producing impulsive SEP events that are too small to resolve as separate events, yet adequate to contribute to the seed population in the pool of impulsive suprathermal ions above a solar active region which may be subsequently sampled and averaged by a shock wave. Thus, no single impulsive event determines the seed population for acceleration by the shock wave in a subsequent gradual SEP event.

Many small jets (i.e. microjets or nanojets?) probably also contribute to the periods of persistent ${}^3$He seen by Wiedenbeck et al. (2008), of long-lived and recurrent sources (Bučík et al. 2014, 2015; Chen et al. 2015), often generated by "gey-sers" that produce multiple jets (e.g. Paraschiv and Donea 2019), and, of course, to the substantial persistent $^3$He abundances below 1 MeV amu$^{-1}$ in the seed population directly observed at 1 AU "upstream" region (labeled period A) in Fig. 2.8 where $^3$He-rich, Fe-rich periods in the seed population are seen on 20 and 21 June 2000 and Fe-rich periods (labeled B) again on 24 and 25 June after the shock wave of interest (e.g. Desai et al. 2003).

Source-plasma temperatures provide a powerful new tool for the comparative study of SEP events.

## 5.9 The Abundance of He and the FIP Effect

The preceding section suggests that He/O, in both impulsive SEP events and in gradual events that re-accelerated impulsive ~3 MK plasma, may be similar to the average value of He/O in the solar wind, while He/O can be much lower in gradual events involving cooler (< 2 MK) ambient source plasma. He variations should not come as a complete surprise since variations of He/O are also seen in the solar wind as functions of time and of solar-wind speed (Collier et al., 1996; Bochsler, 2007; Rakowsky and Laming, 2012). The study of the source-plasma temperatures of gradual SEP events (Reames 2016a) originally assumed the source had He/O = 57 as derived from the SEP average. For some events, however, He fell far from the least-squares fits of enhancements vs. $A/Q$ for all the other elements, as seen in Fig. 5.20a and 5.20c.

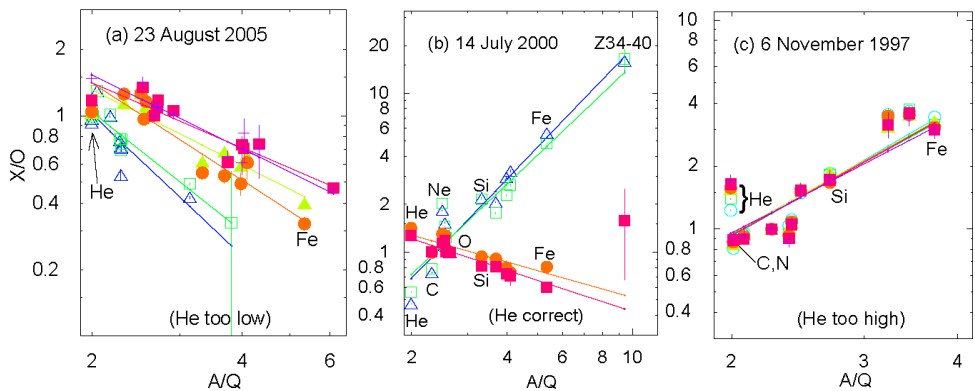

**Fig. 5.20** Best fits to enhancement vs. $A/Q$ assuming the source He/O = 57 at times during three gradual SEP events (Reames 2017b). For (**a**) the He enhancement is too low, for (**c**) it is too high, and for (**b**) it is approximately correct (i.e. in agreement with the fit of other ions) for both enhanced and suppressed periods of Fe/O.

For the SEP event of 23 August 2005 in (a), for which $T = 1.0\pm0.2$ MK, a lower value of source He/O $\approx 45$ would lift the enhancement of He to the fit lines. For the 6 November event (c), with $T = 2.5\pm0.2$ MK, a value of source He/O $\approx 100$ would drive He down to the fit lines. However, for the event of 14 July 2000, with $T = 1.2\pm0.2$, the average value of source He/O $\approx 57$, or slightly less, seems quite reasonable for both ascending and descending enhancement periods (Reames 2017b). He, the element with the highest FIP (24.6 eV), is the slowest element to ionize during its transit of the chromosphere into the corona, so its abundance is most likely to lag that of the other elements (Laming 2009, 2015); thus, low values of He/O may indicate newly arriving coronal ions while high values suggest established abundances.

The distribution vs. $T$ of the values of the source He/O that would be required for the observed He/O to fit the power-law defined by the $Z > 2$ elements in each 8-hour interval is shown in Fig. 5.21. The reaccelerated impulsive-event material at $\approx 3\,\mathrm{MK}$ supplies the higher reference He/O material.

**Fig. 5.21** A histogram for 8-hr intervals in gradual SEP events of the source value of He/O required for the He/O abundance enhancement to fit the power-er-law distribution defined by the elements with $Z > 2$, as a function of the best-fit source-plasma temperature (Reames 2017b © Springer)

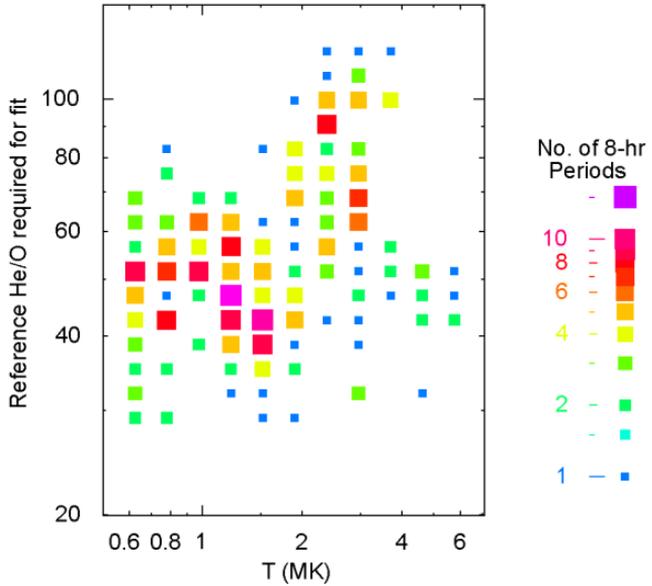

Understanding the possible slow ionization of He helps resolve differences in He/O between SEPs and the solar wind, although the remaining differences, especially in C/O, show us that SEPs and the solar wind have different FIP patterns, hence different origins at the base of the solar corona, as we will find in Chap. 8. This will show that SEPs are *not* merely accelerated solar wind. They are a particle population with unique coronal origin.

## 5.10 Open Questions

This section suggests open questions that might be addressed in future research.

1. What can cause the large non-thermal spread of abundances such as C/He in impulsive SEP events in Fig. 5.19 when both He and C should be fully ionized? Does source depth in the corona matter?

2. How well do SEP-derived temperatures correlate with differently-determined coronal temperatures near the observer's magnetic footpoint early in a gradual SEP event?

3. How can reservoirs contain particles of all energies with such apparently equal efficiency? How do they attain uniformity of intensities across longitudes when the particles upstream of the shock do not? Is diffusion along the turbulent shock front a factor?

4. In principle, a shock could accelerate 1-MK plasma at one longitude and 3-MK plasma at another longitude. Only one clear case is seen. Is there enough lateral transport in and behind the shock to mix SEPs from these sources in the reservoir late in events?

5. Which events have reservoirs and which do not? What is the difference?

6. What happens when the energy in SEPs exceeds the energy in *B* at a shock, especially a quasi-perpendicular shock? Can it do so? Does acceleration cease?

7. Measurements on a spacecraft nearer the Sun can improve SEP onset timing by removing the blurring effect of scattering during transport. How does the SEP onset SPR time at 10-second resolution compare with X-ray and γ-ray-line onsets, type II burst timing, and shock observations? Note that intensities may vary as $\sim r^3$, causing extremely high rates in instruments. To what extent does electron and ion source timing differ in gradual SEP events? In impulsive events?

8. Discrete ionization states affect the assignment of source-plasma temperatures. $^{12}C^{+5}$ is enhanced but $^{12}C^{+6}$ is not; treating *Q* as 5.5 is approximate. $A/\langle Q\rangle$ is not the same as $\langle A/Q\rangle$. Then there is $^{13}C$ which always has high $A/Q$. Can we improve the estimates of *T*?

# References

Ackermann, M., et al., High-energy gamma-ray emission from solar flares: summary of Fermi large area telescope detections and analysis of two M-class flares, Astrophys. J. **787**, 15 (2014) doi: 10.1088/0004-637X/787/1/15

Afanasiev, A., Battarbee, M., Vainio, R., Self-consistent Monte Carlo simulations of proton acceleration in coronal shocks: Effect of anisotropic pitch-angle scattering of particles, Astron. and Astrophys. **584**, 81 (2016) doi: 10.1051/0004-6361/201526750

Afanasiev, A., Vainio, R., Rouillard, A.P., Battarbee, M., Aran, A., Zucca, P., Modelling of proton acceleration in application to a ground level enhancement, Astron.Astrophys **614** A4 (2018) doi: 10.1051/0004-6361/201731343

Ajello, M., et al., Impulsive and long duration high-energy gamma-ray emission from the very bright 2012 March 7 solar flares, Astrophys. J. **789**, 20 (2014) doi: 10.1088/0004-637X/789/1/20

Axford, W.I., Acceleration of cosmic rays by shock waves, *Proc. 17th Int. Cosmic Ray Conf. (Paris)* **12**, 155 (1981)

Bell, A.R., The acceleration of cosmic rays in shock fronts. I, Monthly Notices Roy. Astron. Soc. **182**, 147 (1978a) doi: 10.1093/mnras/182.2.147

Bell, A.R., The acceleration of cosmic rays in shock fronts. II, Monthly Notices Roy. Astron. Soc. **182**, 443 (1978b) doi: 10.1093/mnras/182.3.443

Berdichevsky, D.B., Szabo, A., Lepping, R.P., Viñas, A.F., Mariana, F., Interplanetary fast shocks and associated drivers observed through the 23rd solar minimum by Wind over its first 2.5 years, J. Geophys. Res. **105**, 27289 (2000) doi: 10.1029/1999JA000367

Bochsler, P., Solar abundances of oxygen and neon derived from solar wind observations, Astron. Astrophys. **471** 315 (2007) doi: 10.1051/0004-6361:20077772

Breneman, H.H., Stone, E.C., Solar coronal and photospheric abundances from solar energetic particle measurements, Astrophys. J. Lett. **299**, L57 (1985) doi: 10.1086/184580

Bučík,,R,. Innes, D.E., Mall, U., Korth, A., Mason, G.M., Gómez-Herrero, R., Multi-spacecraft observations of recurrent ³He-rich solar energetic particles, Astrophys. J. **786**, 71 (2014) doi: 10.1088/0004-637X/786/1/71

Bučík, R., Innes, D.E., Chen, N.H., Mason, G.M., Gómez-Herrero, R., Wiedenbeck, M.E., Long-lived energetic particle source regions on the Sun, J. Phys. Conf. Ser. **642**, 012002 (2015) doi: 10.1088/1742-6596/642/1/012002

Chen N.H., Bučík R., Innes D.E., Mason G.M., Case studies of multi-day ³He-rich solar energetic particle periods, Astron. Astrophys. **580**, 16 (2015) doi: 10.1051/0004-6361/201525618

Collier, M.R., Hamilton, D.C., Gloeckler, G., Bochsler, P., Sheldon, R.B., Neon-20, oxygen-16, and helium-4 densities, temperatures, and suprathermal tails in the solar wind determined with WIND/MASS, Geophys. Res. Lett., **23**, 1191 (1996) doi: 10.1029/96GL00621

Daibog, E.I., Stolpovskii, V.G., Kahler, S.W., Invariance of charged particle time profiles at late stages of SCR events from the data of multisatellite observations, Cosmic Research **41**, 128 (2003)

Desai, M.I., Giacalone, J., Large gradual solar energetic particle events, Living Reviews of Solar Physics (2016) doi: 10.1007/s41116-016-0002-5

Desai, M.I., Mason, G.M., Dwyer, J.R., Mazur, J.E., Gold, R.E., Krimigis, S.M., Smith, C.W., Skoug, R.M., Evidence for a suprathermal seed population of heavy ions accelerated by interplanetary shocks near 1 AU, Astrophys. J., **588**, 1149 (2003) doi: 10.1086/374310

Desai, M.I., Mason, G.M., Wiedenbeck, M.E., Cohen, C.M.S., Mazur, J.E., Dwyer, J.R., Gold, R.E., Krimigis, S.M., Hu, Q., Smith, C.W., Skoug, R.M., Spectral properties of heavy ions associated with the passage of interplanetary shocks at 1 AU, Astrophys. J., **661**, 1156 (2004) doi: 10.1086/422211

DiFabio, R., Guo, Z., Möbius, E., Klecker, B., Kucharek, H., Mason, G.M., Popecki, M., Energy-dependent charge states and their connection with ion abundances in impulsive solar energetic particle events, Astrophys. J. **687**, 623 (2008) doi: 10.1086/591833

Ellison, D., Ramaty, R., Shock acceleration of electrons and ions in solar flares, Astrophys. J. **298**, 400 (1985) doi: 10.1086/163623

Gopalswamy, N., Makela, P., Yashiro, S., Lara, A., Akiyama, S., Xie, H., On the shock source of sustained gamma-ray emission from the sun, J. Phys. Conf. Ser. **1332** 1 (2019) doi: 10.1088/1742-6596/1332/1/012004

Gopalswamy, N., Mäkelä, P., Yashiro, S., Akiyama, S., Xie, H., Thakur, N., Source of energetic protons in the 2014 September 1 sustained gamma-ray emission event, Sol. Phys. **295** 18 (2020) doi: 10.1007/s11207-020-1590-8

Hu J, Li G, Ao X, Verkhoglyadova O, Zank G. Modeling particle acceleration and transport at a 2D CME-driven shock. J. Geophys. Res. **122** 10,938 (2017) doi: 10.1002/2017JA024077

Hu, J., Li, G., Fu, S., Zank, G., Ao, X., Modeling a single SEP event from multiple vantage points using the iPATH model, Astrophys. J. Lett. **854** L19 (2018) doi: 10.3847/2041-8213/aaabc1

Jones, F.C., Ellison, D.E., The plasma physics of shock acceleration, Space Sci. Rev. **58**, 259 (1991) doi: 10.1007/BF01206003

Ko Y,-K., Tylka, A.J., Ng C.K., Wang Y.-M., Dietrich W.F., Source regions of the interplanetary magnetic field and variability in heavy-ion elemental composition in gradual solar energetic particle events, Astrophys. J. **776**, 92 (2013) doi: 10.1088/0004-637X/776/2/92

Laming, J.M., Non-WKB models of the first ionization potential effect: implications for solar coronal heating and the coronal helium and neon abundances, Astrophys. J. **695**, 954 (2009) doi: 10.1088/0004-637X/695/2/954

Laming, J.M., The FIP and inverse FIP effects in solar and stellar coronae, Living Reviews in Solar Physics, **12**, 2 (2015) doi: 10.1007/lrsp-2015-2

Laming, J.M., The First Ionization Potential Effect from the Ponderomotive Force: On the Polarization and Coronal Origin of Alfvén Waves Astrophys J. Lett. **844** L153 (2017) doi: 10.3847/1538-4357/aa7cf1

Lario, D., Heliospheric energetic particle reservoirs: Heliospheric energetic particle reservoirs: *Ulysses* and ACE 175-315 keV electron observations, *Proc. 12th Solar Wind Conf.*, AIP Conf. Proc. **1216**, 625 (2010) doi: 10.1063/1.3395944

Lario, D., Decker, R.B., The energetic storm particle event of October 20, 1989, Geophys. Res. Lett. **29** 1393 (2002) doi: 10.1029/2001GL014017

Lee, M.A., Coupled hydromagnetic wave excitation and ion acceleration at interplanetary traveling shocks, J. Geophys. Res., **88**, 6109. (1983) doi: 10.1029/JA088iA08p06109

Lee, M.A., Coupled hydromagnetic wave excitation and ion acceleration at an evolving coronal/interplanetary shock, Astrophys. J. Suppl., **158**, 38 (2005) doi: 10.1086/428753

Lee, M.A., Mewaldt, R.A., Giacalone, J., Shock acceleration of ions in the heliosphere, Space Sci. Rev. **173** 247 (2012) doi: 10.1007/s11214-012-9932-y

Leske, R.A., Cummings, J.R., Mewaldt, R.A., Stone, E.C., von Rosenvinge, T.T., Measurements of the ionic charge states of solar energetic particles using the geomagnetic field, Astrophys J. **452**, L149 (1995) doi: 10.1086/309718

Li, G., Zank, G.P., Rice, W.K.M., Acceleration and transport of heavy ions at coronal mass ejection-driven shocks, J. Geophys. Res. **110** 6104 (2005) doi: 10.1029/2004JA010600

Li G, Shalchi A, Ao X, Zank G, Verkhoglyadova O P. Particle acceleration and transport at an oblique CME-driven shock. Adv Space Res, **49** 1067 (2012) doi: 10.1016/j.asr.2011.12.027

Mason, G.M., Ng, C.K., Klecker, B., Green, G., Impulsive acceleration and scatter-free transport of about 1 MeV per nucleon ions in $^3$He-rich solar particle events, Astrophys. J. **339**, 529 (1989) doi: 10.1086/167315

Mazzotta, P., Mazzitelli, G., Colafrancesco, S, Vittorio, N., Ionization balance for optically thin plasmas: Rate coefficients for all atoms and ions of the elements H to Ni, Astron. Astrophys Suppl. **133**, 403 (1998) doi: 10.1051/aas:1998330

McKibben, R.B., Azimuthal propagation of low-energy solar-flare protons as observed from spacecraft very widely separated in solar azimuth, J. Geophys. Res. **77**, 3957 (1972) doi: 10.1029/JA077i022p03957

Melrose, D.B., *Plasma Astrophysics* New York: Gordon and Breach (1980).

Melrose, D.B., Pope, M. H., Diffusive shock acceleration by multiple shocks, Proc. Astron. Soc. Au. **10**, 222 (1993)

Mewaldt, R.A., Looper, M.D., Cohen, C.M.S., Haggerty, D.K., Labrador, A.W., Leske, R.A., Mason, G.M., Mazur, J.E., von Rosenvinge, T.T., Energy spectra, composition, other properties of ground-level events during solar cycle 23, Space Sci. Rev. **171**, 97 (2012) doi: 10.1007/s11214-012-9884-2

Ng, C.K., Reames, D.V., Focused interplanetary transport of approximately 1 MeV solar energetic protons through self-generated Alfven waves, Astrophys. J. **424**, 1032 (1994) doi: 10.1086/173954

Ng, C.K., Reames, D.V., Pitch angle diffusion coefficient in an extended quasi-linear theory, Astrophys J. **453**, 890 (1995) doi: 10.1086/176449

Ng, C.K., Reames, D.V., Shock acceleration of solar energetic protons: the first 10 minutes, Astrophys. J. Lett. **686**, L123 (2008) doi: 10.1086/592996

Ng, C.K., Reames, D.V., Tylka, A.J., Effect of proton-amplified waves on the evolution of solar energetic particle composition in gradual events, Geophys. Res. Lett. **26**, 2145 (1999) doi: 10.1029/1999GL900459

Ng, C.K., Reames, D.V., Tylka, A.J., Modeling shock-accelerated solar energetic particles coupled to interplanetary Alfvén waves, Astrophys. J. **591**, 461 (2003) doi: 10.1086/375293

Ng, C.K., Reames, D.V., Tylka, A.J., Solar energetic particles: shock acceleration and transport through self-amplified waves, AIP Conf. Proc. **1436**, 212 (2012) doi: 10.1063/1.4723610

Palmer, I.D., Transport coefficients of low-energy cosmic rays in interplanetary space, Rev. Geophys. Space Phys. **20**, 335 (1982)

Paraschiv, A.R., Donea, A., On solar recurrent coronal jets: coronal geysers as sources of elec-tron beams and interplanetary type-III radio bursts, Astrophys. J. **873**, 110 (2019) doi: 10.3847/1538-4357/ab04a6

Parker, E.N.: *Interplanetary Dynamical Processes,* Interscience, New York (1963)

Parker E.N., Nanoflares and the solar X-ray corona, Astrophys J. **330** 474 (1988) doi: 0.1086/166485

Plotnikov, I., Rouillard, A., Share, G., The magnetic connectivity of coronal shocks to the visible disk during long-duration gamma-ray flares, Astron. Astrophys. **608** 43 (2017) doi: 10.1051/0004-6361/201730804 (arXiv:1703.07563)

Post, D.E., Jensen, R.V., Tarter, C.B., Grasberger, W.H., Lokke, W.A., Steady-state radiative cooling rates for low-density, high temperature plasmas, At. Data Nucl. Data Tables **20**, 397 (1977) doi: 10.1016/0092-640X(77)90026-2

Rakowsky, C.E., Laming, J.M., On the Origin of the slow speed solar wind: helium abundance variations, Astrophys. J. **754**, 65 (2012) doi: 10.1088/0004-637X/754/1/65

Raouafi, N.E., Patsourakos, S., Pariat, E., Young, P.R., Sterling, A.C., Savcheva, A., Shimo-jo, M., Moreno-Insertis, F., DeVore, C.R., Archontis, V, et al., Solar coronal jets: observa-tions, theory, and modeling, Space Sci. Rev. **201** 1 (2016) doi: 10.1007/s11214-016-0260-5 (arXiv:1607.02108)

Reames, D.V., Acceleration of energetic particles by shock waves from large solar flares, Astro-phys. J. Lett. **358**, L63 (1990) doi: 10.1086/185780

Reames, D.V., Particle energy spectra at traveling interplanetary shock waves, Astrophys. J. **757,** 93 (2012) doi: 10.1088/0004-637X/757/1/93

Reames, D.V.: The two sources of solar energetic particles, Space Sci. Rev. **175**, 53 (2013) doi: 10.1007/s11214-013-9958-9

Reames, D.V., Element abundances in solar energetic particles and the solar corona, Sol. Phys., **289**, 977 (2014) doi: 10.1007/s11207-013-0350-4

Reames, D.V., Temperature of the source plasma in gradual solar energetic particle events, Sol. Phys., **291** 911 (2016a) doi: 10.1007/s11207-016-0854-9 (arXiv: 1509.08948)

Reames, D.V., The origin of element abundance variations in solar energetic particles, Sol. Phys, **291** 2099 (2016b) doi: 10.1007/s11207-016-0942-x (arXiv: 1603.06233)

Reames, D.V., Spatial distribution of element abundances and ionization states in solar energet-ic-particle events, Sol. Phys. **292** 133 (2017a), doi: 10.1007/s11207-017-1138-8 (arXiv 1705.07471).

Reames, D.V., The abundance of helium in the source plasma of solar energetic particles, Sol. Phys. **292** 156 (2017b) doi: 10.1007/s11207-017-1173-5 (arXiv: 1708.05034)

Reames, D.V., Barbier, L.M., von Rosenvinge, T.T., Mason, G.M., Mazur, J.E., Dwyer, J.R., En-ergy spectra of ions accelerated in impulsive and gradual solar events, Astrophys. J. **483**, 515 (1997a) doi: 10.1086/310845

Reames, D.V., Kahler, S.W., Ng, C.K., Spatial and temporal invariance in the spectra of energet-ic particles in gradual solar events, Astrophys. J. **491**, 414 (1997b) doi: 10.1086/304939

Reames, D.V., Ng, C.K., Berdichevsky, D., Angular Distributions of Solar Energetic Particles , Atrophys. J. **550** 1064 (2001) doi: 10.1086/319810

Reames, D.V., Ng, C.K., Streaming-limited intensities of solar energetic particles on the intensi-ty plateau, Astrophys. J. **722**, 1286 (2010) doi: 10.1088/0004-637X/723/2/1286

Reames, D.V., Ng, C.K., Tylka, A.J., Initial time dependence of abundances in solar particle events, Astrophys. J. Lett. **531**, L83 (2000) doi: 10.1086/319810

Rice, W.K.M., Zank, G.P., Li, G., Particle acceleration and coronal mass ejection driven shocks: Shocks of arbitrary strength, J. Geophys, Res. **108** 1369 (2003) doi: 10.1029/2002JA009756

Roelof, E.C., Propagation of solar cosmic rays in the interplanetary magnetic field, *Lectures in High-Energy Astrophysics*, ed. H. Ögelman and J. R. Wayland, Washington, D. C. NASA SP-199 (1969)

Roelof, E.C., Gold, R.E., Simnett, G.M., Tappin, S.J., Armstrong, T.P., Lanzerotti, L.J., Low-energy solar electrons and ions observed at ULYSSES February-April, 1991 - The inner heliosphere as a particle reservoir, Geophys. Res. Lett. **19**, 1247 (1992) doi: 10.1029/92GL01312

Rouillard, A., Sheeley Jr., N.R., Tylka, A., Vourlidas, A., Ng, C.K., Rakowski, C., Cohen, C.M.S., Mewaldt, R.A., Mason, G.M., Reames, D.V. et al., The longitudinal properties of a solar energetic particle event investigated using modern solar imaging, Astrophys. J. **752**, 44 (2012) doi: 10.1088/0004-637X/752/1/44

Ryan, J.M., Long-duration solar gamma-ray flares, Space Sci. Rev. **93**, 581 (2000)

Sandroos, A., Vainio, R., Simulation results for heavy ion spectral variability in large gradual solar energetic particle events, Astrophys. J. **662**, L127 (2007) doi: 10.1086/519378

Share, G.H., Murphy, R.J., White, S.M., Tolbert, A.K., Dennis, B.R., Schwarz, R.A., Smart, D.F., Shea, M.A., Characteristics of late-phase >100 MeV γ-ray emission in solar eruptive events, Astrophys. J. **869** 182 (2018) doi: 10.3847/1538-4357/aaebf7

Sollitt, L.S., Stone, E.C., Mewaldt, R.A., Cohen, C.M.S., Cummings, A.C., Leske, R.A., Wiedenbeck, M.E., von Rosenvinge, T.T., A novel technique to infer ionic charge states of solar energetic particles Astrophys. J. **679** 910 (2008) doi: 10.1086/587121

Stix, T.H. *The Theory of Plasma Waves* McGraw-Hill, New York (1962)

Stix, T.H., *Waves in Plasmas* AIP New York (1992)

Strauss, R.D., le Roux, J.A., Solar energetic particle propagation in wave turbulence and the possibility of wave generation, Astrophys J. **872** 125 (2019) doi: 10.3847/1538-4357/aafe02

Tan, L.C., Reames, D.V., Ng, C.K., Shao, X., Wang, L., What causes scatter-free transport of non-relativistic solar electrons? Astrophys J **728**, 133 (2011) doi: 10.1088/0004-637X/728/2/133

Terasawa, T., Oka, M., Nakata, K., Keika, K., Nosé, M., McEntire, R.W., Saito, Y., Mukai, T., 'Cosmic-ray-mediated' interplanetary shocks in 1994 and 2003, Adv. Space. Res. **37**, 1408 (2006) doi: 10.1016/j.asr.2006.03.012

Vestrand, W.T., Forrest, D.J., Evidence for a spatially extended component of gamma rays from solar flares, Astrophys. J. Lett. **409**, L69 (1993) doi: 10.1086/186862

Wiedenbeck, M.E, Cohen, C.M.S., Cummings, A.C., de Nolfo, G.A., Leske, R.A., Mewaldt, R.A., Stone, E.C., von Rosenvinge, T.T., Persistent energetic ³He in the inner heliosphere, *Proc. 30th Int. Cosmic Ray Conf.* (Mérida) **1**, 91 (2008)

Zank, G.P., Li, G., Florinski, V., Hu,, Q., Lario,, D., Smith, C.W., Particle acceleration at perpendicular shock waves: Model and observations, J. Geophys. Res.*,* **111**, A6108 (2006) doi: 10.1029/2005JA011524

Zank, G.P., Li, G., Verkhoglyadova, O., Particle Acceleration at Interplanetary Shocks, Space Sci. Rev. **130**, 255 (2007) doi: 10.1007/s11214-007-9214-2

Zank, G.P., Rice, W.K.M., Wu, C.C., Particle acceleration and coronal mass ejection driven shocks: A theoretical model, *J*. Geophys. Res., **105**, 25079 (2000) doi: 10.1029/1999-JA000455

# Chapter 6. High Energies and Radiation Effects

**Abstract** In this chapter we characterize the high-energy spectra of protons that can penetrate shielding and determine the radiation dose to humans and equipment in space. High-energy spectral breaks or "knees", seen in all large SEP events, determine the contribution of highly penetrating protons. The streaming limit, discussed earlier, places an upper bound on particle fluences early in events and the radial variation of intensities is important for near-solar and deep-space missions. The streaming limit is a strong function of radial distance from the Sun. We also consider requirements for a radiation storm shelter for deep space, a mission to Mars, suitability of exoplanets for life, and radiation-induced chemistry of the upper atmosphere of Earth.

We must recognize that solar energetic particles (SEPs) are of more than scientific interest. They can be a serious radiation hazard to astronauts and equipment in space beyond the protection of Earth's atmosphere and magnetic field. Protons of ~150 MeV can penetrate 20 gm cm$^{-2}$ (7.4 cm) of Al or 15.5 cm of water (or human flesh). Such protons are considered to be "hard" radiation, in that they are very difficult to shield, and they are orders of magnitude more intense than the GeV protons that define a ground-level event (GLE). Most of the other radiation risk to humans in space from SEP events comes from protons in the energy region above about 50 MeV, or "soft" radiation. This is where protons begin to penetrate spacesuits and the skin of spacecraft. Further studies of radiation dosage and engineering design and tradeoff are available elsewhere (see Barth et al. 2003; Xapsos et al. 1999, 2007; Cucinotta et al. 2010; Carnell et al. 2016), as is SEP forcasting (e.g. Kahler and Ling 2015; Laurenza et al. 2009). However, we do characterize the high-energy SEP spectra and their limits and spatial variations that affect radiation doses (e.g. Reames and Ng 1998; Reames 1999, 2013; Tylka and Dietrich 2009; Schrijver et al. 2012; Bruno et al. 2018).

## 6.1 High-Energy Spectra

The single most important factor, in the dose of penetrating protons, may be the location of the high-energy spectral break or knee. A comparison of spectra in two events is shown in Fig. 6.1 where the contributions of "hard" and "soft" radiation boundaries are shown. The spectra in the two events are similar in the 10–100 MeV region, partly controlled by the streaming limit (see Sects. 5.1.5 and 6.2). The spectrum of the April 1998 event (green) contributes mostly soft radiation in the region shaded yellow. The *additional* dose from the September 1989 SEP event is shaded red. Even behind 10 g cm$^{-2}$ of material, astronauts would receive a dose of 40 mSv hr$^{-1}$ (~4 rem hr$^{-1}$, rem = Roentgen equivalent for man) at the intensities in the September 1989 event. The *annual* dose limit for a radiation worker in the United States is 50 mSv (see review Cucinotta et al. 2010).

**Fig. 6.1**. Proton spectra in the SEP events of 20 April 1998 (*green*; based on Tylka et al. 2000) and 29 September 1989 (*blue*; based on Lovell, Duldig, and Humble 1998) are compared. Typical energies of "soft" and "hard" radiation are shown. The hazardous portion of the spectrum of the April event is *shaded yellow* and the *additional* hazardous radiation from the September event is *shaded red*. (Reames 2013 © Springer).

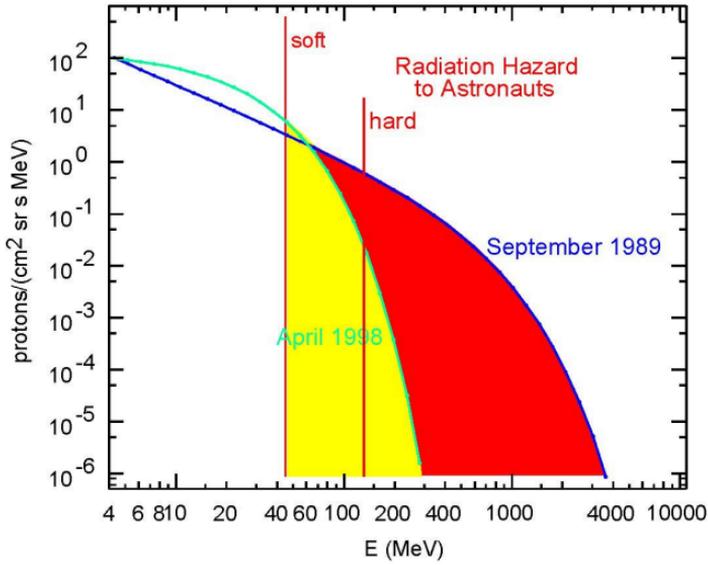

In Fig. 6.1, the proton spectral break or knee for the April 1998 event is about 40 MeV while that for the September 1989 event is nearer 400 MeV. This spectral shape makes an important difference (see Sect. 3.4 and Fig. 3.7).

An extensive study of high-energy spectra in ground-level events (GLEs) has been conducted by Tylka and Dietrich (2009) who merged neutron-monitor data with satellite-based data. Two of these spectra are shown in Fig. 6.2. The authors construct *integral* rigidity spectra using the magnetic cutoff rigidity of the stations. They then correct for the fact that higher-energy protons produce increasingly more secondary neutrons, and they compare with satellite measurements. The spectra are then fit to the empirical double power-law (Band et al. 1993) spectra above 0.137 GV (10 MeV), for which the parameters are stated, and to a single power law in the neutron-monitor region. Note that the Cherenkov-radiation-based GOES/HEPAD instrument and the IMP/GSFC instrument overlap the neutron-monitor measurements extremely well up to rigidities above 1 GV.

Much of the neutron-monitor data have lain idle for 50 years. Tylka and Dietrich have performed a great service to finally find a way to analyze the data, compute spectra, and organize all of these data in a form that is useful for comparing and studying high-energy SEP events. Those responsible for determining the risk of radiation hazards to astronauts should certainly take advantage of this thorough study (see also Raukunen et al. 2018).

**Fig 6.2**.  Integral rigidity spectra are shown for two large GLEs. Cutoff rigidities for individual neutron-monitor stations (listed) are used; the spectra are corrected for neutron production vs. proton energy, and compared with the named satellite measurements.  Fits to double power-law (Band) spectra are shown (Tylka and Dietrich 2009)

The fluence and the power-law fit above 1 GV (430 MeV) for the GLEs from Tylka and Dietrich (2009) are shown in Fig. 6.3. The largest fluence is $4 \times 10^6$ protons cm$^{-2}$ sr$^{-1}$ for the 23 February 1956 event, but the flattest spectra in the high-energy region are for the events of 7 May 1978 and 29 September 1979 (seen also in Figs. 6.1 and 6.5) events with rigidity spectral indices near 4.0. Most of the GLEs have rigidity spectral indices between 5 and 7.

**Fig 6.3**. The *upper panel* shows the proton fluence above 1 GV (430 MeV) vs. time for each GLE. The *lower panel* shows the integral rigidity power-law spectral index also above 1 GV (Tylka and Dietrich 2009; see also Raukunen et al. 2018)

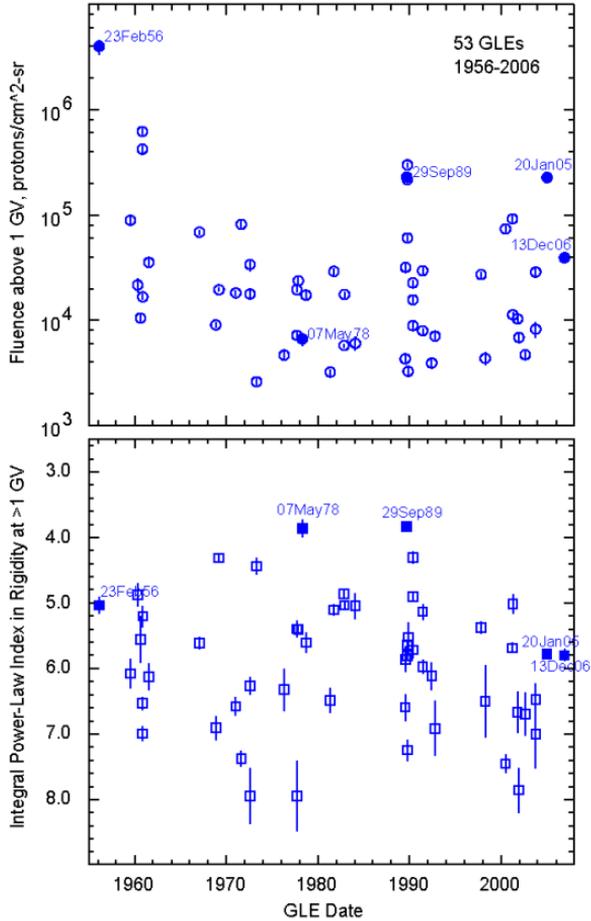

Parameters of these double power-law fits to GLE rigidity spectra have been tabulated by Raukunen et al. (2018) who also discuss interdependence of the fit parameters and fluence models based upon these results.

Recently, measurements of SEP proton spectra between ~80 MeV and a few GeV by the *Payload for Antimatter Matter Exploration and Light-nuclei Astrophysics* (PAMELA) have been reported for events from December 2006 and September 2014 by Bruno et al. (2018). These authors successfully fit proton fluence spectra to the Ellison and Ramaty (1985) form $E^{-\gamma} \exp(-E/E_0)$ for 26 events. Bruno et al. (2018) find no qualitative difference between GLEs, sub-GLEs, and non-GLEs, which form a continuous distribution.

Afanasiev et al. (2018) have applied their shock acceleration model to the GLE of 17 May 2012. They find increased acceleration of GV protons in regions of the shock with high Mach numbers and stress the importance of velocity differences between upstream and downstream scattering centers. Quasi-perpendicular regions of the shock may be involved to produce the observed timing.

## 6.2 The Streaming Limit

Protons streaming out early in a SEP event generate resonant waves that throttle the flow of subsequent particles, trapping them near the source. The streaming limit (see Sect. 5.1.5) is a transport phenomenon placing an upper bound on equilibrium intensities early in events once the waves become established (Reames and Ng 1998, 2010; Ng, Reames, and Tylka 2003, 2012). If we plot the probability of attaining a given intensity, i.e. the number of hours a given intensity is observed in ~11 years, as in Fig. 6.4, we see a sudden drop above the streaming limit. Intensities near shock peaks are not limited by this mechanism since no net streaming is involved there. However, shock peaks occur late, when shocks have weakened and particles have spread spatially.

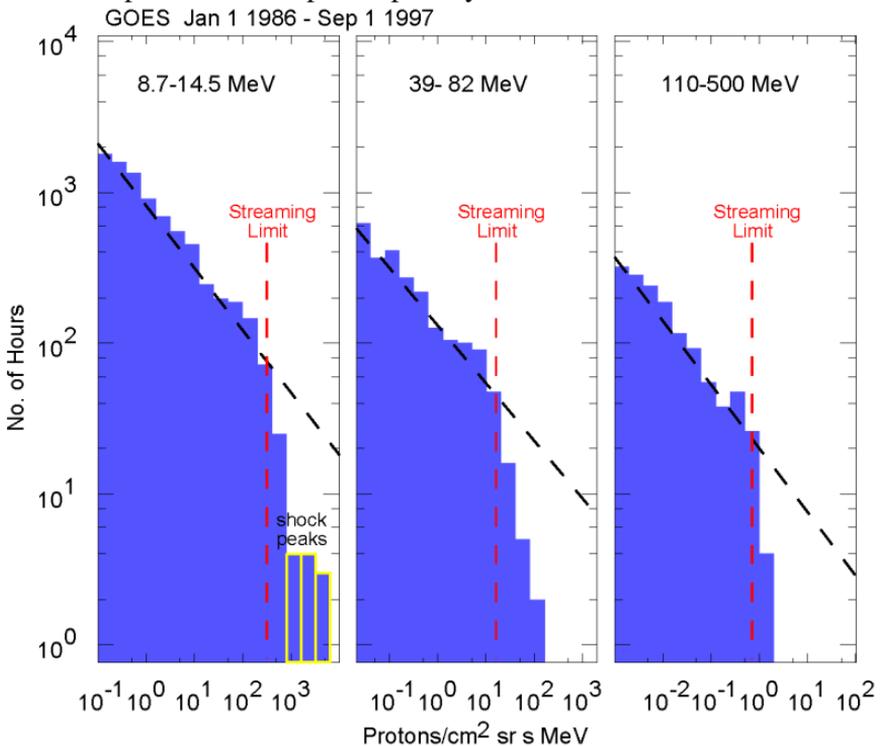

**Fig. 6.4**. The number of hours a given intensity is observed in ~11 years is shown for three different proton energy bins. Only intensities near the rarer shock peaks are seen above the streaming limit (Reames 2013 © Springer).

The black dashed lines in Fig. 6.4 are power-law fits below the streaming limit that decrease as the ~0.4 power of the intensity. This is often expressed as the differential slope, i.e. the rate of change in the number for a given change in the intensity, which decreases as the 1.4 power of the intensity in this case. Cliver et al. (2012) have compared different size measures of SEP events and of hard and soft solar X-ray events.

The streaming limit is not conveniently low so as to prevent excessive radiation exposure to astronauts, but at least it does offer a limit which is in force for a day or so before intensities begin to ramp up as the shock approaches. This allows astronauts time to reenter their vehicles and seek shelter, for example. The intensity level applied as the limits in Fig. 6.4 are shown during several large GLEs in Fig. 6.5. At the lower energies, up to ~80 MeV, the peak at the shock can exceed the streaming limit by an order of magnitude or more.

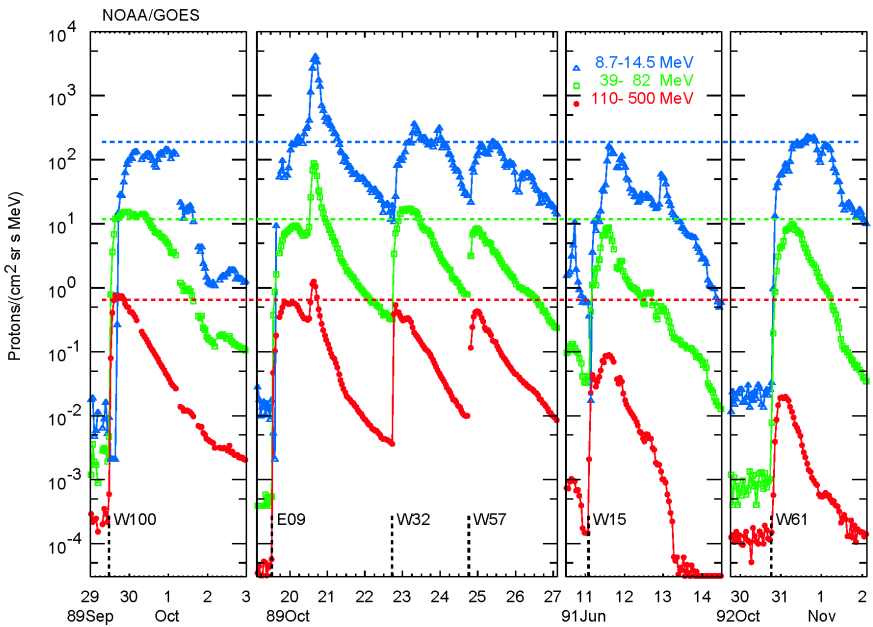

**Fig. 6.5**. Intensity levels are shown in six large SEP events with the corresponding streaming limits (Reames and Ng 1998 © AAS)

The rate of rise of the proton intensity can also be a factor in the *establishment* of equilibrium of the streaming limit as shown in Fig. 6.6. The fast rise of high-energy protons in the SEP event of 20 January 2005 allows the intensity to exceed the equilibrium limit until there has been sufficient wave growth to establish the equilibrium. Most events have slower evolution and do not overshoot the limit. Finally, Lario, Aran, and Decker (2009) have pointed out that trapping might also allow intensities to exceed the streaming limit.

It would be very useful to determine the rate of rare large SEP events and spikes of nitrates in ice cores were once suggested for this. However, these have been found to represent forest fires instead of SEP events (Schrijver et al. 2012).

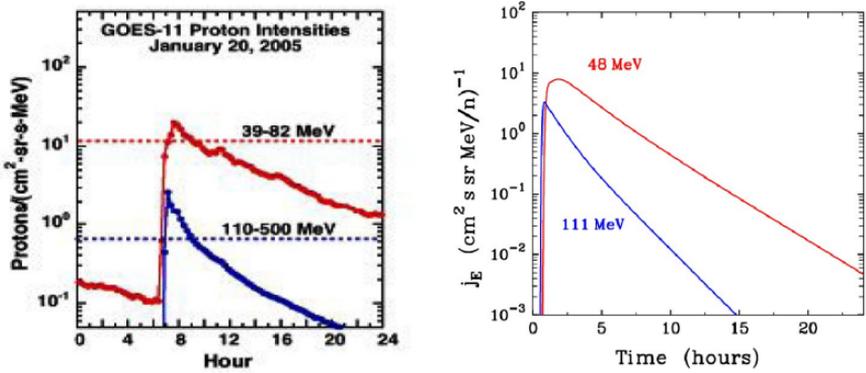

**Fig. 6.6.** The *left panel* shows that intensities in the event of January 20, 2005 briefly exceed the expected streaming limits from Fig. 6.5 (Mewaldt et al. 2007). The *right panel* shows that time-dependent calculations described by Ng, Reames, and Tylka (2012) also exceed these limits because there has not yet been enough proton flow to establish wave equilibrium at the highest energies. The fluence above 1 GV for this event is compared with other events in Fig. 6.3.

## 6.3 Radial Dependence

Radial dependence of SEP intensities can be complex, but is often important for radiation assessment, especially on missions that approach the Sun. There is a wide variation in behavior. We might expect an impulsive injection to diverge like $r^{-3}$ while we have seen that reservoirs have no radial variation at all. Theoretically the dependence on space and time in a large gradual SEP event is shown in the illustrative example in Fig. 6.7.

**Fig. 6.7.** Theoretical intensity of 5.18 MeV protons vs. radius is shown as it varies with time during a large SEP event. Soon after arrival at a given radius, intensities rise to the streaming limit at that radius. At 1 AU, intensities are bounded near ~100 (cm² s sr MeV)⁻¹ until the shock reaches ~0.7 AU (see Ng, Reames, and Tylka 2003, 2012). The tracks of the evolving "peak" and the lower "streaming limit" vs. *r* are indicated.

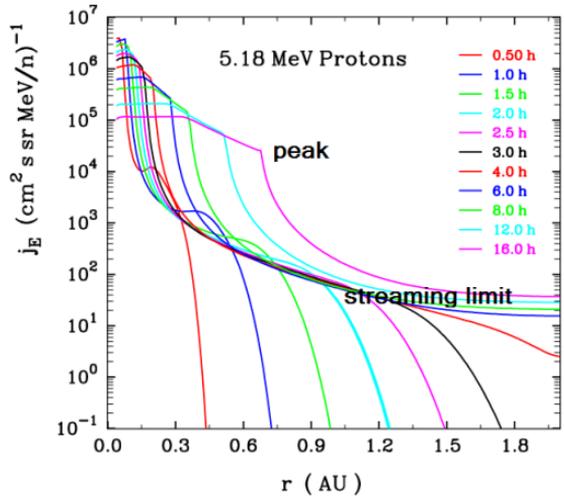

Notice that the streaming limit is itself a strong function of radius and that the peak intensity at the shock follows a different radial track; both are the same at the shock when it is near the Sun. There can be a severe radiation hazard to equipment on a spacecraft that approaches the Sun. However, the probability may be small for occurrence of a large gradual SEP event during a brief passage of space-craft perihelion. Early orbits of the *Parker Solar Probe* have reached inside 0.17 AU, but during the extreme quiet of solar minimum.

A model for calculating the radial dependences has been described by Verkhoglyadova et al. (2012).

## 6.4 Radiation Hazards and an SEP Storm Shelter

The NASA (2007; Carnell et al. 2016) permissible radiation exposure limits for humans in space are shown in Table 6.1. Entries in mGy-Eq (milligray equivalent) have been multiplied by the relative biological effectiveness of the radiation (1.5 for protons, 2.5 for heavy ions) and are for short-term or career non-cancer effects (1 Gy = 1 joule kg$^{-1}$ = 100 rad = $6.24 \times 10^{12}$ MeV kg$^{-1}$ deposited energy). The NASA standard is an excess risk of exposure induced death from cancer of no more than 3% at the 95% confidence limit. Short term limits are used to prevent clinically significant non-cancer effects, such as acute radiation syndrome response, and degradations in crew performance (Townsend et al. 2018).

**Table 6.1** Permissible Radiation Exposure Limits for Humans

BFO = Blood-forming organs, CNS = Central nervous system

| Organ | 30 days (mGy-Eq) | 1 year (mGy-Eq) | Career (mGy-Eq) |
|---|---|---|---|
| Lens | 1000 | 2000 | 4000 |
| Skin | 1500 | 3000 | 6000 |
| BFO | 250 | 500 | NA |
| Heart | 250 | 500 | 1000 |
| CNS | 500 mGy | 1000 mGy | 1500 mGy |
| CNS (Z≥10) | – | 100 mGy | 250 mGy |

For human missions into deep space for extended periods, such as missions to the Moon or Mars, large SEP events become a significant radiation risk. Since it is impractical to shield an entire spacecraft from a SEP event, the idea of a temporary shelter that could be entered during a large SEP event has emerged as a possible alternative. It has even been suggested that such a shelter could be assembled from components that are already onboard (e.g. drinking water or even waste), but, of course, it must be possible to maintain and use this shelter for periods of several days. Water-filled vests have been proposed, but water-filled helmets may be as essential.

As a design standard, it has been proposed (Townsend et al. 2018) that the shelter should protect humans against a radiation dose based upon that encountered in the 19 October 1989 SEP event series (see Fig. 6.5 or 1.5B) as is shown in Fig. 6.8 and tabulated in Townsend et al. (2018).

**Fig. 6.8** The "Design Basis" SEP proton fluence spectrum proposed for shelter design (data from Townsend et al. 2018). This fluence may be distributed over several days (e.g. see Fig.1.5B).

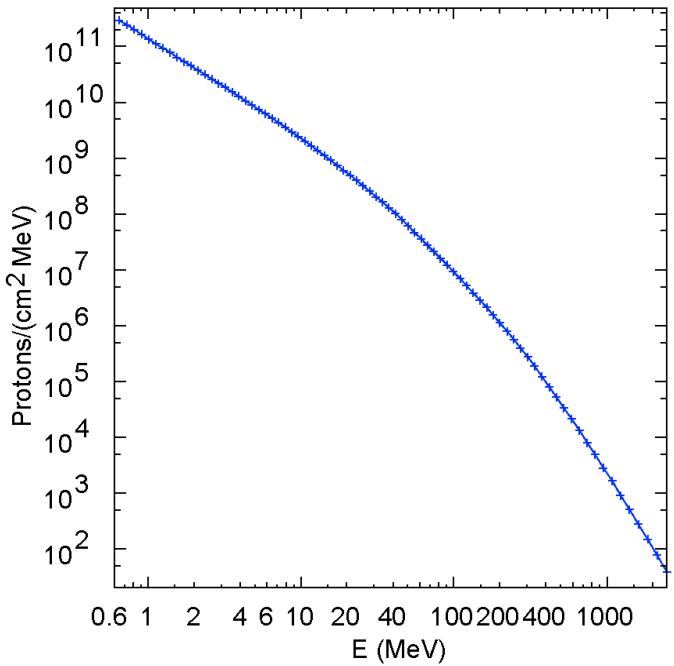

Typical BFO dose as a function of shielding thickness for Al and polyethylene are shown, for example, for female crew members in Fig. 6.9; similar curves for male crew members are also shown in Xapsos et al. (1999) and Townsend et al. (2018).

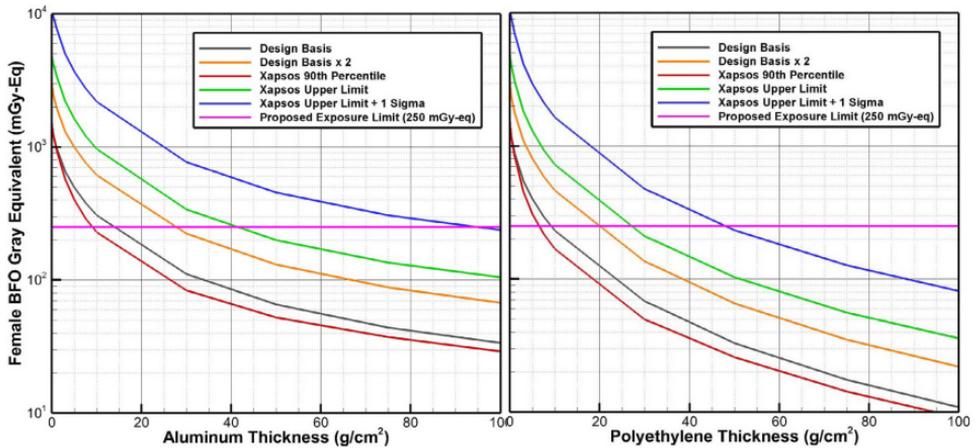

**Fig. 6.9** The BFO radiation dose to female crew members is shown as a function of thickness of Al (*left panel*) and of polyethylene (*right panel*) for the energy spectra including the "Design Basis" spectrum in Fig. 6.8 (see Townsend et al. 2018; Xapsos et al. 1999). The figure shows that low-Z polyethylene is much more efficient than Al.

On planets or moons it may be possible to build efficient shelters out of local materials. In some cases, including the Earth's Moon, there may be caverns or caves, similar to lava tubes from volcanoes on Earth, which can provide a safe refuge.

## 6.5 A Mission to Mars

A mission to Mars beginning 26 November 2011 carried instruments that led to an estimate that the radiation dose during a ~1-yr round-trip mission would be $660\pm12$ mSv (Zeitlin et al. 2013). Of course, this mission occurred during the notoriously weak Solar Cycle 24 that did not contain a SEP event like that of 23 February 1956 (see Fig. 6.3). Fortunately those events are very rare.

For the timing of a manned mission to Mars, one can go during solar maximum when SEP events are more probable but GCR intensities are reduced, or during solar minimum when SEPs are reduced but GCRs are at maximum (see Fig. 1.8). The continuous radiation of GCRs causes cancer risk in astronauts, while the SEPs pose a small risk of serious radiation sickness or even a fatal exposure. Most planning assumes a trip to Mars during solar maximum to reduce the cancer risk. It is assumed that SEP risk can be reduced somewhat by a safe-haven shelter with shielding of 20–40 g cm$^{-2}$, combined with an adequate warning system. GCR radiation is *not* reduced by shielding; it is actually increased by production of secondary nuclear-reaction products, including highly penetrating neutrals (Carnell et al. 2016).

Little effort is presently expended on studying the hazard from SEP events, i.e. assessing their risk of occurrence, and ensuring an appropriate structure is in place to provide adequate warning. In addition, there is little planning for contingencies in case of an extreme event. The probability of an extremely hazardous event occurring during a specific mission, even a one-year mission, is relatively small, perhaps less than a few percent. The problem actually comes when there is a continuous human presence outside the Earth's magnetosphere; then it is not a question of if, but when.

## 6.6 The Upper Atmosphere of Earth

Ionization of the upper atmosphere during large SEP events can have significant long-term effects on the chemistry of the Earth's polar atmosphere. SEP ionization produces $HO_x$ and $NO_y$ in the mesosphere and stratosphere and the lifetime of the $NO_y$ allows it to affect ozone for months to years. Mesospheric ozone depletions of 50% can last for hours or days. Significant ozone depletions of >10% can last a few months after SEP events. However, interference with the Cl- and Br-loss cycles actually caused an increase in total ozone, for example in 1992–1994, a few years after the October 1989 series of SEP events shown in Fig. 6.5 (Jackman, Fleming, and Vitt, 2000; Jackman et al. 2006). More recent events from January and March 2012 have also produced effects (von Clarmann et al. 2013) from this weaker solar cycle.

## 6.7 SEPs and Exoplanets

The possible effects of SEPs, *stellar* energetic particles, from other stars may also be an issue for the habitability and the development of life on exoplanets, or their moons (e.g. Lingham et al. 2018; Airapetian et al. 2019). SEPs could either constitute a radiation hazard in an otherwise habitable environment or they could induce reactions that produce the organic chemicals necessary for life. SEPs can have a significant effect on planetary atmospheres, depending upon planetary magnetic fields. In any case they are another factor that must be considered in the definition of a habitable zone.

   The energy and intensity of energetic particles scales with the speed of CMEs and, while CMEs cannot be observed on other stars, their speed depends upon the intensity of the stellar magnetic field. That field is likely to be increased in young rapidly-rotating stars. Also, X-ray flares can be measured for stars and, while they are not directly related to SEPs, their number and intensity is related to the stellar magnetic field reconnection. Actually, the Sun is relatively inactive among stars of its class; it would be most interesting to see the range of SEP variation and its dependence upon stellar properties.

   Fu et al. (2019) used the improved PATH model (Hu et al. 2017) to study the dependence of SEPs on the rotation rate of the star. They found that within 0.8 AU for three times the solar rotation rate, the maximum particle energy at the shock front increased with the rotation rate of the star.

   Detection of type II radio bursts could provide additional evidence of stellar SEPs. Unfortunately, radio emission below about 10 MHz cannot be measured from the Earth's surface because of ionospheric interference.

## References

Afanasiev, A., Vainio, R., Rouillard, A.P., Battarbee, M., Aran, A., Zucca, P., Modelling of proton acceleration in application to a ground level enhancement, Astron.Astrophys **614** A4 (2018) doi: 10.1051/0004-6361/201731343

Airapetian, V.S., Barnes, R., Cohen, O., Collinson, G.A., Danchi, W.C., Dong, C.F., Del Genio, A.D., France, K., Garcia-Sage, K., Glocer, A et al., Impact of space weather on climate and habitability of terrestrial type exoplanets, arXiv: 1905.05093

Band, D., Matteson, J., Ford, L., Schaefer, B., Palmer, D., Teegarden, B., Cline, T., Briggs, M., Paciesas, W., Pendleton, G., et al., BATSE observations of gamma-ray burst spectra. I - Spectral diversity, Astrophys. J., **413**, 281 (1993) doi: 10.1086/172995

Barth, J.L., Dyer, C.S., Stassinopoulos, E.G., Space, atmospheric, and terrestrial radiation environments, IEEE Trans. Nucl. Sci. **50**, 466. (2003) doi: 10.1109/TNS.2003.813131

Bruno, A., Bazilevskaya, G.A., Boezio, M., Christian, E.R., de Nolfo, G.A., Martucci, M., Merge', M., Mikhailov, V.V., et al., Solar energetic particle events observed by the PAMELA mission, Astrophys. J. **862**, 97 (2018) doi: 10.3847/1538-4357/aacc26

Carnell, L., Blattnig, S, Hu, J. Huff, J., Kim, M.-H., Norman, R., Patel, Z., Simonsen, L., Wu, H., *NASA 1 evidence report: Risk of Acute Radiation Syndromes Due to Solar Particle*

*Events* (2016)        https://humanresearchroadmap.nasa.gov/evidence/reports/Acute.pdf? rnd=0.543557888150009

Cliver, E.W., Ling, A.G., Belov, A., Yashiro, S., Size distributions of solar flares and solar energetic particle events, Astrophys. J. Lett. **756,** L29 (2012) doi: 10.1088/2041-8205/756/2/L29

Cucinotta, F.A., Hu, S., Schwadron, N.A., Kozarev, K., Townsend, L.W., Kim, M.-H. Y., Space radiation risk limits and Earth-Moon-Mars environmental models, Space Weather **8** S00E09 (2010) doi: 10.1029/2010SW000572

Fu, S., Jiang, Y., Airapetian, V., Hu, J., Li, G., Zank, G., Effect of star rotation rate on the characteristics of energetic particle events, Astrophys. J. **878** 36 (2019) doi: 10.3847/2041-8213/ab271d

Hu J, Li G, Ao X, Verkhoglyadova O, Zank G. Modeling particle acceleration and transport at a 2D CME-driven shock. J. Geophys. Res. **122** 10,938 (2017) doi: 10.1002/2017JA024077

Jackman, C.H., Deland, M.T., Labow, G.J., Fleming, E.L., López-Puertas, M., Satellite measurements of middle atmospheric impacts by solar proton events in Solar Cycle 23, Space Sci. Rev. **125**, 381 (2006) doi: 10.1007/s11214-006-9071-4

Jackman, C.H., Fleming, E.L., Vitt, F.M., Influence of extremely large solar proton events in a changing stratosphere, J. Geophys. Res. **105** 11659 (2000) doi: 10.1029/2000JD900010

Kahler, S.W., Ling, A. Dynamic SEP event probability forecasts, Space Weather **13**, 665 (2015) doi: 10.1002/2015SW001222

Lario, D., Aran, A., Decker, R.B., Major solar energetic particle events of solar cycles 22 and 23: Intensities close to the streaming limit, Sol. Phys. **260**, 407 (2009) doi: 10.1007/s11207-009-9463-1

Laurenza, M., Cliver, E.W., Hewitt, J., Storini, M., Ling, A.G., Balch, C.C., Kaiser, M.L., A technique for short-term warning of solar energetic particle events based on flare location, flare size, and evidence of particle escape, Space Weather, **7** S04008 (2009) doi: 10.1029/2007SW000379

Lingam, M., Dong, C., Fang, X., Jakosky, B. M., and Loeb, A., The propitious role of solar energetic particles in the origin of life. Astrophys. J. **853** 10 (2018) doi: 10.3847/1538-4357/aa9fef

Lovell, J.L., Duldig, M.L., Humble, J.E., An extended analysis of the September 1989 cosmic ray ground-level enhancement, J. Geophys. Res., **103**, 23,733 (1998) doi: 10.1029/98-JA02100

Mewaldt, R.A., Cohen, C.M.S., Haggerty, D.K., Mason, G.M., Looper, M.L., von Rosenvinge, T.T.. Wiedenbeck, M.E., Radiation risks from large solar energetic particle events, AIP Conf. Proc. **932**, 277 (2007) doi: 10.1063/1.2778975

NASA (2007). *NASA Space Flight Human System Standard. Volume 1: Crew Health,* NASA-STD-3001 (National Aeronautics and Space Administration, Washington). https://standards.nasa.gov/standard/nasa/nasa-std-3001-vol-1

Ng, C.K., Reames, D.V., Tylka, A.J., Modeling shock-accelerated solar energetic particles coupled to interplanetary Alfvén waves, Astrophys. J. **591**, 461 (2003) doi: 10.1086/375293

Ng, C.K., Reames, D.V., Tylka, A.J., Solar energetic particles: shock acceleration and transport through self-amplified waves, AIP Conf. Proc. **1436**, 212 (2012) doi: 10.1063/1.4723610

Raukunen, O., Vainio, R., Tylka, A.J., Dietrich, W.F., Jiggens, P., Heynderickx, D., Dierckxsens, M., Crosby, N., Ganse, U., Siipola, R., Two solar proton fluence models based on ground level enhancement observations, J. Spa. Wea. Spa. Clim. **8**, A04 (2018) doi: 10.1051/swsc/2017031

Reames, D.V., Solar energetic particles: Is there time to hide? Radiation Measurements **30**/3, 297 (1999) doi: 10.1016/S1350-4487(99)00066-9

Reames, D.V.: The two sources of solar energetic particles, Space Sci. Rev. **175**, 53 (2013) doi: 10.1007/s11214-013-9958-9

Reames, D.V. Ng, C.K., Streaming-limited intensities of solar energetic particles, Astrophys. J. **504**, 1002 (1998) doi: 10.1086/306124

Reames, D.V., Ng, C.K., Streaming-limited intensities of solar energetic particles on the intensity plateau, Astrophys. J. **722**, 1286 (2010) doi: 10.1088/0004-637X/723/2/1286

Schrijver, C.J., Beer, J., Baltensperger, U., Cliver, E. W., Güdel, M., Hudson, H.S., McCracken, K.G., Osten, R.A., Peter, T., Soderblom, D.R., Usoskin, I.G., Wolff, E.W., Estimating the frequency of extremely energetic solar events, based on solar, stellar, lunar, and terrestrial records, J. Geophys. Res. **117**, A08103 (2012) doi: 10.1029/2012JA017706

Townsend, L.W., Adams, J.H., Blattnig, S.R., Clowdsley, M.S., Fry, D.J., Jun, I., McLeod, C.D., Minow, J.I., Moore, D.F., Norbury, J.W., Norman, R.B., Reames, D.V., Schwadron, N.A., Semones, E.J., Singleterry, R.C., Slaba, T.C., Werneth, C.M., Xapsos, M.A., Solar particle event storm shelter requirements for missions beyond low Earth orbit, Life Sci. in Space Res. **17** 32 (2018) doi: 10.1016/j.lssr.2018.02.002

Tylka, A.J., Boberg, P.R., McGuire, R.E., Ng, C.K., Reames, D.V., Temporal evolution in the spectra of gradual solar energetic particle events, in AIP Conf. Proc. **528**, *Acceleration and Transport of Energetic Particles Observed in the Heliosphere*, ed. R. A. Mewaldt, J. R. Jokipii, M.A. Lee, E. Möbius, T.H. Zurbuchen (Melville: AIP), 147 (2000) doi: 10.1063/1.1324300

Tylka, A.J., Dietrich, W.F., A new and comprehensive analysis of proton spectra in ground-level enhanced (GLE) solar particle events, in *Proc. 31st Int. Cos. Ray Conf* , Lódz (2009) http://icrc2009.uni.lodz.pl/proc/pdf/icrc0273.pdf

Verkhoglyadova, O.P., Li, G., Ao, X., Zank, G.P., Radial Dependence of Peak Proton and Iron Ion Fluxes in Solar Energetic Particle Events: Application of the PATH Code, Astrophys. J. **757**, 75 (2012) doi: 10.1016/j.physrep.2014.10.004

von Clarmann, T., Funke, B., López-Puertas, M., Kellmann, S., Linden, A., Stiller, G.P., Jackman, C.H., Harvey, V.L., The solar proton events in 2012 as observed by MIPAS, Geophy. Res. Lett. **40** 2339 (2013) doi: 10.1002/grl.50119

Xapsos, M.A., Stauffer, C., Jordan, T., Barth, J.L., Mewaldt, R.A., Model for cumulative solar heavy ion energy and linear energy transfer spectra, IEEE Trans. Nucl. Sci. **54**, 1985 (2007) doi: 10.1109/TNS.2007.910850

Xapsos, M.A., Summers, G.P., Barth, J.L., Stassinopoulos, E.G., Burke E.A. Probability model for worst case solar proton event fluences. IEEE Trans. Nucl. Sci. 46 (6): 1481 (1999) doi: 10.1109/23.819111

Zeitlin, C, Hassler, D.M., Cucinotta, F.A., Ehresmann, B., Wimmer-Schweingruber, R.F., Brinza, D.E., et al., 2013, Measurements of energetic particle radiation in transit to Mars on the Mars Science Laboratory, Science **340**, 1080 (2013) doi: 10.1126/science.1235989

# Chapter 7. Measurements of SEPs

**Abstract**   Those who study solar energetic particles (SEPs) should be aware of the basic types of experiments that have contributed most of the observations studied in this book, and especially the tradeoff of their strengths and weaknesses, and how they fail.  However, this is *not* a comprehensive review, only an introduction. We focus on d$E$/d$x$ vs. $E$ instruments that are the workhorses of SEP studies, and also study time-of-flight vs. $E$ instruments that dominate precision measurements below 1 MeV amu$^{-1}$.  Single-detector instruments and high-energy techniques are discussed briefly as are supplementary data and CME lists.

Nearly every experimenter who builds instruments thinks he has made the best tradeoff within the triple constraints of weight, power, and expense, to maximize the scientific return.   Many instruments are designed to extend coverage to a previously unmeasured region: energy coverage, isotope resolution, heavy elements. Others hitchhike on spacecraft going to a new and interesting region of space. There are usually tradeoffs among resolution, geometry factor, background elimination, element or isotope coverage, and energy coverage.  Higher-energy ions are much easier to resolve, but high intensities at lower energies sample rarer species and rarer events.

The rate of energy loss of an ion in a detector material is approximately

$$\frac{dE}{dx} \approx \frac{4\pi e^2 n_e}{mc^2 \beta^2}\left(\frac{Q^2}{A}\right)\left[\ln\left(\frac{2mc^2\beta^2\gamma^2}{I}\right) - \beta^2\right] \tag{7.1}$$

where $m$ and $e$ are the mass and charge of an electron, $n_e$ is the electron density, $I$ is the "mean ionization potential" of the stopping material, and $\beta$ and $\gamma$ are the relativistic velocity and Lorentz factor of the ion as defined in Sect. 1.5.4. Here we have again used $E = \mathcal{E}/A = M_u (\gamma - 1) \approx \frac{1}{2} M_u \beta^2$, a function of velocity alone, to show that the only dependence on the stopping ion is $Q^2/A$ and its velocity $\beta$.  $M_u = m_u c^2 = 931.494$ MeV.

Equation 7.1 is derived from the electron-ion scattering cross section (Rutherford scattering) where we view incoming electrons of the stopping material being scattered by the electric field of the ion.  Energy transfers to the electrons are integrated from a minimum of $I$ to a maximum of $2mc^2\beta^2\gamma^2$, which is approximately the maximum energy that can be transferred to a scattered electron.  Note that, when $Q \approx Z$, the dominant energy dependence of $dE/dx$ is $\sim\beta^{-2}$, or nonrelativistically, $\sim E^{-1}$, sometimes a useful approximation.

At relativistic energies, $dE/dx$ reaches a broad minimum at ~2.5 GeV amu$^{-1}$ then rises slightly from density effects not included here. At low energies, $dE/dx$ actually peaks, because $Q$ decreases, but $Q \rightarrow Z$ at moderate energies. A simple approximation sometimes used for this is $Q \approx Z [1-\exp(-\beta/\beta_0)]$. For capture into the K orbital, $\beta_0 \approx Z/137$; for the Fermi-Thomas model $\beta_0 \approx Z^{2/3}/137$. Modern empirical tables use more complex expressions and tabulate both stopping power and range (Hubert, Bimbo, and Gauvin 1990). The particle range $R = \int dE (dE/dx)^{-1}$. For energies down to 1 keV amu$^{-1}$, the tables of Paul and Schinner (2003) are available.

## 7.1 Single-Element Detectors

Conceptually, the simplest detector is that with a single sensitive element. Modern "solid-state" detectors are a Si wafer biased as a capacitor that collects the electron-hole pairs produced when an ionizing particle penetrates, loses energy, or stops within its volume. The charge collected, proportional to the energy loss, is measured as a pulse height by analog-to-digital converters. Single-element detectors are generally shielded to define the access geometry for low-energy particles.

Measuring the energy of each arriving particle works at low energies, but penetrating particles contribute as if they had a much lower energy (Fig. 7.1). When the SEPs have a steep energy spectrum, the contribution of high-energy particles may be small, but early in SEP events nearly all particles are penetrating and single-detector instruments falsely appear to show low-energy particles arriving from the Sun much earlier than they possibly could. This effect is sometimes called "punchthrough," it occurs in *every* SEP event, and can cause serious misconceptions. These detectors also confuse heavier ions similarly, even though they deposit an increasing amount of energy. Single-element telescopes should never be used for SEP onset timing. They are more appropriate for the study of steep energetic particle spectra at interplanetary shock waves. Their obvious advantage is low weight, power, and cost.

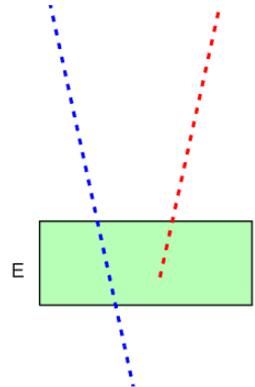

**Fig. 7.1** A single-detector telescope measures the total kinetic energy of stopping ions (*red*) and the energy loss of penetrating ions (*blue*), which is much lower. The latter are (incorrectly) assumed to be rare. Access geometry is somewhat controlled by shielding (not shown) and a permanent magnet may be included to sweep away electrons, or to measure electrons by comparing detectors with and without magnets.

Electrons are particularly difficult to measure since they do not travel in straight lines in detectors, and can suffer numerous large-angle scatters. The best remedy is extensive instrument calibration before launch.

Single-element and other limited telescopes are sometimes flown on deep-space or planetary missions where weight and power are severely limited and where SEPs are not the primary objective. Unfortunately, these low-priority hitchhikers may even be turned off during transit to the mission destination to save resources, further decreasing their limited value.

## 7.2 ∆E vs. E Telescopes

These telescopes consist of at least three active detector elements. Particles enter the first detector, penetrate into the second, and stop before entering the third anti-coincidence detector. The separation of the first two detectors, and their areas, determine the instrument geometry-factor. The detector thicknesses determine the minimum and maximum energy according to the range-energy relation in Si (e.g. Hubert, Bimbo, and Gauvin 1990). Front detector thicknesses of 10–20 μm set a lower bound of ~1 MeV amu$^{-1}$, depending upon species. Total thicknesses of D+E of up to 10 cm of Si are used for energies ~200 MeV amu$^{-1}$. The energy range can be extended to above ~400 MeV amu$^{-1}$ by observing the *change* in d$E$/d$x$ between D and E, if penetrating ions are measured. The concept of a two-element telescope is shown in Fig. 7.2.

**Fig. 7.2**. A minimal ∆$E$ vs. $E$ or two-dimensional telescope requires coincidence of signals from the D and E elements and no signal from the A element to define stopping (*red*) ions. "Matrix" plots of pulse-heights of D vs. E are used to resolve elements and measure their energies (see Fig. 7.3). The anti-coincidence element or inert shielding may surround the telescope. Note that the energy deposit in D depends upon the angle of the particle trajectory with the telescope axis; this energy spread and thickness variations control charge and isotope resolution.

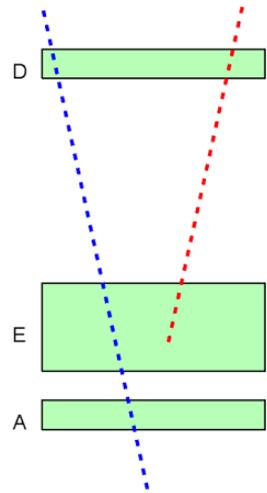

Most of the particle telescopes flown in space are of this general type, although multiple detectors may be used in place of the D and E elements. Early telescopes used plastic scintillator or even gas drift chambers, but most telescopes of the last 25 years are "solid state" Si detectors which have extremely high resolution and *stability, i.e. their response does not change at all during decades of operation*. Gas detectors soon suffer from out-gassing, contamination, or leakage.

Anti-coincidence detectors were sometimes wrapped around the whole telescope. However, at the high rates in a large SEP event these may be recording particles nearly all the time, and insure that the telescope is effectively turned off.

A geometry factor defined by surfaces $S_1$ and $S_2$ is

$$A\Omega = \iint\limits_{S_1} dS_1 \iint\limits_{S_2} dS_2 \, \frac{(\mathbf{n}_1 \cdot \mathbf{r})(\mathbf{n}_2 \cdot \mathbf{r})}{r^4} \tag{7.2}$$

where $\mathbf{n}_1$ and $\mathbf{n}_2$ are unit vectors normal to the surface elements $dS_1$ and $dS_2$, respectively and $\mathbf{r}$ is the vector distance between them. Geometry factors are usually calculated numerically.

### 7.2.1 An Example: LEMT

Response of a telescope with a thin front detector, the *Low-Energy Matrix Telescope* (LEMT) on the *Wind* spacecraft (von Rosenvinge et al. 1995), is shown in Fig. 7.3. LEMT has three important virtues, large geometry (51 cm² sr), broad element coverage (H–Pb at ~2–20 MeV amu⁻¹), and, equally important, the author is familiar with it. Each LEMT consists of a domed array of 16 D-detectors 18 µm thick, followed by a large 1-mm-thick E-detector with coarse $5 \times 5$ position sensing and an anticoincidence detector (see von Rosenvinge et al. 1995).

**Fig 7.3**. Response of the LEMT telescope (diagram shown in inset) to ions from a small ³He-rich event in 1995 is shown with "tracks" of species indicated. The telescope has only modest resolution of He isotopes. The track of O is heavily populated by anomalous cosmic rays during this period near solar minimum (see Reames et al. 1997 © AAS)

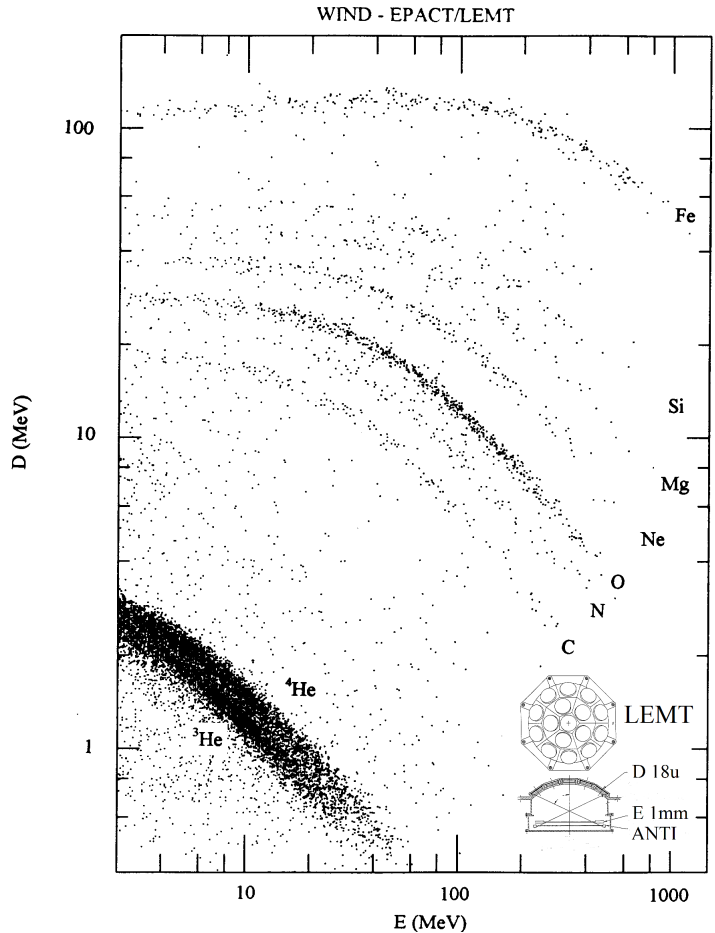

Particles entering LEMT are corrected for angle of entry, mapped in a log $D$ vs. log $E$ space, like that of Fig. 7.3, and binned onboard according to particle species and energy interval (see Reames, Ng, and Berdichevsky 2001).   The right-hand ends of the particle tracks, especially noticeable for C, N, and O, occur just before the ions have enough energy to begin to penetrate into the anticoincidence detector.

In the region of the rarer elements with $Z \geq 34$, "priority" measurements of individual ions are rare enough to be telemetered for later analysis.   The performance of LEMT at high $Z$ is shown in Fig. 7.4.

**Fig. 7.4**. High-$Z$ response of LEMT is shown where resolution (i.e. track width) is comparable with that at Fe, but track locations are well calibrated using beams of Fe, Ag, and Au before launch. Energy varies along each calibration curve from *left* to *right*, from 2.5 to 10 MeV amu$^{-1}$.

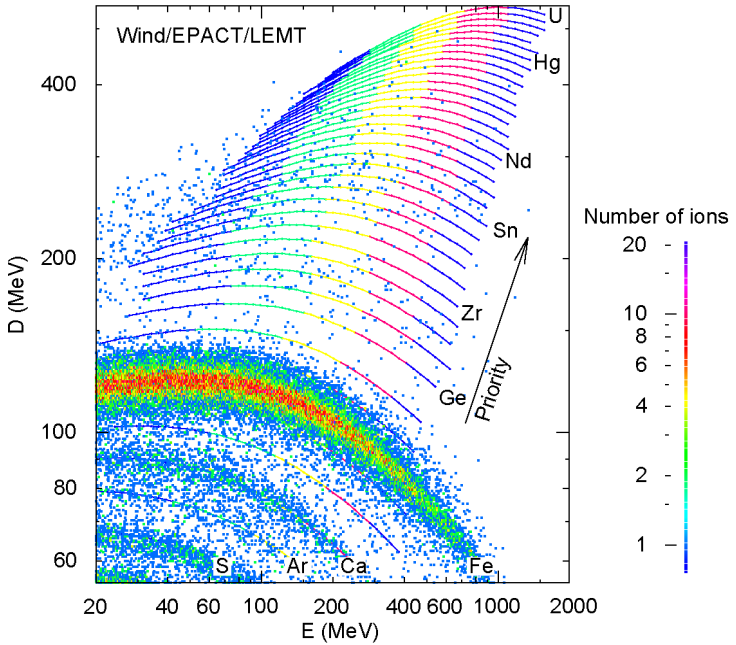

While the error at high $Z$ is $\pm 2$ to 3 units, the resolution is adequate to show bands of enhanced abundances, such as that between Ge and Zr and the band near Sn, that reflect an abundance maximum at $50 \leq Z \leq 56$. The absolute locations of the reference curves of the elements were well calibrated prior to launch using accelerator beams of He, C, O, Fe, Ag, and Au (see von Rosenvinge et al. 1995).   By measuring at low energy with a fairly large geometry factor, LEMT can move up the steep energy spectra to get a rough measure of the abundances of the rare elements with $34 \leq Z \leq 82$.  For results of these measurements see Figs. 4.7 and 4.9.

### 7.2.2 Isotope Resolution: SIS

Accuracy can be affected by thickness variations and sec $\theta$ variations by particle trajectories inclined by an angle $\theta$ to the telescope axis. Both of these may be reduced by accurately measuring sec $\theta$ using two sets of $x$ and $y$ strip detectors (e.g. Stone et al. 1998).  The additional detector thickness required for these measurements raises the energy threshold above ~10 MeV amu$^{-1}$, depending upon particle species, but also permits isotope resolution up to Fe – an important tradeoff.  Fig.

7.5 shows the resolution of Ne isotopes by the *Solar Isotope Spectrometer* (SIS) on the *Advanced Composition Explorer* (ACE) in two different SEP events. Isotopic abundances show the same *A/Q* variations we have seen in element abundance enhancements in both impulsive and gradual SEP events. Here, however, there is no question about the average value of *Q*, which we assume to be the same for all isotopes of an element.

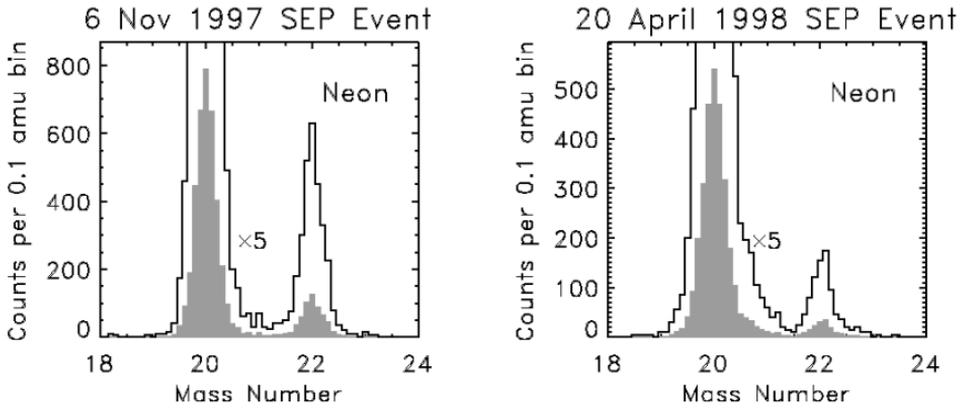

**Fig. 7.5**. Panels show the resolution of Ne isotopes by the SIS telescope in two SEP events. Histograms are also shown enhanced by a factor of 5 to clarify $^{22}$Ne measurement (Leske et al. 2007 © Springer). Isotope measurements show *A/Q* enhancements like those in element abundances.

### 7.2.3 Angular Distributions

Most early missions outside the magnetosphere, such as IMP 8, ISEE 3, and even *Wind*, were spin stabilized and could choose the spin axis normal to the ecliptic, allowing easy measurement of each particle's direction of incidence as a time in the spin phase relative to a Sun pulse or a directional pulse from the magnetic field instrument (see Figs. 2.3, 2.6, 3.11, 5.6, and 9.17). Since the magnetic field often lies in the ecliptic plane, this allows easy measurement of pitch-angle distributions. Instruments with apertures above and below the ecliptic could even have 3D coverage.

More-recent missions, like *Voyager*, SOHO, STEREO, PSP, and even ACE are controlled by stabilized or sun-pointing instruments such as cameras or coronagraphs, so that the particle instruments have great difficulty measuring meaningful distributions, even with multiple telescopes and look angles. Particle transport can be complex and the ability of older instruments to compare distributions of H, $^3$He, $^4$He, O, and Fe (e.g. Reames, Ng, and Berdichevsky 2001) can often help answer modern questions of ion transport when they arise (e.g. Reames 2019).

### 7.2.4 Onboard Processing

Early instruments telemetered pulse-height measurements of each particle to the ground for processing. Often a new particle could not be sampled until the current one was telemetered. Onboard sampling criteria for H, He, and heavies were soon invented to insure that rarer species were sampled, but telemetry was limited to about one particle s$^{-1}$. Onboard microprocessors can now handle up to ~10,000 particles s$^{-1}$, correcting for geometry and identifying particle species and energy by lookup in 64×64 element log-log tables (one for $^3$He-$^4$He, one for $6 \leq Z \leq 26$) that would overlay regions of Fig. 7.3, for example. This is a major statistical improvement; the telemetry is now filled with bin counts by particle species and energy with some space reserved for pulse heights of rare species and samples for quality control.

Buffering is also possible, so that rates can be sampled for longer or shorter periods, some sampled vs. spin phase, and all are queued in a buffer for readout. However, for SEPs near 1 AU, we rarely need less than 5-min averages. Typically, for an energy interval from $E_1$ to $E_2$, it is not worthwhile to study time intervals less than $t_1 - t_2$, where $t_i$ is the typical travel time from the Sun for particles of energy $E_i$. Higher time resolution can be useful for more local sources, e.g. for spacecraft nearer the Sun or for study of local shocks. For the broad energy intervals on some SEP monitors, the mean energy can decrease markedly with time.

## 7.3 Time-of-Flight vs. E

Measurement of a particle's time of flight over a fixed distance determines its velocity. If the particle subsequently stops in a Si detector its total kinetic energy can be measured, and the pair of measurements determines the particle mass. The design of the *SupraThermal Energetic Particle* (STEP) system flown on the *Wind* and STEREO spacecraft is shown in Fig. 7.6.

**Fig. 7.6**. The STEP telescope measures time of flight vs. energy (see text; von Rosenvinge et al. 1995 © Springer).

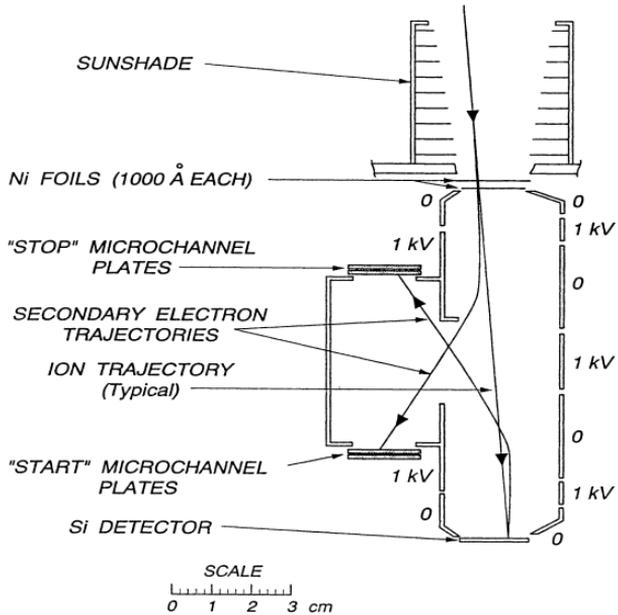

SUNSHADE

Ni FOILS (1000 Å EACH)

"STOP" MICROCHANNEL PLATES

SECONDARY ELECTRON TRAJECTORIES

ION TRAJECTORY (Typical)

"START" MICROCHANNEL PLATES

Si DETECTOR

SCALE

0   1   2   3   cm

A particle penetrating the entrance Ni foil in STEP may knock off ~4–30 electrons that are accelerated and deflected by the 1 kV electric field into the "start" microchannel plates that multiply the signal by ~100. If the particle then enters the Si detector, backscattered electrons are accelerated into the "stop" microchannel plates, and energy is measured in the Si detector. The time between the start and stop signals, 2–100 ns, is processed by a time-to-amplitude converter (TAC). The TAC and energy signals are combined into a weighted analog sum that assigns a priority that controls further processing. Heavies, with $A > 4$, are assigned the highest priority, He next, and then H. The response of STEP to a small $^3$He-rich SEP event is shown in Fig. 7.7.

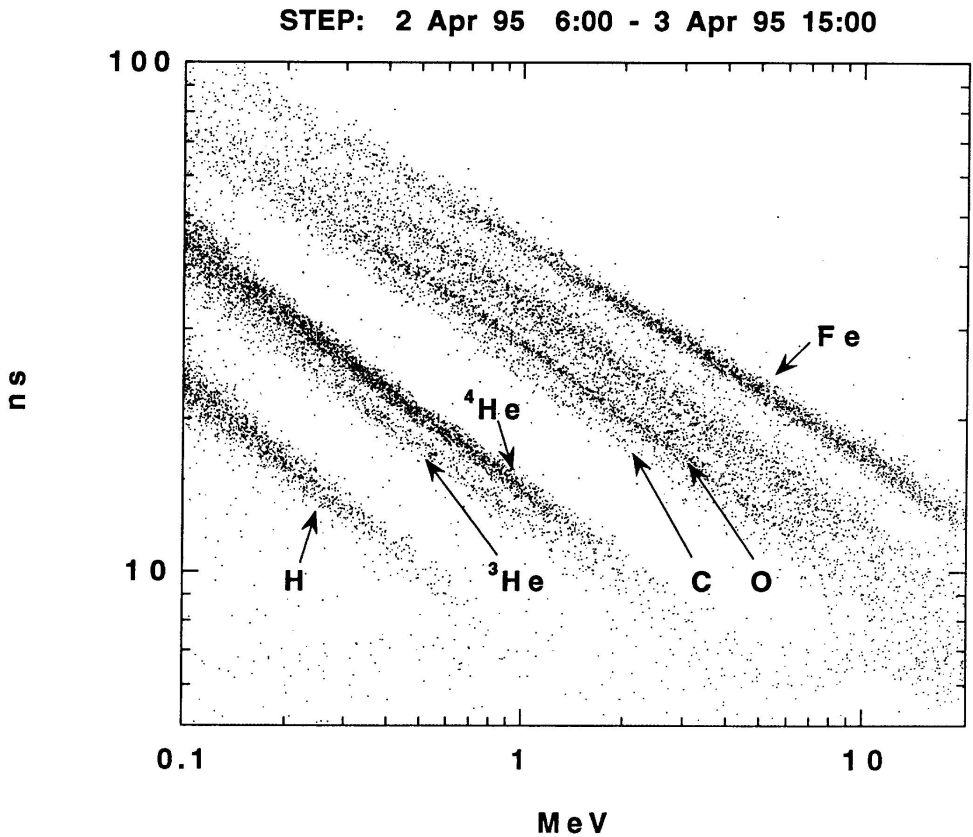

**STEP: 2 Apr 95 6:00 - 3 Apr 95 15:00**

**Fig. 7.7.** The response of the STEP telescope shows the time-of-flight (ns) vs. the total kinetic energy (MeV) for a sample of ions during a small $^3$He-rich SEP event (see von Rosenvinge et al. 1995; Reames et al. 1997 © AAS).

The resolution using this technique can be greatly improved by adding an additional timing plane, using electrostatic mirrors to reflect the electrons, and using microchannel plates with position-sensing anodes. This was done for the ULEIS instrument on the ACE spacecraft (Mason et al. 1998). This instrument produced the resolution seen in Fig. 4.19.

## 7.4 NOAA/GOES

The *Geostationary Operational Environmental Satellites* (GOES), operated by the *National Oceanic and Atmospheric Administration* (NOAA), are a series of satellites intended to give continuous time coverage of the space environment. A new GOES spacecraft with equivalent capabilities is launched every few years.

Energies of interest for SEP observations are proton energies in five channels from 4 to 500 MeV measured by two-element telescopes behind different thicknesses of shielding in the *Energetic Particle Sensor* (EPS). In addition, the *High Energy Proton and Alpha Detector* (HEPAD) adds a Cherenkov detector to measure protons in the intervals 350–420, 420–510, 510–700, and >700 MeV. These are extremely useful high-energy measurements. GOES data since 1986 are available at https://www.ngdc.noaa.gov/stp/satellite/goes/index.html (although the web site has been known to change). Note that the low-energy channels of the EPS should *not* be used for onset timing since they are contaminated by higher-energy particles. Geometry factors for high-energy particles are too uncertain to allow channel differences to exclude all contamination in EPS. However, GOES provides an excellent synoptic summary of SEP events (see Fig. 5.1) and the >700 MeV channel may be a better indicator of high-energy protons than neutron monitors (Thakur et al. 2016).

GOES also provides 1–8 Å soft X-ray peak intensities that is a classic measure of heating in solar flares. The X-ray "CMX class" specifies the decade of X-ray peak intensity with C$n$ for $n \times 10^{-6}$ W m$^{-2}$, M$n$ for $n \times 10^{-5}$ W m$^{-2}$, and X$n$ for $n \times 10^{-4}$ W m$^{-2}$ (e.g. lower panel of Fig 7.8; also see Fig. 4.15).

A very useful data base is provided by the SOHO/LASCO CME catalog which lists CME data from January 1996 on (Gopalswamy et al. 2009; https://cdaw.gs-fc.nasa.gov/CME_list/). Plots of GOES particle and X-ray data are also shown correlated with CME height-time plots as shown in the example in Fig. 7.8. Note the steep height-time plots (i.e. fast CMEs) in the middle panel near the onsets of the two large gradual SEP events in the upper panel. The SEP event on 7 January has a prominent associated X-ray increase (lower panel), while that on 6 January does not, even though both SEP events have similar intensities of >100 MeV protons. Other data in the catalog include movies of CME evolution and EUV flares along with radio data from *Wind*/WAVES.

There are also data bases of CMEs produced automatically such as the *Solar Eruptive Event Detection System* (SEEDS; http://spaceweather.gmu.edu/seeds/; Olmedo et al. 2008) which generates listings of CMEs for both SOHO/LASCO and STEREO/SECCHI and CME movies. The SEEDS web site contains links to other data bases.

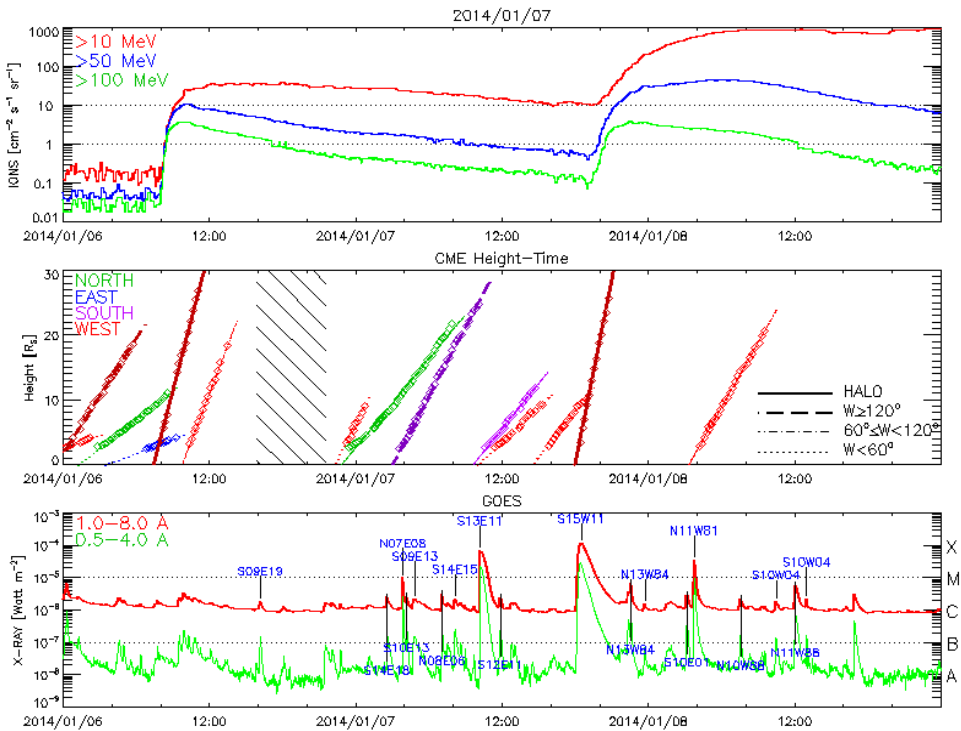

**Fig 7.8**. CME height-time plots (*center*) are shown together with GOES X-ray (*below*) and particle (*above*) data in the SOHO/LASCO CME catalog ( see Gopalswamy et al . 2009)

## 7.5 High-Energy Measurements

Ground-level neutron monitors have provided the historic information on SEPs above ~0.5 GeV by observing the products that rain down from nuclear interactions of energetic protons with atomic nuclei of the upper atmosphere. When the signal from the SEPs can be seen above the background produced by galactic cosmic rays we have a ground-level event (GLE). However, most GLEs rise less than 10% above GCR background, providing rather poor information on timing.

As noted previously, high-energy protons are often strongly beamed along the interplanetary magnetic-field line, so a particular neutron monitor on Earth sees an intensity maximum when its asymptotic look direction is aligned with that field. Since the field direction can vary, neutron monitors often see sudden increases or decreases, or even multiple peaks and valleys of intensity as their look direction scans across the pitch-angle distribution as the interplanetary magnetic-field direction swings around. Nevertheless, integrating over an event at multiple stations can produce creditable spectra, that compare well with those from GOES and IMP, as obtained by Tylka and Dietrich (2009) and shown in Figs. 6.2 and 6.3. This was a significant advance in high-energy spectra.

Two newer instruments, the *Payload for Antimatter Exploration and Light-nuclei Astrophysics* (PAMELA) mission and the *Alpha Magnetic Spectrometer* (AMS) are large complex instruments that were justified and funded for particle physics and cosmology, which may also prove useful for high-energy SEP measurements. These instruments use transition-radiation detectors, time-of-flight detectors, a permanent magnet and tracking system, Cherenkov systems, and calorimeters to measure each incident particle. They were designed to search for antimatter, such as anti-helium, strange quark matter, and dark matter.

The PAMELA satellite (2006–2015) is in a near-polar, 70°-inclination, orbit. It can measure protons and He above about 80 MeV amu$^{-1}$ and reported spectra for the 13 December 2006 SEP event (Adriani et al. 2011), for events in 2010–2012 (Bazilevskaya et al. 2013), and for 26 events from 2006 – 2014 (Bruno et al. 2018). AMS is on the *International Space Station*. It can measure protons and isotopes of light ions above about 200 MeV amu$^{-1}$.

While these instruments must deal with geomagnetic-field limitations, as neutron monitors do, they can directly measure spectra and abundances, and represent a great improvement in the accuracy of measurements at high energies.

## 7.6 Problems and Errors

The single most difficult problem in measuring SEPs is exploring rare species and small events while still dealing with the high intensities in large events. Most high-resolution instruments fail or degrade during periods of *high SEP intensity*.

Early instruments sampled particles randomly and sent the measurements to the ground for analysis. However, since telemetry was slow and the H/O ratio can exceed $10^4$ at fairly high energies, H and He consumed all the telemetry and heavy ions were almost never seen. Later instruments incorporated priority schemes to distinguish H, He, and "heavies" and selectively telemetered them at different priorities, keeping track of the number received onboard for re-normalization. Most modern instruments determine particle species and energy and bin them onboard. The higher onboard processing rates have allowed geometry factors to profitably expand, improving statistics and observing rare species.

As rates increase, the first problem to solve involves "dead-time corrections." An instrument cannot process a new particle while it is still busy processing the previous one. Knowing the processing times, these corrections are usually already included while calculating intensities. However, it does make a difference whether the telescope has become busy because too many high-energy particles traverse the anticoincidence detector, or because too many low-energy particles are striking the front detector. Some instruments can determine coincidence and priority at high rates before they decide to perform the slower pulse-height analysis; they can handle much higher throughput. Instruments that must pulse-height analyze every above-threshold signal in every detector are more limited in speed, by factors of 10 or more, since many of the pulse heights are not of interest; perhaps because they do not even meet the coincidence conditions.

Eventually, problems come from multiple particles in the telescope within the resolution time. A proton stops in the back detector and triggers the coincidence while a low energy Fe stops in the front detector, or while an energetic He or heavier ion crosses the front detector at some large angle. Background in LEMT during the first day of the Bastille-Day SEP event on 14 July 2000 is shown in Fig. 7.9. Background stretches all the way up the ordinate in the 2< E < 20 MeV band. Calibration curves that are shown have omitted the 2.5–3.3 MeV amu$^{-1}$ interval which would extend into this band (also compare Fig. 7.3). The added background not only contaminates measurements but also reduces the time available for real particles. Fortunately this is a rare problem for LEMT and it fails quite gracefully in this case, i.e. abundances and spectra above 3.3 MeV amu$^{-1}$ are still quite valid.

**Fig. 7.9**. Sampled response of LEMT is shown during the first day of the Bastille-Day event, 14 July 2000. Calibration curves are only shown from 3.3–10 MeV amu$^{-1}$, to emphasize the band of background covering the region where the 2.5–3.3 MeV amu$^{-1}$ interval would be. The low-energy SEP ions are throttled by the streaming limit (Sect. 5.1.5) and only background survives. Compare the region 2 < E < 20 MeV with that in Fig. 7.3.

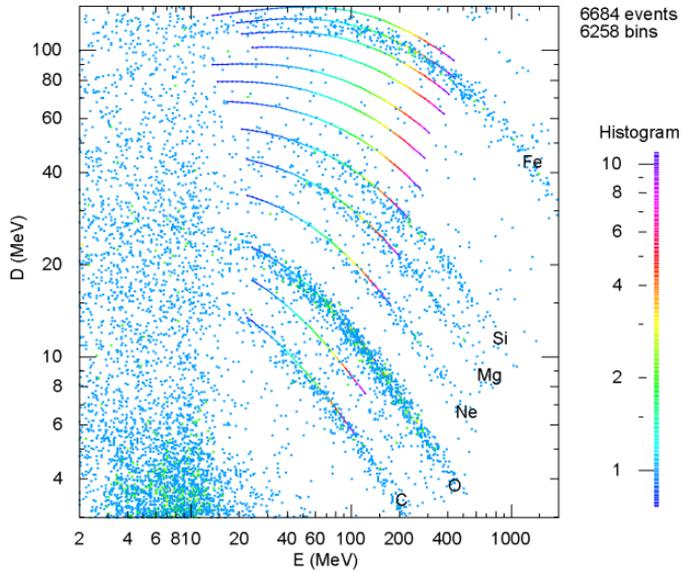

The upper limit of E of the background band in LEMT occurs because it is difficult for a proton, the most abundant species, to deposit more than 10 or 15 MeV into the E detector before penetrating into the anticoincidence detector. Instruments that require three energy measurements on each ion can use consistency to eliminate much background, but they have higher energy thresholds than LEMT.

One easy way to detect background is to check for unrealistic abundances, such as measurable ratios of F/O or B/O. *If you discover something really unusual, it is wise to check the pulse-height matrix before publishing your amazing new finding*.

Different instruments have different problems and some have interesting solutions. Some early instruments suffered gain changes in large events so the particle tracks moved around with time. *Many* instruments saturate at high particle rates, the smaller, faster instruments on GOES and *Helios* do not. ULEIS has a restricting aperture that can be rotated into place to reduce intensities. Other telescopes turn off detector elements to reduce their effective geometry factor.

The data base for many measurements from many spacecraft, including SEP intensities, is http://cdaweb.gsfc.nasa.gov/sp_phys/. Generally, however, pulse-height data are not widely available, since the more-extensive data and the specialized processing and software required are only developed by the instrument teams. This software is generally not modified to keep up with evolution of computer hardware and operating systems. Software longevity can become a general problem for instruments that last over two solar cycles.

Last, but not least, good calibrations of instruments before launch are important. A wide variety of beam species from particle accelerators are now commonly available over a vast energy range. Not only does this validate the nominal particle response derived from the range-energy relation, but multifaceted telescopes like LEMT were mounted in a rotatable fixture that allowed each detector to be exposed to an $^{56}$Fe beam at normal incidence to measure thickness distributions that could be used in corrections. This, and confirmation of viewing angles, could only be done with terrestrial monoenergetic beams and not with "inflight" calibrations sometimes used.

## References

Adriani, O., Barbarino, G.C., Bazilevskaya, G.A., Bellotti, R., Boezio, M., Bogomolov, E.A.; Bonechi, L., Bongi, M., Bonvicini, V., Borisov, S., et al., Observations of the 2006 December 13 and 14 solar particle events in the 80 MeV n$^{-1}$ – 3 GeV n$^{-1}$ range from space with the PAMELA detector, Astrophys. J. **742**, 102 (2011) doi: 10.1088/0004-637X/742/2/102

Bazilevskaya, G.A., Mayorov, A.G., Malakhov, V.V., Mikhailov, V.V., Adriani, O., Barbarino, G.C.; Bellotti, R., Boezio, M., Bogomolov, E.A., Bonechi, L., et al., Solar energetic particle events in 2006-2012 in the PAMELA experiment data, J. Physics: Conf. Series, **409**, 012188 (2013) doi: 10.1088/1742-6596/409/1/012188

Bruno, A., Bazilevskaya, G.A., Boezio, M., Christian, E.R., de Nolfo, G.A., Martucci, M., Merge', M., Mikhailov, V.V., et al., Solar energetic particle events observed by the PAMELA mission, Astrophys. J. **862**, 97 (2018) doi: 10.3847/1538-4357/aacc26

Gopalswamy, N., Yashiro, S., Michalek, G., Stenborg, G., Vourlidas, A., Freeland, S., Howard, R., The SOHO/LASCO CME catalog, Earth, Moon, Planets **104** 295 (2009) doi: 10.1007/s11038-008-9282-7

Hubert, F., Bimbot, R., Gauvin, H., Range and stopping-power tables for 2.5–500 MeV/nucleon heavy ions in Solids, Atom. Dat. Nucl. Dat. Tables **46**, 1 (1990) doi: 10.1016/0092-640X(90)90001-Z

Leske, R.A., Mewaldt, R.A., Cohen, C.M.S., Cummings, A.C., Stone, E.C., Wiedenbeck, M.E., von Rosenvinge, T.T., Solar isotopic composition as determined using solar energetic particles, Space Sci. Rev. **130**, 195 (2007) doi: 10.1007/s11214-007-9185-3

Mason, G.M., Gold, R.E., Krimigis, S.M., Mazur, J.E., et al., The Ultra-Low-Energy Isotope Spectrometer (ULEIS) for the ACE spacecraft, Space Sci. Rev. **86**, 409 (1998) doi: 10.1023/A:1005079930780

Olmedo, O., Zhang, J., Wechsler, H., Poland, A. and Borne, K., Automatic detection and tracking of coronal mass ejections in coronagraph time series, Sol. Phys., **248**, 485 (2008) doi:. 10.1007/s11207-007-9104-5

Paul, H., Schinner, A., Empirical stopping power tables for ions from $^3$Li to $^{18}$Ar and from 0.001 to 1000MeV/nucleon in solids and gases, Atom. Dat. Nucl. Dat. Tables **85**, 377 (2003) doi: 10.1016/j.adt.2003.08.003

Reames, D.V. Hydrogen and the abundances of elements in impulsive solar energetic-particle events, Sol. Phys. **294** 37(2019) doi: 10.1007/s11207-019-1427-5

Reames, D.V., Barbier, L.M., von Rosenvinge, T.T., Mason, G.M., Mazur, J.E., Dwyer, J.R., Energy spectra of ions accelerated in impulsive and gradual solar events, Astrophys. J. **483**, 515 (1997) doi: 10.1086/304229

Reames, D.V., Ng, C.K., Berdichevsky, D., Angular Distributions of Solar Energetic Particles , Atrophys. J. **550** 1064 (2001) doi: 10.1086/319810

Stone, E.C., Cohen, C.M.S., Cook, W.R., Cummings, A.C., Gauld, B., Kecman, B., Leske, R.A., Mewaldt, R.A., Thayer, M.R., Dougherty, B.L., et al., The Solar Isotope Spectrometer for the Advanced Composition Explorer, Space Sci. Rev. **86** 357 (1998) doi: 10.1023/A:1005027929871

Thakur, N., Gopalswamy, N., Mäkelä, P., Akiyama, S., Yashiro, S., Xie, H., Two exceptions in the large SEP events of solar cycles 23 and 24, Sol. Phys. **291**, 513 (2016) doi: 10.1007/s11207-015-0830-9

Tylka, A.J., Dietrich, W.F., A new and comprehensive analysis of proton spectra in ground-level enhanced (GLE) solar particle events, in *Proc. 31st Int. Cos. Ray Conf* , Lódz (2009), http://icrc2009.uni.lodz.pl/proc/pdf/icrc0273.pdf

von Rosenvinge, T.T., Barbier, L.M., Karsch, J., Liberman, R., Madden, M.P., Nolan, T., Reames, D.V., Ryan, L., Singh, S., Trexel, H., The Energetic Particles: Acceleration, Composition, and Transport (EPACT) investigation on the *Wind* spacecraft, Space Sci. Rev. **71** 152 (1995) doi: 10.1007/BF00751329

# Chapter 8. Element Abundances and FIP: SEPs, Corona, and Solar Wind

**Abstract**  We have used abundance measurements to identify the sources and the physical processes of acceleration and transport of SEPs.  Here we study energetic particles themselves as samples of the solar corona that is their origin, distinguishing the corona from the photosphere and the SEPs from the solar wind.  Theoretically, differences in the first ionization potential "FIP effect" may distinguish closed- and open-field regions at the base of the corona, which may also distinguish SEPs from the solar wind.  There is not a single coronal FIP effect, but two patterns, maybe three.  Are there variations?  What about He?

There are many different opportunities to sample solar and coronal material.  Meteorites are a sample of the non-volatile material left over from the time when the solar system was forming, and can be compared with abundances from spectral line intensities from the photosphere (e.g. Lodders, Palme, and Gail 2003).  The fast and slow solar wind (Bochsler 2009) and gradual SEP events (Reames 2018a, b, 2020) each provide unique samples of the solar corona, which all seem to differ.  Separately, the shock-accelerated solar wind provides a comparison with the direct measurements (Sect. 8.3; Reames 2018c).  Spectroscopic extreme-ultraviolet, X-ray (Feldman and Widing 2008; Sylwester et al. 2014), and γ-ray line (Ramaty, Mandzhavidze, and Kozlovsky 1996) measurements provide localized measurements in the corona and in the heated plasma of solar flares.

In a modern theory of the FIP effect (Laming 2015, Laming et al. 2019), the ponderomotive force of Alfvén waves can preferentially drive ions, but not neutral atoms, across the chromosphere and into the corona.  The Alfvén waves can resonate with the loop length of magnetic loops that are closed in this region but not with open field lines, producing two fundamentally different FIP patterns.

## 8.1 Element Abundances in the Sun

Element abundances in the Sun are presumed to have changed little, apart from some gravitational settling, from their values in the pre-solar molecular cloud from which the Sun was formed by gravitational collapse.  Material left over from that time period was left as meteorites or eventually collapsed further to form the planets.  As the collapsing Sun began to heat and radiate, the surrounding material was heated then condensed into pea-sized molten drops called chondrules (after Greek chondros for sphere) which would later clump to form chondritic meteorites.  With the onset of the solar wind, the volatile elements tended to be blown out of the inner heliosphere, denuding it of volatiles and leading to the inner rocky planets.  Most of the H and He was swept out to Jupiter and beyond.

   Thus the chondritic meteorites provide a measure of the abundances of the non-volatile elements in the pre-solar and early-solar nebula and hence of the Sun. In particular, the carbonaceous chondrites or CI chondrites (Carbonaceous of the Ivuna type), of which there are actually few examples, show the least depletion of volatiles, giving the most complete set of abundances. When these meteoritic abundances are compared with those from spectral line measurements of the photosphere, nearly 40 elements now show agreement within about 10% (Lodders, Palme, and Gail 2009).

   For the spectral-line measurements, recent years have seen a change from assumed local thermal equilibrium to three-dimensional modeling of the emitting plasma (Grevesse et al. 2013; see also Schmelz et al. 2012). This has led to the abundances of Asplund et al. (2009) shown in Table 1.1, who have concentrated on well-resolved and isolated spectral lines, but also led to the abundance measurements of the dominant volatile elements, especially C, N, O, S, and Fe, by Caffau et al. (2011), who have resolved elements from blended spectral lines and provided the photospheric abundances listed in Table 8.1. However, the new abundances do conflict with helioseismic determinations of the sound speed, the He abundance, and depth of the convection zone (Grevesse et al. 2013).

## 8.2 The Solar Wind

Unlike the conceptually simple $dE/dx$ vs. $E$ measurement of SEPs (Chap. 7), measurements of the solar wind abundances must resolve the charge state, mass, and energy of each ion. Instruments use stages of collimation, electrostatic deflection, post acceleration, time of flight, and energy deposit (e.g. Gloeckler et al. 1992). The need to measure all of the individual ionization states of each element leads to some cases of overlap, which are difficult to resolve, e.g. $C^{+6}$ and $O^{+8}$, or some charge states of Ne or S. However, abundance measurements have now been made of solar wind abundances emitted spatially over three dimensions around the Sun (von Steiger et al. 2000) and abundance measurements have been collected and summarized for us by Bochsler (2009). An alternative technique that has contributed to the results is the return and analysis of foils that have been exposed to collect solar wind ions.

   The solar wind consists of fast ($500 > V_S > 800$ km s$^{-1}$) solar wind or high-speed streams emerging from coronal holes and slow ($200 > V_S > 500$ km s$^{-1}$) wind or interstream wind diverging (Wang and Sheeley 1990) around closed structures such as streamers which contain magnetic field reversals and pseudo-streamers which do not. Much of the solar wind also consists of solar ejecta and magnetic clouds. The fast wind contains cooler plasma with $O^{+7}/O^{+6} < 0.1$ while the slow wind and magnetic clouds tend to contain hotter plasma with $O^{+7}/O^{+6} > 0.1$. A more complete scheme for distinguishing different regions of the solar wind has been developed by Xu and Borovsky (2015). We will see that the amplitude of the FIP-level difference, i.e. the FIP *bias*, for the slow solar wind is similar to that for the SEPs, yet the location of the crossover from high to low FIP differs. The fast solar wind has a smaller FIP bias (≈70%), but not for all elements (Bochsler 2009). Abundances for the slow solar wind are shown in Table 8.1

Variations of He/O are seen in the solar wind as functions of time and of solar-wind speed (Collier et al. 1996; Bochsler 2007; Rakowsky and Laming 2012) and large variations are seen in H/He with phase in the solar cycle (Kasper et al. 2007). This makes it difficult to include H and He in FIP studies of the solar wind, as is also the case for SEPs (see Sect. 5.9 and Chap. 9). However, variations of H and He in SEPs seem to be unrelated to those in the solar wind.

## 8.3 Corotating Interaction Regions: Accelerated Solar Wind

Corotating interaction regions (CIRs) occur when high-speed solar-wind streams, emerging from coronal holes, overtake and collide with slow solar wind emitted earlier in the solar rotation. Two shock waves can form from this collision, a forward shock propagates outward into the slow solar wind and a stronger, reverse shock propagates sunward into the fast wind (Belcher and Davis 1971; Richardson 2004). Near the Sun, the fast wind flows nearly parallel to the slow wind, but as the distance increases, the fast wind begins to "bite" into the slow wind, often beyond 1 AU, and the shock waves strengthen. *Pioneer* spacecraft observations have shown that the maximum particle acceleration by the shock waves occurs at about 5 AU (McDonald et al. 1976; Van Hollebeke, McDonald, and von Rosenvinge 1978). The energetic particle intensity seen at 1 AU is usually a combination of acceleration from the solar wind at the reverse shock and of transport, streaming sunward, in the solar-wind frame, along the field lines to the observers (Marshall and Stone 1978) and displaying unique abundances (e.g. Reames et al. 1991; Richardson et al. 1993; Mason et al. 1997, 2008). CIRs allow us to measure element abundances of shock-accelerated solar wind with the same instruments that measure element abundances of shock-accelerated SEPs.

As we found for SEP events, scattering mean free paths of particles can often be expressed as a power law in particle rigidity so that high-rigidity particles propagate more easily. This can be seen in Fig. 8.1 where scattering has confined low-energy He from a CIR much more closely to the source, when it is near 1 AU, than high-energy He which is seen relatively undiminished a week later when the spacecraft is magnetically connected to a point on the shock that is far out from the Sun. Whereas SEP events show velocity dispersion with the fastest particles arriving first, for CIRs, the low energies dominate early but decrease in importance rapidly. Of course, this rigidity dependence affects element abundances as well. If we can accommodate the corresponding *A/Q* dependence, as we did with SEPs, the energetic ions at CIRs give us an alternate way to measure the FIP dependence when the source is the solar wind. The two sources of fast and slow wind are poorly resolved by the energetic ion abundances.

**Fig. 8.1** The *lower pan-el* shows intensities of He at various energies given in MeV amu[-1] observed near Earth by the *Wind* spacecraft in a CIR event produced by the high-speed solar-wind stream shown in the *upper panel*. Increased scattering causes low-rigidity ions to be confined near the shock while high-rigidity ions spread widely from the distant shock even as it moves far out in the heliosphere.

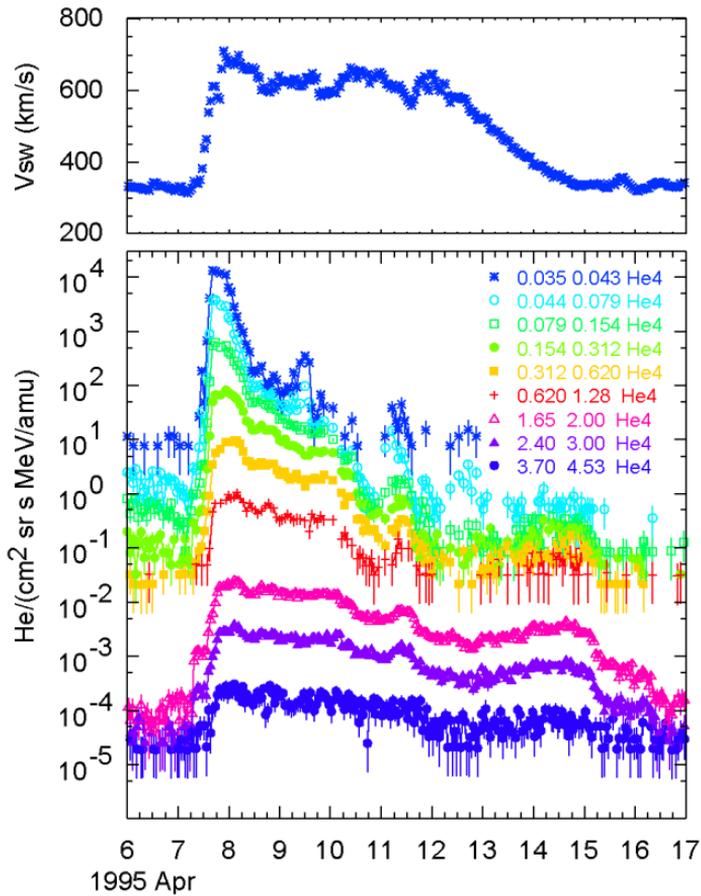

CIRs form a steady spiral spatial pattern that rotates with the Sun and can persist for several rotations, on each pass accelerating ions that have been seen to last 17 days and span 225° of solar longitude at MeV energies (Reames et al. 1997). The lagging intensities of higher-energy particles are fed by the much higher intensities from the shock wave that is stronger in the outer heliosphere at ~ 5 AU, not by the weaker shock near us that produced the low-energy peak.

Figure 8.2 shows time variations of element abundances during passage of a CIR and during three impulsive SEP events. Like any other shock wave, a CIR shock wave can accelerate a seed population from any source, including pre-accelerated ions from impulsive and gradual SEP events. Even a very small impulsive SEP event, which could occur near coincidence with a CIR passage, would contaminate the otherwise small Fe abundances of the CIR, but would appear to affect little else.

**Fig. 8.2.** The *lower panel* shows intensities of $^4$He, C, O, and Fe at 2.5 – 3.2 MeV amu$^{-1}$ vs. time early in the year 2000. A CIR event and three impulsive SEP events are indicated. The SEP events have C/O < 0.5 and Fe/O $\approx$ 1, while the CIR event has C/O $\approx$ 1 but very low Fe/O. The *upper panel* shows the solar-wind speed (Reames 2018c, © Springer).

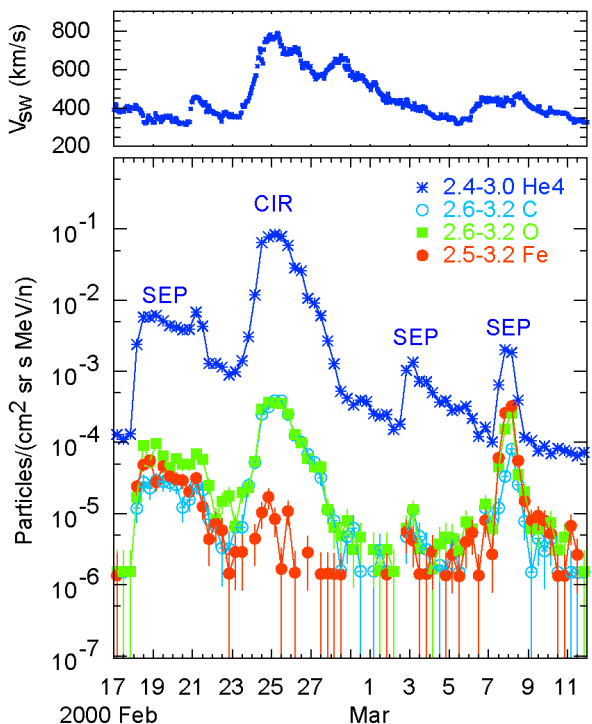

Energy spectra of ions are shown for a sample of four CIR events in Fig. 8.3. The intensity spectra of $^4$He, C, and O are well matched to the form

$$j(v) = j_{0i} (A/Q)^a\, v^b\, exp(-v/v_0) \qquad (8.1)$$

where $v$ is the ion speed and $j_{0i}$, $a$, $b$, and $v_0$ are adjustable constants, $j_{0i}$ varying with species as injected at the shock. This is the form derived by Fisk and Lee (1980) with the addition of the explicit $A/Q$ dependence. In general, $a \neq b$, and both values vary from event to event. In principle, simple shock acceleration should yield $b = a - 4$ since shocks have a correlated affect on both spectra and abundances, but transport can complicate and disrupt this simple relationship as discussed in detail in Sect. 9.9. Also, the observed spectra flatten with time (see Reames et al. 1997) because the distant shock strengthens with time but low rigidity ions are increasingly suppressed during transport.

The spectra of $^4$He, C, and O show similar shapes for all elements in the events in the lower panel of Fig 8.3 and for the normalized Fe and $^4$He spectra for most events in the upper panel. However, the event numbered 3 shows either possible low-energy contamination of Fe or steepening of high-rigidity Fe relative to lower-rigidity He and O. Fe spectra in the same range of $E$ have much higher rigidity $P$ and transport depends upon $P$.

**Fig. 8.3**. Energy spectra of four CIR events compare C and O with He in the *lower panel* and Fe with He in the *upper panel*. Spectral shapes for He, C, and O (and other species) generally agree well, but Fe in event 3 shows either (i) steepening at Fe because of its higher rigidity, or (ii) Fe background injected by impulsive SEPs at low energies (Reames 2018c, © Springer).

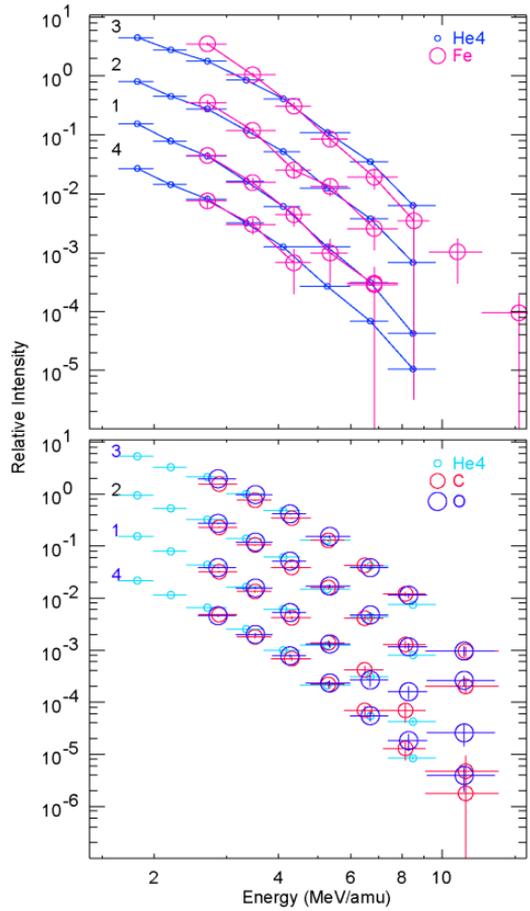

We can now use the same analysis technique for these CIR events that we used for gradual SEP events in Sect. 5.6. Figure 8.4 shows best-fit power-law lines for the relative abundances, i.e. the normalization factors for the spectra relative to O divided by the corresponding fast solar wind (FSW) abundances relative to O (Bochsler 2009), vs. *A/Q*. The *A/Q* values correspond to plasma temperatures of 1.0 – 1.3 MK which are appropriate for the FSW and are determined from the best-fit power law.

**Fig. 8.4**. Best-fit power laws (*blue*) and element abundance enhancements (*black*) relative to O, divided by corresponding abundances in the FSW (Bochsler 2009), are plotted vs. $A/Q$ for each of our four CIR events. The atomic number $Z$ is shown at the position of each element and successive measured elements are joined. The reference FSW abundance of S ($Z = 16$) seems to be consistently too large in all of the events and Fe may be too small.

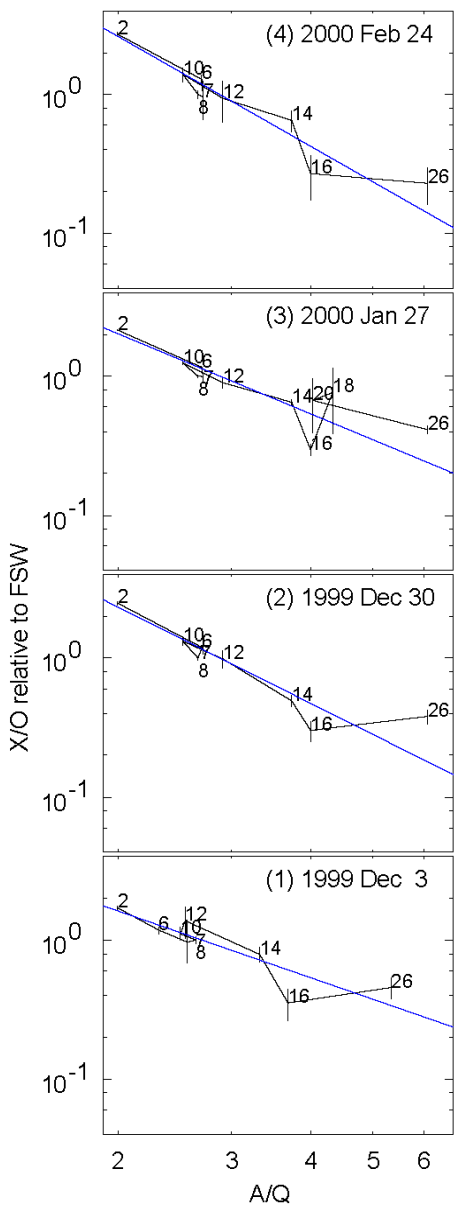

All of the 12 recent CIR events studied by Reames (2018c) showed negative power-law dependence of abundance enhancements on $A/Q$; for these events the shock waves began outside 1 AU. However, in an earlier event in May 1982 (Reames, Richardson, and Barbier 1991) the shock began locally and the relative abundances seemed to be independent of $A/Q$, directly providing an alternative measure of element abundances of the solar wind are shown as CIR in Table 8.1.

It seems that when the shock forms near 1 AU, we see a direct sample of the source ions locally (both slow and fast), as accelerated, but when the shocks form far outside 1 AU, ions (from the FSW) with high $A/Q$ are suppressed as they spread widely by transport. These energetic CIR ions give us another measure of the FIP effect in the solar corona, presumably similar to the combined solar wind.

## 8.4 Comparing FIP Patterns of SEPs and the Solar Wind

Assuming the higher value of 91 for source He/O for SEPs (Sects. 5.8 and 5.9), we compare the FIP plot for SEPs with that for the slow solar wind (SSW; Bochsler 2009) in Fig. 8.5. The upper panel compares these abundances relative to those in the photosphere from Caffau et al. (2011) and Lodders, Palme, and Gail (2009). Here we renormalized the solar wind values by a factor of 1.2 to improve the agreement at both high and low FIP. The lower panel compares SEP and slow-solar-wind abundances directly, showing the alternate normalization as a dashed line. The alternative normalization means that O is not the best choice for a reference, i.e. O is more enhanced in the SSW than in SEPs.

The FIP bias, i.e. the ratio of the levels at low and high FIP, is quite similar for SEPs and the slow solar wind. The FIP bias of the fast solar wind is smaller (Bochsler 2009). However, the difference in Fig. 8.5 is in the FIP value of the crossover between low- and high-FIP regions, ~10 eV for SEPs and ~14 eV for the slow solar wind. Thus, the elements C, S, and P, seem to behave more like neutral atoms for the SEPs and like ions for the slow solar wind, as if the SEPs are derived from regions of cooler plasma where C, S, and P are not sufficiently ionized. Theory suggests that it may be more likely that SEPs are accelerated from closed field lines while the solar wind must come from open field lines near the base of the corona (Reames 2018a; Laming 2015; Laming et al. 2019). Spectral line measurements of the corona also show suppressed values of S/O (Schmelz et al. 2012) presumably on closed field lines.

**Fig. 8.5** The *upper panel* shows the SEP/photospheric and 1.2 times the slow solar wind (SSW)/photospheric abundance ratios as a function of FIP. The curves are empirical curves used to show the trend of the data. The *lower panel* shows the direct ratio of the coronal abundances from SEPs to those of the SSW (Bochsler 2009), as a function of FIP. The dashed line suggests the alternate normalization factor of 1.2 (Reames 2018a, © Springer).

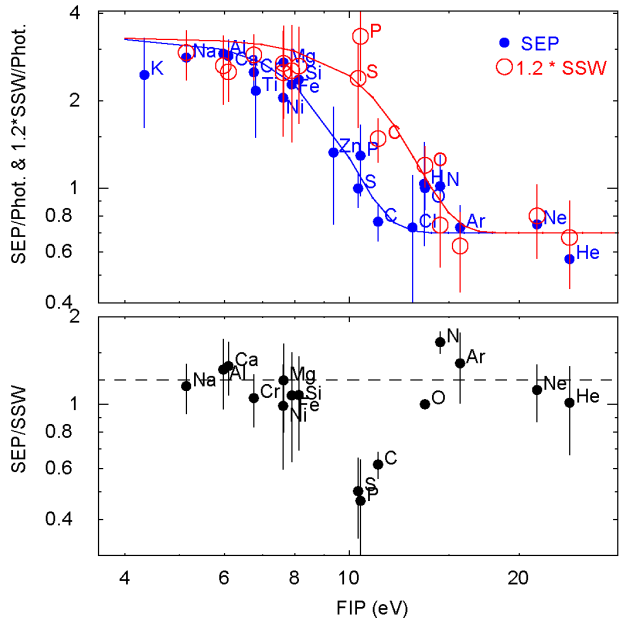

In any case, the differences in FIP behavior seen in Fig 8.5 strongly suggest that SEPs provide a unique sample of coronal material. FIP patterns are determined near and below the base of the corona, *not* at the height or time of SEP acceleration. Thus, at energies above a few MeV amu$^{-1}$, SEPs *cannot* be merely reaccelerated solar wind; they are an *independent* sample of coronal material (Reames 2018a). This irreconcilable difference in FIP patterns was first noted by Mewaldt et al. (2002) and Desai et al. (2003) also noted that SEPs were not merely reaccelerated solar wind. This is why SEPs exhibit the same abundances whether they are found in fast or in the slow wind (Kahler and Reames 2003; Kahler, Tylka, and Reames 2009).

**Table 8.1** Photospheric, reference SEP, CIR, and SSW abundances.

|  | Z | FIP [eV] | Photosphere[1] | SEPs[2] | CIRs[3] | SSW[4] |
|---|---|---|---|---|---|---|
| H | 1 | 13.6 | $(1.74\pm0.04)\times10^6$ * | $(1.6\pm0.2)\times10^6$ | $(1.81\pm0.24)\times10^6$ | – |
| He | 2 | 24.6 | $1.46\pm0.07\times10^5$ | 91000±5000 | 159000±10000 | 90000±30000 |
| C | 6 | 11.3 | 550±76* | 420±10 | 890±36 | 680±70 |
| N | 7 | 14.5 | 126±35* | 128±8 | 140±14 | 78±5 |
| O | 8 | 13.6 | 1000±161* | 1000±10 | 1000±37 | 1000 |
| Ne | 10 | 21.6 | 195±45 | 157±10 | 170±16 | 140±30 |
| Na | 11 | 5.1 | 3.47±0.24 | 10.4±1.1 | – | 9.0±1.5 |
| Mg | 12 | 7.6 | 60.3±8.3 | 178±4 | 140±14 | 147±50 |
| Al | 13 | 6.0 | 5.13±0.83 | 15.7±1.6 | – | 11.9±3 |
| Si | 14 | 8.2 | 57.5±8.0 | 151±4 | 100±12 | 140±50 |
| P | 15 | 10.5 | 0.501±0.046* | 0.65±0.17 | – | 1.4±0.4 |
| S | 16 | 10.4 | 25.1±2.9* | 25±2 | 50±8 | 50±15 |
| Cl | 17 | 13.0 | 0.55±0.38 | 0.24±0.1 | – | – |
| Ar | 18 | 15.8 | 5.5±1.3 | 4.3±0.4 | – | 3.1±0.8 |
| K | 19 | 4.3 | 0.224±0.046* | 0.55±0.15 | – | – |
| Ca | 20 | 6.1 | 3.72±0.60 | 11±1 | – | 8.1±1.5 |
| Ti | 22 | 6.8 | 0.138±0.019 | 0.34±0.1 | – | – |
| Cr | 24 | 6.8 | 0.759±0.017 | 2.1±0.3 | – | 2.0±0.3 |
| Fe | 26 | 7.9 | 57.6±8.0* | 131±6 | 97±11 | 122±50 |
| Ni | 28 | 7.6 | 2.95±0.27 | 6.4±0.6 | – | 6.5±2.5 |
| Zn | 30 | 9.4 | 0.072±0.025 | 0.11±0.04 | – | – |
| Se–Zr | 34–40 | – | ≈0.0118 | 0.04±0.01 | – | – |
| Sn–Ba | 50–56 | – | ≈0.00121 | 0.0066±0.001 | – | – |
| Os–Pb | 76–82 | – | ≈0.00045 | 0.0007±0.0003 | – | – |

[1]Lodders et al. (2009). Ratios to O of elements from Lodders et al. (2009) are taken before correction of O by Caffau et al. (2011).

* Caffau et al. (2011).

[2] Reames (1995, 2014, 2020).

[3] Reames, et al. (1991); Reames (1995).

[4] Bochsler (2009).

## 8.5 FIP Theory: the Sources of SEPs and the Solar Wind

Recent theory of FIP fractionation (Laming 2015; Laming et al. 2019) involves extensive numerical calculations of ionization states of the elements as a function of altitude across the chromosphere where ions are guided along magnetic fields in the presence Alfvén waves, which neutral atoms cannot feel. Comparing with the results of that theory, it is shown in the lower panel of Fig. 8.6 that the SEPs best fit the theoretical FIP pattern derived for closed field lines with adiabatic invariant conservation (Laming et al. 2019), where the elements C, P, and S are suppressed like high-FIP elements, while, in the middle and upper panel of Fig. 8.6, the CIRs and solar wind fits the open-field pattern (Laming 2019), where C, P, and S are elevated like low-FIP ions. The pattern of CIR abundances vs. FIP in Fig 8.6 clearly resembles the SSW in that C, P, and S are elevated. Here the principal photospheric abundances are those of Caffau et al. (2011) supplemented by those of Lodders, Palme, and Gail (2003). The differences between the panels in the abundances of C, P, and S in the crossover region are highlighted in the light blue band.

As atoms cross the chromosphere, densities and ionized fractions change, as do the collision frequencies of ions and neutrals with the background H plasma. Low in the chromosphere, the plasma $\beta_P > 1$, and turbulence prevents fractionation. Higher, when $\beta_P < 1$, ions flow along $B$ under the ponderomotive force of Alfvén waves; increased ionization of the background H tends to reduce that flow. On closed loops, where Alfvén waves resonate with the loop length, fractionation is concentrated near the top of the chromosphere where ionization of H limits fractionation to ions with FIP < 10 ev. On open field lines, lack of resonance allows fractionation to take place more broadly throughout the loop where H is not ionized, leading to additional enhancements of C, S, and P. These conclusions are the result of extensive calculations of the wave patterns and the ionization states of the elements as a function of altitude (Laming 2015; Laming et al. 2019). The pattern of waves required for the observed abundances suggest (Laming 2017) that the Alfvén waves actually descend from the corona where they may be created by nanoflares.

**Fig. 8.6** Average abundances of (**a**) SEPs, (**b**) CIR ions and (**c**) slow solar wind (SSW; Bochsler 2009), relative to solar photospheric abundances, (*solid blue*) are shown as a function of the FIP of each element and compared with theoretical calculations (*open red*) by Laming et al. (2019) for loop structures (**a**), and for the open field SSW (**b** and **c**). All abundances are normalized at O. The *light blue* band compares elements C, P, and S that are suppressed like high-FIP elements in SEPs, but are elevated like low-FIP elements in the solar wind. However, decreasing the photospheric C/O ratio by 20% would greatly improve the agreement of observation and theory for all three samples: SEPs, CIRs, and SSW (Reames 2020).

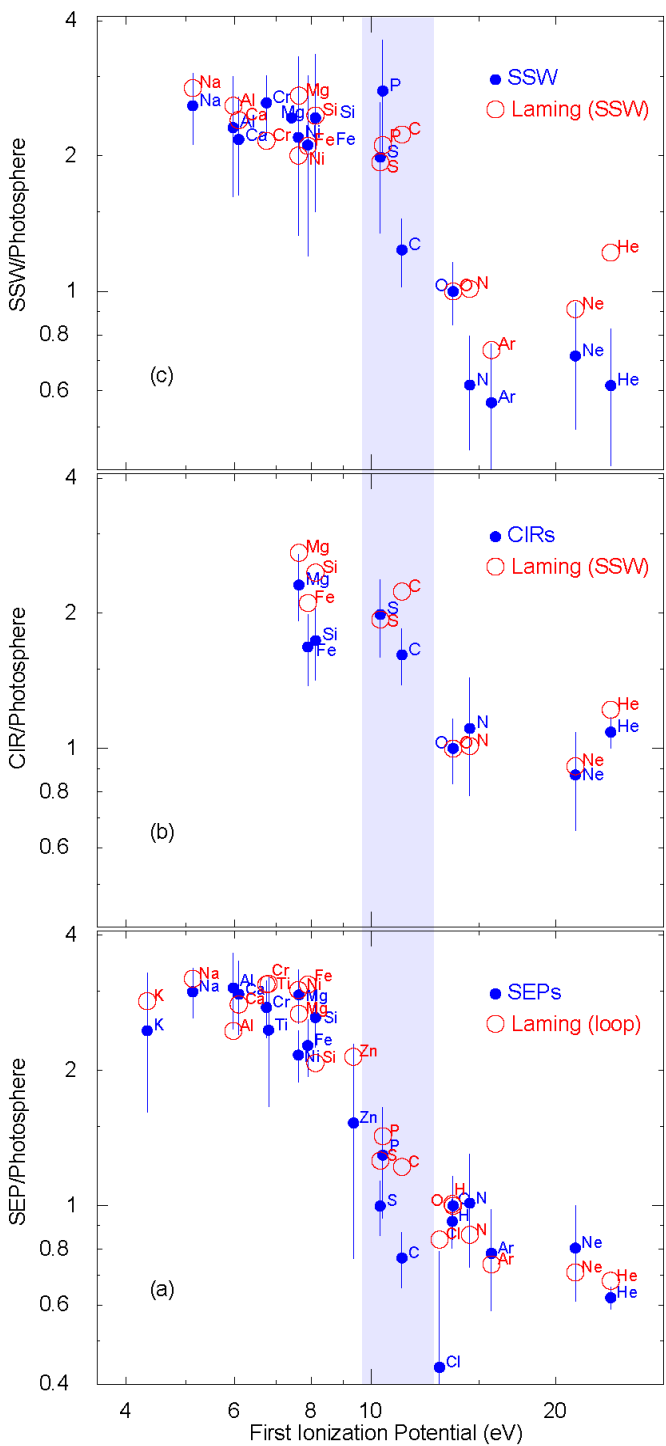

While there have been some small changes in CIR abundances (Reames 2018c) since publication of the comparisons in Fig. 8.6, the apparent difference based upon open- and closed-field geometry persists. Recently, this theory has been updated and has been compared with the SEP, SW, and CIR abundance measurements by Laming et al. (2019).

The most significant difference between all of the observed open- and closed-field patterns is in C/O as shown in Table 8.2. Actually the closed-field measurements are significantly *below* the recent photospheric values, contrary to any expectations. Agreement with theory for all three samples, SEPs, CIRs, and slow solar wind, would improve greatly if the photospheric C/O ratio were decreased 20%.

**Table 8.2** Various C/O values

| Photosphere | C/O = 0.550 ± 0.076 | Caffau et al. (2011) |
|---|---|---|
| | C/O = 0.550 ± 0.063 | Asplund et al. (2009) |
| Closed Fields | SEPS C/O = 0.430 ± 0.010 | Reames (2014) |
| | γ-rays 0.35 < C/O <0.44 | Ramaty et al. (1996) |
| | FIP Theory C/O ≈0.671  (1.22×phot) | Laming et al. (2019) |
| Open Fields | FSW C/O = 0.68 ± 0.07 | Bochsler (2009) |
| | SSW C/O = 0.68 ± 0.07 | Bochsler (2009) |
| | CIRs C/O = 0.860 ± 0.054 | Reames (2018c) |
| | FIP Theory C/O ≈1.232  (2.24×phot) | Laming et al. (2019) |

## 8.6 A Full-Sun Map of FIP

Since it now is possible to make full-disk images the Sun in the light of a single spectral line, it has become possible to compare spectral lines of high-FIP and low-FIP elements. Thus, Brooks, Ugarte-Urra, and Warren (2016) have constructed a map of FIP derived from the Si/S ratio, specifically Si X 258.37 Å/S X 264.22 Å that is shown on the right in Fig. 8.7. From the evidence above, it is not certain that this is a map of FIP for the slow solar wind, as the authors suggest, since both Si and S seem to behave as low-FIP elements for the solar wind. However, this may be an excellent map of distribution of FIP for the source of the SEPs, where Si is low-FIP and S is definitely high-FIP (Fig. 8.6a).

The image on the left in Fig. 8.7 shows the Sun in the Fe XIII 202.044 Å line formed at 2 MK which highlights active regions. The locations of the active regions on the left can be compared with the regions of high FIP bias on the right, suggesting that these regions have the appropriate FIP bias to be the sources of the SEPs.

The technique in Fig. 8.7 is an extremely powerful way of exploring the distribution of element abundances in the solar corona when the ionization-state distributions of the ions are understood. However, it is important to insure that differences measure FIP and not density or ionization state (temperature).

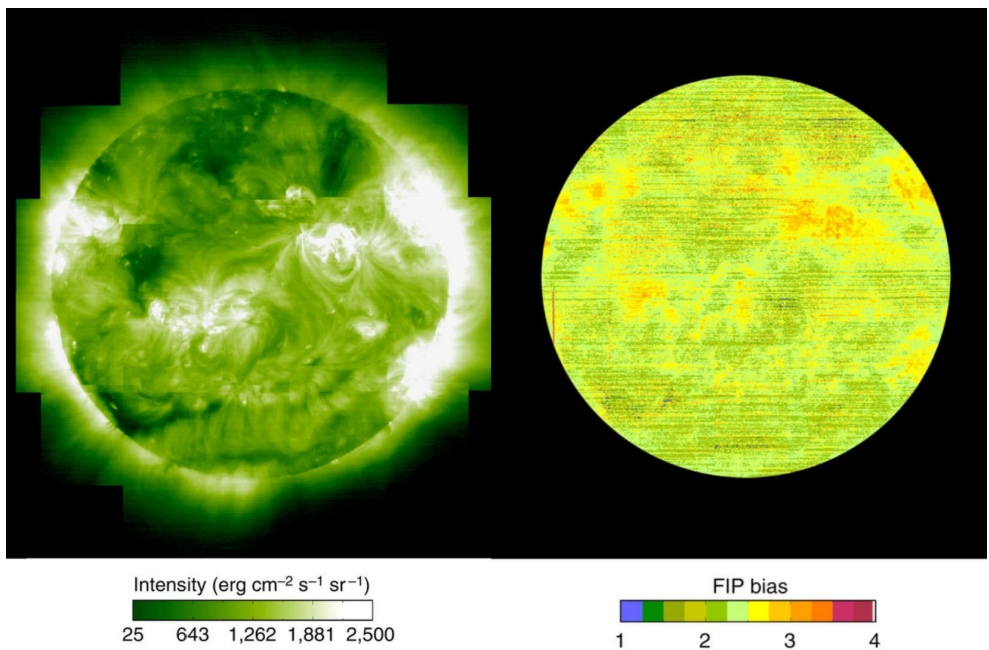

**Fig. 8.7** The solar image on the *left* in the 2 MK Fe XIII 202.044 Å line shows the location of active regions, while that on the *right* shows the ratio of Si X 258.37 Å/S X 264.22 Å that is a measure of FIP bias for SEPs (Brooks, Ugarte-Urra, and Warren 2016, CC BY 4.0).

If we measure the corresponding FIP bias using Si/S from the 8-hr averages in gradual SEP events, we find a mean of $2.44\pm0.04$ with a $\pm20\%$ spread in the individual measurements. We limited to SEPs to measurements with $0.8 \leq T \leq 1.5$ MK to avoid possible bias from the steep positive power-law increase vs. *A/Q* for the events with $T \geq 2$ MK from impulsive suprathermal seeds. Perhaps the FIP bias of ~3 MK near active regions shown in Fig. 8.7 is diluted somewhat by acceleration of surrounding material to produce the average bias of 2.4 in SEPs.

## 8.7 A Possible SW-SEP Model

The foregoing observations suggest a model that distinguishes the origins of the SEPs and the solar wind sketched in Fig. 8.8. The *solar wind speed varies inversely with the divergence of the magnetic field lines from the corona* (Wang and Sheeley 1990) so high wind speeds come from slowly diverging fields (blue) and the slow wind (green, yellow) comes from highly divergent regions as shown in Fig. 8.8. Both fast and slow solar winds come from field lines that were never closed to the currently emerging plasma, but always open from the chromosphere, where FIP fractionation occurs, outward. Solar jets (brown) can occur at magnetic reconnection sites involving open and closed loops of active regions (red), and producing impulsive SEP events and narrow CMEs.

As the slow solar wind diverges past closed field lines above active regions there is the possibility of some exchange in their abundances through magnetic re-connection. Abundances produced on closed field lines are not forever trapped or only limited to ejection in a CME.

**Fig 8.8** The *lower panel* shows a possible configuration of the solar magnetic field where the fast solar wind flows from coronal holes (*blue*), the slow wind from highly divergent open fields (*green, yellow*) and jets (*brown*) emerge from active regions (*red*). In the *upper panel*, a CME-driven shock (*gray*) accelerates SEPs from weakly closed loops and from suprathermal ions and plasma residue from jets, when it is present (Reames 2018b).

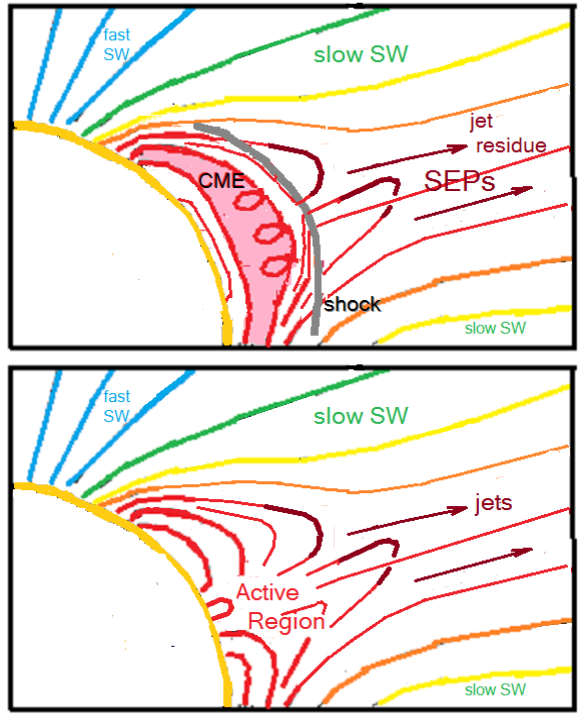

How can SEPs come from closed-field regions?

(1) The gradual SEP events with source temperatures of T > 2 MK are probably accelerated by shock waves that sample both the residual suprathermal impulsive SEPs from pools formed by numerous small jets from an active region and some recently ejected local plasma. Gradual SEP events with ions from T < 2 MK plasma may simply fail to encounter any regions with impulsive suprathermal ions.

(2) When plasma, which is contained near loop tops with $\beta_P \approx 1$, is suddenly shock accelerated, ion rigidities are boosted by an order of magnitude or more, so the ions are no longer trapped; thus, shock waves can also accelerate SEPs from ambient plasma on cooler less-active loops where $T < 2$ MK.

Therefore, a single CME-driven shock, shown in the upper panel of Fig. 8.8 can reaccelerate material from the impulsive SEPs and jet ejecta above an active region and can also accelerate cooler plasma from the closed loops on its flanks. SEP source composition and temperature can change across the shock face.

## 8.8 FIP-Dependent Variations in He

The abundances of many elements vary from event to event in SEP events, but He is the only example where we believe those variations may depend upon FIP, i.e.

they occur as material transitions from the chromosphere to the corona, not later during acceleration or transport. Source abundance variations in He were discussed in Sect. 5.9. In principle, these variations could occur because He, with the highest FIP value of 24.6 eV, is slow to ionize as it is transported across the chromosphere as calculated by Laming (2009).

However, the order-of-magnitude suppression of He in some small impulsive SEP events shown in Fig. 4.12 presents a greater challenge to the theory, but the jet model for these events shown in Fig. 4.16 may involve emerging flux which may suddenly project only the ionized chromospheric material up into the corona. Calculation of the He ionization in such a dynamic situation has not yet been performed, but it is possible that assumed slow ionization of He may conflict with the fast transport of material that is an important factor in this case.

Nevertheless, it is the impulsive SEP events with shock reacceleration (see Fig. 9.6) and the gradual events that reaccelerate residual impulsive suprathermal material (see Sect. 5.9; Reames 2017; 2018b) that have the highest ratios of He/O $\approx$ 90. Most of the larger gradual SEP events that sample ambient coronal material have He/O $\approx$ 40 – 60. However, photospheric models have He/O $\approx$ 150 – 170.

## 8.9 Open Questions

1. What do differences in the FIP-effect for SEPs and for slow solar wind say about the specific coronal locations of origins of the two populations?

2. Why is C/O so low in SEPs, even lower than in the photosphere?

3. What is the level of FIP processing for the chromospheric filament ejected within a CME?

4. What does He/O tell about the locations of SEP acceleration and the physical processes?

## References

Asplund, M., Grevesse, N., Sauval, A.J., Scott, P., The chemical composition of the sun, Ann. Rev. Astron. Astrophys. **47**, 481 (2009) doi: 10.1146/annurev.astro.46.060407.145222

Belcher J.W., Davis L. Jr., Large-amplitude Alfvén waves in the interplanetary medium, 2., J Geophys Res **76**, 3534 (1971) doi 10.1029/JA076i016p03534

Bochsler, P., Solar abundances of oxygen and neon derived from solar wind observations, Astron. Astrophys. **471** 315 (2007) doi: 10.1051/0004-6361:20077772

Bochsler, P., Composition of matter in the heliosphere, Proc. IAU Sympos **257**, 17 (2009) doi: 10.1017/S1743921309029044

Brooks, D.H., Ugarte-Urra, I., Warren, H.P., Full-Sun observations for identifying the source of the slow solar wind, Nature Comms. **6**, 5947 (2016) doi: 10.1038/ncomms6947

Caffau, E., Ludwig, H.-G., Steffen, M., Freytag, B., Bonofacio, P., Solar chemical abundances determined with a CO5BOLD 3D model atmosphere, Sol. Phys. **268**, 255. (2011) doi: 10.1007/s11207-010-9541-4

Collier, M.R., Hamilton, D.C., Gloeckler, G., Bochsler, P., Sheldon, R.B., Neon-20, oxygen-16, and helium-4 densities, temperatures, and suprathermal tails in the solar wind determined with WIND/MASS, Geophys. Res. Lett., **23**, 1191 (1996) doi: 10.1029/96GL00621

Gloeckler, G., Geiss, J., Balsiger, H., Bedini, P., Cain, J.C., Fischer, J., Fisk, L.A., Galvin, A.B., Gliem, F., Hamilton, D. C., et al., The solar wind ion composition spectrometer, Astron. Astrophys. Suppl. **92**, 267 (1992)

Grevesse, N., Asplund, M., Sauval, A.J., Scott, P., "Old" versus "new" solar chemical composition, in ASP Conf. Proc., Vol. **479**. San Francisco: Astron. Soc. Pacif., 481 (2013)

Kahler, S.W., Reames, D.V., Solar energetic particle production by coronal mass ejection-driven shocks in solar fast-wind regions, Astrophys. J. **584**, 1063 (2003) doi: 10.1086/345780

Kahler, S.W., Tylka, A.J., Reames, D.V., A comparison of elemental abundance ratios in SEP events in fast and slow solar wind regions, Astrophys. J. **701**, 561 (2009) doi: 10.1088/0004-637X/701/1/561

Kasper, J.C., Stevens, M.L., Lazarus, A.J., Steinberg, J.T., Ogilvie, K. W., Solar wind helium abundance as a function of speed and heliographic latitude: variation through a solar cycle, Astrophys. J. **660**, 901 (2007) doi: 10.1086/510842

Laming, J.M., Non-WKB models of the first ionization potential effect: implications for solar coronal heating and the coronal helium and neon abundances, Astrophys. J. **695**, 954 (2009) doi: 10.1088/0004-637X/695/2/954

Laming, J.M., The FIP and inverse FIP effects in solar and stellar coronae, Living Reviews in Solar Physics, **12**, 2 (2015) doi: 10.1007/lrsp-2015-2

Laming, J.M., Vourlidas, A., Korendyke, C., et al., Element abundances: a new diagnostic for the solar wind. Astrophys. J. **879** 124 (2019) doi: 10.3847/1538-4357/ab23f1 arXiv: 19005.09319

Lodders, K., Palme, H., Gail, H.-P., Abundances of the elements in the solar system, In: Trümper, J.E. (ed.) *Landolt-Börnstein, New Series VI/4B*, Springer, Berlin. Chap. **4.4**, 560 (2009) doi: 10.1007/978-3-540-88055-4_34

Marshall, F., Stone, E.C., Characteristics of sunward flowing protons and alpha particle fluxes of moderate intensity, J. Geophys. Res., **83**, 3289 (1978) doi: 10.1029/JA083iA07p03289

Mason, G.M., Mazur, J.E., Dwyer, J.R., Reames, D.V., von Rosenvinge, T.T., New spectral and abundance features of interplanetary heavy ions in corotating interaction regions, Astrophys, J. **486**, 149 (1997) doi: 10.1086/310845

Mason, G.M., Leske, R.A., Desai, M.I., Cohen, C.M.S., Dwyer, J. R., Mazur, J.E., Mewaldt, R.A., Gold, R.E., Krimigis, S.M., Abundances and Energy Spectra of Corotating Interaction Region Heavy Ions Observed during Solar Cycle 23, Astrophys. J. **678,** 1458 (2008) doi: 10.1086/533524

McDonald, F.B., Teegarden, B.J., Trainor, J.H., von Rosenvinge, T.T., Webber, W.R., The interplanetary acceleration of energetic nucleons, Astrophys. J. Lett., **203**, L149 (1976) doi: 10.1086/182040

Mewaldt, R.A., Cohen, C.M.S., Leske, R.A., Christian, E.R., Cummings, A.C., Stone, E.C., von Rosenvinge, T.T., Wiedenbeck, M.E., Fractionation of solar energetic particles and solar wind according to first ionization potential, Advan. Space Res. **30**, 79 (2002) doi: 10.1016/S0273-1177(02)00263-6

Rakowsky, C.E., Laming, J.M., On the origin of the slow speed solar wind: helium abundance variations, Astrophys. J. **754**, 65 (2012) doi: 10.1088/0004-637X/754/1/65

Ramaty, R., Mandzhavidze, N., Kozlovsky, B., Solar atmospheric abundances from gamma ray spectroscopy, in AIP Conf. Proc. 374, High Energy Solar Physics, ed. R. Ramaty, N. Mandzhavidze, and X.-M.Hua (NewYork,AIP), 172 (1996) doi 10.1063/1.50953

Reames, D.V., Coronal abundances determined from energetic particles, Adv. Space Res. **15** (7), 41 (1995)

Reames, D.V., Element abundances in solar energetic particles and the solar corona, Sol. Phys., **289**, 977 (2014) doi: 10.1007/s11207-013-0350-4

Reames, D.V., The abundance of helium in the source plasma of solar energetic particles, Sol. Phys. **292** 156 (2017) doi: 10.1007/s11207-017-1173-5 (arXiv: 1708.05034)

Reames, D.V., "The "FIP effect" and the origins of solar energetic particles and of the solar wind, Sol. Phys. **293** 47 (2018a) doi: 10.1007/s11207-018-1267-8 (arXiv 1801.05840 )

Reames, D.V., Abundances, ionization states, temperatures, and FIP in solar energetic particles, Space Sci. Rev. **214**, 61 (2018b) doi: 10.1007/s11214-018-0495-4 (arXiv 1709.00741 )

Reames, D.V., Corotating shock waves and the solar-wind source of energetic ion abundances: power laws in *A/Q*, Sol. Phys. **293** 144 (2018c) doi: 10.1007/s11207-018-1369-3 (arXiv 1808.06132 )

Reames, D. V., Four distinct pathways to the element abundances in solar energetic particles, Space Science Rev. **216** 20 (2020) doi: 10.1007/s11214-020-0643-5 (arXiv 1912.06691)

Reames, D.V., Ng, C.K., Mason, G.M., Dwyer, J.R., Mazur, J.E., von Rosenvinge, T.T., Late-phase acceleration of energetic ions in corotating interaction regions Geophys. Res. Lett. **24** 2917 (1997) doi: 10.1029/97GL02841

Reames, D.V., Richardson, I.G., Barbier, L.M., On the differences in element abundances of energetic ions from corotating events and from large solar events, Astrophys. J. Lett. **382**, L43 (1991) doi: 10.1086/186209

Richardson, I.G., Energetic particles and corotating interaction regions in the solar wind, Space Sci. Rev. **111,** 267 (2004) doi: 10.1023/B:SPAC.0000032689.52830.3e

Richardson, I.G., Barbier, L.M., Reames, D.V., von Rosenvinge, T.T., Corotating MeV/amu ion enhancements at ≤1 AU from 1978 to 1986, J. Geophys. Res. **98** 13 (1993) doi: 10.1029/92-JA01837

Schmelz , J.T., Reames, D.V., von Steiger, R., Basu, S., Composition of the solar corona, solar wind, and solar energetic particles, Astrophys. J. **755**, 33 (2012) doi: 10.1088/0004-637X/755/1/33

Sylwester, B., Sylwester, J., Phillips, K.J.H., Kepa, A., Mrozek, T., Solar flare composition and thermodynamics from RESIK X-ray spectra, Astrophys. J. **787**, 122 (2014) doi 10.1088/0004-637X/787/2/122

Van Hollebeke, M.A.I., McDonald, F.B. and von Rosenvinge, T.T.: 1978, The radial variation of corotating energetic particle streams in the inner and outer solar system, *J. Geophys. Res*. **83**, 4723, doi: 10.1029/JA083iA10p04723

von Steiger, R., Schwadron, N.A., Fisk, L.A., Geiss, J., Gloeckler, G., Hefti, S., Wilken, B., Wimmer-Schweingruber, R.F., Zurbuchen, T.H., Composition of quasi-stationary solar wind flows from Ulysses/Solar Wind Ion Composition Spectrometer , J. Geophys. Res. **105** 27217 (2000) doi: 10.1029/1999JA000358

Wang, Y.-M., Sheeley, Solar wind speed and coronal flux-tube expansion, Astrophys J. **355**, 726 (1990) doi: 10.1086/168805

Xu, F., Borovsky, J.E., A new four-plasma categorization scheme for the solar wind J. Geophys. Res., Space Phys. **120** 70 (2015) doi: 10.1002/2014JA020412

# Chapter 9. Hydrogen Abundances and Shock Waves

**Abstract**  How well do protons fit into the abundance patterns of the other elements? Protons have $Q = 1$ and $A/Q = 1$ at all temperatures of interest. When does their relative abundance fit on the power law in $A/Q$ defined by the elements with $A/Q > 2$? For small "pure" impulsive events, protons fit well, but for larger CME-associated impulsive events, where shock waves boost the intensities, protons are enhanced a factor of order ten by addition of seed protons from the ambient plasma. During most large gradual SEP events with strong shock waves, protons again fit the power law, but with weaker or quasi-perpendicular shock waves, dominated by residual impulsive seed particle abundances at high $Z$, again protons are enhanced. Proton enhancements occur when moderately weak shock waves happen to sample a two-component seed population with dominant protons from the ambient coronal plasma and impulsive suprathermal ions at high $Z$; thus proton-enhanced events are a surprising new signature of shock acceleration in jets. $A/Q$ measures the rigidity dependence of both acceleration and transport but does not help us distinguish the two. Energy-spectral indices and abundances are correlated for most gradual events but not when impulsive ions are present; thus we end with powerful new correlations that probe both acceleration and transport.

There are substantial variations in the abundances of elements from event to event in SEPs. It was a leap of faith to assume that most of those variations could be explained by differing source plasma temperatures plus a smooth power-law dependence upon $A/Q$. Only the element He seems to vary, because of its FIP, from the average coronal abundance underlying SEPs. However, we have not yet considered protons. Should we expect protons to fit the power law?

   Proton abundances in the solar wind, as measured by He/H, vary by a factor of about five with the solar cycle depending on the wind speed (Kasper et al. 2007). Most of this variation is in H, since more modest variations of He/O are seen in the solar wind as functions of time and of solar-wind speed (Collier et al. 1996; Bochsler 2007; Rakowsky and Laming 2012). However, in Chap. 8 we have seen that SEP abundances are not related to those of the solar wind. Different physics.

   With the exception of protons, elements in SEPs are "test particles" which are influenced by the electromagnetic fields they encounter but are too rare to alter those fields significantly. When streaming protons reach sufficient intensities, they can amplify or generate resonant waves of sufficient intensities to alter the behavior of all ions that follow behind. The power-law of enhancements vs. $A/Q$ might be expected to break down when different particle species encounter different interplanetary scattering conditions at different resonant frequencies that have been varying with space and time (Chap. 5; Reames, Ng, and Tylka 2000; Ng,

Reames, and Tylka 1999, 2003, 2012). For what circumstances does the power law from high Z predict the intensity of protons?

## 9.1 Impulsive SEP Events

Figure 9.1 begins the study with two very small impulsive SEP events. For these small events, an extension, down to protons at $A/Q = 1$, of the power law fit of enhancement vs. $A/Q$ from the elements with $Z \geq 6$, fits extremely well for the energies shown in the figure, despite some scatter of the elements that define the fit. These cases seem typical for smallest impulsive SEP events (Reames 2019b).

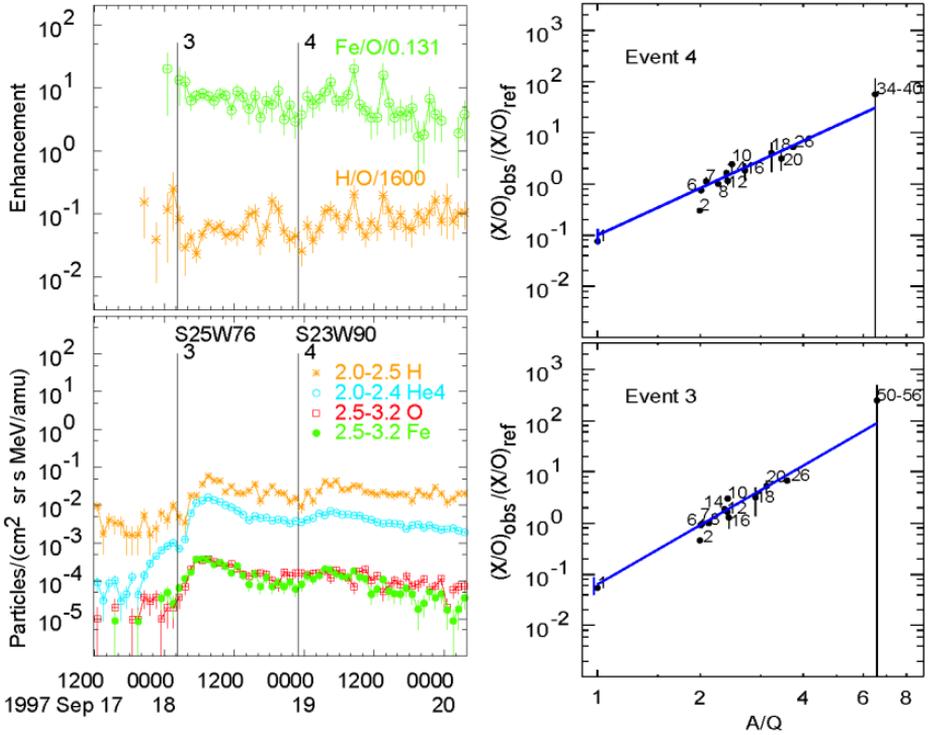

**Fig. 9.1** Intensities of H, He, O, and Fe (*lower-left*) and normalized abundance enhancements H/O and Fe/O (*upper-left*) are shown versus time for two impulsive SEP events. Event numbers 3 and 4, marking the event onset times, refer to the list of Reames, Cliver, and Kahler (2014a); the solar source coordinates are also listed. The *right panels* show enhancements, relative to average SEP coronal abundances, labeled by $Z$, versus $A/Q$ for each event, with best-fit power law for elements with $Z \geq 6$ (*blue line*) extrapolated down to protons at $A/Q = 1$ (Reames 2019b).

Of the 111 impulsive SEP events studied and listed by Reames, Cliver, and Kahler (2014a), 70 had measurable proton intensities not buried in background. It is interesting that even for the "He-poor" impulsive events (see Fig 4.12; Reames, Cliver, and Kahler 2014a; Reames 2019a) we find that the fit lines often predict the proton abundances accurately, as shown in Fig. 9.2. Here, the protons fit the power law in Event 35 while the suppression of He seems completely unrelated. This is the case for most of the small He-poor impulsive SEP events. How-ever, as a counter-example, protons exceed the expected value in Event 34, perhaps because of background from a small prior event in this case.

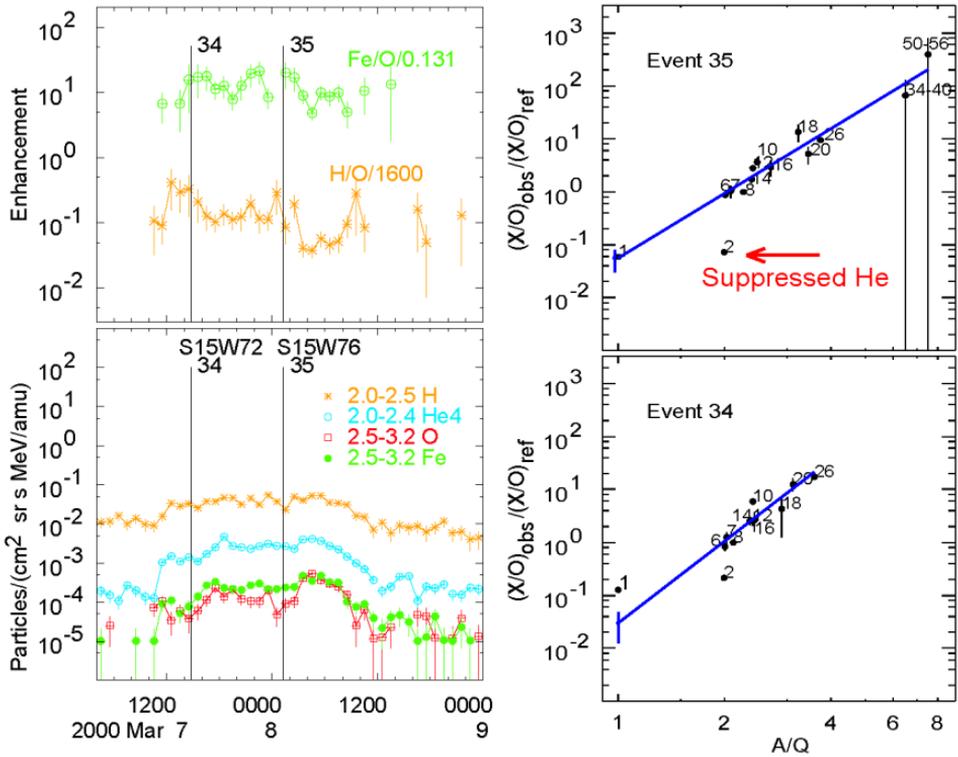

**Fig. 9.2** Intensities of H, He, O, and Fe (*lower-left*) and normalized abundance enhancements H/O and Fe/O (*upper-left*) are shown versus time for two He-poor impulsive SEP events. Event numbers 34 and 35, marking the event onset times, refer to the list of Reames, Cliver, and Kahler (2014a); the solar source coordinates are also listed. The *right panels* show enhancements, relative to SEP coronal abundances, labeled by *Z*, versus *A/Q* for each event, with best-fit power law for elements with $Z \geq 6$ (*blue line*) extrapolated down to protons at $A/Q = 1$ (Reames 2019b).

Figure 9.3 shows two well-known large impulsive SEP events (see e.g. Fig. 3.2). In both events, protons exceed the values predicted by the power-law fit by more than an order of magnitude. For Event 49 the pre-event background of protons might contribute, in principle, but they do not explain the proton excess. For Event 37, the very low intensity of pre-event protons could certainly not be a factor. Events 37 and 49 have associated CMEs with speeds of 1360 and 840 km s$^{-1}$, respectively (Reames, Cliver, and Kahler 2014a).

In the case of Event 37, particle angular distributions present some evidence that protons undergo additional scattering, which may depend upon rigidity, in comparison with other ions of the same velocity (Reames 2019b). However, other events show no evidence of this, so it is not a common factor that causes proton excesses. Most impulsive SEP events are nearly scatter-free. The presence of significant CME speeds suggests that shock waves are the important factor.

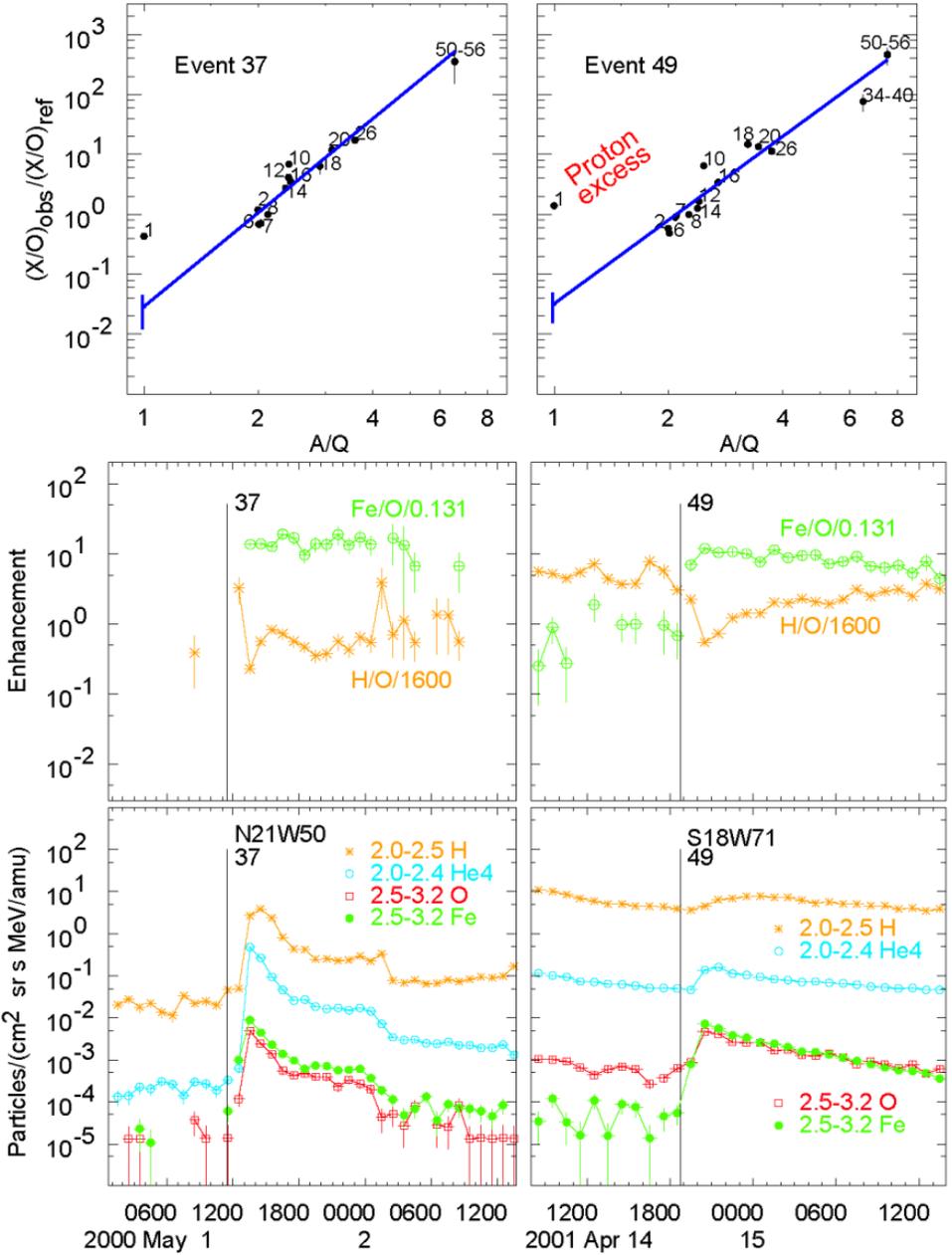

**Fig. 9.3**. Intensities of H, He, O, and Fe (*lower-panels*) and normalized abundance enhancements H/O and Fe/O (*center-panels*) are shown versus time for two large impulsive SEP events. Event numbers 37 and 49, marking the event onset times, refer to the list of Reames, Cliver, and Kahler (2014a); the solar source locations are also listed. The *upper panels* show enhancements, relative to SEP coronal abundances, labeled by $Z$, versus $A/Q$ for each event, with best-fit power law for elements with $Z \geq 6$ (*blue line*) extrapolated down to protons at $A/Q = 1$ (Reames 2019b).

It is when we consider the size or intensity of events that differences begin to become clearer as seen in Fig. 9.4. The ratio of observed proton enhancement to that expected from the least-squares fit, i.e. the proton excess, is shown on the abscissa. A value of 1.0 implies that the protons agree with the fit from $Z > 2$ ions. Figure 9.4a shows that, for small events with higher anisotropy, protons are more likely to fit. Figure 9.4b shows that small events tend to fit while increasingly larger ones do not. Figure 9.4c shows a histogram of the overall distribution for impulsive SEP events.

**Fig. 9.4**. Panel (**a**): the front/back directional anisotropy of protons during the first six hours of impulsive SEP events is shown versus the observed enhancement of H relative to that expected from the power-law fit for elements $Z \geq 6$ for 70 impulsive SEP events; the symbol size and color show the peak 2-MeV proton intensity of each event. Panel (**b**) shows the peak proton intensity versus the observed/expected H ratio for each event, and panel (**c**) shows a histogram of the distribution of observed/expected H enhancement ratio. An observed/expected enhancement of H = 1 is shown as a solid line in panels (**a**) and (**b**) (Reames 2019b).

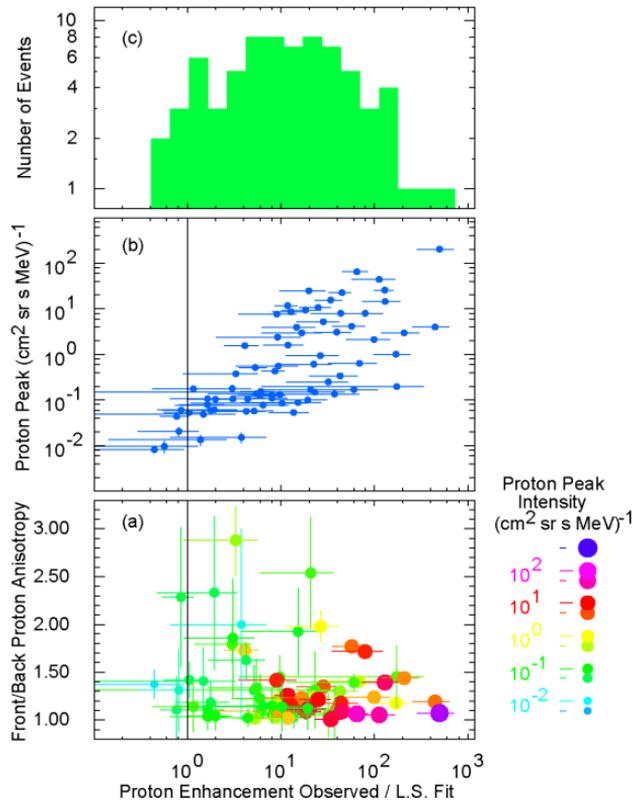

Statistically, for 24% of the events in Fig. 9.4 protons are within one standard deviation of the least-squares fit line. However, a possible explanation for the proton excess in larger impulsive SEP events comes when we consider the speed of associated CMEs, if any, found by Reames, Cliver, and Kahler (2014) as shown in Fig. 9.5. Fast CMEs tend to be associated with those intense events where protons fall a factor of ten or more above the fitted line. Events with CME speeds above $\approx 500$ km s$^{-1}$ are likely to drive shock waves that can boost the energies and intensities of the SEPs from the impulsive event and also dip into ambient coronal protons. Excess protons may be evidence of a shock wave driven by the CME associated with the same jet that produced energetic ions, i.e. jets with fast CME-driven shock waves accelerate protons from the ambient plasma along with the reaccelerated $Z > 2$ impulsive seed particles with $T \approx 3$ MK from the reconnection.

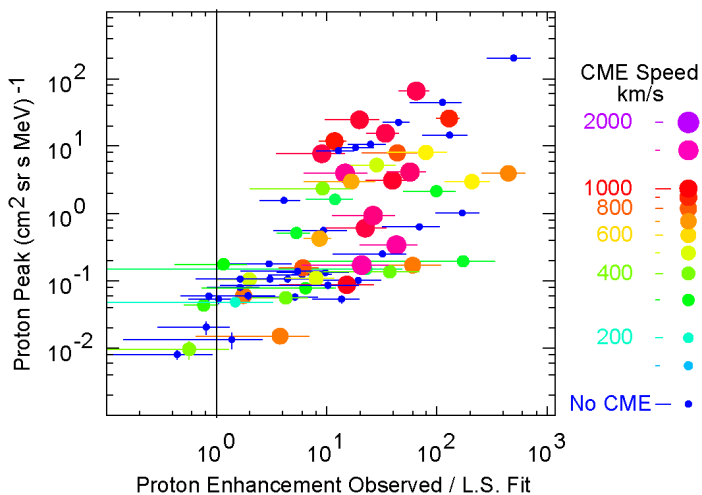

**Fig. 9.5** The peak proton intensity of impulsive SEP events is shown versus the observed/expected H ratio for each event. The color and size of the circle show the speed of the associated CME, if any is seen. An observed/expected H enhancement =1 is shown as a solid line.

In impulsive SEP events, at a source temperatures of ≈3 MK, He and C are fully ionized and O is nearly so. Thus variations in the abundances of these ions, such as He/C, reflect variations in the source itself, rather than in acceleration or transport. Figure 9.6 shows how various properties of impulsive events associate with relative O/C vs. He/C variations. The larger impulsive events with fast CMEs and proton excesses have higher average reference He/O. These events are less [3]He-rich, perhaps because larger events have depleted the available [3]He (Sect.2.5.2). CMEs with speeds above 500 km s[-1] are easily capable of accelerating ions that are pre-accelerated in the magnetic reconnection region of the impulsive jet event. Smaller events do not have proton excesses, but may or may not have suppressed He.

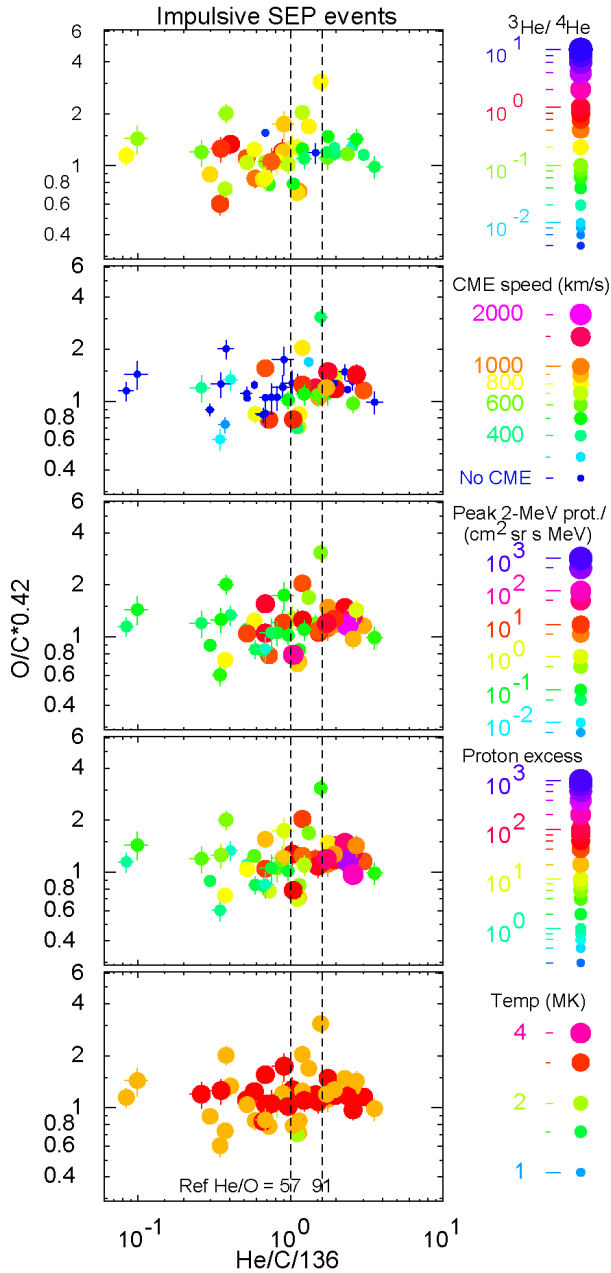

**Fig. 9.6** *Each panel* shows normalized abundances of O/C versus He/C with errors <20% for impulsive SEP events with temperature, proton excess, 2 MeV proton intensity, CME speed, and ³He/⁴He ratio highlighted by color and size of the points, as indicated. Dashed lines indicate reference abundances of He/O = 57 and 91. Note that the larger events with faster CMEs have higher average He/O and have limited ³He/⁴He (Reames 2019d).

## 9.2 Gradual SEP Events

For gradual SEP events, the power-law fit to the abundance enhancements versus *A/Q* often varies with time during an event, so we consider 8-h intervals which usually provide adequate statistics for abundance measurements.

Figure 9.7 shows the analysis of two events of the "Halloween" series in October 2003 (Reames 2019c). This analysis is similar to that described for Fig. 5.13.

Least-squares power-law fits of abundance enhancements are shown for each time interval in Fig. 9.7e using $A/Q$ values for the temperature minimum of $\chi^2/m$ determined in Fig. 9.7d and shown in Fig 9.7c. As in all similar fits, the observed abundances are divided by the reference coronal abundances to determine enhancements, all relative to O. Fit lines determined for the elements with $Z > 2$ are extended to $A/Q = 1$ and compared with protons in Fig 9.7e. Here the agreement is reasonably good, even when there are sudden large changes in slope as in the case between the last two time periods near the bottom of panel 9.7e.

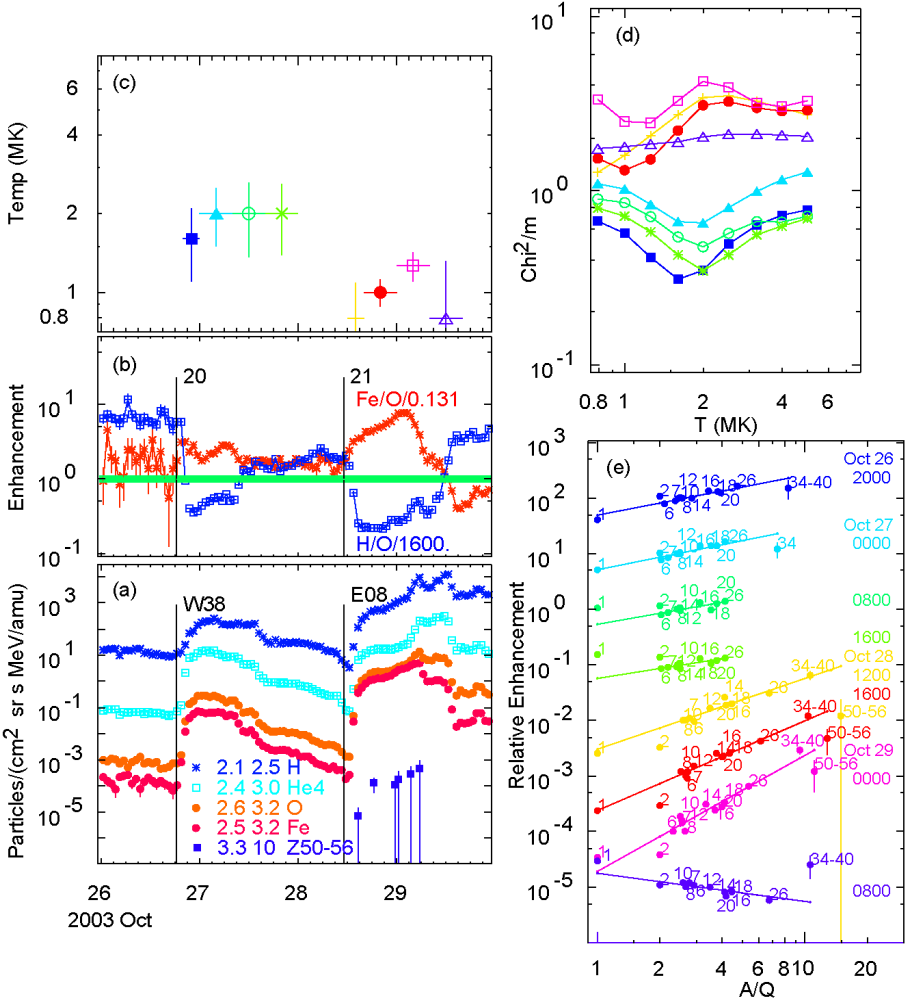

**Fig. 9.7.** Panel (**a**) shows intensities of H, He, O, Fe, and $50 \leq Z \leq 56$ ions (energies in MeV amu[-1]), (**b**) normalized abundance enhancements H/O and Fe/O, and (**c**) derived source temperatures are shown versus time for the 26 and 28 October 2003 SEP events. Panel (**d**) shows $\chi^2/m$ versus $T$ for each 8-h interval while (**e**) shows enhancements, labeled by $Z$, versus $A/Q$ for each 8-h interval shifted ×0.1, with best-fit power law for elements with $Z \geq 6$ extrapolated down to H at $A/Q = 1$. Colors correspond for the eight intervals in (**c**), (**d**), and (**e**) and symbols in (**c**) and (**d**); times are also listed in (**e**). Event onsets are flagged with solar longitude in (**a**) and event number from Reames (2016a) in (**b**) (Reames 2019c). Event 21 is a GLE.

A typical gradual event where protons exceed the prediction is an event with impulsive suprathermal seed ions and $T \approx 3$ MK, as shown in Figure 9.8. Here, the dashed lines down to $A/Q = 1$ in Fig. 9.8e show a broken power law with significant proton excesses.

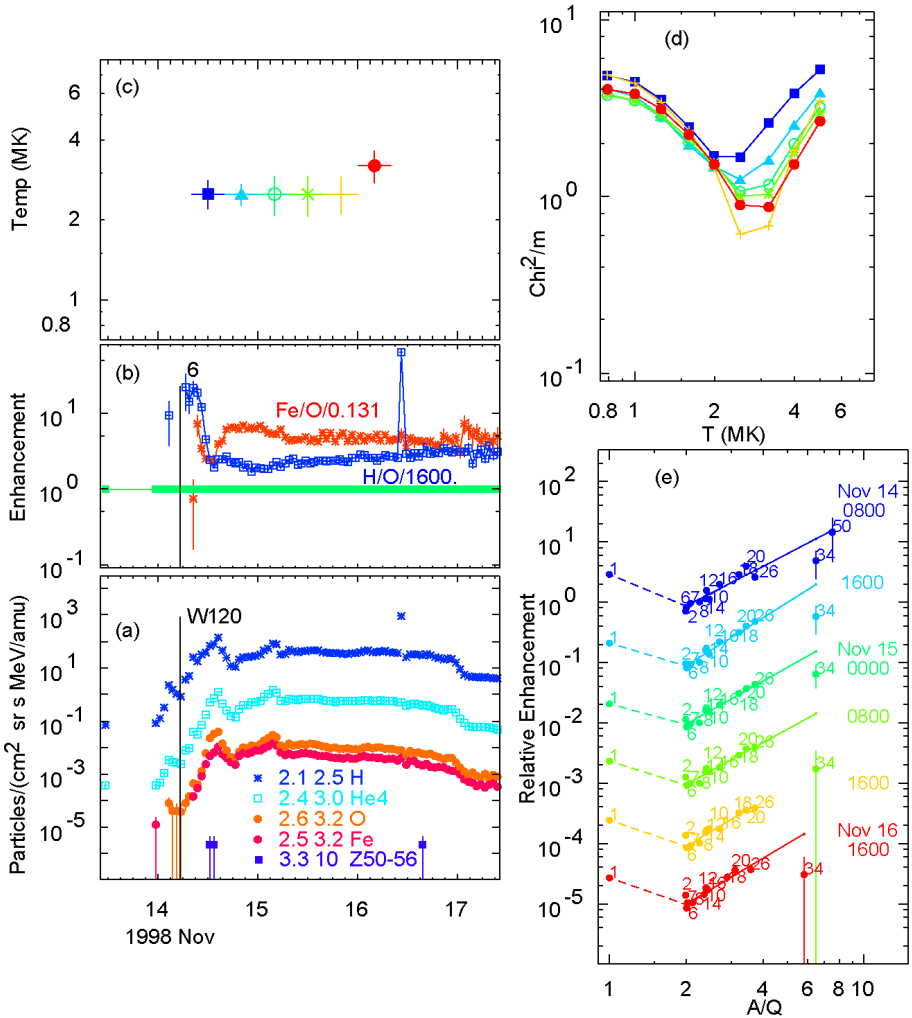

**Fig. 9.8**. Panel (**a**) shows intensities of H, He, O, Fe, and $50 \leq Z \leq 56$ ions, (**b**) normalized abundance enhancements H/O and Fe/O, and in (**c**) temperatures are shown versus time for the 14 November 1998 SEP event. Panel (**d**) shows $\chi^2/m$ versus T for each 8-h interval while (e) shows enhancements, labeled by Z, versus A/Q for each 8-h interval shifted ×0.1, with best-fit power law for elements with $Z \geq 6$ (*solid*) joined to H by *dashed* lines. *Colors* correspond for the six intervals in (**c**), (**d**), and (**e**) and *symbols* in (**c**) and (**d**); times are also listed in (**e**). *Dashed* lines join H with its associated elements in panel (**e**). Event onset is flagged with solar longitude in (**a**) and event number from Reames (2016a) in (**b**) (Reames 2019c).

We see the distribution of all 8-h gradual-event intervals in Figure 9.9. The *upper panel* shows a histogram of the distribution of proton excess versus slope or power of the *A/Q* of the fit while the lower panels show the distribution for different source temperatures.

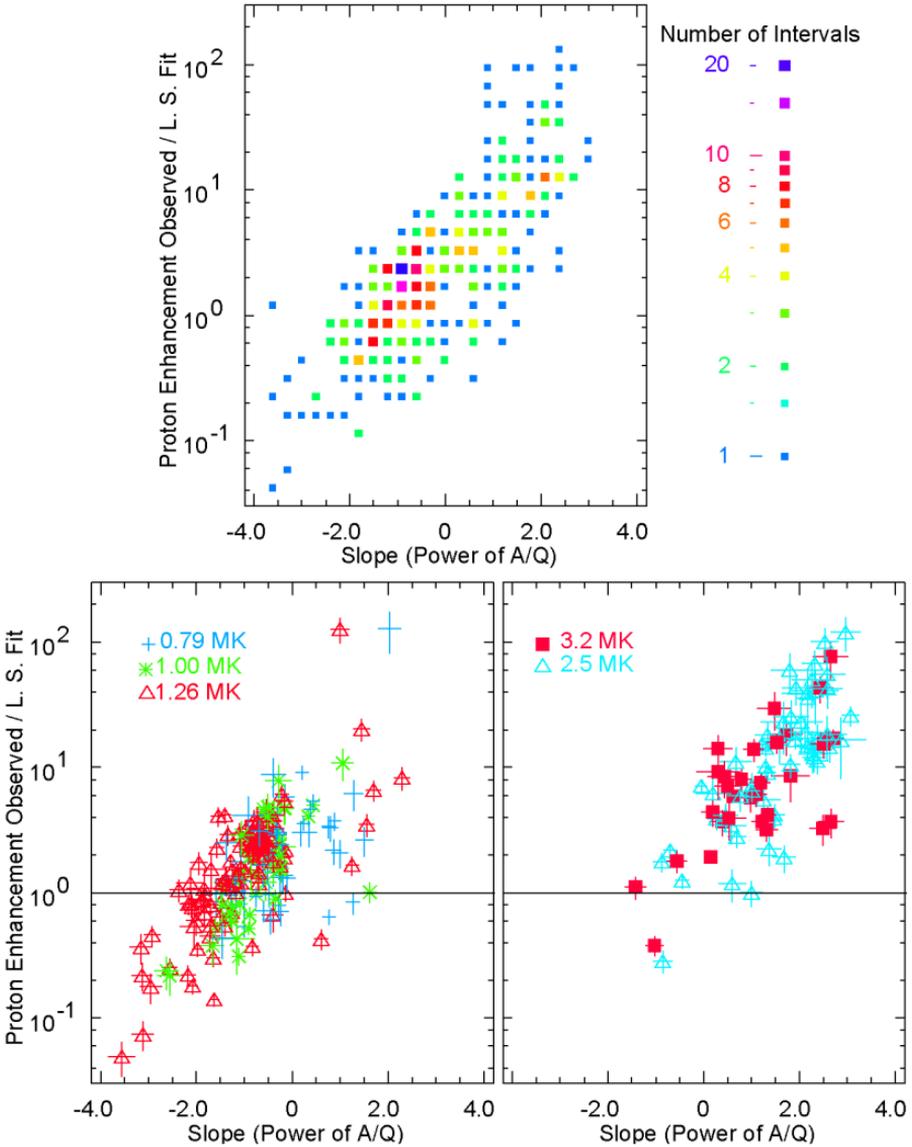

**Fig. 9.9**. In all panels, the enhancement of H relative to that expected from the power-law fit of elements $Z \geq 6$ is shown versus the "slope" or power of $A/Q$ from the fit of elements $Z \geq 6$. The *upper panel* shows a histogram of the distribution of all 398 8-h intervals in this space with symbol color and size showing the number at each location. The *lower left panel* shows the distribution of intervals with $T = 0.79$, 1.0, and 1.26 MK. The *lower right panel* shows the distribution of intervals with $T = 2.5$ and 3.2 MK (Reames 2019c).

For most gradual-event periods, with source plasma temperatures below 2 MK and declining slope versus *A/Q*, the proton intensities are predicted within a factor of order 2 or 3, by the ions with $Z > 2$. However, the ~25% of gradual events with $T > 2$ MK and positively sloping intensities versus $A/Q$, have persistent large proton excesses; these are the events dominated by impulsive seed ions.

The lower panel in Fig. 9.10 shows source temperature in a plot of O/C vs. He/C for gradual SEP events at the same scale as that for impulsive SEP events in Fig. 9.6. Upper panels show temperature, proton excess, and 20-MeV proton intensity on panels of normalized Fe/O vs. He/C.

**Fig. 9.10** The *lower panel* shows normalized abundances of O/C versus He/C for 8-h intervals during gradual SEP events with temperature as the size and color of the points. The *upper panels* show Fe/O vs. He/C with temperature, proton excess, and 20-MeV proton intensity for each interval highlighted by color and size of the points. Dashed lines indicate reference abundances of He/O = 57 and 91

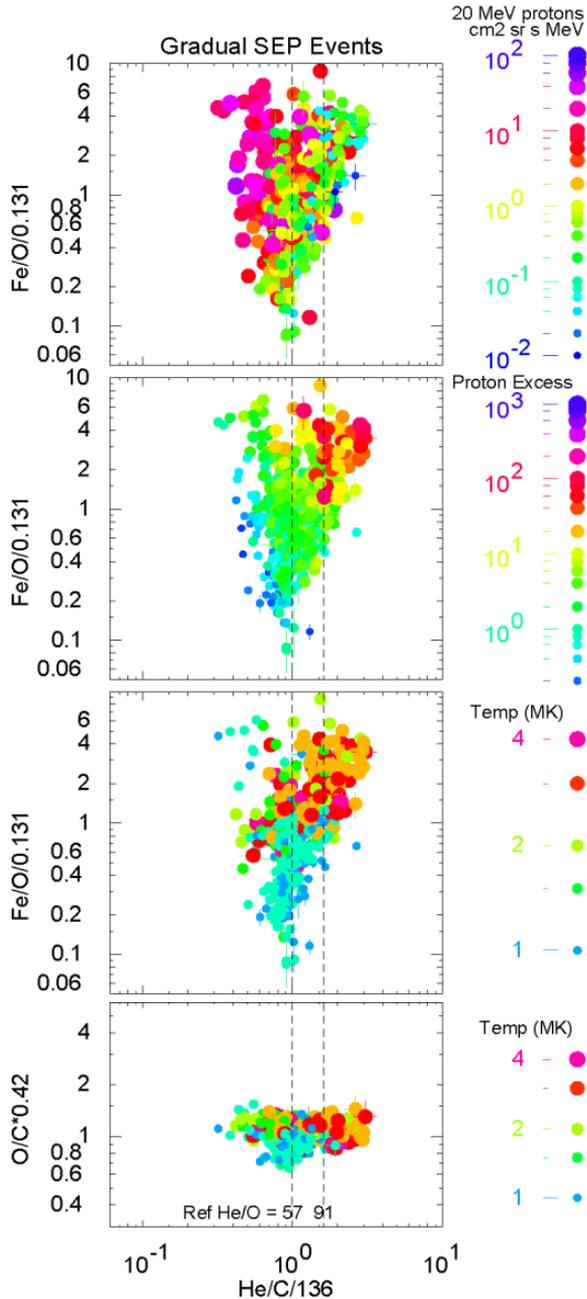

The upper panels in Fig. 9.10 show that events in the higher temperature range ≈3 MK, which involve reaccelerated impulsive seed ions, also have the large proton excesses as we saw in Fig. 9.9. These events have modest intensities of 20-MeV protons and lower average shock speeds (Reames 2019d). Like the impulsive events with CMEs in Fig. 9.6 they have higher He/C ratios. In contrast, the biggest gradual events, with the highest intensities of 20-MeV protons, lie on the low-He/C side of Fig 9.10 and they lack proton excesses. These measurements drift to lower values of Fe/O with time during each event at constant He/C.

The lowest panels of Figs. 9.6 and 9.10 compare impulsive and gradual events, respectively, at the same scale. The gradual events have much smaller intrinsic abundance variations, such as He/C, especially if we restrict the sample to $T > 2$ MK impulsive suprathermal seed particles. This comparison was already seen in Fig. 5.19 and will be discussed below. Sampling seed ions from a pool fed by many small impulsive events reduces the abundance variation.

## 9.3 Waves Coupling Proton Velocity with *A/Q*

When we compare ions at the same velocity, as we do when we study the power-law dependence on *A/Q*, those ions interact with different parts of the ambient or proton-generated wave spectrum. Neglecting pitch angle variations, for example, 2.5 MeV protons resonate with waves generated by streaming 2.5 MeV protons. However, 2.5 MeV amu$^{-1}$ He, C, and O with $A/Q = 2$ resonate with waves generated by streaming 10 MeV protons, and 2.5 MeV amu$^{-1}$ Fe at $A/Q = 4$ resonates with the wave spectrum generated by streaming 39 MeV protons; at $T = 1$ MK, 2.5 MeV amu$^{-1}$ Fe has $A/Q = 6.1$ and resonates with protons near 90 MeV. Thus the shape of the *A/Q* dependence is related to the shape of the proton spectrum and its time dependence, since high-energy protons arrive earlier to modify the wave spectrum that resonates with the ions with high *A/Q*. The time behavior in large events is complicated and its *A/Q* dependence has not been modeled extensively.

There are some gradual events that have excess protons early in the events but the expected numbers of protons later. In these events, the initial abundance ratios are affected by hard proton spectra as described in the large gradual event of 30 September 1998 discussed in Sect. 5.1.4 and shown in Fig. 5.2. Here high-energy protons arrive first and create resonant waves that scatter He and the heavier ions of a given velocity while the protons of that velocity are just beginning to be scattered by self-generated waves. This process suppresses He/H initially in Fig 5.2, i.e. it effectively increases H/He and creates excess protons. An excess of protons actually means that for a given proton intensity the heavier ions are suppressed (see Reames 2020a).

## 9.4 Compound Seed Particles

The SEP events with shock acceleration may sample a complex seed population. Are the protons sampled from the same component of this seed population as the other ions? When the heavy-ion abundances increase with *A/Q*, they may be sam-pled from pre-accelerated impulsive suprathermal ions that have $T \approx 3$ MK. The protons in that population are already suppressed, but ambient coronal material is also available for shock acceleration. This situation has been described by Reames (2019b, 2019d, 2020a) and shown in Fig. 9.11a. Here, impulsive Event 54 on 20 February 2002 (from the list in Reames, Cliver, and Kahler 2014) with a CME of 954 km s$^{-1}$ may include shock-accelerated protons predominantly from the ambient corona (red source) plus ions with $Z \geq 2$ mainly from a local impulsive jet magnetic-reconnection source (blue source labeled SEP1). The red components show a decreasing slope in $A/Q$, typical of shocks accelerated coronal plasma, while the original slope of the blue component retains most of its steep positive dependence on $A/Q$. The lower temperature of the (red) coronal material in Fig. 9.11a decreases $Q$ and raises the $A/Q$ for heavier ions like C, O, and etc.

**Fig. 9.11.** (**a**) Element enhancements, labeled by $Z$, versus $A/Q$ for impulsive Event 54 (20 February 2002), together with possible sources for shock acceleration from pre-accelerated impulsive ions (*blue*, SEP1) and ambient corona or pre-event plasma (*red*). Helium may receive comparable contributions from both sources in this event (Reames 2019b; 2019d). Panel (**b**) shows possible selection of seed particle H (*solid*) and O (*dashed*) from ambient (*red*) and harder pre-accelerated impulsive (*blue SEP1*) spectra by shock waves with different threshold energies (Reames 2020a).

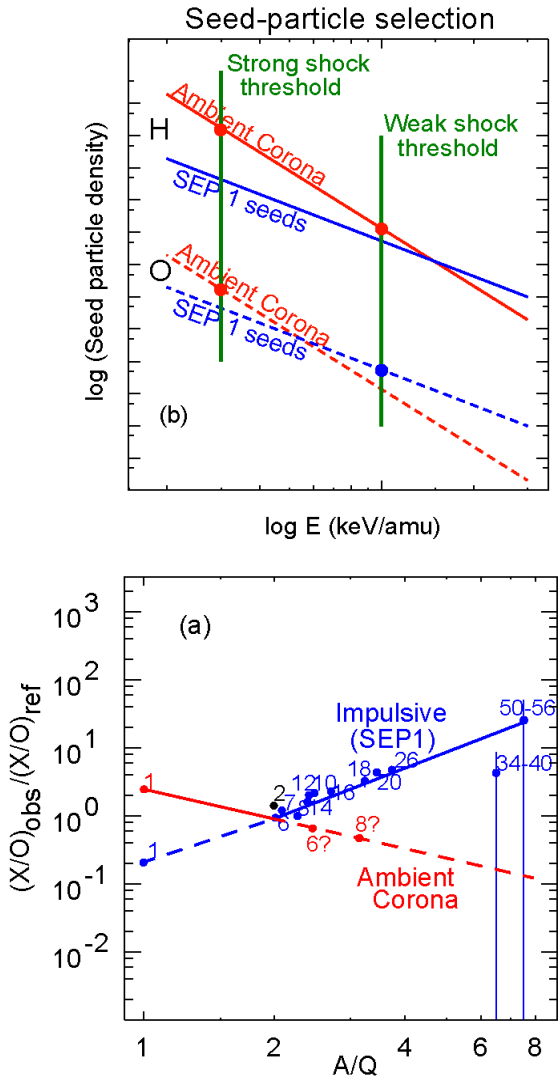

Figure 9.11b shows possible seed-particle spectra of H and O (representing other heavy ions) with appropriate abundances for ambient thermal ions and for an assumed harder spectrum of pre-accelerated impulsive suprathermal ions (labeled as SEP1). The combination of spectral hardness and abundances of the seeds allow weaker shocks to sample both populations while strong shocks are dominated by the more-abundant ambient ions throughout.

This is a possible explanation for the impulsive events with observed proton excesses, but Fig. 9.11 applies to gradual SEP events as well. Of course, for gradual SEP events the shocks must just happen to encounter the pools of impulsive suprathermal ions in order to include them. Also, the extremely fast, wide shocks in most gradual events may sweep up ambient coronal material so efficiently that the impulsive suprathermal ions become negligible, and all ions, including protons, fit on the same power law of the SEP ions. However, the smaller gradual events with weaker shock waves favor the residual impulsive suprathermal ions swept up from a large region by the wide shock; only protons from the ambient coronal plasma are able to predominate. The preference for the higher-velocity seed particles is enhanced when quasi-perpendicular shock waves are involved where ions downstream must be fast to overtake the shock to continue the acceleration (Tylka et al. 2005; Tylka and Lee 2006).

In Fig 9.10 (and also in Fig 9.6) the proton excess shows a tendency to increase as a function of He/C. It is possible that events with high proton excess accelerated from ambient coronal material also have a component of He from the same source as the protons.

Figure 9.12 shows the distinction between the similar process in impulsive and gradual events. Both involve moderately weak shocks. In both, impulsive suprathermal seed ions with $T \approx 3$ MK dominate high $Z$ while ambient plasma dominates the protons and occasionally the He. The essential difference is that each impulsive event probably involves a single jet source, for which seed abundances may vary locally from event to event, while the wide shock in the gradual events sweeps up suprathermal residue from a pool of many (N) impulsive jet sources that has been accumulating, reducing abundance variations by a factor of $\sqrt{N}$. The residue from many jets has been observed to collect in large regions for substantial periods of time, so that these $^3$He-rich, Fe-rich pools of suprathermal ions are often seen (see regions labeled A and B in Fig. 2.8, for example; Desai et al., 2003; Bučík et al., 2014, 2015, 2018a, 2018b; Chen et al., 2015). The number of small flares increases logarithmically with decreasing size, leading Parker (1988) to propose that nanoflares were sufficiently numerous to heat the solar corona. Jets, being the open-field version of flares, may increase similarly, so that many small jets, perhaps we should call them microjets or even nanojets, contribute the seed population for the gradual SEP events with high-$Z$ enhancements and $T \approx 3$ MK.

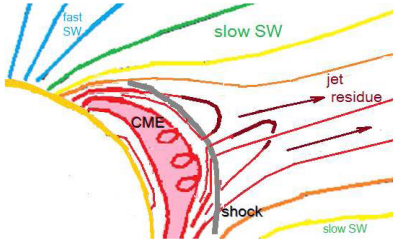

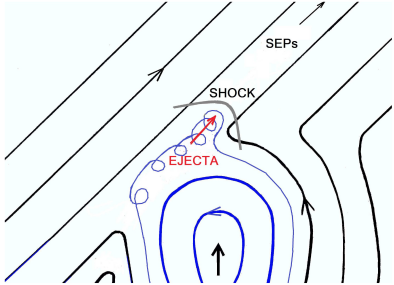

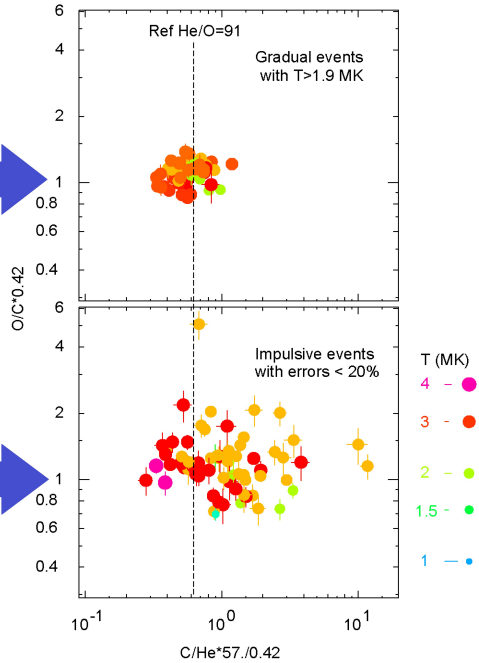

**Fig. 9.12** Suggested explanations for (**a**) the broad distribution of source abundances in impulsive events from variations in single localized jet events where ejecta may (SEP2) or may not (SEP1) drive shocks, and (**b**) the narrow distribution in high-$T$ gradual events where pre-accelerated impulsive seed populations from a pool of many small individual jets are averaged by a large shock (SEP3). Abundance distributions shown (from Fig 5.19) are equivalent to those that were shown in the lower panels of Figs. 9.6, and 9.10.

While the model impulsive SEP event involves a single jet, we can certainly imagine a region of magnetic reconnection that is extensive enough to involve several individual jets at a given time. Element abundances from these compound regions might tend toward average impulsive SEP abundances as the gradual events do. Perhaps this is why the smallest impulsive SEP events tend to have the largest abundance variations – they involve smaller regions with less averaging.

## 9.5 CME Associations of Impulsive and Gradual Events

Figure 9.13 reviews the CME associations of impulsive and gradual SEP events. Simply by selecting events based upon the Fe/O ratio to initially define our impulsive events (Fig. 4.1), we have found significant differences in the nature and properties of the associated CMEs involved.

**Fig. 9.13** Histograms compare the speed (*left panels*) and width (*right panels*) distributions of CMEs associated with impulsive (*upper panels*) and gradual (*lower panels*) SEP events. The "?" indicates events with no known CME association, often for lack of coverage (Reames 2019d).

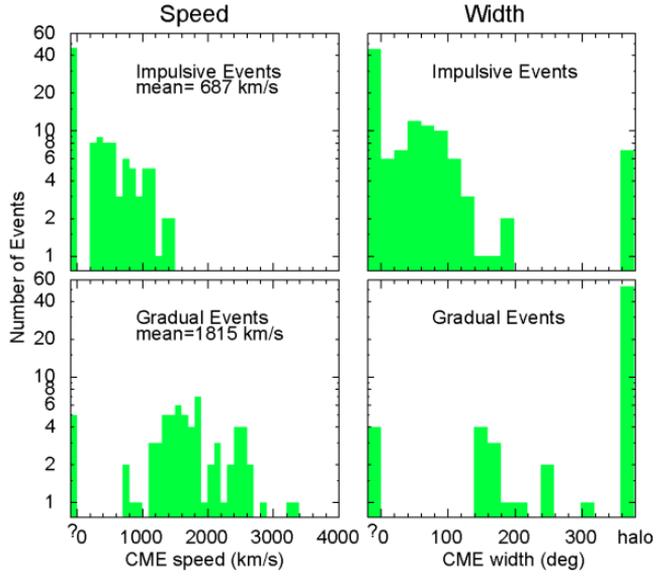

In Figure 9.14, we examine the events based upon source plasma temperature *T*. The events with *T* >1.9 MK that involve reaccelerated impulsive ions tend to involve slower, weaker CMEs. Events with the fastest CMEs and shocks accelerate the cooler ambient plasma.

**Fig. 9.14** CME speed is shown for gradual SEP events as a function of time, with source plasma temperatures indicated in the *lower panel*. CME speed distributions for gradual SEP events, with T < 1.9 MK (SEP4) and T > 1.9 MK (SEP3), are shown in the *middle* and *upper panels*, respectively (Reames 2019d).

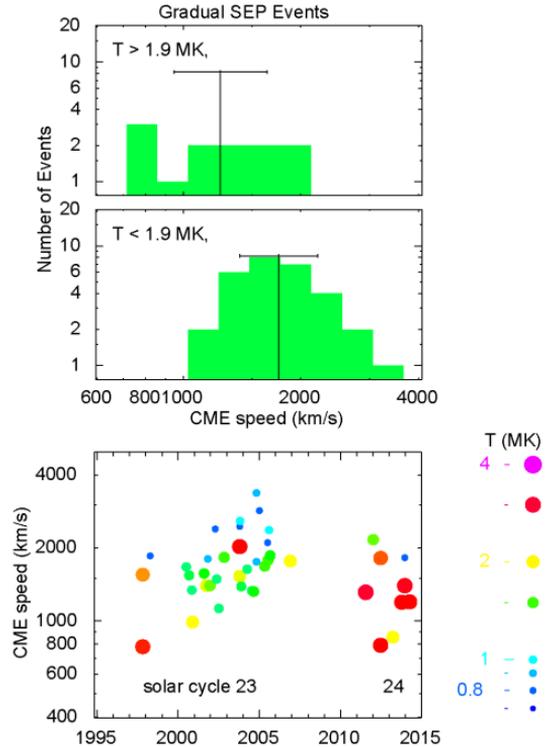

The smaller gradual events with the hotter ($\approx 3$ MK) reaccelerated impulsive-SEP source plasma not only involve slower, weaker CMEs, but may also involve quasi-perpendicular shock waves (not shown). These events with $T \approx 3$ MK tend to arrive early in solar cycle 23 and the weaker solar cycle 24. Faster CMEs that occur later in cycle 23 tend to be dominated by cooler ambient coronal plasma of which they sample deeply – these more powerful events have less need for pre-accelerated ions. Like all SEP events, all solar cycles are not the same.

## 9.6 Four Subtypes of SEP Events

Thus, inclusion of protons in SEP abundance patterns leads to the suggestion of four types of SEP events (after Reames 2019d, 2020a):

i) SEP1 – Pure Impulsive: "Pure," shock-free impulsive SEP events accelerate ions in islands of magnetic reconnection in solar jets. Element abundance enhancements increase as a power law in $A/Q$ from H to elements as heavy as Pb derived from $T \approx 3$ MK plasma; they can be distinguished initially by Fe/O that is over four times the coronal value and confirmed by the lack of any proton excess. He/O is normally high, but may be greatly suppressed in occasional events perhaps by a rapid rise of ionized material that may be too fast to allow much ionization of high-FIP He. Abundant electrons streaming out from the event generate waves that are resonantly absorbed by ${}^3$He; these electrons produce a type III radio burst. Plasma may also be ejected from the event, producing a narrow CME that is too slow to drive a significant shock wave.

ii) SEP2 – Impulsive + Shock: An impulsive event occurs when the narrow CME from a jet, that of an otherwise pure impulsive SEP1 event, drives a fast shock wave. The shock wave samples all available ions, those from the ambient plasma and residual energetic ions from the SEP1 event. The abundant ambient protons form the wave structure at the shock and dominate at $Z = 1$, but the pre-enhanced, pre-accelerated, $T \approx 3$ MK ions dominate the $Z > 2$ region because they are favored by the weak shock. This appears as a large ten-fold proton excess.

iii) SEP3 – Weak Gradual with Impulsive Seeds: A moderately-fast, wide CME from an eruptive event drives a moderately-fast shock wave producing a gradual SEP event. The sampled shock region may be quasi-perpendicular (or just weak) so its dominant contribution from sampling of ambient plasma is limited mainly to protons, while preferring the faster residual impulsive suprathermal ions surviving in pools of a dozen or more previous impulsive SEP events from small jets that combine to produce well-defined average impulsive-SEP abundances for $Z > 2$. These gradual events have substantial proton excesses plus the $T \approx 3$ MK, high-$Z$ signature of the impulsive seed particles.

iv) SEP4 – Pure Gradual: A fast, wide CME from an eruptive event drives a fast shock wave that expands broadly, producing an energetic gradual SEP event that lasts many days. If the shock is quasi-parallel or samples deeply into the tail of the thermal distribution of the ambient plasma with $T < 2$ MK, it will produce a "pure" gradual event with dominant coronal ion abundances modified by a power-law dependence on $A/Q$ that may be enhanced or suppressed during ion transport. Protons generally fit with other ions although some regions of unusual transport may produce modest local excesses or depletions of protons. Any impulsive suprathermal ions present are also accelerated by the shock, but their contribution is overwhelmed by the accelerated ambient coronal ions; these shocks do not need pre-accelerated ions. Small, weak shocks may also join this category when they find no suprathermal ions to accelerate.

The inclusion of protons in abundance studies has provided surprising new information on the underlying physics in SEP events. More generally, the study of power-law patterns in the $A/Q$-dependence of element abundance enhancements has provided an important new source of information on the difficult-to-obtain temperature and origin of the ions accelerated as SEPs, on the physical processes involved, and on the nature of the solar corona.

## 9.7 Spatial Distributions

Proton abundances can clarify the pattern of abundances and their spatial distribution, even when H is not included in the power-law fit of the abundances. Fig. 9.15 shows power-law fits as a function of time for three widely separated spacecraft for the 23 January 2012 SEP event. Protons were not included in the original study of this event (Reames 2017) and the power-law fits seemed to be disrupted by a spectral break, and of poor quality. However, when we include H in Fig 9.15, it is clear that the power-law fits of high-Z elements point directly toward the protons in most cases. H validates the power-law behavior and marks this as an SEP4-class event – viewed from any longitude.

Early in the event in Fig 9.15d, the power-law fits are extremely flat at the well-connected *Wind* spacecraft, i.e. the SEP abundances are nearly coronal, so it is difficult to determine a plasma temperature. Later, as the higher-rigidity heavy ions at higher $A/Q$ leak away, the increasing slope defines a temperature of $1.3 \pm 0.3$ MK (see Fig. 3 of Reames 2019c).

Intensities are considerably lower at the STEREO spacecraft than at *Wind* and, unfortunately, geometry factors are also an order-of-magnitude smaller for the STEREO instruments, so the STEREO abundances are more poorly defined, especially STEREO B. However, the trends in the data are clear.

By the time the shock (S in Figs. 9.5a and 9.5b) has reached about 2 AU, STEREO A shares a magnetic reservoir (Sect. 5.7) with *Wind*, and all ion intensities become spatially uniform, declining with time as the magnetic "bottle" containing them expands. Reservoirs treat differing energies and rigidities of ions invariantly. STEREO B does not sample this reservoir and intensities there are much lower late in the event.

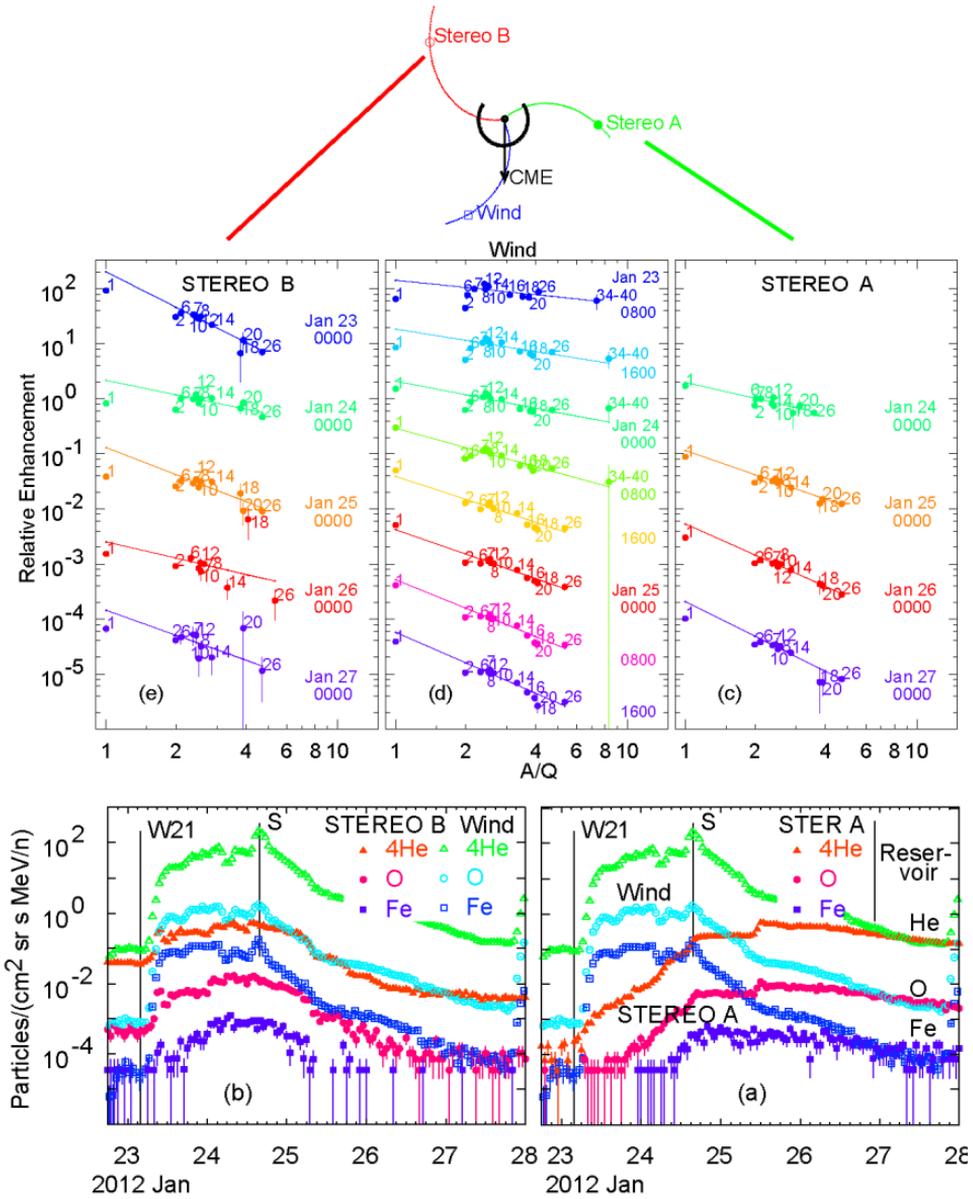

**Fig. 9.15** Power-law abundances distributions are shown for the three-spacecraft configuration of STEREO B, *Wind,* and STEREO A shown at the top. (**a**) Time profiles of few MeV/n $^4$He, O, and Fe are compared for STEREO A and *Wind* during the 23 January 2012 SEP4 event. (**b**) Similar time profiles of $^4$He, O, and Fe are compared for STEREO B and *Wind* during the event. (**c–e**) shows abundance enhancements, labeled by *Z*, versus *A/Q* for relative time intervals beginning at the time listed, with best-fit power law for elements with $Z \geq 6$ extrapolated down to H at $A/Q = 1$ for (**c**) daily intervals on STEREO A, (**d**) 8-h intervals on *Wind*, and (**e**) daily intervals on STEREO B. Abundances at STEREO A and B are for $4 – 6$ MeV amu$^{-1}$ ions, and those at *Wind* are typically for $3 – 5$ MeV amu$^{-1}$ ions. In (**a**), Wind and STEREO A enter an equal-intensity, magnetic reservoir on 27 January, long after shock passage.

## 9.8 Rigidity-Dependence: Acceleration or Transport?

By now, one might conclude that the average acceleration by shocks has minimal dependence upon rigidity (i.e. upon $A/Q$). If we exclude reacceleration of impulsive suprathermal ions, which have a built-in dependence on $A/Q$, the dependence we find in SEP4 events, for example, could be explained easily by transport from the shock. Scattering is one big adjustable parameter. However, there does seem to be a net negative slope to the SEP4 power-law fits for the small and moderate events. This is shown in Fig. 9.16, where the lower panel shows source temperature vs. the power-law slope of $A/Q$, averaged over each event, with proton excess as point size and color. If we eliminate the obvious SEP3 events by requiring $T < 2$ MK, we obtain the event-size distribution in the upper panel of Fig 9.16. This distribution suggests that small events, with minimal wave generation, tend to have negative slopes of $A/Q$, while large events, with significant wave generation, have average slopes that are net neutral or positive (Reames 2020b).

**Fig. 9.16** The *lower panel* shows $T$ vs. the power-law slope of enhancement vs. $A/Q$, averaged over each gradual SEP event, with the proton excess shown as the point size and color. Impulsive-sourced SEP3 events stand out as orange or red in the upper-right corner. The *upper panel* shows the fluence of >30 MeV protons vs. the average slope of $A/Q$ only for SEP4 events with $T < 2$ MK (Reames 2020b, © Springer)

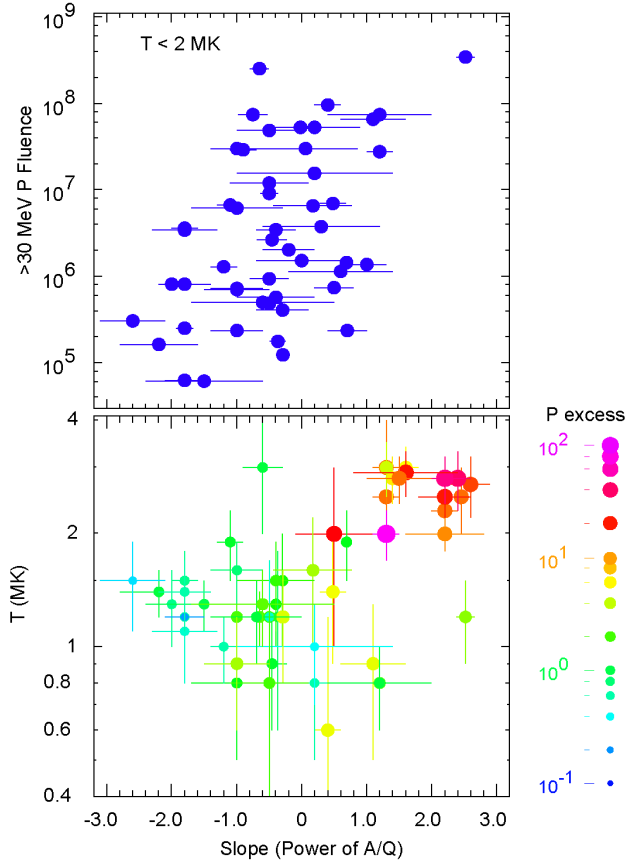

At the outset we should say that this discussion may seem a bit circular since we actually derived the reference coronal abundances by averaging over gradual SEP events. However, when we compared these average SEP abundances with those in the photosphere (Chap. 8) we found a FIP effect that was compatible with current theory and which could not support any significant residual power-law in *A/Q*. However they were obtained, the reference abundances are not arbitrary and we cannot simply redefine them. For example, solar wind abundances do not work as a reference for SEPs; temperature minima are not formed.

Naively, we would expect that ions scatter against Alfvén waves, back and forth across the shock, with a mean free path proportional to the ion's rigidity. This would surely produce rigidity dependence in the acceleration, making the acceleration time vary inversely with rigidity until an equilibrium spectrum is attained. Jones and Ellison (1991) discuss the issue of reduced rigidity dependence in reference to the Earth's bow shock, where Monte Carlo calculations (Ellison, Möbius, and Paschmann 1990) showed that shock smoothing could compensate for the expected rigidity dependence. Shock smoothing allows ions with longer mean free paths to encounter a larger shock-velocity difference, compensating for the slower acceleration otherwise.

In fact, shock waves are not simple planar structures; they are complex surfaces, modulated by waves that vary in space and time. Particle-in cell (PIC) simulations show these variations (e.g. Trotta et al. 2020) and small-scale variations have been observed in interplanetary shocks by the *Cluster* spacecraft (e.g. Kajdič et al. 2019). These variations include upstream waves produced by reflected particles, some including variations in $\theta_{Bn}$. While these considerations have not been applied to coronal shocks or to the rigidity dependence of SEPs, they are interesting and could be important. *A/Q* dependence of abundances may, in fact, allow some measure of shock structure, with some shocks relatively enhancing heavier-element abundances and others suppressing them. Perhaps future simulations of shock acceleration of SEPs will aid the study these effects.

The power law in *A/Q* that we observe in SEPs is a result of the combined rigidity dependence of both acceleration and transport. In an effort to separate these processes, we first consider very small gradual events, where the proton intensities do little to disrupt the transport. Perhaps SEPs from small gradual events even travel scatter-free like those from small impulsive events. Figure 9.17 shows properties of the small gradual SEP event of 17 June 1998. During the first day of this event the ions show very little scattering in their angular distributions for H and He shown in panels (d) and (c), respectively. The best power-law fit for the abundance enhancements vs. *A/Q* measured on 17 June are shown in Fig 9.17f has a slope of $-1.07 \pm 0.14$. Perhaps this represents the rigidity dependence of the acceleration by this shock that is unmodified by transport. Small SEP4 events in Fig. 9.16 have slopes in the region of -1 to -2, which may be the rigidity dependence from acceleration alone in these events.

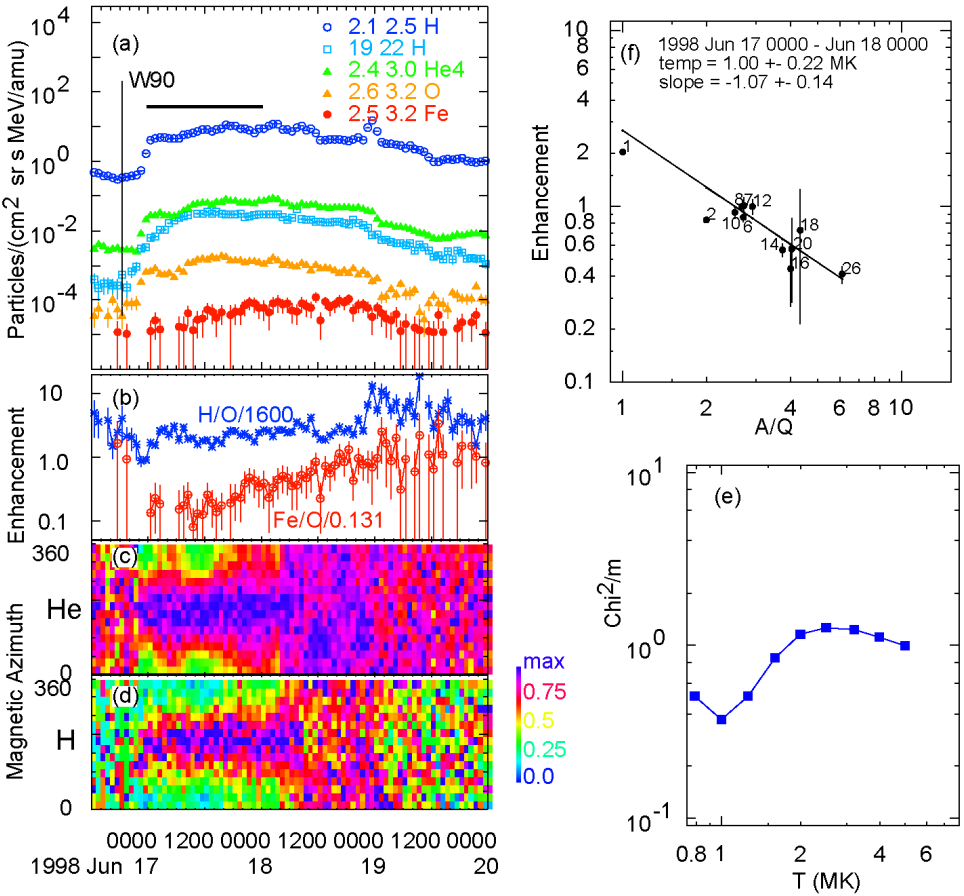

**Fig. 9.17** Selected particle intensities (**a**) and enhancements (**b**) are shown vs. time, along with angular distributions (in °) relative to the magnetic field **B** for H (**d**) and He (**c**) in the small 17 June 1998 gradual SEP event. The plot of $\chi^2/m$ vs. $T$ in (**e**) selects $T = 1$ MK as the best fit for abundance enhancements vs. $A/Q$, shown in (**f**), with each element noted by $Z$. The time period measured is listed and is shown by the bar above the curves in (**a**) (Reames 2020b, © Springer).

At the opposite extreme, we can consider huge gradual SEP events where proton intensities are so high as to reach the streaming limit (Sect. 5.1.5), impeding the flow of low-energy ions and causing flattened or rounded low-energy spectra during the early plateau period (Sect.5.1.5; Reames and Ng 2010; Ng, Reames, and Tylka 2012).

Figure 9.18 shows the behavior of element abundances during two of the large events studied by Reames and Ng (2010). For these events and other streaming-limited events the power of $A/Q$ changes markedly from positive to negative after shock passage. Early in the events, low-rigidity ions are strongly scattered but, at a given velocity, higher-rigidity ions like Fe can penetrate the turbulence more easily than low-rigidity ions like H. However, a small but consistent proton excess in the 5 November 2001 event, on the left in Fig. 9.18, may indicate that the wave spectrum is not a perfect power law.

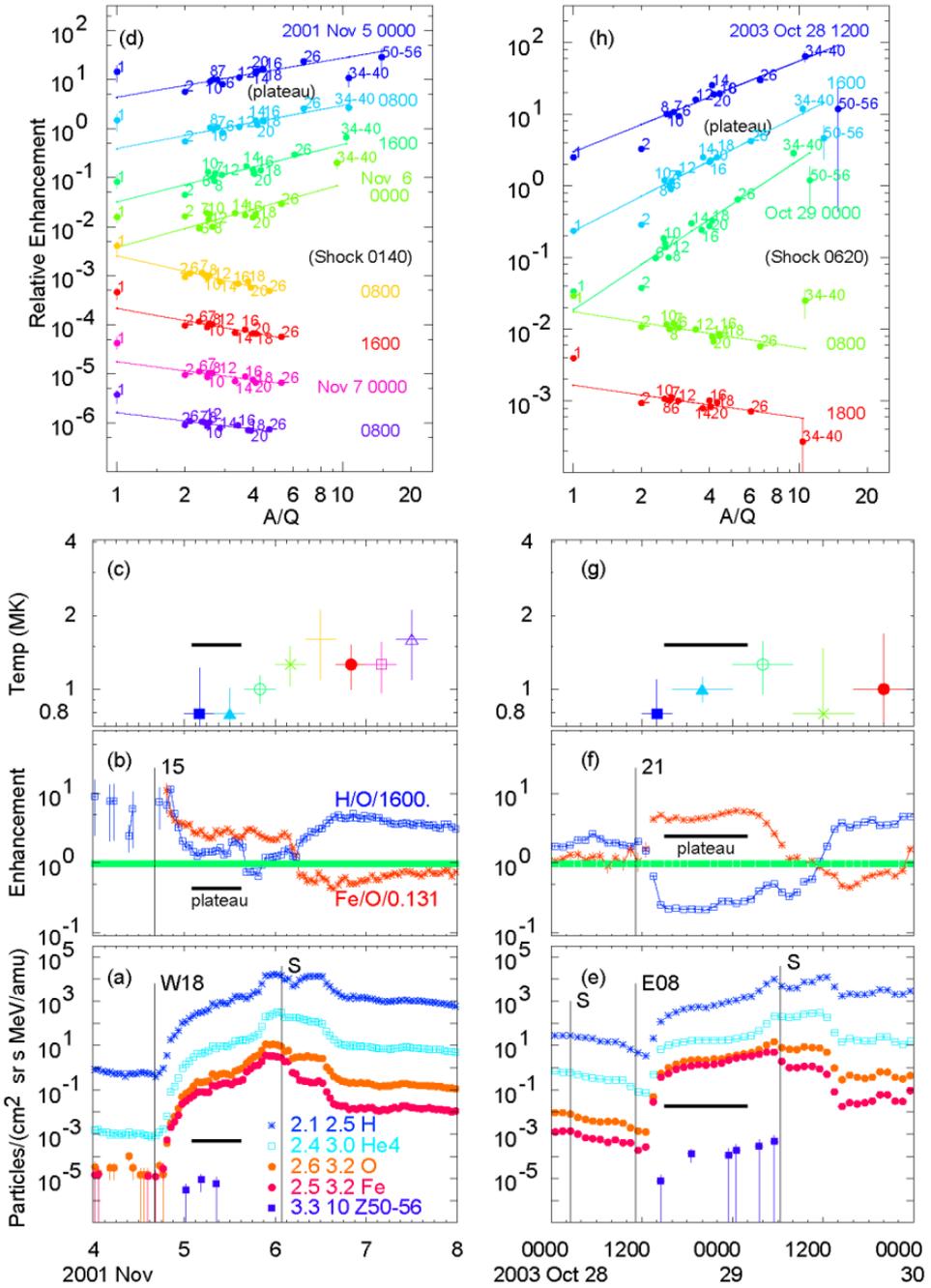

**Fig. 9.18** Time evolution of intensities, (**a**) and (**e**), H and Fe enhancements, (**b**) and (**f**), and derived temperatures, (**c**) and (**g**), respectively, is shown for the large gradual events of 4 November 2001 (*left*) and 28 October 2003 (*right*). Panels (**d**) and (**h**) show the respective power-law fits of enhancements vs. *A/Q* with *Z* shown and time indicated by color. Plateau regions are shown and times of shock passage are indicated. Powers (slopes) of the *A/Q* dependence change markedly from positive to negative after shock passage (Reames 2020b, © Springer).

For the 5 November 2001 event the power or slope is positive 0.8 – 1.2 before the shock and -0.6 – -1.0 after. For the 28 October 2003 event the power is 1.3 – 2.1 before and -0.4 – -0.5 after. There is little doubt that scattering during transport increases this power early in these events, trapping the lower-rigidity ions near the Sun, but the suppressed values after the shock passage might represent, not the source, but also the depletion of the high-rigidity ions that escaped previously. In a study of SEP abundances, Reames (2014) found that abundances were very stable in reservoirs late in events, but they showed a strong energy dependence which was not present in the overall abundance averages used for FIP studies. Self-consistent models of wave generation and scattering of ions during acceleration and transport are not yet available to help resolve these issues.

## 9.9 Correlations between Spectra and Abundances

When shock waves sweep up coronal material for acceleration, the pattern of abundances at a given velocity is related to spectral indices of the ions. Both features result from the same rigidity-dependent scattering, as does the interplanetary transport, which can also maintain or disrupt the relationship. Non-relativistically, spectra of form $E^y$ with enhancements of $(A/Q)^x$ has been found to obey $y = x/2 - 2$ (Reames 2020b). The power of velocity is $x - 4$. The origin of this fundamental relationship is not understood theoretically. It relates to the "injection problem" and describes the way a shock selects ions of different $A/Q$ from the solar plasma.

The analysis of a moderate-sized, well-behaved event is shown in Fig. 9.19. For events of this size, the spectra seem to be determined early in the event, subsequent action of the shock only maintains the same spectral shape and abundance pattern; the spectrum is no harder at the shock peak (also true for the event in Fig. 9.15). A weakening shock only maintains the spectrum of a previous stronger shock, as we saw in Eq. 5.9. Apparently transport has little effect on the properties of this event.

Analysis of a more complex example is shown in Fig. 9.20. Here the spectral index and the power of $A/Q$ change rapidly throughout the event, only roughly tracking the expected relationships. For this event O spectra vary from $E^{-1}$ to $E^{-5}$ while abundances vary from $(A/Q)^{+1}$ to $(A/Q)^{-2}$. Fe spectra vary less, probably because of their higher rigidity. The abundances follow the classic pattern of high-Z enhancement early and suppression later, because Fe scatters less than O during transport, for example. This is a large event and wave amplification is certainly a factor. However, it is also true that its longitude suggests that we are connected to the strong nose of the shock early and to the weaker flanks later. Spatial vs. temporal effects, acceleration vs. transport, it is not easy to tell, but comparing smaller and larger events suggests that transport is the new factor that dominates this behavior, since smaller events show no significant longitude dependence (e.g. Fig. 9.15). Differences in the spectral indices of O and Fe are also a clue. We lack event simulations that explore these abundance variations. Other examples are shown by Reames (2020c), including SEP3 events and events dominated by the streaming limit.

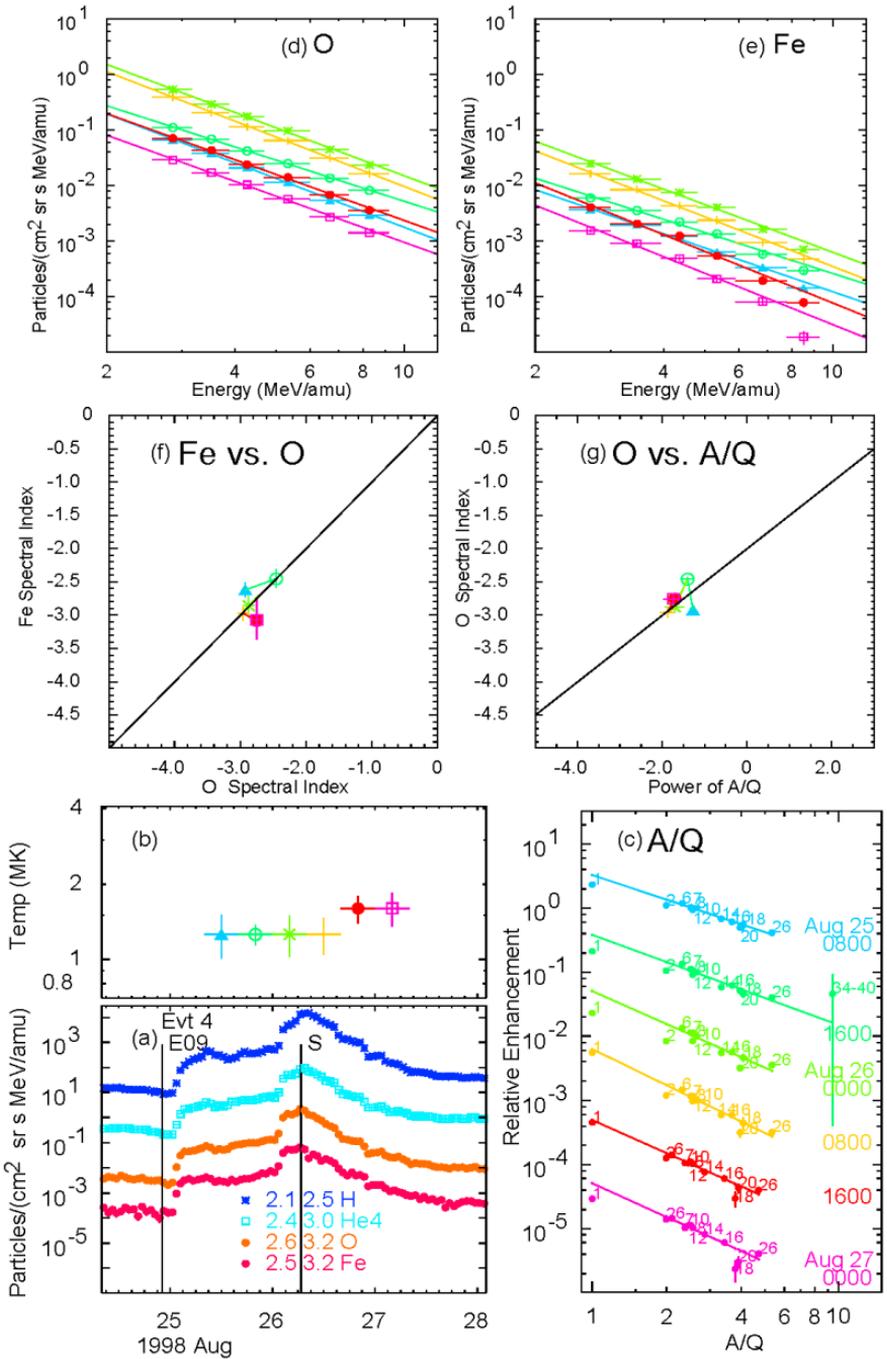

**Fig. 9.19** (**a**) Intensities, for listed ions and energies in MeV amu⁻¹ and (**b**) derived source temperatures are shown versus time for gradual SEP Event 4 (list of Reames, 2016), on 24 August, 1998. Fits are shown for (**c**) enhancements of elements, listed by Z, versus A/Q, and for energy spectra of (**d**) O and (**e**) Fe. Correlation plots are shown for spectral indices of (**f**) Fe versus O and of (**g**) O versus A/Q. Colors for time intervals correspond in (**b**), (**c**), (**d**), (**e**), (**f**), and (**g**). In (**f**) the solid line is diagonal, y = x, in (**g**) it is y = x/2 − 2 (Reames 2020c).

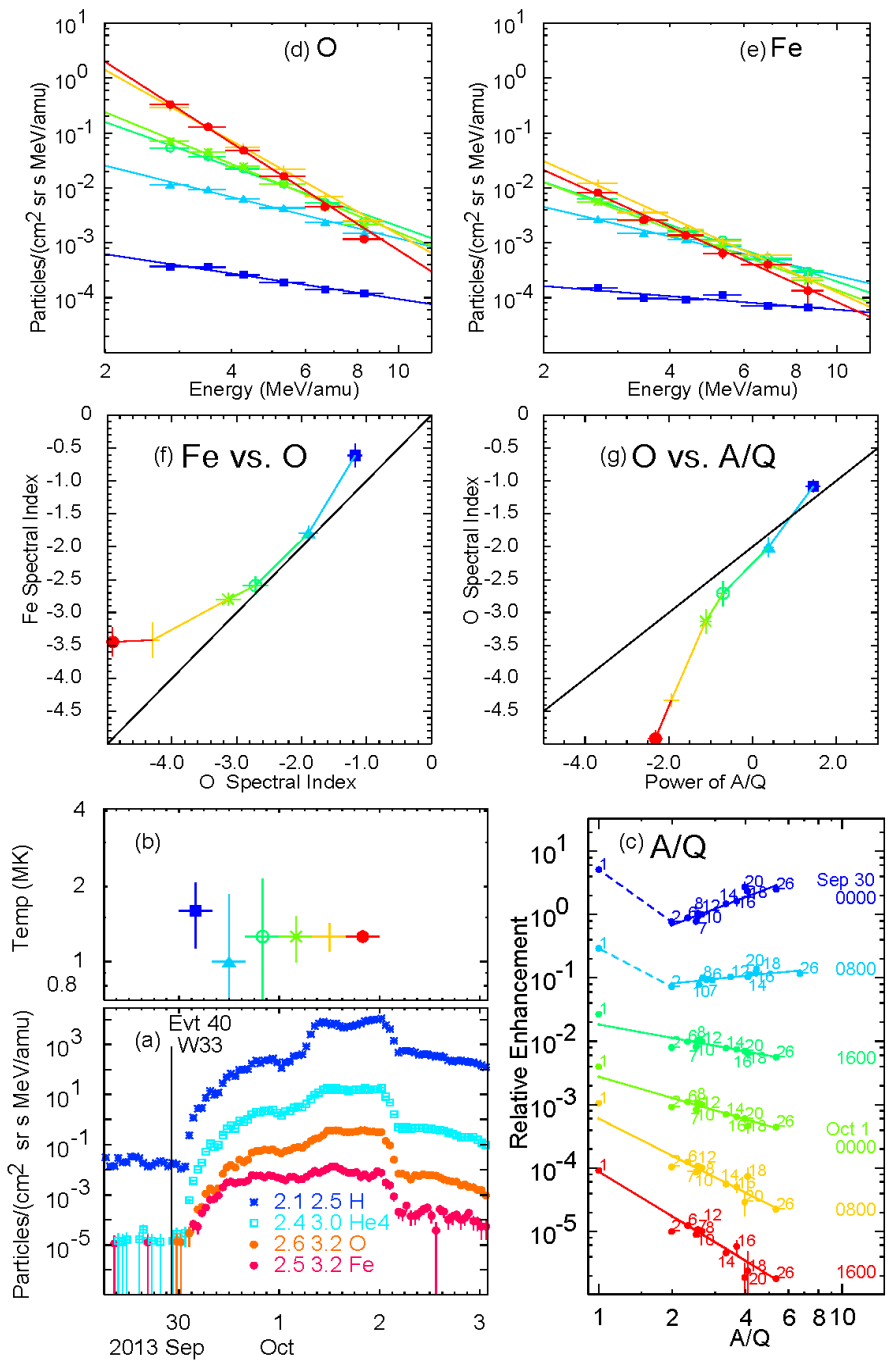

**Fig. 9.20** (**a**) Intensities and (**b**) derived source temperatures are shown versus time for gradual SEP Event 40, of 29 September, 2013. Fits are shown for (**c**) enhancements of elements, listed by Z, versus A/Q, and for energy spectra of (**d**) O and (**e**) Fe. Correlation plots are shown for spectral indices of (**f**) Fe versus O and of (**g**) O versus A/Q. Colors for time intervals correspond in (**b**), (**c**), (**d**), (**e**), (**f**), and (**g**). In (**f**) the solid line is diagonal, y = x, in (**g**) it is y = x/2 − 2 (Reames 2020c).

For the large SEP4 events, we can use 2-h time intervals and improve the resolution of the complex temporal evolution. Spectral and abundance correlations are shown for two SEP4 GLEs in Fig. 9.21. Eight-hour spectral fits for these events are shown in Reames (2020c). Event 15 was also shown in Fig. 9.18 (see also Fig. 5.1) as an example of an event with a streaming-limited plateau formed when Alfvén waves generated by streaming protons limit the intensities of low-rigidity ions (Sects. 5.1.2 and 5.1.5; Reames and Ng 2010; Ng, Reames, and Tylka 2012).

The O and Fe spectra we are using have similar energy amu$^{-1}$ but rigidities of O span the range of about 180 – 360 MV and Fe spans about 370 – 700 MV at $T \approx$ 1.3 MK. Thus the proton-generated waves preferentially trap the low-rigidity O near the shock, making the early O spectra flatter, but these waves have less affect on Fe spectra. Persistently flatter slopes of O than Fe are clearly seen in Fig. 9.21h and also in Fig. 9.21d to a lesser extent. The spectra recover just before the shock arrives.

Behind the shock in Event 15 we see evidence of adiabatic trapping in a magnetic reservoir (Sect. 5.7) where all of the red points in Figs. 9.21c and 9.21d pile together, indicating invariant spectra. Spectral shapes are invariant in the reservoir. These points fall near the lines of expected correlation.

Strictly speaking, the complexity in the disrupted correlation of spectra and abundances like those in Fig. 9.21 come from a breakdown in our power-law assumption for these quantities because of proton-generated waves at high intensities. The O and Fe spectra are still observed to be power laws over the observed range of each species, but they have different powers, whereas for small and medium SEP4 events, with no significant transport component, a single power spans from O through Fe. The proton-generated waves are rigidity dependent, but they vary with time and a single power no longer spans from O to Fe.

Time variation is an important factor. At a given time the wave spectrum near the shock may produce a power-law wave-spectral modification. However, this wave spectrum varies with time (Ng, Reames, and Tylka 2003). The relative scattering delay and trapping of the low-energy ions during the evolution of the event can mix low-rigidity ions produced at one time and place with higher-rigidity ions from another time and place involving much different wave spectra. This causes the complexity we see in Fig. 9.21.

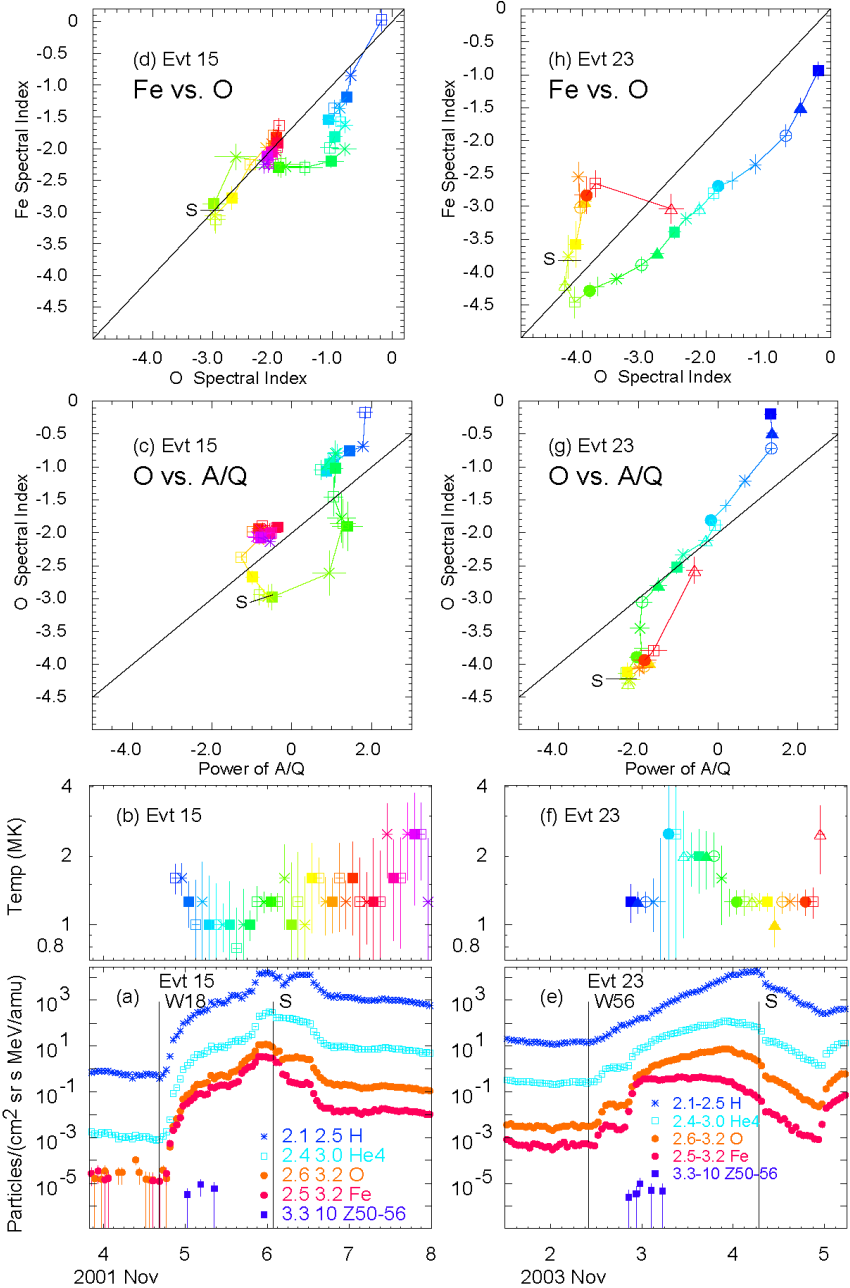

**Fig. 9.21** shows two GLEs, Event 15 of 4 November, 2001 on the *left*, and Event 23 of 2 November, 2003 on the *right* with 2-hour time resolution. Panels (**a**) and (**e**) show intensities of listed species and energies (in MeV amu⁻¹) versus time; (**b**) and (**f**) show derived temperatures versus time, providing time-tagged symbols and colors for the upper panels; (**c**) and (**g**) show time evolution of O spectral index versus power of A/Q; (**d**) and (**h**) show time evolution of Fe versus O spectral indices. S denotes the time of shock passage (Reames 2020c).

In Fig. 9.22 we compare the spectra and abundances of the first four 8-h periods in 45 gradual SEP events of all kinds (listed in Reames 2016). SEP3 events with $T > 2$ MK are indicated by red. Fig. 9.22a shows that He spectra tend to be a bit harder than O spectra. In Fig. 9.22b, Fe spectra show larger variation vs. O, especially for SEP4 events. In Fig. 9.22c, the orange and red ($T > 2$ MK) SEP3 events have high powers of $A/Q$ that are uncorrelated with spectral indices, since those powers are largely determined by the strong positive enhancements of the impulsive seeds prior to shock acceleration. If we remove the red and orange SEP3 events from Fig. 9.22c, the remaining SEP4 events follow broadly along the expected correlation; of course, some are distorted along the lines we would expect from the rather extreme examples in Figs. 9.20 and 9.21.

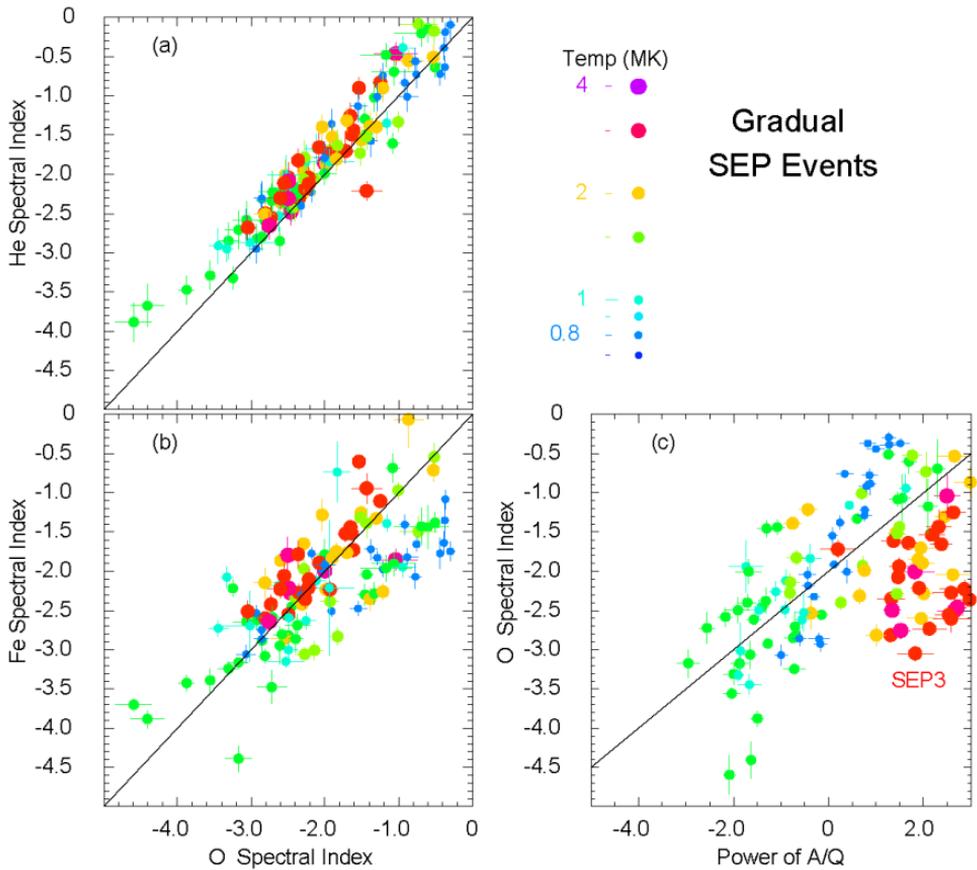

**Fig. 9.22** Shown are (**a**) spectral indices of He versus O, (**b**) spectral indices of Fe versus O, and (**c**) spectral indices of O versus powers of $A/Q$ for the first four 8-h periods in 45 gradual SEP events listed by Reames (2016). The size and color of each point is determined by the source plasma temperature $T$ as shown by the scale. Events with $T > 2$ MK are SEP3-class events dominated by reaccelerated impulsive suprathermal ions with enhanced abundances already determined before shock acceleration. In (**a**) and (**b**) the solid line is diagonal, y = x, in (**c**) it is y = x/2 – 2 (Reames 2020c).

## 9.10 Open Questions

1. What physical parameters determine the magnitude of the large excess of protons in some impulsive and gradual SEP events?

2. Why is reduced He/C associated with the largest gradual SEP events?

3. What causes a few $^3$He-rich events to have greatly suppressed $^4$He/C?

4. How does the rigidity dependence of SEP4 shock acceleration depend upon shock structure? Can we measure it?

5. Will the next generation of SEP simulations include abundances of $Z > 1$ ions? Only protons are a significant hazard, but getting the abundances right for GLEs is a powerful test of the quality of the underlying physics.

## References

Bochsler, P., Solar abundances of oxygen and neon derived from solar wind observations, Astron. Astrophys. **471** 315 (2007) doi: 10.1051/0004-6361:20077772

Bochsler, P., Composition of matter in the heliosphere, Proc. IAU Sympos **257**, 17 (2009) doi: 10.1017/S1743921309029044

Bučík, R., Innes, D.E., Chen, N.H., Mason, G.M., Gómez-Herrero, R., Wiedenbeck, M.E., Long-lived energetic particle source regions on the Sun, J. Phys. Conf. Ser. **642**, 012002 (2015) doi: 10.1088/1742-6596/642/1/012002

Bučík,,R,. Innes. D.E., Mall, U., Korth, A., Mason, G.M., Gómez-Herrero, R., Multi-spacecraft observations of recurrent $^3$He-rich solar energetic particles, Astrophys. J. **786**, 71 (2014) doi: 10.1088/0004-637X/786/1/71

Bučík, R., Innes, D.E., Mason, G.M., Wiedenbeck, M.E., Gómez-Herrero, R., Nitta, N, $^3$He-rich solar energetic particles in helical jets on the sun, Astrophys, J. **852** 76 (2018) doi: 10.3847/1538-4357/aa9d8f

Bučík, R., Wiedenbeck, M.E., Mason, G.M., Gómez-Herrero, R., Nitta, N.V., Wang, L., $^3$He-rich solar energetic particles from sunspot jets, Astrophys. J. Lett. **869** L21 (2018b) doi: 10.3847/2041-8213/aaf37f

Chen N.H., Bučík R., Innes D.E., Mason G.M., Case studies of multi-day $^3$He-rich solar energetic particle periods, Astron. Astrophys. **580**, 16 (2015) doi: 10.1051/0004-6361/201525618

Collier, M.R., Hamilton, D.C., Gloeckler, G., Bochsler, P., Sheldon, R.B., Neon-20, oxygen-16, and helium-4 densities, temperatures, and suprathermal tails in the solar wind determined with WIND/MASS, Geophys. Res. Lett., **23**, 1191 (1996) doi: 10.1029/96GL00621

Ellison, D.C., Möbius, E., Paschmann, G., Particle injection and acceleration at Earth's bow shock: comparison of upstream and downstream events, Astrophys. J. **352** 376 (1990) doi: 10.1086/168544

Jones, F.C., Ellison, D.C., The plasma physics of shock acceleration, Space Sci. Rev. **58**, 259 (1991) doi: 10.1007/BF01206003

Kajdič, P., Preisser, L., Blanco-Cano, X., Burgess, D., Trotta, D., First observations of irregular surface of interplanetary shocks at ion scales by *Cluster*, Astrophys. J. Lett. **874** L13 (2019) doi: 10.3847/2041-8213/ab0e84

Kasper, J.C., Stevens, M.L., Lazarus, A.J., Steinberg, J.T., Ogilvie, K. W., Solar wind helium abundance as a function of speed and heliographic latitude: variation through a solar cycle, Astrophys. J. **660**, 901 (2007) doi: 10.1086/510842

Ng, C.K., Reames, D.V., Tylka, A.J., Effect of proton-amplified waves on the evolution of solar energetic particle composition in gradual events, Geophys. Res. Lett. **26**, 2145 (1999) doi: 10.1029/1999GL900459

Ng, C.K., Reames, D.V., Tylka, A.J., Modeling shock-accelerated solar energetic particles coupled to interplanetary Alfvén waves, Astrophys. J. **591**, 461 (2003) doi: 10.1086/375293

Ng, C.K., Reames, D.V., Tylka, A.J., Solar energetic particles: shock acceleration and transport through self-amplified waves, AIP Conf. Proc. **1436**, 212 (2012) doi: 10.1063/1.4723610

Parker, E.N.: Nanoflares and the solar X-ray corona, Astrophys. J. **330** 474 (1988) doi: 10.1086/166485

Rakowsky, C.E., Laming, J.M., On the origin of the slow speed solar wind: helium abundance variations, Astrophys. J. **754**, 65 (2012) doi: 10.1088/0004-637X/754/1/65

Reames, D.V., Element abundances in solar energetic particles and the solar corona, Sol. Phys., **289**, 977 (2014) doi: 10.1007/s11207-013-0350-4

Reames, D.V., Spatial distribution of element abundances and ionization states in solar energetic-particle events, Sol. Phys. **292** 133 (2017), doi: 10.1007/s11207-017-1138-8 (arXiv 1705.07471).

Reames, D.V., Temperature of the source plasma in gradual solar energetic particle events, Sol. Phys., **291** 911 (2016a) doi: 10.1007/s11207-016-0854-9 (arXiv: 1509.08948 )

Reames, D.V., Helium suppression in impulsive solar energetic-particle events, Sol. Phys. **294** 32 (2019a) doi: 10.1007/s11207-019-1422-x (arXiv: 1812.01635)

Reames, D.V. Hydrogen and the abundances of elements in impulsive solar energetic-particle events, Sol. Phys. **294** 37(2019b) doi: 10.1007/s11207-019-1427-5

Reames, D.V., Hydrogen and the abundances of elements in gradual solar energetic-particle events, Sol. Phys. **294** 69 (2019c) doi: 10.1007/s11207-019-1460-4

Reames, D.V., Excess H, suppressed He, and the abundances of elements in solar energetic particles, Sol. Phys. **294**, 141 (2019d) doi: 10.1007/s11207-019-1533-4

Reames, D. V., Four distinct pathways to the element abundances in solar energetic particles, Space Science Rev. **216** 20 (2020a) doi: 10.1007/s11214-020-0643-5

Reames, D.V., Distinguishing the rigidity dependences of acceleration and transport in solar energetic particles, Sol. Phys. **295** 113 (2020b) doi: 10.1007/s11207-020-01680-6 (arXiv 2006.11338 )

Reames, D.V., The correlation between energy spectra and element abundances in solar energetic particles, Sol. Phys. submitted (2020c) arXiv: 2008.06985

Reames, D.V., Ng, C.K., Streaming-limited intensities of solar energetic particles on the intensity plateau, Astrophys. J. **722**, 1286 (2010) doi: 10.1088/0004-637X/723/2/1286

Reames, D.V., Cliver, E.W., Kahler, S.W., Abundance enhancements in impulsive solar energetic-particle events with associated coronal mass ejections, Sol. Phys. **289**, 3817, (2014) doi: 10.1007/s11207-014-0547-1

Reames, D.V., Ng, C.K., Tylka, A.J., Initial time dependence of abundances in solar particle events, Astrophys. J. Lett. **531**, L83 (2000) doi: 10.1086/312517

Trotta, D., Burgess, D., Prete, G., Perri, S., Zimbardo, G., Particle transport in hybrid PIC shock simulations: A comparison of diagnostics, Mon. Not. Roy. Astron. Soc. **491** 580 (2020) doi: 10.1093/mnras/stz2760

Tylka, A.J., Lee, M.A., Spectral and compositional characteristics of gradual and impulsive solar energetic particle events, Astrophys. J. **646**, 1319 (2006) doi: 10.1086/505106

Tylka, A.J., Cohen, C.M.S., Dietrich, W.F., Lee, M.A., Maclennan, C.G., Mewaldt, R.A., Ng, C.K., Reames, D.V., Shock geometry, seed populations, and the origin of variable elemental composition at high energies in large gradual solar particle events, Astrophys. J. **625**, 474 (2005) doi: 10.1086/429384

# Chapter 10. Summary and Conclusions

**Abstract**  In this chapter we summarize our current understanding of SEPs, of properties of the sites of their origin and of the physical processes that accelerate or modify them.  These processes can leave an indelible mark on the abundances of elements, isotopes, ionization states, anisotropies, energy spectra and time profiles of the SEPs.  Transport of the ions to us along magnetic fields can impose new variations in large events or even enhance the visibility of the source parameters as the SEPs expand into the heliosphere.  We lack physical models that can follow the complexity of SEP abundance variations.

What is our current understanding of solar energetic particles (SEPs)?

1. All acceleration of the SEPs that we see in space occurs on magnetic field lines that are *open* to particles of that magnetic rigidity.  We also see γ rays and neutrons from nuclear reactions of similar SEPs on closed field lines in solar flares, but no charged products of those nuclear reactions are seen in space.  Neither the primary nor the secondary ions can escape, instead they heat the flares.

2. There are two acceleration sites for the SEPs we see in space: *solar jets* and *CME-driven shock waves*. A) "Impulsive" SEP events, accelerated at solar jets, appear to involve two physical mechanisms, magnetic reconnection and resonant wave-particle absorption.  Both produce striking, and identifiable, enhancements of abundances of chemical elements and isotopes.  B) For "gradual" SEP events the dominant mechanism is acceleration by CME-driven shock waves, but the seed population may be complex, and abundances are also modified by pitch-angle scattering during transport in large events.

3. *Impulsive SEP events* are small and brief.  Solar jets, where acceleration occurs, are associated with slow, *narrow CMEs*.  *Magnetic reconnection* in jets, sampling ions of 2–4 MK plasma in active regions, cause abundance enhancements rising as a steep *power law in A/Q* by factors up to ~1000 from H to Pb. *Wave-particle resonance* causes large, highly variable, *enhancements in $^3He/^4He$* by factors up to 10,000 that vary strongly with the ion energy and may sometimes cause rounded, steep, low-energy spectra of ions with gyro-frequencies near the second harmonic of the $^3$He gyro-frequency.  The waves may be generated by the copious streaming electrons that also produce type III radio bursts.  Acceleration may occur near 1.5 $R_S$ and ions traverse enough material for electron stripping to attain equilibrium velocity-dependent *Q*, but not enough energy loss to disrupt the strong high-*Z* enhancements that are seen. "Pure" impulsive events (SEP1), lack shock acceleration. Local shocks can reaccelerate SEP1 ions plus "excess" protons from the ambient corona (SEP2).

**4.** *Gradual SEP events* are large, energetic, and intense. They sometimes acceler­ate multi-GeV protons, and they have long durations and broad spatial extent, often exceeding ~180°. They are associated with *fast*, *wide CMEs* that drive shock waves that accelerate ions from ambient coronal plasma of ~0.8–1.6 MK in ~69% of the events (SEP4). In 24% of gradual events the shock waves pass through solar active regions where they sample a seed population that includes ambient plasma laced with residual suprathermal ions from pools fed by *multi­ple* small solar jets (SEP3). The seed population and source-plasma tempera­ture can vary across the face of a shock. The location of high-energy spectral breaks or knees depends upon both shock properties and *A/Q* of the ion species, causing complex abundance variations at high energies. Shock waves begin to form near 1.5 $R_S$, first accelerating electrons that produce type II radio bursts; acceleration of SEPs can begin above the tops of magnetic loops by 2–6 $R_S$, de­pending upon longitude around the CME.

**5.** *Self-amplified Alfvén waves become increasingly important in larger gradual SEP events.* Pitch-angle scattering by proton-amplified waves limits particle in­tensities at the *streaming limit*, alters initial element abundance ratios after on­set, rapidly broadens angular distributions, and flattens low-energy spectra dur­ing the early intensity-plateau period. Preferential scattering of ions with lower *A/Q* during transport causes regions of relative *A/Q*-dependent abundance en­hancements or depletions in space that evolve with time. This Q-dependence allows determination of the source plasma temperature. In contrast, non-rela­tivistic electrons and particles from small impulsive SEP events travel scatter free. Larger events are increasingly dominated by self-generated waves, but SEPs become scatter-free again later in the reservoir behind the CME (see 7. below). Wave growth and scattering depend upon rigidity, spatial location, and time during a large SEP event.

**6.** Can we always distinguish impulsive and gradual events? Usually, but not al­ways. Shocks often reaccelerate residual impulsive suprathermal ions with pre-enhanced abundances. Some SEP events, called "impulsive" because of their high Fe/O enhancement, for example, may have also undergone reacceleration by a shock wave. These (SEP2) events may be distinguished by their abun­dance of excess protons sampled from the ambient plasma. Impulsive seed ions reaccelerated by wide, fast shock waves from pools (SEP3) are usually distin­guished by lower Fe/O and fewer abundance fluctuations.

**7.** Reservoirs are large volumes of adiabatically-trapped SEPs seen late in gradual events. Particles are magnetically trapped between the CME and the Sun with negligible leakage. Intensities of all species and energies are spatially uniform but all decrease with time as the trapping volume expands. Early workers mis­took this slow decline as slow *spatial* diffusion. Actually, particles in reservoirs propagate nearly scatter-free since waves most have been absorbed. Reservoirs often provide the high-energy particles that slowly precipitate to produce long-duration, spatially-extensive, energetic γ-ray events when they scatter into the magnetic loss cone and interact in the denser corona below.

**8.** In large events, CMEs capture the largest share of the magnetic energy released at the Sun and SEPs can acquire as much as ~15% of a CME's energy. In flares, SEPs capture 30–60% of the energy of magnetic reconnection, but those SEPs do not escape from closed loops, they scatter into the loss cone and dump their energy into the footpoints of the loops. Trapping creates hot, bright flares.

**9.** SEPs, at energies above a few MeV amu⁻¹, and ions of the slow solar wind, show differences in the pattern of their element abundances relative to corresponding photospheric abundances as a function of first ionization potential, FIP, specifically for the elements C, P, and S. SEPs show a source where C, P, and S is less likely to be ionized crossing the chromosphere. Such differences are determined near the base of the corona, long before acceleration, so that SEPs and the solar wind must be derived from *different* coronal regions. Thus, SEPs are *not* merely accelerated solar wind but an *independent* sample of the solar corona. Theory, based upon the ponderomotive force of Alfvén waves, suggests that material that will become SEPs is transported up into the corona along closed field lines as in active regions, while that forming the solar wind arrives on open field lines. It is ironic that the SEP ions accelerated on open field lines probably entered the corona on closed field loops.

**10.** The properties discussed above distinguish four patterns of element abundances shown in Table 10.1 below.

**Table 10.1** Properties of the four SEP element abundance patterns.

|      | **Observed Properties** | **Physical Association** |
|------|-------------------------|--------------------------|
| **SEP1** | Power-law enhancement vs. $A/Q$ (Fe/O >×4) with $T \approx 3$ MK including $Z=1$ and $Z>2$ | Magnetic reconnection in solar jets with no fast shock |
| **SEP2** | Power-law enhancement vs. $A/Q$ (Fe/O > ×4) with $T \approx 3$ MK for Z >2; ~30% scatter in He/C, etc. Proton excess ~×10 | Jets with fast, narrow CMEs drive shocks that reaccelerate SEP1 ions plus excess protons from ambient plasma |
| **SEP3** | Power-law enhancement vs. $A/Q$ (Fe/O < ×4) with $T \approx 3$ MK for Z >2; <10% scatter in He/C, etc. Proton excess ~×10 | Moderately fast, wide CME-driven shocks accelerate SEP1 residue left by many jets in pools, plus excess protons from ambient plasma |
| **SEP4** | Power-law enhancement vs. $A/Q$ with $0.8< T < 1.8$ MK for $Z=1$ and $Z > 2$ | Extremely fast, wide CME-driven shocks accelerate all ions so that ambient plasma dominates. |

We have seen that some processes depend upon particle velocity and others depend upon magnetic rigidity. Early studies could not distinguish these processes using proton spectra alone.  Ions highlight rigidity dependence of abundances upon $A/Q$ at constant velocity, giving us new leverage on the underlying physics, as well as the nature of the source plasma and even an estimate of its temperature. For the SEP4 events, the abundances and energy spectral indices can be correlated.  With spectra of the form $E^y$ and abundance enhancements of the form $(A/Q)^x$ we find $y = x/2 - 2.$; it is not yet clear why.  This relationship provides new information on the "injection problem," i.e. on the way shocks select ions from the plasma.  For SEP2 and SEP3 events, the power-law abundances of ions from impulsive events are so distinctive that this signature can be followed through reacceleration by a strong shock.

Thus, some of the early mysteries of SEP origin seem to be resolved, even though many new questions have arisen.  The progress has come almost entirely from the direct measurement of SEPs in space, especially from their abundances. The story is complex.  It involves acceleration and reacceleration of ions that, nevertheless, carry measurable properties of their convoluted histories.  We have identified the physical mechanisms that contribute to particle acceleration and developed new tools to explore them.  What remains is to understand their detailed interplay.  What parameters determine when and where each mechanism operates, and how can we predict their onset, their magnitude and their outcome?  Other questions abound.  How is it that reservoirs are so uniform and so well maintained?  What causes He-poor events? We anticipate the next generation of understanding.

Theories and models of acceleration often treat element abundances as adjustable parameters – or not at all.  However, we now know absolute coronal abundances sampled at the source and we need theories that follow ion injection and map those abundances through acceleration and transport into observations. Spectra and abundances are correlated, at least for many SEP4 events.  How does the shock-shape matter?  Element abundances probe the physics of SEPs.  Only proton predictions are required for astronaut safety, but a model that could predict the complex abundance variations in a large SEP event could gain a powerful badge of quality and reliability.  We have no such model today.